\documentclass{aa}  
\usepackage{graphicx}
\usepackage{subfig}
\usepackage{txfonts}
\usepackage{siunitx} 
\usepackage[version=4]{mhchem}
\usepackage{hyperref}
\makeatletter
\renewcommand*\aa@pageof{, page \thepage{} of \pageref*{LastPage}}
\makeatother
\bibpunct{(}{)}{;}{a}{}{,} 
\usepackage{natbib}
\newcommand{\hii}{H{\sc ii}}
\newcommand{\hi}{H{\sc i}}

\begin{document} 

\title{High-mass star formation across the Large Magellanic Cloud}

\subtitle{I. Chemical properties and hot molecular cores observed with ALMA at 1.2 mm}

\author{Roya Hamedani Golshan\inst{1}\fnmsep\thanks{rhgolshan@ph1.uni-koeln.de}, 
        {\'A}lvaro S{\'a}nchez-Monge \inst{2,3},
        Peter Schilke \inst{1}, 
        Marta Sewi\l{}o\inst{4,5,6},
        Thomas M{\"o}ller\inst{1},
        V.\ S.\ Veena \inst{1,7} and
        Gary A.\ Fuller \inst{1,8}
        }

\institute{I. Physikalisches Institut, Universit{\"a}t zu K{\"o}ln, Z{\"u}lpicher Stra{\ss}e 77, 50937 Cologne, Germany
        \and
            Institut de Ci\`encies de l'Espai (ICE, CSIC), Carrer de Can Magrans s/n, E-08193, Bellaterra, Barcelona, Spain
        \and
            Institut d'Estudis Espacials de Catalunya (IEEC), Barcelona, Spain
        \and
            Exoplanets and Stellar Astrophysics Laboratory, NASA Goddard Space Flight Center, Greenbelt, MD 20771, USA
        \and
            Center for Research and Exploration in Space Science and Technology, NASA Goddard Space Flight Center, Greenbelt, MD 20771, USA
        \and
            Department of Astronomy, University of Maryland, College Park, MD 20742
        \and      
           Max-Planck-Institut für Radioastronomie, Auf dem H{\"u}gel 69, D-53121 Bonn, Germany
        \and
            Jodrell Bank Centre for Astrophysics, Department of Physics \& Astronomy, The University of Manchester, Oxford Road, Manchester M13 9PL, UK
         }

\date{}

\authorrunning{Hamedani Golshan et al.}
 
\abstract
{The formation of massive stars passes through a so-called hot molecular core phase, where the temperature of molecular gas and dust rises to above 100 K within a size scale of approximately 0.1 pc. The hot molecular cores are rich in chemical compounds found in the gas phase, which are a great probe of ongoing star formation.}  
{To study the impact of the initial effects of metallicity (i.e., the abundance of elements heavier than helium) on star formation and the formation of different molecular species, we searched for hot molecular cores in the sub-solar metallicity environment of the Large Magellanic Cloud (LMC).}
{We conducted Atacama Large Millimeter/submillimeter Array (ALMA) Band 6 observations of 20 fields centered on young stellar objects (YSOs) distributed over the LMC in order to search for hot molecular cores in this galaxy.}
{We detected a total of 65 compact 1.2~mm continuum cores in the 20 ALMA fields and analyzed their spectra with XCLASS software. The main temperature tracers are \ce{CH3OH} and \ce{SO2}, with more than two transitions detected in the observed frequency ranges. Other molecular lines with high detection rates in our sample are \ce{CS}, \ce{SO}, \ce{H^{13}CO^+}, \ce{H^{13}CN}, \ce{HC^{15}N}, and \ce{SiO}. More complex molecules, such as \ce{HNCO}, \ce{HDCO}, \ce{HC3N}, \ce{CH3CN}, and \ce{NH2CHO}, and multiple transitions of \ce{SO} and \ce{SO2} isotopologues showed tentative or definite detection toward a small subset of the cores. According to the chemical richness of the cores and high temperatures from the XCLASS fitting, we report the detection of four hot cores and one hot core candidate. With one new hot core detection in this study, the number of detected hot cores in the LMC increases to seven.}
{Six out of seven hot cores detected in the LMC to date are located in the stellar bar region of this galaxy. These six hot cores show emission from complex organic molecules (COMs), such as \ce{CH3OH}, \ce{CH3CN}, \ce{CH3OCHO}, and \ce{CH3OCH3}. The only known hot core in the LMC with no detection of COMs is located outside the bar region. The metallicity in the LMC presents a shallow gradient increasing from outer regions toward the bar. Various studies emphasize the interaction between the LMC and the Small Magellanic Cloud, which resulted in the mixing and inhomogeneity of the interstellar medium of the two galaxies. These interactions triggered a new generation of star formation in the LMC. We suggest that the formation of hot molecular cores containing COMs ensues from the new generation of stars forming in the more metal-rich environment of the LMC bar.}

\keywords{Astrochemistry, Galaxies: Magellanic Clouds, star formation, ISM: molecules, Stars: protostars, Radio lines: ISM}

\maketitle
%

%
\section{Introduction}\label{sec:introduction}

Hot molecular cores are compact (d $\leq$ 0.1 pc) and dense ($\geq$ 10$^6$ cm$^{-3}$) regions that are heated up to temperatures above 100 K, either internally or externally, in the immediate surroundings of 
formation sites of massive stars \citep[e.g.][]{2022MNRAS.511.3463Q}. At these temperatures, the molecules that are formed and developed in the ice mantles on the surfaces of dust grains are released into the gas due to thermal evaporation or sputtering with shock waves, UV photons, or cosmic rays. Once in the gas phase, more complex chemical compounds can be formed \citep[e.g.,][]{2008ApJ...682..283G}. This results in the formation and development of complex organic molecules \citep[COMs, molecules with six or more atoms including carbon;][]{2009ARA&A..47..427H}. Eventually, most complex molecules are dissociated by strong UV radiation from the newly formed massive stellar object(s) and the associated \hii\ region. In the time between the evaporation of the ice mantels to the destruction of the complex molecular species in the gas phase, the rich molecular gas chemistry is accompanied by the excitation of a plethora of rotational transitions observable at (sub)millimeter wavelengths \citep[e.g.][]{2013A&A...559A..47B, 2023A&A...676A.121M}.

The development, chemical richness, and molecular composition of hot molecular cores may be affected by environmental conditions. One important environmental factor is the metallicity of the gas where stars are forming. Metallicity is the abundance of elements heavier than helium, and it is known to increase over cosmic time with every new generation of stars due to the nucleosynthesis processes in the stellar interiors, supernova explosions, and neutron star mergers. This triggers a fundamental question in astrochemistry regarding how the difference in the elemental abundances influences the formation, survival, and destruction of molecular species. To answer this question, searching for the sites of rich chemistry in different (e.g., sub-solar) metallicity conditions is a crucial step, and hot molecular cores are prominent candidates for providing a complete molecular inventory in the gas phase. Thanks to the high sensitivity and resolving power of the Atacama Large Millimeter/submillimeter Array (ALMA), detecting and pinpointing massive star-forming regions and hot molecular cores down to sub-parsec scales at 1~mm is feasible for relatively short integration times in our Galaxy and the immediate galactic neighborhood (i.e., the Magellanic Clouds). In this work, we aim to search for and characterize hot molecular cores in the lower metallicity gas of the Large Magellanic Cloud (LMC).
%
%

The LMC is a nearby dwarf galaxy at a distance of $D\approx49.59\pm0.09$ (statistical) $\pm0.54$ (systematic) kpc \citep{2019Natur.567..200P}. This galaxy, with an average metallicity\footnote{Different works have estimated the metallicity of the LMC to be between 0.3 and 0.5 of the metallicity in the Solar neighborhood interstellar gas (Z$_\mathrm{\odot}$). In this work, we use the recent estimate by \cite{2021MNRAS.507.4752C}, who derives [Fe/H] = $-0.42$ dex, with [Fe/H] = $\log\left(\frac{Z_{\mathrm{gal}}}{Z_{\odot}}\right)$), and results in Z$_\mathrm{LMC}$ $\sim$ 0.38 Z$_\mathrm{\odot}$.} of Z$_\mathrm{LMC}\sim$ 0.38 Z$_\mathrm{\odot}$ 
is one of the closest laboratories to study the physics and chemistry of a low-metallicity environment. Low metallicity means less dust and hence diminished shielding effects, allowing UV radiation to penetrate deeper into clouds. The UV field strength in the LMC is ten to 100 times larger than in the Milky Way \citep[MW, e.g.,][]{2006ApJS..165..138W}. As a result, the dust temperature is expected to be higher, likely affecting the development of the more complex molecular species. Moreover, the metallicity has been found to vary across the LMC, showing a moderate gradient that increases toward the bar region in this galaxy \citep[e.g.,][]{2021MNRAS.507.4752C}. 

The development of ALMA enabled the detection of the first hot molecular cores in the LMC in recent years. The six hot cores detected so far \citep[see][]{2016ApJ...827...72S, 2018ApJ...853L..19S, 2020ApJ...891..164S, 2022ApJ...931..102S} present a picture of a diverse chemistry. While metallicity-scaled abundances for detected COMs in four hot cores are comparable to those observed in the Galaxy, COMs are either absent or underabundant in the remaining two hot cores. Hot cores in the N\,113 star-forming region show emission from molecules containing up to nine atoms, with abundances that match the lower end of Galactic values after scaling for metallicity \citep{2018ApJ...853L..19S}. The two bona fide hot cores in the N\,105 star-forming region \citep{2022ApJ...931..102S} also exhibit a rich chemistry, with abundances that follow the trend of the ones in N\,113, although with lower values. On the COM-poor side, a hot core with CH$_3$OH and CH$_3$CN molecular lines but no larger molecules has been detected by \cite{2020ApJ...891..164S}. This hot core, which shows an underabundance of complex species compared to Galactic hot cores, is associated with the massive young stellar object (YSO) ST16. Even more extreme is the hot core associated with the YSO ST11 \citep{2016ApJ...827...72S}. There are no emission lines from COMs in this source, and the temperature above 100 K is obtained from \ce{SO2} and its isotopologues. \cite{2016ApJ...827...72S} have suggested high dust temperatures that prohibit effective freeze-out and subsequent hydrogenation of CO on the dust surface to be the likely cause of the suppression of \ce{CH3OH} formation, the building block for more complex COMs \citep[e.g.][]{2008ApJ...682..283G}. The existence of hot cores with metallicity-scaled COM abundances similar to the Galactic values shows that higher dust temperatures in the cold collapse phase are not ubiquitous in the LMC and may exist only in specific regions. Based on these results, a statistically significant sample needs to be analyzed to better characterize the hot cores in the LMC and their relation to environmental conditions (e.g., proximity to star-forming regions having higher UV radiation levels or variations in metallicity).

To expand the search for hot molecular cores from the LMC to other sub-solar metallicity environments, recent works have studied objects in the outer regions of the MW \citep[e.g.][]{2021ApJ...922..206S, 2022A&A...660A..76F} as well as in the Small Magellanic Cloud \citep[SMC][]{2023ApJ...946L..41S}. The Galaxy's outer regions harbor an interstellar medium with a diminished metallicity compared to the inner Galactic regions \citep[e.g.,][]{2021MNRAS.502..225A} as well as more quiescent radiation fields. Studies in the outer Galaxy are crucial, as they can bridge our knowledge of the higher metallicity (and more active) star-forming environment of the inner Galaxy to the low metallicities of the LMC and SMC. Recently, a hot molecular core has been discovered in the outskirts of the MW in the WB\,89$-$789 star-forming region at a distance of 10.7~kpc (galactocentric distance of 19 kpc) \citep{2021ApJ...922..206S}. The metallicity at this position is estimated to be Z $\sim$ 0.25 Z$_\mathrm{\odot}$. Interestingly, the chemical composition of the hot core in WB\,89$-$789 resembles that of hot cores in the inner Galactic regions and not of those in the LMC. On the other hand, the SMC is a star-forming dwarf galaxy at a distance of $62.1\pm1.9$~kpc \citep{2014ApJ...780...59G} and has the lowest metallicity in the nearby universe where we are able to search for hot cores with currently available facilities. The metallicity is Z $\sim$ 0.2 Z$_\mathrm{\odot}$, with a shallow but significant gradient \citep{2018MNRAS.475.4279C}. \cite{2023ApJ...946L..41S} have reported the detection of two hot cores in the SMC, S07 and S09, coinciding with the position of two massive YSOs. Notably, \ce{SO2} traces the hot compact ($\sim0.1$~pc) component with temperatures higher than 100~K in these two hot cores, while \ce{CH3OH} is extended ($\sim0.2$--0.3~pc) and much cooler ($\le40$~K for S07).

In summary, the detection of hot cores in the LMC, the SMC, and the outskirts of the MW, added to the multiple Galactic examples, suggests that the development of hot molecular cores is common during the formation of stars in the Z = 0.2--1~Z$_\mathrm{\odot}$ metallicity range \citep{2023ApJ...946L..41S}. Despite this, the first studies in low-metallicity environments have found a diverse population of objects with molecular abundances sometimes consistent with Galactic-like metallicity environments and others with lower abundances. Furthermore, some hot cores appear to be poor in COMs abundance, while others have abundances similar to those in the MW disk. All in all, it is noticeable that a larger sample of sources observed and analyzed with a common methodology is still necessary to provide insight into the formation and properties of hot cores in low-metallicity regions. With this in mind, we studied 20 different fields in the LMC, the largest sample to date, in order to search for hot cores and improve our current understanding of the physics and chemistry of high-mass star-forming cores in this nearby dwarf galaxy. 

This is the first paper of a series that is accompanied by a publication by Hamedani Golshan et al.\ (2024, hereafter Paper~II). Paper~II presents the observed sample and studies the dense cores and star-forming clusters identified in the observed regions. In the current paper, we focus on the chemical content of the identified cores and their connection to the LMC's global metallicity trend and large-scale properties. In Section~\ref{sec:observations} we introduce the sample and the observations. In Section~\ref{sec:results}, we present the results and main analysis procedures. We discuss the results in the context of our knowledge about the hot cores in the LMC, SMC, and MW in Section~\ref{sec:discussion}. Several appendixes show our dataset or elaborate on specific concepts or methodologies.
%
\begin{figure*}[h]
\centering
\includegraphics[width=0.95\textwidth]{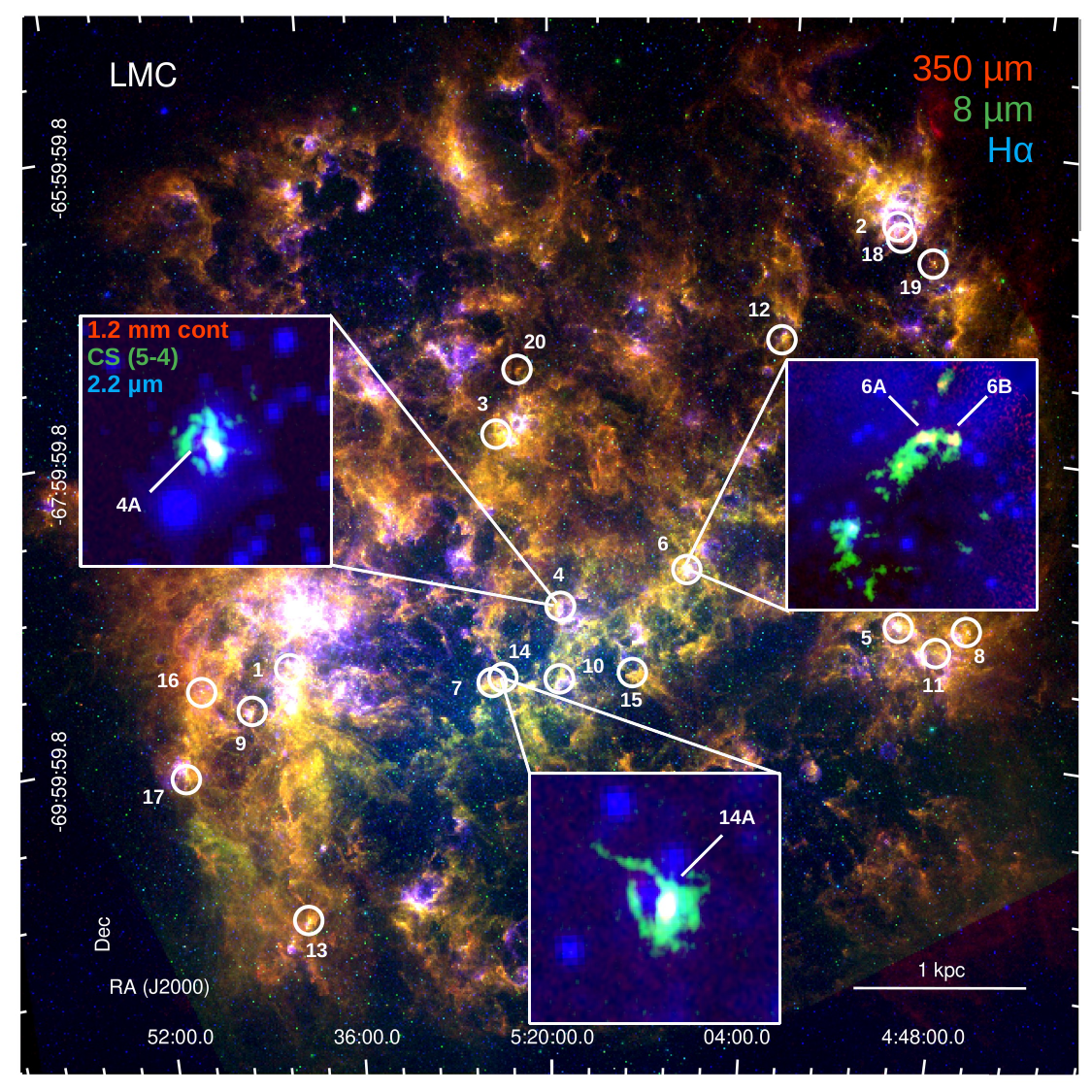}
\caption{Color-composite image of the LMC combining the H$\alpha$ \protect\citep[blue,][]{2000ASPC..221...83S}, SAGE/IRAC 8~$\mu$m \protect\citep[green,][]{2006AJ....132.2268M}, and HERITAGE/SPIRE 350~$\mu$m \protect\citep[red,][]{2013AJ....146...62M} images. The positions of the 20 ALMA fields are marked with white circles 15 times larger than the field of view of our ALMA observations and the field number. The color-composite images of the fields with hot core detection combining the VMC K-band \protect\citep[blue,][]{2011A&A...527A.116C}, ALMA CS\,(5--4) peak intensity (green, this work), and 1.2~mm ALMA continuum (red, this work and Paper~II) images are shown as insets. The hot cores reported in this paper, 4A \citep[ST16;][]{2020ApJ...891..164S}, 6A \protect\citep[N105-2 B;][]{2022ApJ...931..102S}, 6B \protect\citep[N105-2 A;][]{2022ApJ...931..102S}, and 14A are highlighted in the insets images.}
\label{fig:LMCRGB}
\end{figure*}
\begin{table}
\centering
\caption{Coordinates and names of our target fields.}
\begin{tabular}{ccc}
\hline\hline \noalign{\smallskip} 
Field
& Target ID
& Alternative
\\
Number
& \citep{2009ApJS..184..172G}
& name

\\
\hline \noalign{\smallskip}
     1 & 053941.12-692916.8 & ST6 \tablefootmark{a} \\ 
     2 & 045640.79-663230.5 & ST10 \tablefootmark{a} \\ 
     3 & 052343.48-680033.9 &  \\ 
     4 & 051912.27-690907.3 & ST16 \tablefootmark{a,b} \\ 
     5 & 045358.57-691106.7 &  \\ 
     6 & 050953.89-685336.7 & N105-2 E \tablefootmark{c,d} \\ 
     7 & 052423.39-693904.7 &  \\ 
     8 & 044854.41-690948.3 &  \\ 
     9 & 054248.90-694446.3 &  \\ 
    10 & 051916.87-693757.5 &  \\ 
    11 & 045100.16-691934.4 &  \\ 
    12 & 050355.87-672045.1 &  \\ 
    13 & 053952.11-710930.7 &  \\ 
    14 & 052333.40-693712.2 &  \\ 
    15 & 051344.99-693510.6 &  \\ 
    16 & 054629.32-693514.2 &  \\ 
    17 & 054826.21-700850.2 &  \\ 
    18 & 045622.61-663656.9 &  \\ 
    19 & 045406.43-664601.4 &  \\ 
    20 & 052210.08-673459.6 &  \\ 
\hline 
\end{tabular}
\tablefoot{
\tablefoottext{a}{\cite{2016A&A...585A.107S}}\\
\tablefoottext{b}{\cite{2020ApJ...891..164S}}\\
\tablefoottext{c}{\cite{2022ApJ...931..102S}}\\
\tablefoottext{d}{\cite{1956ApJS....2..315H}}
}
\label{tab:pointings_noise}
\end{table}
\begin{table*}
\centering
\caption{Spectral cube parameters for primary beam corrected images.}
\begin{tabular}{cccccc}
\hline\hline \noalign{\smallskip} 
Spectral & Frequency range\tablefootmark{a} & Synthesized beam ($\Theta_{B}$)\tablefootmark{a} & Channel width & Cube rms\tablefootmark{a,b} \\ 
window & (GHz) & (" $\times$ ") & (MHz / kms$^{-1}$) & (mJy~beam$^{-1}$ / K)\\ 
\hline \noalign{\smallskip} 
    241~GHz & 241.37--241.84 & 0.497 $\times$ 0.436 & 0.122 / 0.15 & 2.58 / 0.28 \\ 
    244~GHz & 243.67--245.54 & 0.421 $\times$ 0.362 & 0.488 / 0.6 & 1.50 / 0.21\\ 
    257~GHz & 256.72--258.59 & 0.405 $\times$ 0.347 & 0.488 / 0.6 & 1.61 / 0.22\\ 
    259~GHz & 258.76--260.63 & 0.405 $\times$ 0.360 & 0.488 / 0.6 & 2.01 / 0.25\\ 
\hline 
\end{tabular}
\tablefoot{
\tablefoottext{a}{The average value for the 20 fields.}\\
\tablefoottext{b}{The rms noise per channel width is estimated in a line-free channel.}
}
\label{tab:spectral_setup}
\end{table*}
%
\section{Observations and sample}\label{sec:observations}
%
%
\subsection{Sample selection}\label{sec:sample}

We selected 20 massive YSOs identified with Spitzer \citep[e.g.,][]{2009ApJS..184..172G} distributed throughout the LMC to be observed with ALMA (see Table~\ref{tab:pointings_noise} for their names and coordinates and see Fig.~\ref{fig:LMCRGB} for their distribution across the LMC). We selected YSOs without ice features or strong fine-structure lines in their Spitzer IRS spectra \citep{2009ApJ...699..150S} and with \textit{Herschel} counterparts at submillimeter wavelengths \citep{2014AJ....148..124S} to maximize the chance of selecting the evolutionary stage of hot cores and not earlier or later evolutionary stages. These sources are located away from prominent massive star-forming sites \citep[based on the absence of strong H$\alpha$ emission in the H$\alpha$ image from the MCELS survey,][]{2000ASPC..221...83S}, making the cold initial conditions (which favor methanol formation) more likely. The selected sources are all bright at 160~$\mu$m to ensure the presence of large amounts of dust and large column densities, favoring the detection of molecular lines. We also added two regions centered on the YSOs ST6 and ST10, for which \cite{2016A&A...585A.107S} detected methanol ice. A more detailed description of the sample and its properties, including the continuum images and the compact sources detected with ALMA can be found in Paper~II.
%
%
\subsection{ALMA observations}\label{sec:ALMA}

The sample of 20 regions (see Fig.~\ref{fig:LMCRGB}) in the LMC was observed with the ALMA 12-m Array during its Cycle~5 between November 11 and December 3, 2018 (project number 2017.1.00696.S). The observations were carried out as single pointings centered at the coordinates listed in Table~\ref{tab:pointings_noise} with a duration of about 19~minutes on-source per field. The observations were executed during eight different execution blocks that were later merged to reach the requested sensitivity. We used the main array of ALMA with the 12-m antennas in two configurations (C43-4 and C43-5) and baselines ranging from 15 to 1400~m, which resulted in an angular resolution of $\approx0\farcs4$ (corresponding to $\sim0.09$~pc at the distance of the LMC) and a maximum recoverable angular scale of $\approx5$\arcsec\ (corresponding to $\sim1.2$~pc at the distance of the LMC). This maximum recoverable scale, although not enough to map large extended structures, allowed us to detect dense cores and clumps with typical sizes $\leq0.5$~pc. The total field of view (or primary beam) of the ALMA observations is $\sim44$\arcsec\ (corresponding to $\sim10$~pc). The spectral setup (see Table~\ref{tab:spectral_setup}) covered four spectral windows, including a 469~MHz-wide band centered at 241.6~GHz and three 1875~MHz-broad bands centered at 244.6, 257.6, and 259.5~GHz. All four spectral windows have 3840 channels, resulting in channel widths of 0.122~MHz and 0.488~MHz, respectively. Throughout the paper, we refer to these four spectral windows as 241, 244, 257, and 259~GHz.

\begin{figure*}[h!]
\begin{tabular}{cc} 
\includegraphics[width=0.48\textwidth]{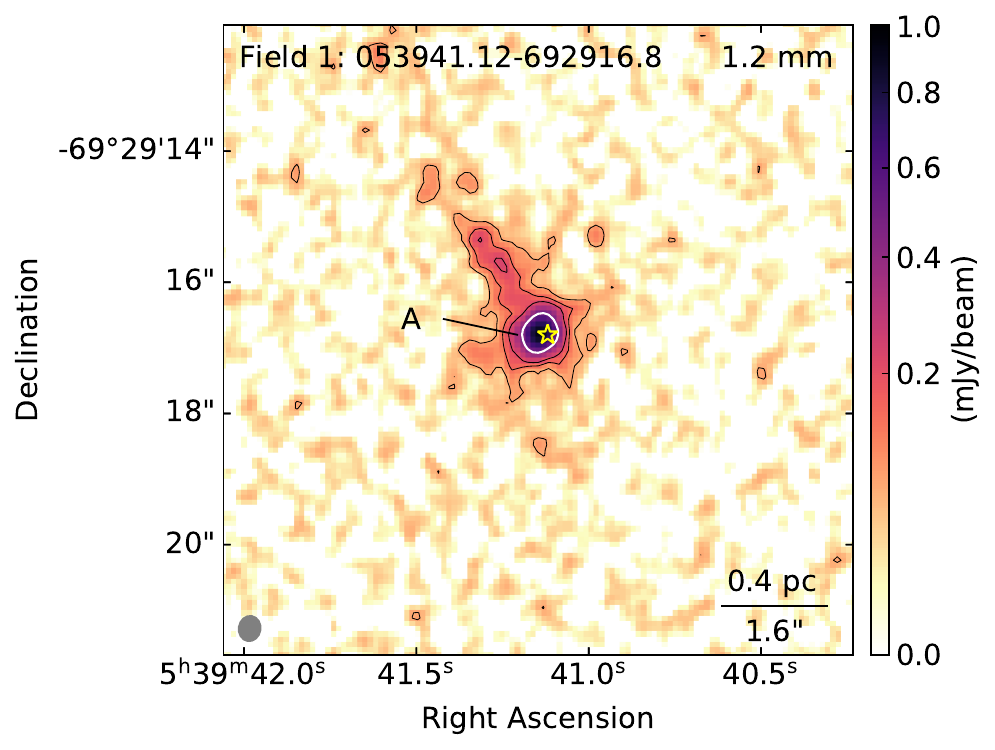} &
\includegraphics[width=0.48\textwidth]{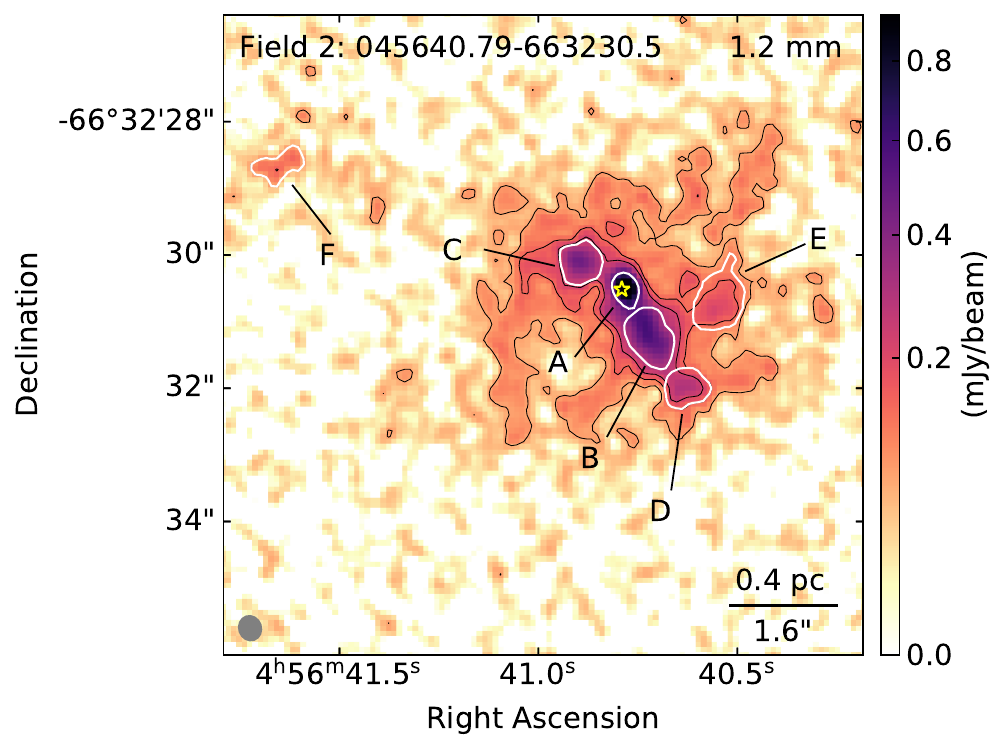} \\
\includegraphics[width=0.48\textwidth]{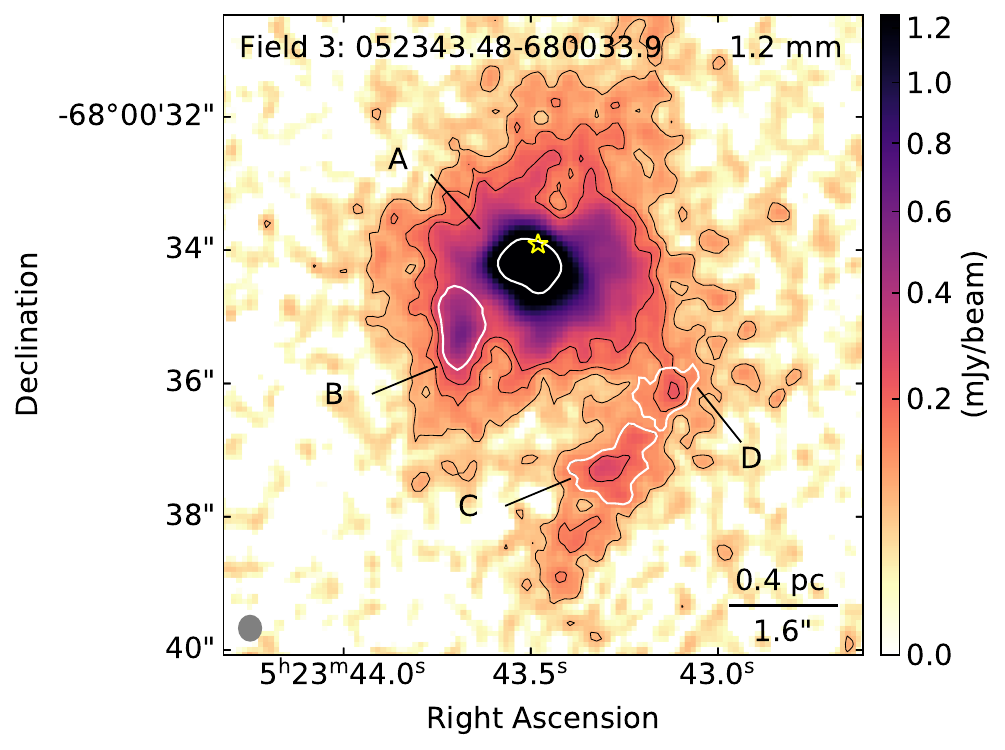} &
\includegraphics[width=0.48\textwidth]{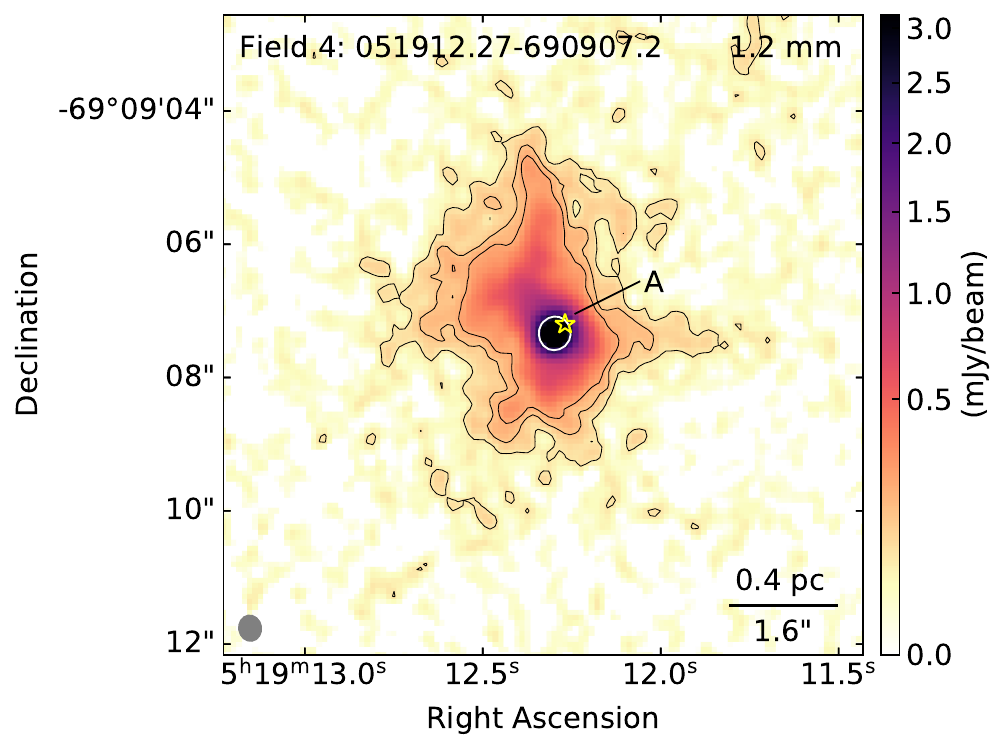} \\
\end{tabular}
\caption{Examples of the 1.2~mm continuum emission toward four regions of the LMC sample.  
The contours show the three, six, and nine times the $\sigma_{rms}$.
The identified compact cores are highlighted with white contours at 50\% of the peak emission. The synthesized beam of the images and a scale bar are shown in the bottom-left and bottom-right corners of each panel, respectively. See Paper~II for more details on the 20 fields observed across the LMC (see Table~\ref{tab:pointings_noise}).}
\label{fig:1st_cont_maps}
\end{figure*}
\begin{figure*}[h!]
\centering
\includegraphics[width=0.88\textwidth]{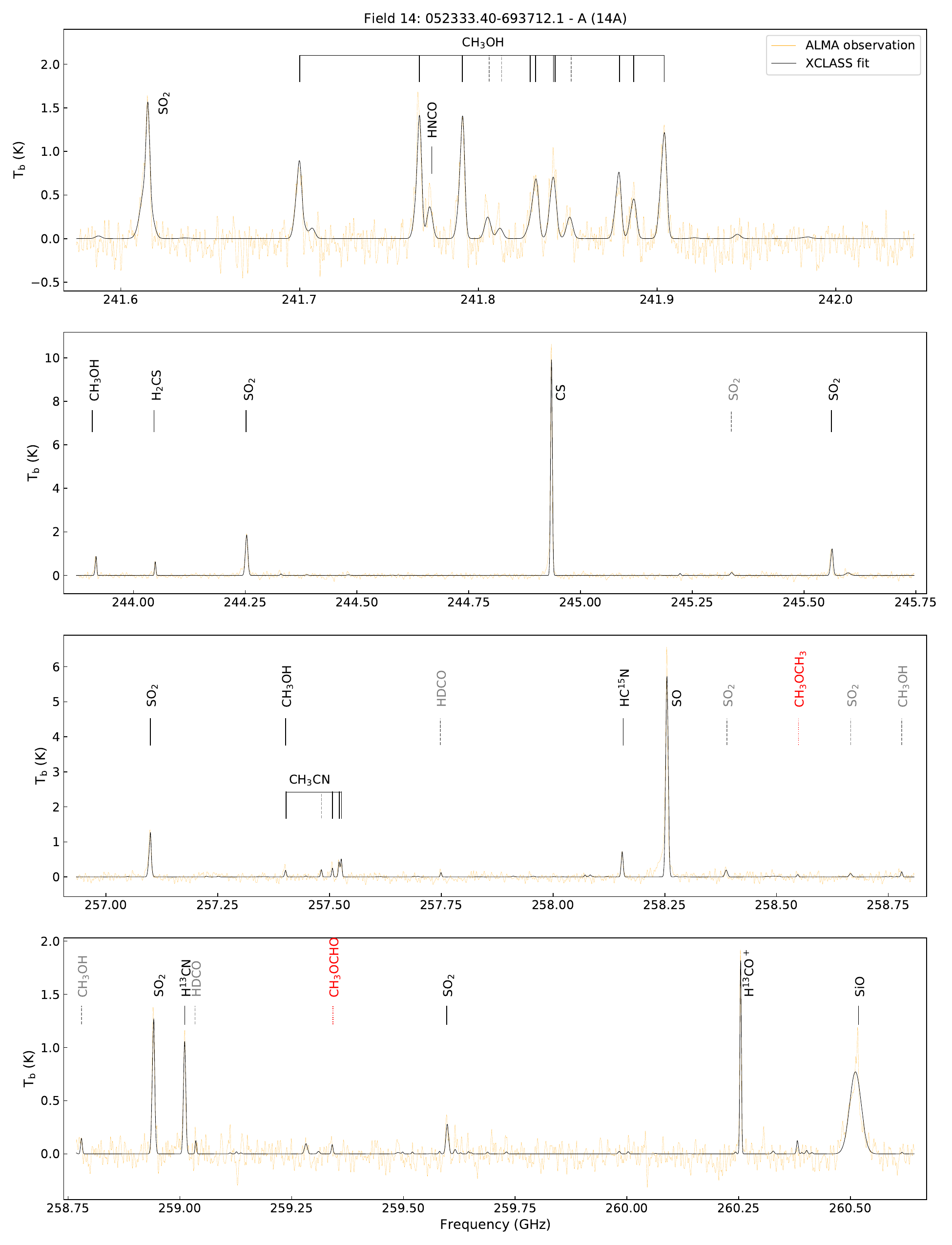}
\caption{ALMA Band 6 observed spectra toward source 14A overlaid with the best-fit XCLASS model. Molecular transitions with definite detections ($\mathrm{S/N} > 5$) are represented by solid black lines, while those with tentative detections ($3 < \mathrm{S/N} < 5$) are denoted by dotted gray lines. Dotted red lines indicate transitions of molecules for which abundance upper limits are provided.}
\label{fig:14A_spec}
\end{figure*}
%
The data were calibrated using the ALMA pipeline in CASA \citep[Common Astronomy Software Applications;][]{2007ASPC..376..127M} version 5.4.0-68. The amplitude-bandpass calibrator was chosen to be J0635$-$7516 or J0519-4546 in different execution blocks, with the complex gains being corrected with the calibrators J0529$-$7245 or J0601$-$7036. Continuum subtraction, determined from line-free channels, was done in the \text{u,v}-domain using the CASA task \texttt{uvcontsub}. The images were produced with the CASA task \texttt{tclean}, using a standard gridding and the \textit{Briggs} weighting with the robust parameter set to 0.5 as a compromise between sensitivity and resolution. The final images, both continuum and spectral data cubes, have been primary beam corrected using the \texttt{impbcor} task in CASA. Typical errors for the continuum images are about 65~$\mu$Jy~beam$^{-1}$ (10~mK) and about 1.92~mJy~beam$^{-1}$ (0.23~K) for the cubes (see Table~\ref{tab:spectral_setup}). More details on the observations and the continuum images can be found in Paper~II. As an example, Fig.~\ref{fig:1st_cont_maps} shows the 1.2~mm ALMA continuum emission toward four fields.
%
%
\section{Results and analysis}\label{sec:results}

A total of 65 compact sources were identified in the continuum images of the 20 fields in the LMC, defined by the 50$\%$ intensity contour level around well-defined intensity peaks (see Paper~II). These objects have peak continuum fluxes above five times the image rms noise and constitute clear detections of compact sources. We used the temperatures estimated in the current paper to calculate the dust and gas mass of each core, assuming that the 1.2~mm continuum emission originates in optically thin dust with an opacity of 1.04~cm$^2$~g$^{-1}$ \citep[for dust grains coated by thin ice mantels after \num{e5}~yr of coagulation in a hydrogen density of \num{e6}~cm$^{-3}$,][]{1994A&A...291..943O} and a gas-to-dust mass ratio of 316 \citep{2022ApJ...931..102S}. The cores have masses\footnote{According to the tentative detection of radio recombination lines toward one of the cores (see Section~\ref{sec:spectra}), we note that there might be a certain level of contamination from thermal free-free emission in the 1.2~mm continuum fluxes for at least the most massive core. Therefore, the higher-end value of the masses, \ce{H2} column, and volume densities may be overestimated.} in the range of 20--1000~M$_\odot$, H$_2$ densities in the range \num{1.2E5}--\num{4.0E6}~cm$^{-3}$, and H$_2$ column densities ranging from \num{6.0e+22}--\num{1.5e+24}~cm$^{-2}$. In this paper, we focus on the chemical properties of these 65 compact objects. In Section~\ref{sec:spectra}, we describe the spectral analysis performed toward these sources, while in Section~\ref{sec:XCLASS} we analyze their spectral properties, such as temperature, line width, and chemical content.
%
%
\subsection{Spectral analysis of the continuum cores}\label{sec:spectra}

In this section, we present the steps to analyzing the extracted spectra from the 65 compact sources identified in the ALMA 1.2~mm continuum images (see Paper~II). The spectra of each source were extracted from the primary beam corrected cube images of the four spectral windows (see Table~\ref{tab:spectral_setup}). The average spectra were computed using the \texttt{spectral-cube} python package \citep{2016ascl.soft09017R} with the pixels enclosed within the 50\% peak-intensity contour of the 1.2~mm continuum emission maps. 
Figure~\ref{fig:14A_spec} shows an example spectrum (see Appendix~\ref{app:spectra} for the rest of the spectra).
%
%

As a first step in the analysis of the extracted spectra, we identified the molecular lines toward the most chemically rich sources (i.e., spectra with the largest number of line detections; see Fig.~\ref{fig:14A_spec} for an example) using the CDMS catalog (Cologne Database for Molecular Spectroscopy; \citealp{Mueller2005}). For this, we estimated the noise level ($\sigma_{\rm_{rms}}$) in each spectrum (i.e., for each source and spectral window) by performing a sigma clipping analysis. Based on this noise-level determination and the intensity of each spectral line feature, we sorted the detected lines. We considered the lines as a definite detection if their peak value exceeded 5$\sigma_{\rm_{rms}}$ and as tentative features if the line strength is between 3 and 5$\sigma_{\rm_{rms}}$. The definite and tentative detections are respectively labeled with black and gray in Fig.~\ref{fig:14A_spec} and Figures~\ref{fig:spectraField01} to \ref{fig:spectraField20}. 
%
\begin{table}
\centering
\caption{Definite and tentative detections of molecular species along with the corresponding number of sources exhibiting these emissions.}

\begin{tabular}{l c | l c}
\hline\hline \noalign{\smallskip}
Molecular 
& Number 
& Molecular 
& Number \\

species
& of sources\tablefootmark{a}
& species
& of sources\tablefootmark{a}
\\
\hline \noalign{\smallskip} 
\ce{CS}       & 58 / -- & \ce{HC^15N}  & 6 / 3 \\
\ce{SO}       & 43 / 2 & \ce{H2CS}    & 23 / 7 \\
\ce{^33SO}    & 3 / -- & \ce{HNCO}    & 3 / 1  \\ 
\ce{SiO}      & 8 / 12 & \ce{HDCO}    & -- / 1  \\
\ce{SO2}      & 16 / 11 & \ce{HC3N}    & -- / 3  \\
\ce{^33SO2}   & 1 / 1  & \ce{CH3OH}   & 45 / 2 \\
\ce{^34SO2}   & 1 / 2  & \ce{CH3CN}   & 4 / 1  \\
\ce{H^13CO^+} & 32 / 3 & \ce{NH2CHO}  & -- / 1 \\
\ce{H^13CN}   & 15 / 8 & & \\
\hline 
\end{tabular}
\tablefoot{
\tablefoottext{a}{Definite / tentative detection(s)}
}
\label{tab:detected_species}
\end{table}
%
With a complete set of molecules presented in Table~\ref{tab:detected_species}, including definite and tentative detections, we fit the spectra using the XCLASS toolbox\footnote{\url{https://xclass.astro.uni-koeln.de/}} \citep[eXtended CASA Line Analysis Software Suite;][with additional extensions, T. M\"{o}ller, priv. comm.]{2017A&A...598A...7M, 2023A&A...676A.121M}. XCLASS solves the radiative transfer equation in one dimension in local thermodynamic equilibrium (LTE) or non-LTE based on the request. It can solve the radiative transfer equation for an unlimited number of molecules simultaneously, taking line blending into account and producing modeled spectra. XCLASS provides the basis for finding the best-fit model to the observed spectra by encompassing a large set of fitting algorithms available within the MAGIX interface \citep{2013A&A...549A..21M}. The fitting procedure works with the $\chi^2$ minimization. The contribution of a certain molecule can be described by more than one component. This is useful when the detected molecular emission line may come from different unresolved structures with different sizes, temperatures, or even distances to the observer. Each component is described by source size, temperature, column density, line width, and velocity offset. These can be fixed or free to vary to get the best-fit spectra and the best set of parameters.  

Using XCLASS, we analyzed the spectra assuming that all the considered molecular species are located within a core layer at the position of the source (i.e., we did not consider foreground clouds located between the source and the observer). In the absence of optically thick transitions, column density and source size are degenerate and cannot be determined separately. Therefore, we fixed the source size for all the molecules to a very large value to impose a beam-filling factor of one.\footnote{We note that a smaller source size results in higher column densities.} The temperature, column density, line width, and velocity shift are in general free parameters to fit. While parameters such as line width and velocity shift can be determined from Gaussian fits to the detected spectral lines, a good determination of the temperature requires the detection of several transitions from the same molecular species. Based on this, we grouped the molecules into two different categories: molecules with more than two transitions detected within the four spectral windows and molecules with only one or two transitions. The first group includes \ce{CH3OH}, \ce{SO2}, and \ce{CH3CN}, and we fit the temperature for them using XCLASS when multiple transitions were detected.

Out of 65 sources, \ce{CH3OH} was detected in 48, while 37 sources show clear detections of three or more transitions. One or two transitions were detected toward the other 11 sources. Seventeen sources have no \ce{CH3OH} features with line intensities higher than 3$\sigma_{rms}$. In the first iteration, we fit \ce{CH3OH} using an LTE approximation, which resulted in low temperatures (5--17~K; see Table~\ref{tab:XCLASS_LTE}) and some line features to be underfit or overfit (see Fig.~\ref{fig:residual_spectra}). Based on these inaccuracies, we switched to non-LTE for fitting \ce{CH3OH}, which resulted in better fits. In the non-LTE description, we have two additional parameters to consider: the ratio of A and E \ce{CH3OH} and the collision partner\footnote{XCLASS uses a similar prescription as RADEX \citep{2007A&A...468..627V} for its non-LTE fitting. The collision parameters are from the Leiden Atomic and Molecular Database \citep[LAMDA,][]{2005A&A...432..369S}.} volume density. We fixed the A to E \ce{CH3OH} ratio to one and fit the collision partner density in the cores with more than two \ce{CH3OH} lines. The fit collision partner densities are $\approx$~\num{e6}~cm$^{-3}$ for most of the cores and about $\approx$~\num{e7}~cm$^{-3}$ for those cores with bright \ce{CH3OH} lines. For the sources with one or two \ce{CH3OH} lines, we fixed the density of collision partners to the median value of the rest of the cores: \num{2.0e6}~cm$^{-3}$. With this, we then fit the temperature, column density, line width, and velocity offset. The derived gas kinetic temperatures for \ce{CH3OH} with the non-LTE description are generally in the range of 20--60~K (see Table~\ref{tab:XCLASS_params}). Finally, for those sources with no \ce{CH3OH} detection, we fixed the temperature to the median value of the other cores and the line width and velocity offset to values derived for \ce{CS}, if this was detected, and we derived an upper limit for the \ce{CH3OH} column density. 

We found \ce{SO2} to be present in 12 sources with enough lines for temperature determination under LTE conditions. However, only in five sources were there enough \ce{SO2} lines to fit the spectra using non-LTE. For these sources, we compared the LTE and non-LTE fits, finding no major differences in the output temperatures and fit quality. Therefore, we fit \ce{SO2} for all the fields using an LTE approach. For the rest of the molecular species, we fixed the temperature to the best-fit value for \ce{CH3OH}. If \ce{CH3OH} was not detected or properly fit in the spectra of a source, we fixed the temperature of other molecules to the median of the best-fit value for \ce{CH3OH} from other sources. 

For four sources (4A, 6A, 6B, and 14A), a one-component model for \ce{CH3OH} and \ce{SO2} did not give satisfactory results. A two-layered morphology with a hot central component and a colder envelope as the foreground provided the most convincing results. The core component's source size was fixed to a sub-beam size (not beam filling), while the foreground component was set to fill the beam. In these sources, we detected several transitions of \ce{CH3CN} with a high enough S/N, which provided an additional temperature estimate. As done for the other sources in the sample, for the rest of the molecules, we fixed the temperature to what was derived from the hot component of \ce{CH3OH}. Moreover, some isotopologues were also detected in these sources. We used the flexibility of XCLASS to link the isotopologues to the main species, giving them the same temperature and velocity profile to fit the isotopologue ratio from the ratio of column densities. We also detected \ce{HNCO} and tentatively detected \ce{HDCO}, \ce{HC3N}, and \ce{NH2CHO} in one or more cores among these four cores. Furthermore, we estimated an upper limit for \ce{HDCO}, \ce{NH2CO}, \ce{CH3CHO}, and \ce{CH3OCH3} in these four cores when they appeared in the spectra, but they were not fit properly due to having a low S/N (S/N < 3). To this purpose, we fixed the temperature and line width and velocity shift to the hot component of \ce{CH3OH}. These transitions are marked with red in the figures of the corresponding spectra.
%
\begin{figure*}[t!]
\centering 
\includegraphics[scale = 0.7]{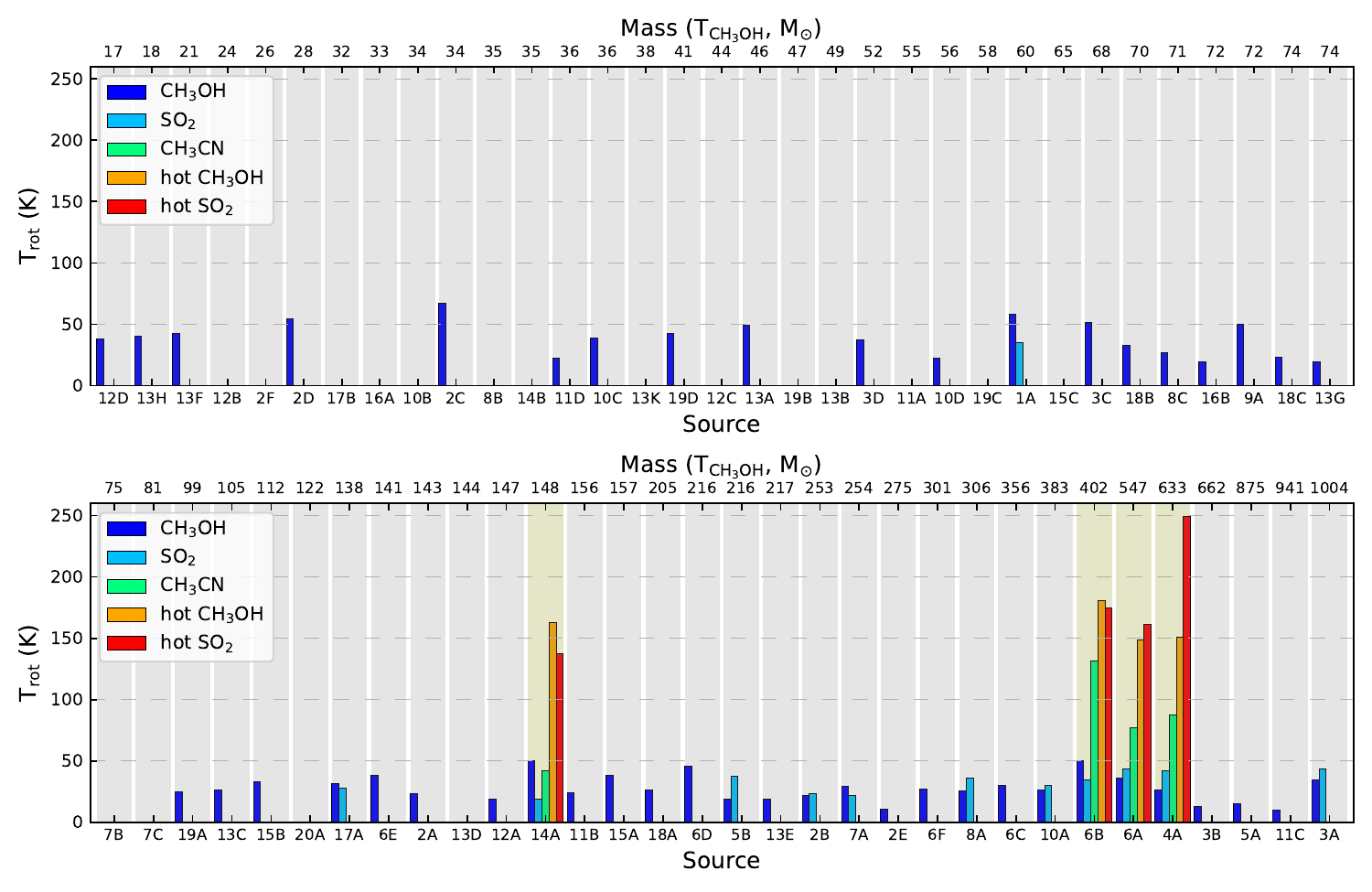}
\caption{Best-fit XCLASS temperature for different molecular tracers. In case of definite detection in more than two transitions, \ce{CH3OH}, \ce{SO2}, and \ce{CH3CN} were free for fitting (for more details, see Section \ref{sec:spectra}). There is an uncertainty of 40\%--60\% for the temperatures. The sources are ordered by increasing mass (from Paper~II), calculated with the temperature of the \ce{CH3OH} (cold) component to account for the total mass in the core. The hot cores are highlighted with yellow bars.}
\label{fig:T_Sorted_M}
\end{figure*}
\begin{figure*}[t!]
\centering 
\includegraphics[trim={2cm 0.5cm 2.8cm 1.5cm}, scale = 0.72]{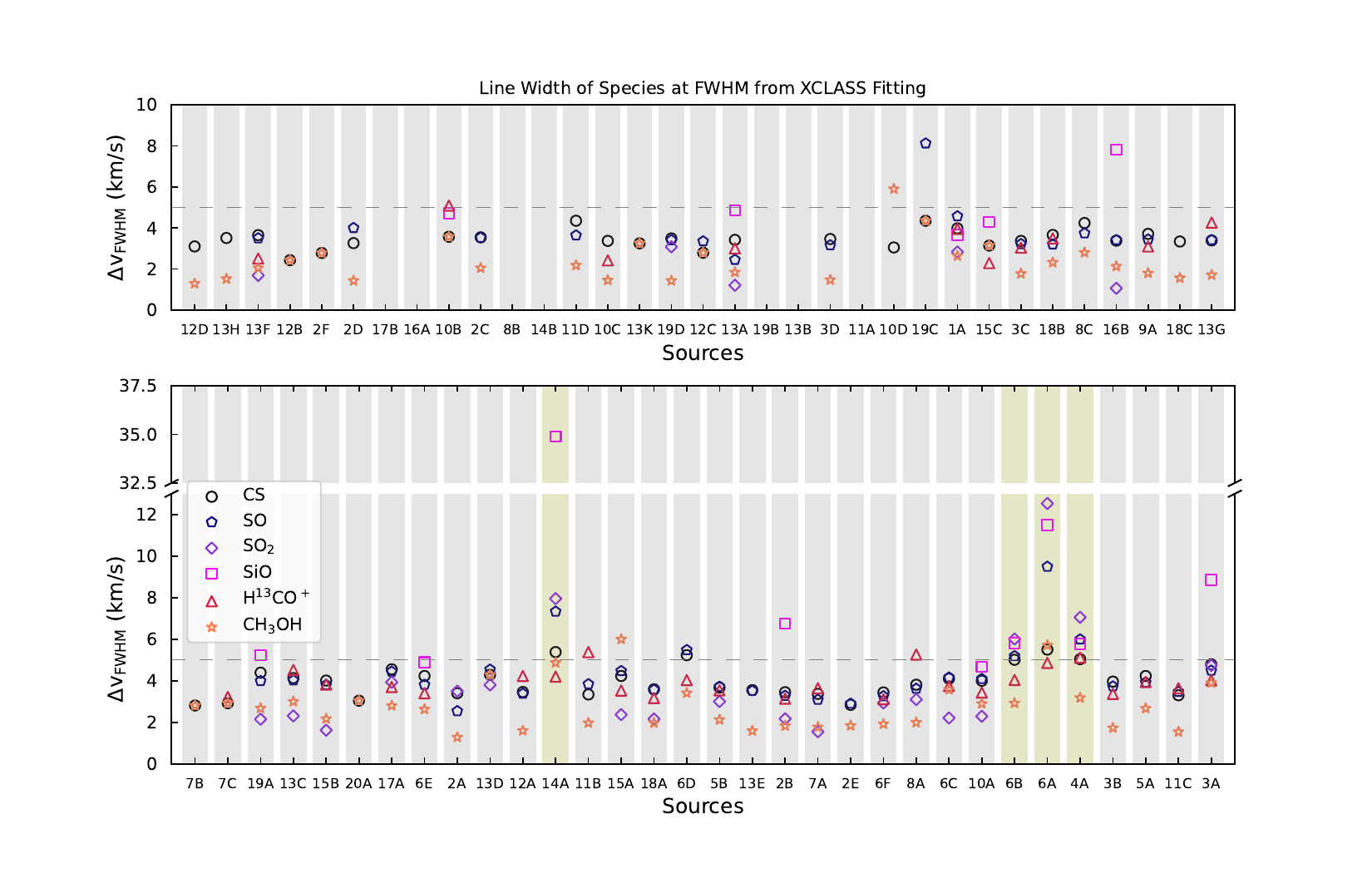} 
\caption{Best-fit XCLASS line width (full width at half maximum, FWHM) for different molecular species. The line width of 5 km/s is marked with a dashed line as a reference. There is an uncertainty of 10\%--30\% for the line widths. The sources are ordered by increasing mass (from Paper~II) with the temperature of \ce{CH3OH} (cold) component to account for the whole mass in the core. The hot cores are highlighted with yellow bars.}
\label{fig:LW_all}
\end{figure*}
%
In addition, we searched for hydrogen recombination lines (RLs) in all sources. Within the four observed spectral ranges, there are two potentially detectable RLs (H36$\beta$ and H41$\gamma$). However, we only found one tentative detection of H36$\beta$ toward source 3A (see Fig.~\ref{fig:spectraField03}). We used XCLASS to fit the RL emission. For the fit, we fixed the electron temperature to \num{e4}~K \citep[e.g.][]{2012A&A...545A..29S, 2023ApJ...958..179J} and fit the other three parameters, obtaining an emission measure estimate of \num{2.5e6}~pc~cm$^{-6}$, a line width of 18.2~km~s$^{-1}$, and a velocity offset of 284.4~km~s$^{-1}$, consistent with the velocity derived for the molecular species.

The fitting procedure in XCLASS uses an algorithm chain from the MAGIX interface, including the genetic algorithm as a global minimizer and the Levenberg-Marquart algorithm as a local minimizer. We also ran XCLASS with a Markov chain Monte Carlo (MCMC) algorithm on a few spectra to estimate typical errors on the different fitting parameters. The error values range from 40\%-60\% for the temperatures, 20\%-40\% for the column densities, and 10\%-30\% for the line widths and velocity shifts. In the non-LTE solution, there is a degeneracy in defining the temperature, column density, and collision partner density that makes it hard to agree on an error value. The outcome of the fitting procedure is shown as modeled spectra (in black) overlaid on observed spectra (in colors) in Fig.~\ref{fig:14A_spec} and Figures.~\ref{fig:spectraField01} to \ref{fig:spectraField20}. The best-fit parameters are listed in Table~\ref{tab:XCLASS_params}, and the upper-limit estimations are marked with a less than ($<$) sign. 
%
%
\subsection{Results from spectral analysis}\label{sec:XCLASS}
%
%
\subsubsection{Temperature of compact cores}\label{sec:temperature}

As discussed in Section~\ref{sec:spectra}, we fit the temperature for three molecular tracers, \ce{CH3OH}, \ce{SO2}, and \ce{CH3CN}, when there were firm detections. This resulted in 47 sources with a temperature determination using \ce{CH3OH}, 12 sources with \ce{SO2}, and only four sources with \ce{CH3CN}-derived temperatures. Figure~\ref{fig:T_Sorted_M} shows the derived temperatures for the different molecular tracers for all 65 compact cores. The cores are ordered based on their mass (as determined in Paper~II).

As stated earlier, out of the 65 cores, only four have a reliable temperature determined using \ce{CH3CN}, and it is only for these four cores that we needed more than one component to fit \ce{SO2} and \ce{CH3OH} (see Table~\ref{tab:XCLASS_params}). Furthermore, we only obtained temperatures above 100~K for the hottest component of these four sources, with the temperature remaining consistently below 60~K for the rest of the sources. Based on the definition of a hot molecular core to have a temperature higher than 100~K, we can classify these four sources (4A, 6A, 6B, and 14A) as such. These hot cores are likely embedded in a colder envelope with temperatures about 30--40~K, similar to the temperatures derived for the other cores in the sample. 
%
%
\subsubsection{Line-width of different molecules}\label{sec:deltav}

The line width of a molecular species provides useful information about the kinematics of the gas. Figure~\ref{fig:LW_all} compares the line widths derived for different molecular species from fitting the Gaussian line profile with XCLASS (see Section~\ref{sec:spectra}). Six molecules, including the ubiquitous ones, \ce{CS}, \ce{SO}, \ce{CH3OH}, and two other abundant species, \ce{SO2} and \ce{H^13CO^+}, together with the shock tracer \ce{SiO} were utilized. The line widths vary between 1 and 35~km~s$^{-1}$, with methanol having the lowest value in 49\% of the cores. In this plot, we present the line width value for the hot component of \ce{SO2} and \ce{CH3OH} when there are two temperature components. A value of 5~km~s$^{-1}$ is shown with a dashed line as a reference. The number of cores with line-width values larger than 5~km/s for at least one of the molecular species is 16. In nine of these sources, SiO is one of the molecules with a large line width suggesting the presence of molecular outflows, which will be studied in a forthcoming paper. Among them, we can identify the four cores (4A, 6A, 6B, and 14A) with temperatures higher than 100~K (see Section~\ref{sec:temperature} and Fig.~\ref{fig:T_Sorted_M}).
%
\begin{figure*}
\centering
\includegraphics[scale=0.7]{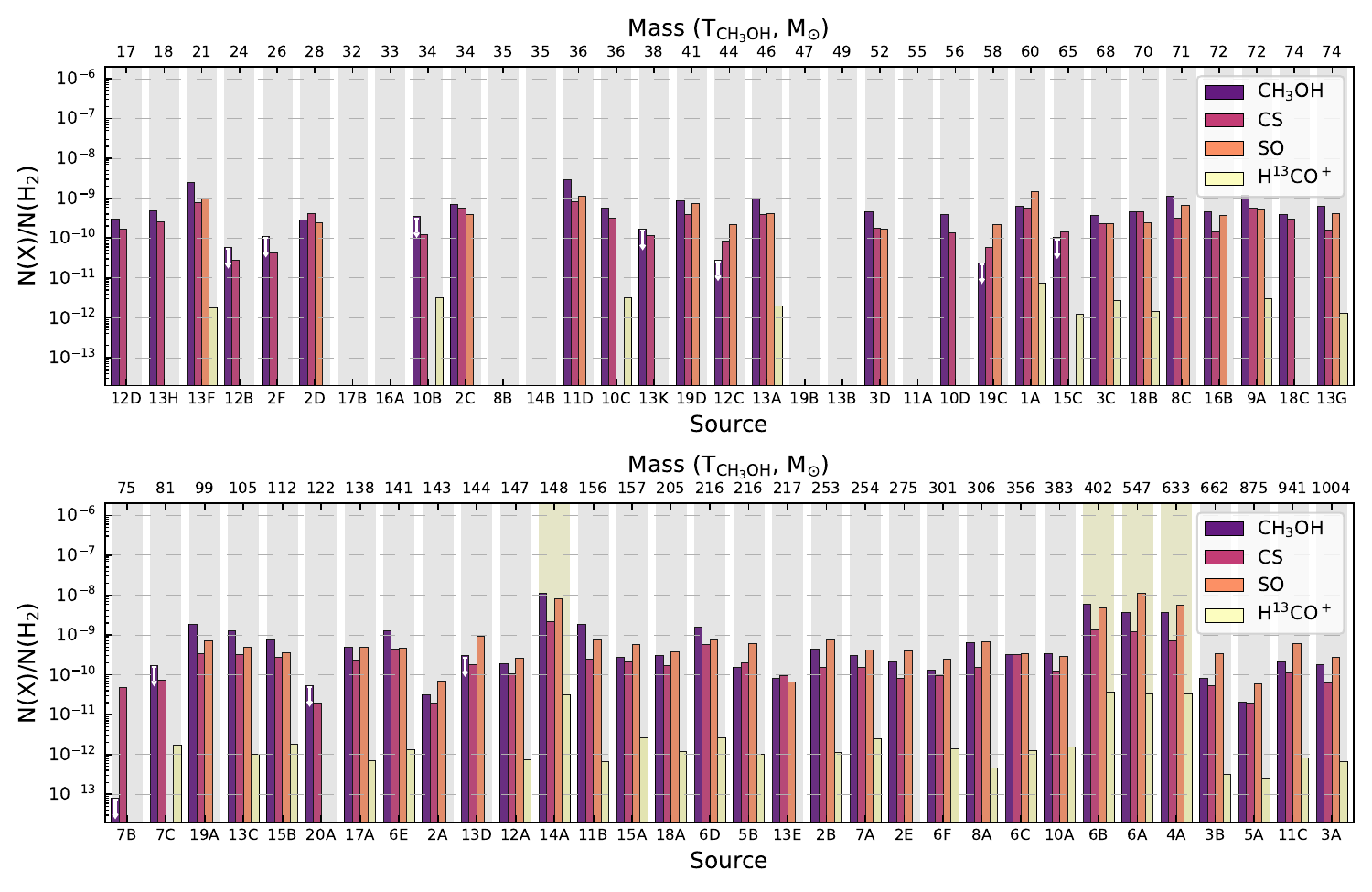} 
\caption{Abundances of molecular species with a high detection rate in our sample. There is an uncertainty of 20\%--40\% on the column densities. The sources are ordered by increasing mass (from Paper~II) with the temperature of \ce{CH3OH} (cold) component to account for the whole mass in the core. The hot cores are highlighted with yellow bars. The arrows show an upper-limit estimation.}
\label{fig:Molecular_Abundances_HDR}
\end{figure*}
\begin{figure*}
\centering
\includegraphics[scale=0.7]{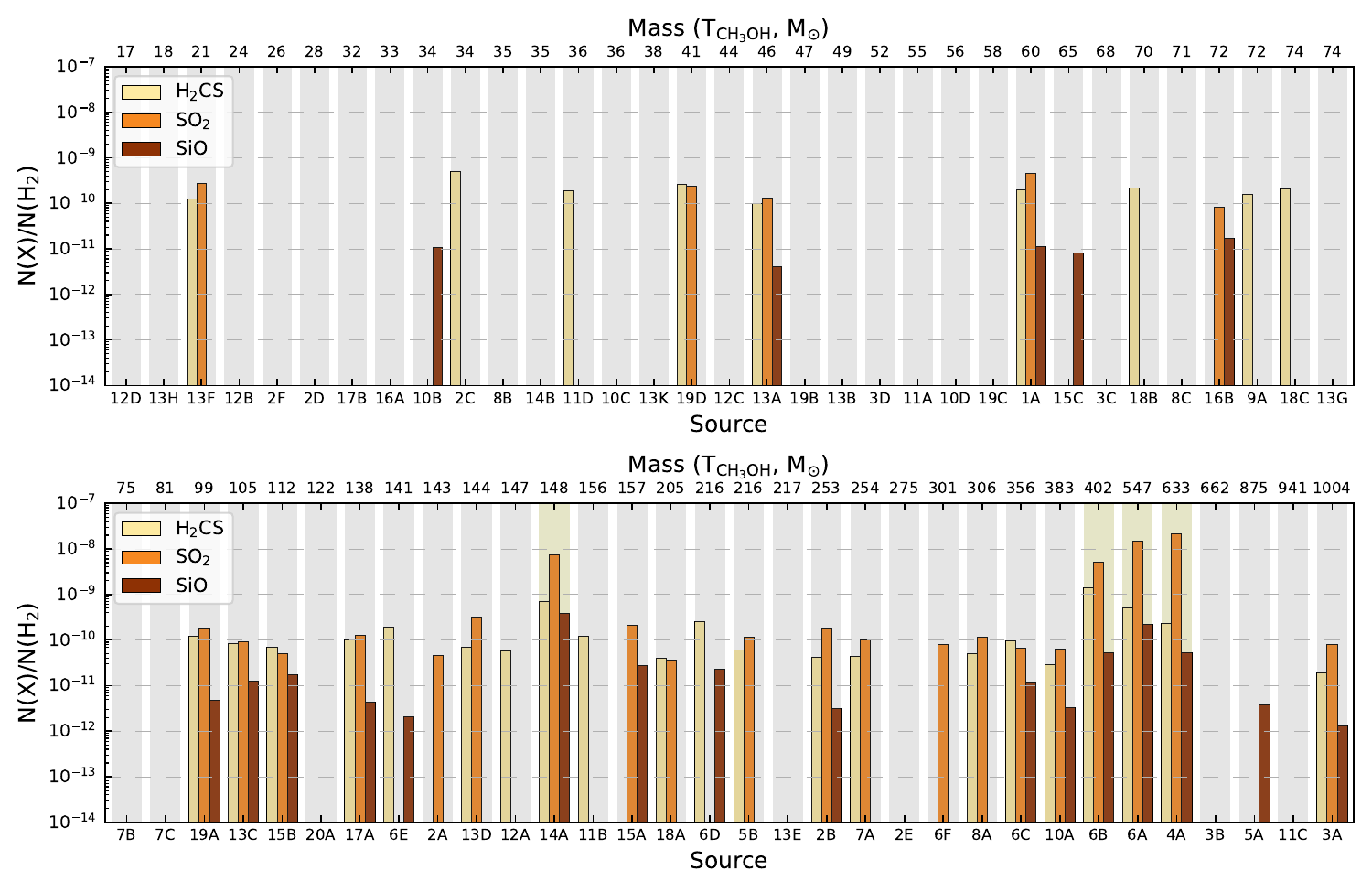} 
\caption{The same as Fig.~\ref{fig:Molecular_Abundances_HDR} but for S-bearing species, \ce{H2CS}, \ce{SO2}, and \ce{SiO}.}
\label{fig:Molecular_Abundances_S-bear}
\end{figure*}
\begin{figure*}
\centering
\includegraphics[scale=0.7]{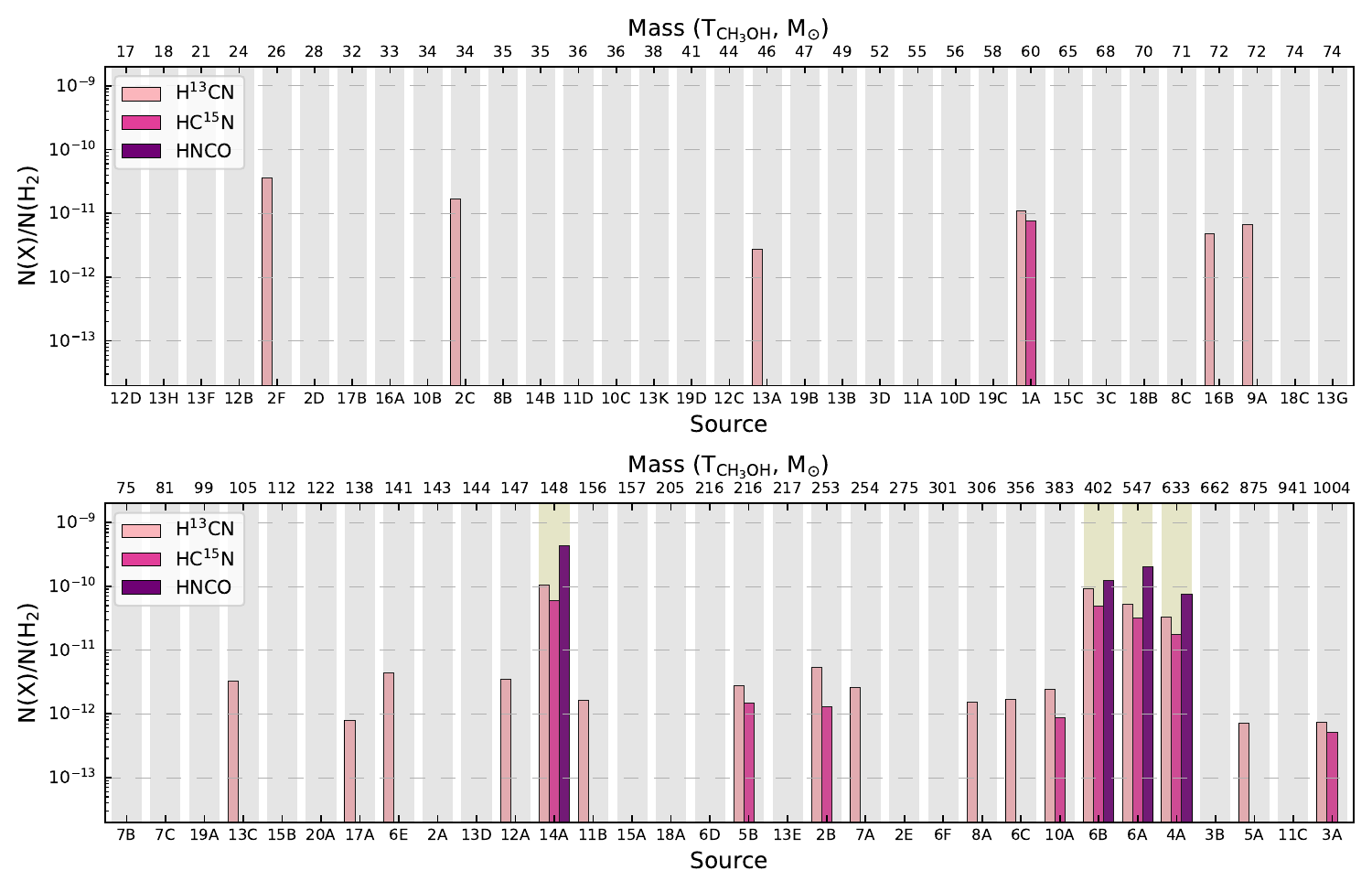} 
\caption{Same as Fig.~\ref{fig:Molecular_Abundances_HDR} but for N-bearing species, \ce{H^13CN}, \ce{HC^15N}, and \ce{HNCO}.}
\label{fig:Molecular_Abundances_N-bear1}
\end{figure*}
\begin{figure*}
\centering
\includegraphics[scale=0.7]{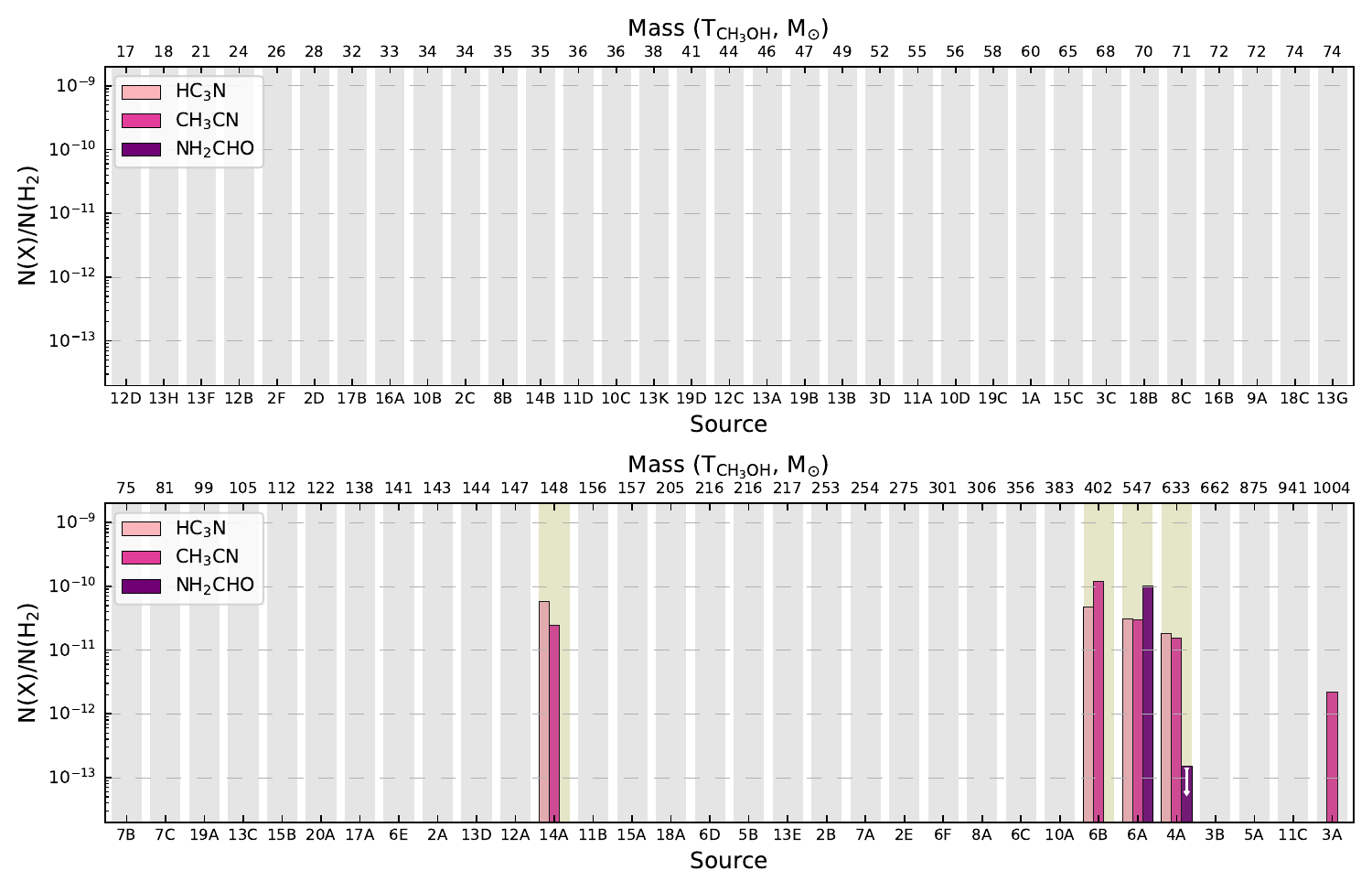} 
\caption{Same as Fig.~\ref{fig:Molecular_Abundances_HDR} but for N-bearing species, \ce{HC3N}, \ce{CH3CN}, and \ce{NH2CHO}.}
\label{fig:Molecular_Abundances_N-bear2}
\end{figure*}
\begin{figure*}
\centering
\includegraphics[scale=0.7]{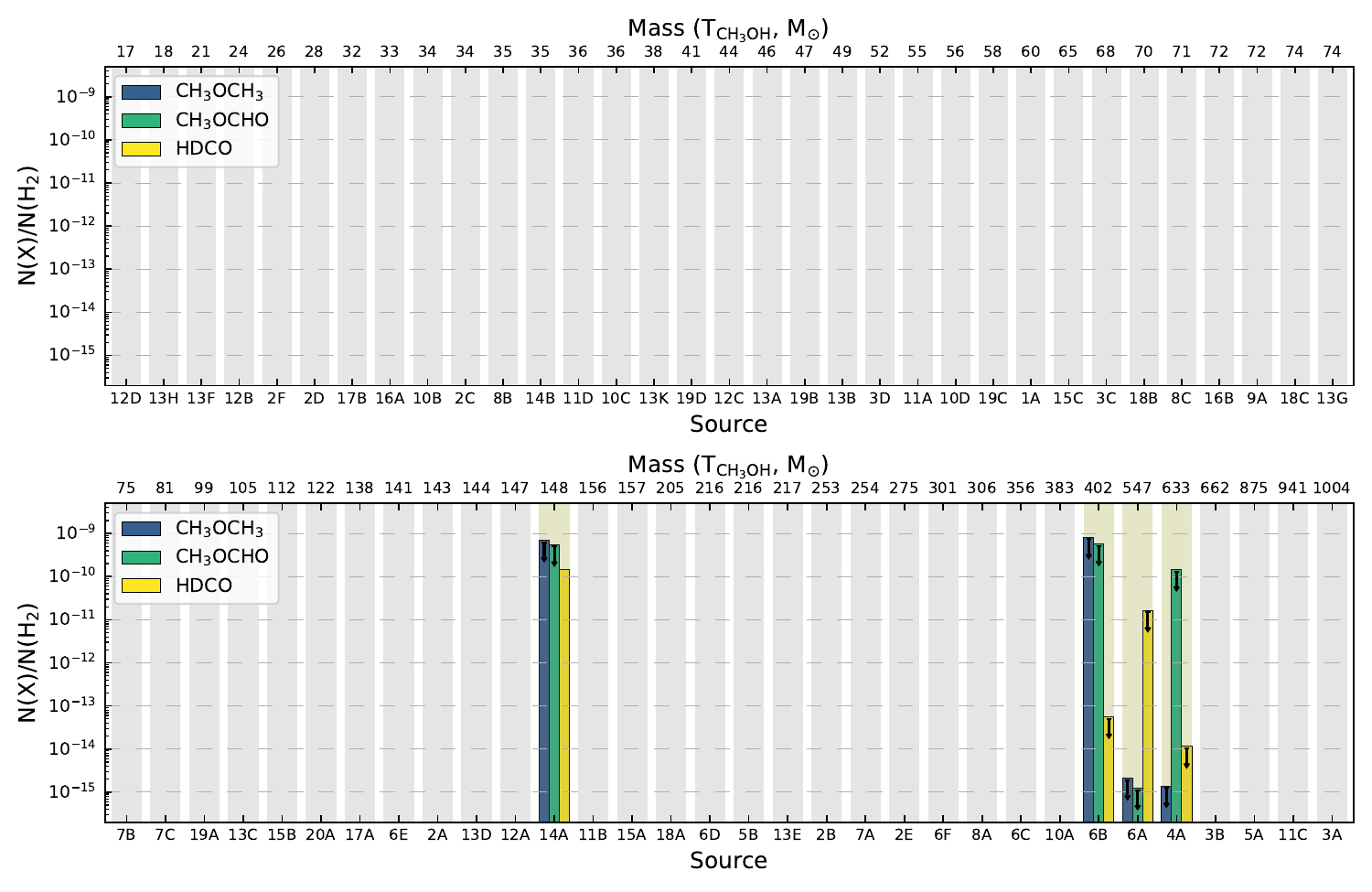} 
\caption{Same as Fig.~\ref{fig:Molecular_Abundances_HDR} but for O-bearing species, \ce{CH3OCH3}, \ce{CH3OCHO}, and \ce{HDCO}.}
\label{fig:Molecular_Abundances_O-bear}
\end{figure*}
%
%
\subsubsection{Chemical contents of the cores}\label{sec:chemicalcontent}

Several molecular lines were detected in the spectra of each core. Among all the species, methanol was detected toward 74\% of the cores (48/65). Among 17 cores with no-\ce{CH3OH} line detection, seven cores showed no emission lines at all (8B, 11A, 13B, 14B, 16A, 17B, and 19B). Other molecular species with high detection rates include \ce{CS}, \ce{SO}, \ce{SO2}, \ce{H^13CO+}, \ce{H^13CN}, \ce{HC^15N}, \ce{H2CS}, and \ce{SiO}, while more complex molecules, such as \ce{HDCO}, \ce{HNCO}, \ce{HC3N}, and \ce{CH3CN} or isotopologues of \ce{SO} and \ce{SO2}, were detected only toward a small subset of objects (about 8\% of the sample). We also report a tentative detection of \ce{NH2CHO} in one of the cores (6A). Moreover, \ce{CH3OCHO} or \ce{CH3OCH3} emerged in the modeled spectra after fitting an upper limit column density for them toward the cores 4A, 6B, and 14A. In addition, a hydrogen RL, H36$\beta$, was detected tentatively toward the most massive core in our sample (3A). In addition, H41$\gamma$ is also in the observed bands, and we mark it with red on the spectra of this core, as it was detected with an S/N less than three. The detection of this RL suggests the presence of \hii\ regions that likely contribute to the 1.2~mm continuum emission. Preliminary data at 6~GHz obtained with ATCA (Hamedani Golshan, priv.\ communication) favors this interpretation. The mass derived for this source is likely overestimated in the current analysis. Interestingly, we have a tentative detection of \ce{CH3CN} towards the source 3A. The detection of \ce{CH3CN}, which is only clearly detected toward the four cores 4A, 6A, 6B, and 14A in the sample (see Section~\ref{sec:discussion}), together with the detection of RLs (likely related to an \hii\ region) in 3A suggests the presence of a hot core near an expanding \hii\ region, a scenario that is commonly found in several hot cores of the MW \citep[e.g.,][]{2000prpl.conf..299K, 2013ApJ...766..114S, 2017A&A...602A..59C}.

We compared the abundances of different molecules in all the sources. The abundances were derived by dividing the column density of every molecule by the molecular hydrogen column density (N({\ce{H2})}, derived in Paper~II). For this purpose, we assumed that the dust and gas at the position of the source are well mixed and have the same temperature as derived from the XCLASS fits (see Table~\ref{tab:XCLASS_params}). Figure~\ref{fig:Molecular_Abundances_HDR} shows the abundance of the molecules with the highest detection rate for all the cores, whereas in Figures~\ref{fig:Molecular_Abundances_S-bear} to \ref{fig:Molecular_Abundances_O-bear}, the molecules are grouped into three different groups: S-bearing species (\ce{SO}, \ce{SO2}) and \ce{SiO}; N-bearing species (\ce{H^13CN}, \ce{HC^15N}, and \ce{HNCO}) and (\ce{HC3N}, \ce{CH3CN}, and \ce{NH2CHO}); and O-bearing species (\ce{HDCO}, \ce{CH3OCHO}, \ce{CH3OCH3}). We ordered the cores based on their mass, but we did not see any clear correlation between mass and abundance for any of the species. When comparing the different groups of molecules, we observed that S-bearing species are, on average, more abundant than N- and O-bearing species. 

Finally, comparing all the abundance plots, it was evident that four sources show a rich chemistry with the presence of various molecular species. These sources would classify as hot molecular cores based on their chemistry, as well as their temperatures, since these objects are the only cores with temperatures above 100~K and average line widths above 5~km~s$^{-1}$. Interestingly, these four sources are those with the largest \ce{SO2} abundances, thus supporting the interpretation by \cite{2023ApJ...946L..41S} that bright \ce{SO2} lines are a good tracer of hot molecular cores in the low-metallicity environment of the LMC and SMC.
%
%
\section{Discussion}\label{sec:discussion}

%
\subsection{Hot cores in our survey}\label{sec:hotcores}

Based on the spectral analysis of the 65 continuum sources, we report the detection of four hot molecular cores and one hot core candidate from the ALMA observations of 20 fields hosting YSOs in the LMC. As highlighted in Section~\ref{sec:results}, four cores (4A, 6A, 6B, and 14A) have temperatures above 100~K, large line widths, and the highest chemical richness among all the cores in the sample. This, together with the fact that the spatial size of these cores where the emission reaches half of the peak value in the continuum map, $\mathrm{\Omega}_\textrm{obs}$ ($\mathrm{\Omega}_\textrm{deconv}$), is between 0.12 (0.08) and 0.14 (0.11) pc (see Paper II) suggest that they are in the hot molecular core phase. This result confirms the previous detection of hot cores in ST16 YSO \citep{2020ApJ...891..164S} and two hot cores N105-2 A and N105-2 B in the N105 star-forming region \citep{2022ApJ...931..102S}. In addition, we present a new hot core detection in field 14: 0523333.40$-$693712.1, 14A. With this new detection, the number of known hot cores in the LMC increases to seven.

Isotopologues of \ce{SO} and \ce{SO2} were detected toward the hot cores in our sample. The hot cores 4A and 6A show emission lines from \ce{^33SO}, \ce{^33SO2}, and \ce{^34SO2}. Hot core 6B contains \ce{^33SO} and \ce{^34SO2} but no \ce{^33SO2}, and the core 14A shows no reliable detection of these isotopologues. We derived the isotopic ratios of \ce{^32S}/\ce{^33S} and \ce{^32S}/\ce{^34S} toward hot cores 4A and 6A and found them to vary between 12 and 21 and 9 and 19. These values can be compared with previous estimates in the LMC and other galaxies. \cite{2023A&A...679L...6G} used the APEX 12-m telescope to measure sulfur isotopic ratios in the LMC and compiled a set of measurements toward the LMC, different regions of the MW, and starburst galaxies (see their Table~2). For the LMC, \cite{2023A&A...679L...6G} found the ratios of \ce{^32SO}/\ce{^33SO} and \ce{^32SO}/\ce{^34SO} to range between 27 and 53 and 15 and 19, respectively. Combined with our results, which are similar to or slightly lower than previous estimates, the range for these isotopic ratios increases to 12 to 53 for \ce{^32SO}/\ce{^33SO} and 9 to 19 for \ce{^32SO}/\ce{^34SO}. The sulfur isotopic ratios in the LMC are lower than in different regions throughout the MW but more consistent with measurements in starburst galaxies. Gong et al.\ (2023) suggest that the low values in the LMC might be attributed to a combination of age, low metallicity, and star-formation history.

Source 3A may be a potential hot core candidate based on the presence of COMs in the analyzed spectra. This source is associated with a common hot core tracer, \ce{CH3CN}. However, the temperatures that we derived for this source are below 100~K. We note that source 3A is also associated with RLs in addition to \ce{CH3CN}, suggesting a scenario where a compact hot core may be located in the near vicinity of an \hii\ region. Future observations of these objects at higher resolution as well as covering additional frequency ranges may help in unambiguously determining the presence of hot cores in this region.  
%
\begin{figure*}
\centering
\begin{tabular}{ccc}
\includegraphics[trim={0cm 0cm 2.5cm 1.3cm}, scale=0.4]{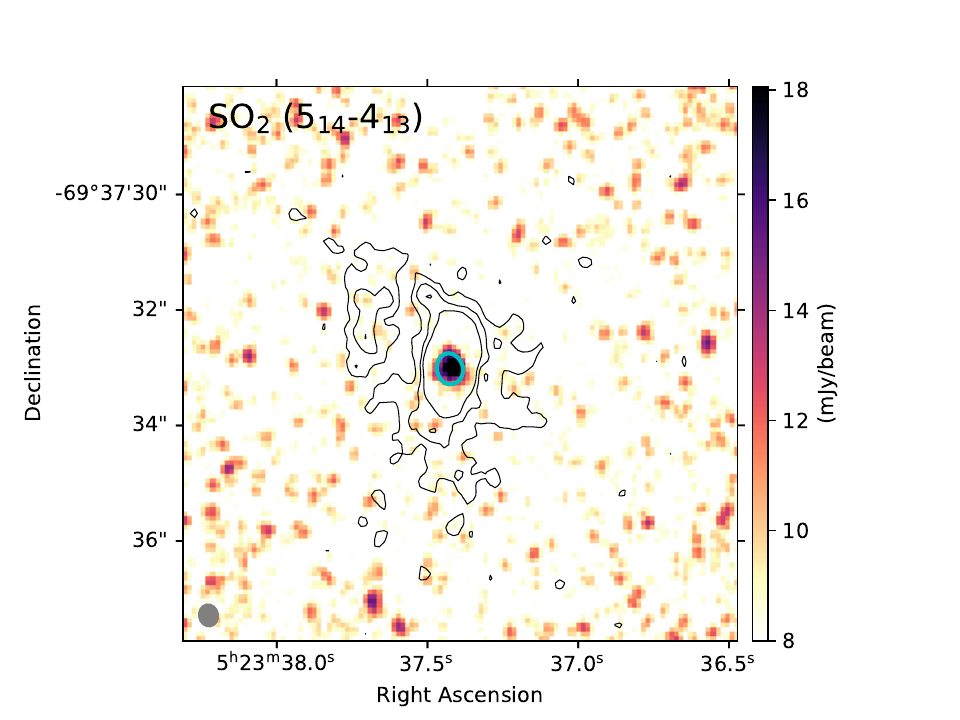} &
\includegraphics[trim={0cm 0cm 2.5cm 1.3cm}, scale=0.4]{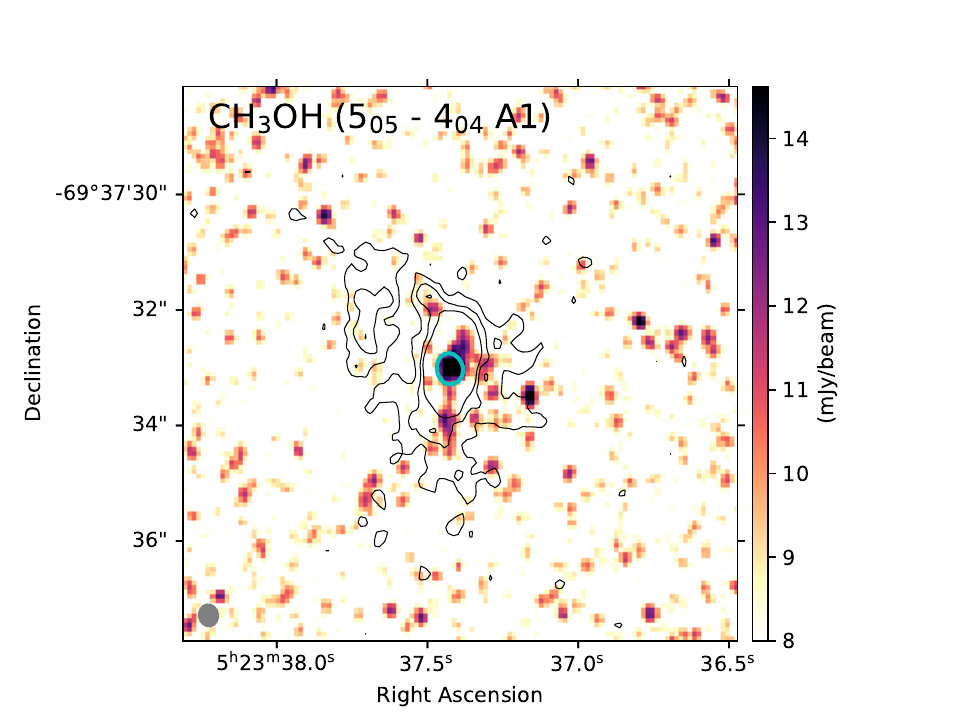} & 
\includegraphics[trim={0cm 0cm 2.5cm 1.3cm}, scale=0.4]{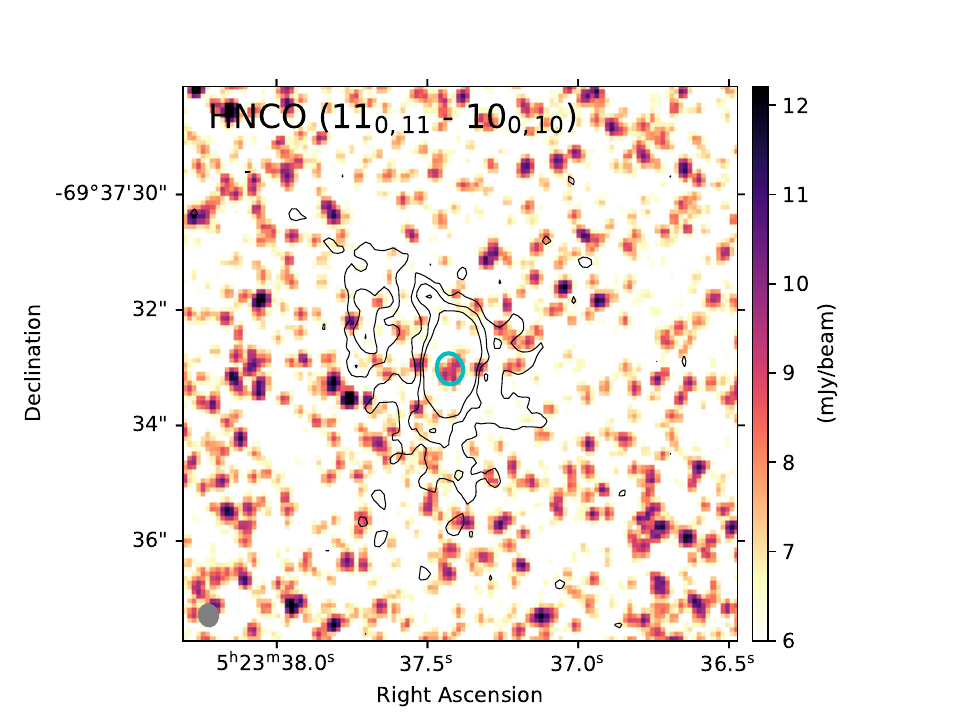} \\
\includegraphics[trim={0cm 0cm 2.5cm 0.5cm}, scale=0.4]{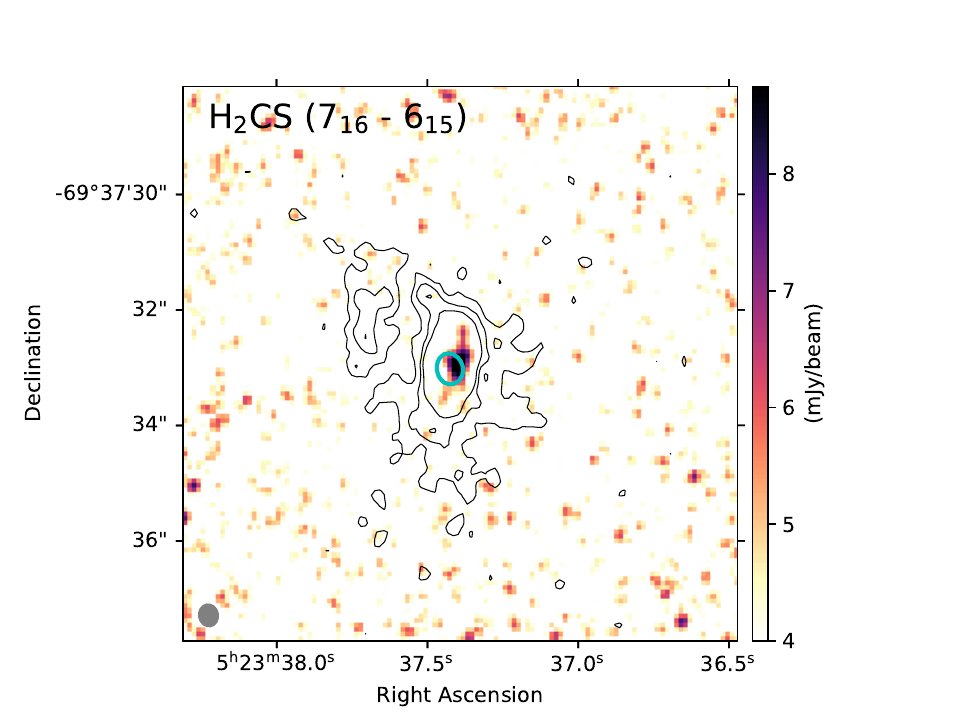} &
\includegraphics[trim={0cm 0cm 2.5cm 0.5cm}, scale=0.4]{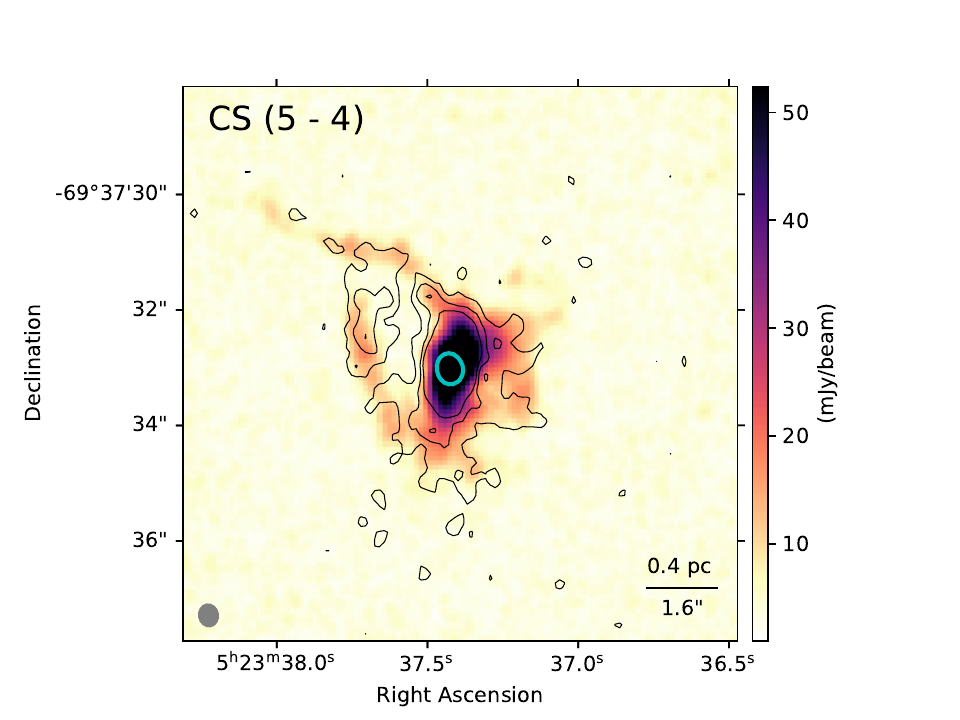} &
\includegraphics[trim={0cm 0cm 2.5cm 0.5cm}, scale=0.4]{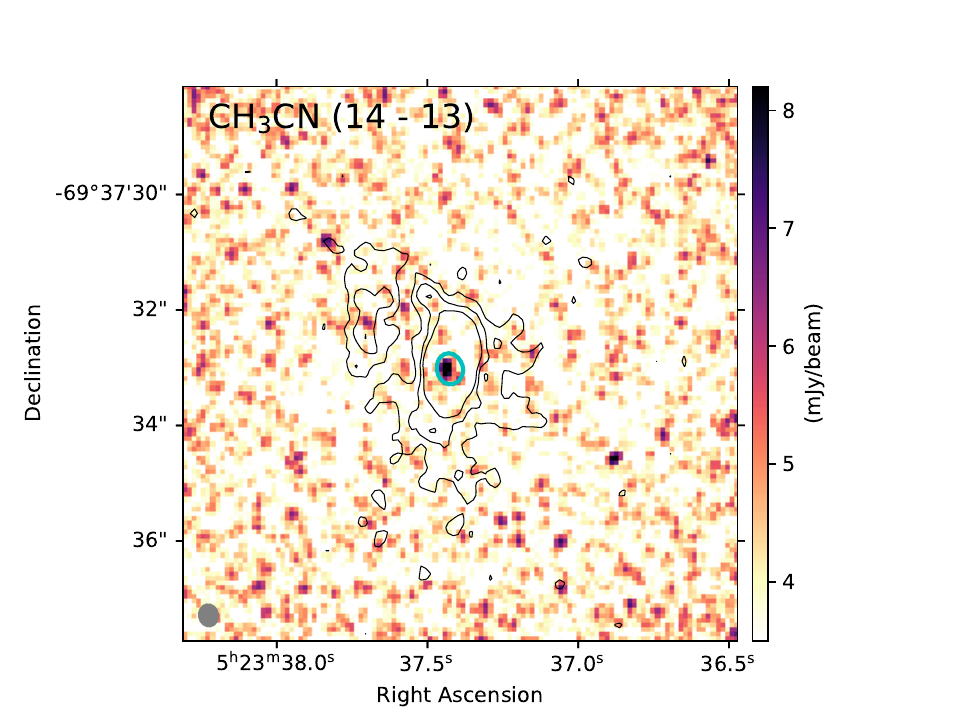} \\
\includegraphics[trim={0cm 0cm 2.5cm 0.5cm}, scale=0.4]{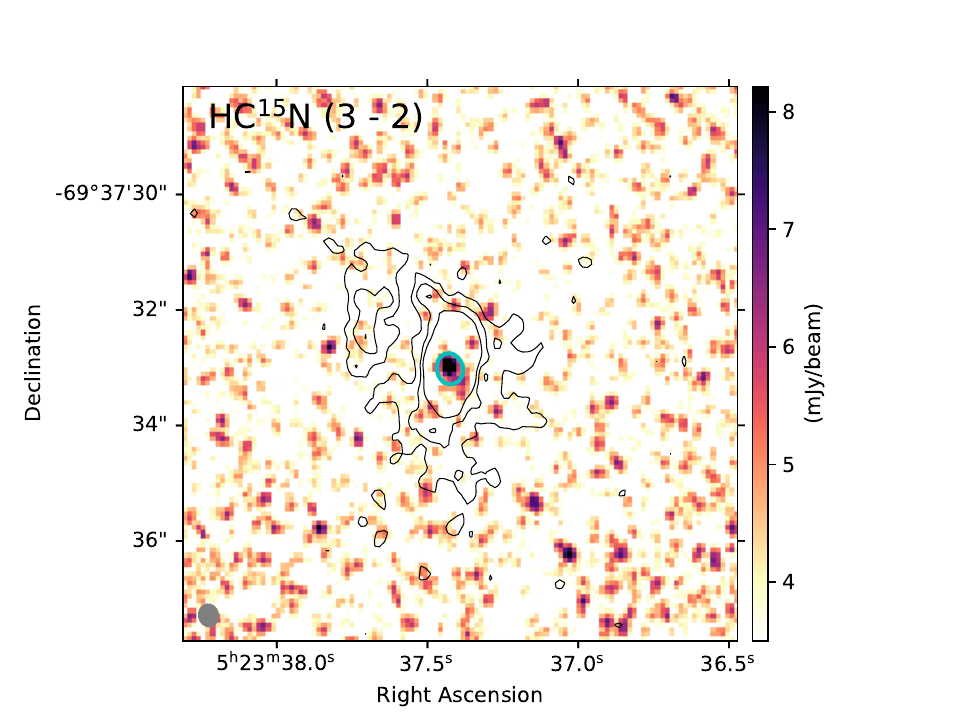} &
\includegraphics[trim={0cm 0cm 2.5cm 0.5cm}, scale=0.4]{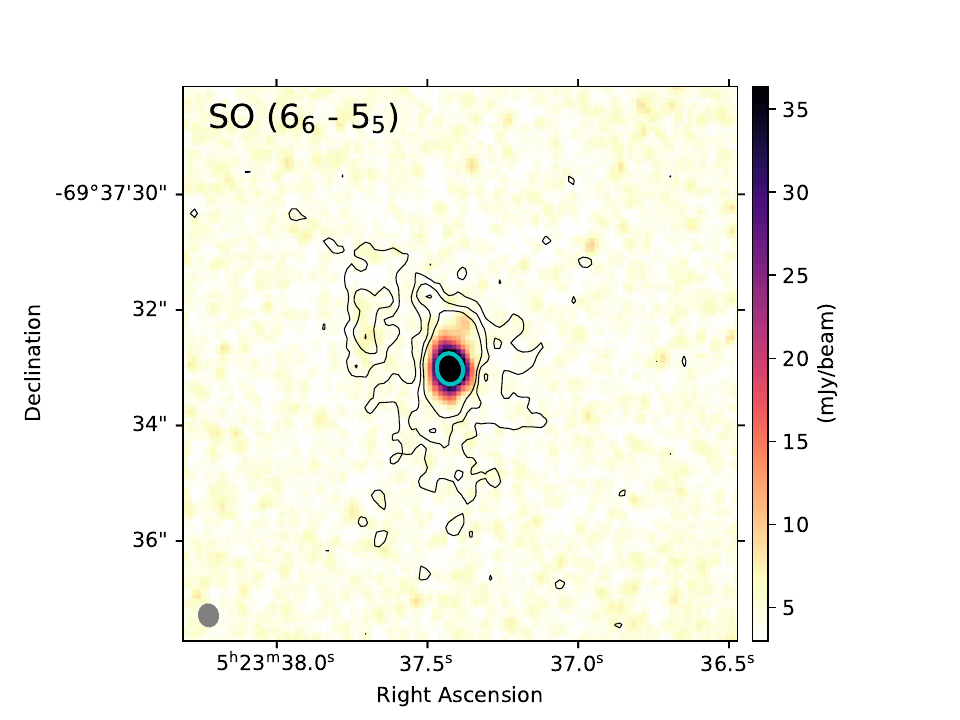} &
\includegraphics[trim={0cm 0cm 2.5cm 0.5cm}, scale=0.4]{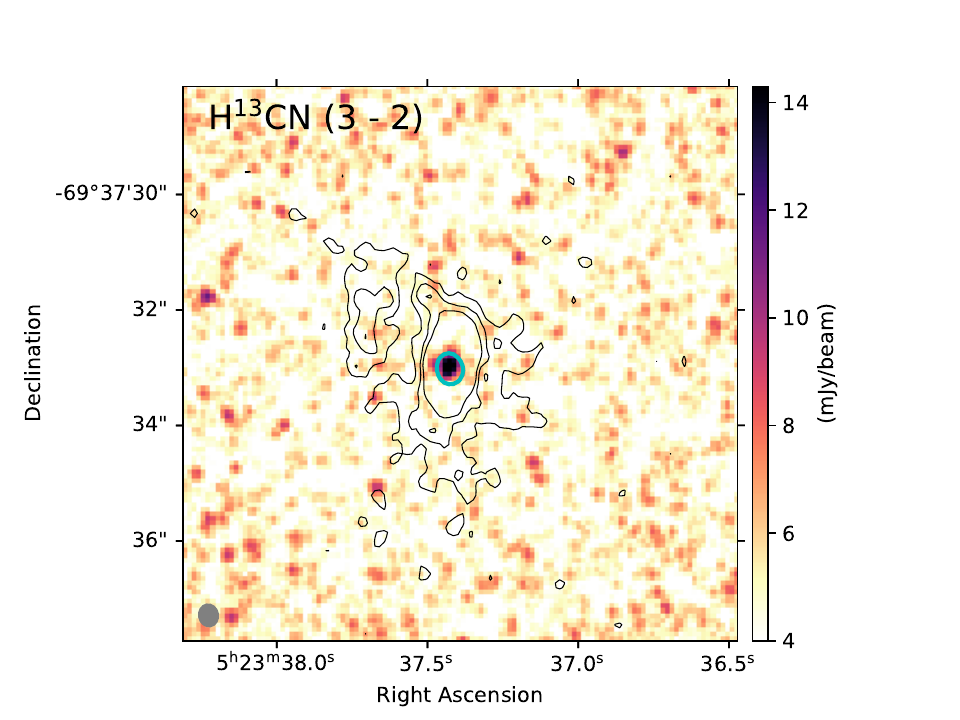} \\
\includegraphics[trim={0cm 0cm 2.5cm 0.5cm}, scale=0.4]{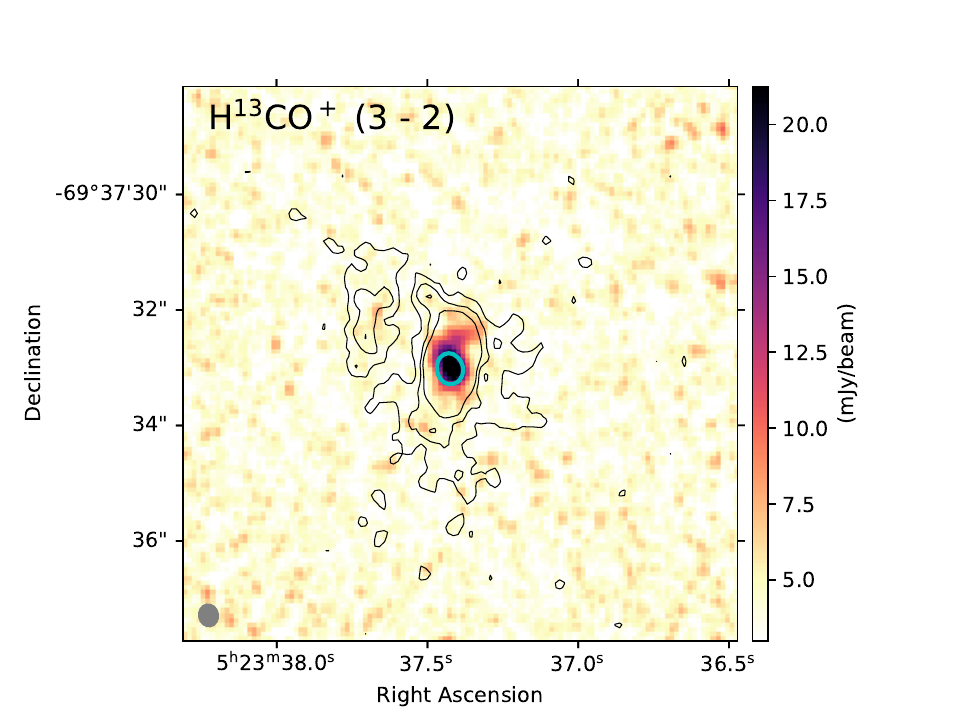} &
\includegraphics[trim={0cm 0cm 2.5cm 0.5cm}, scale=0.4]{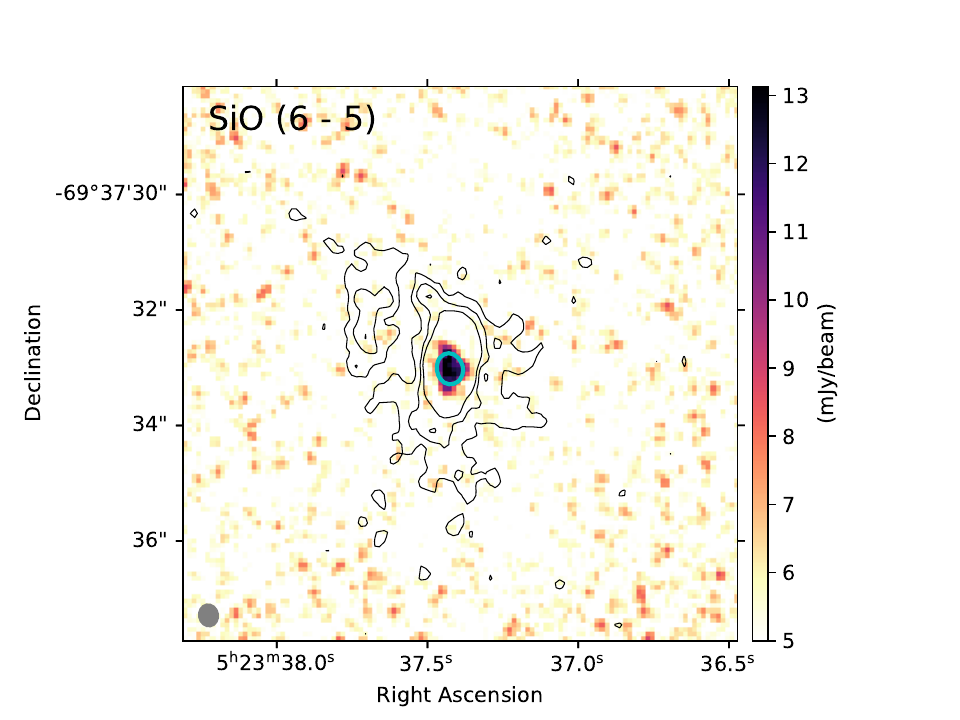} &
\includegraphics[trim={0cm 0cm 2.5cm 0.5cm}, scale=0.4]{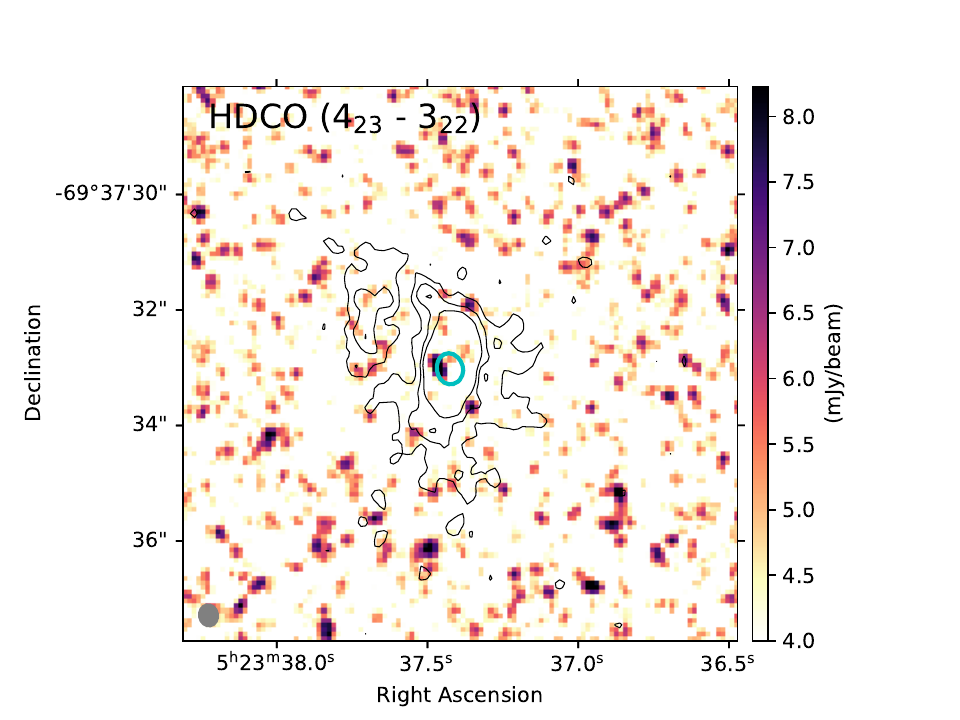} \\
\end{tabular}
\caption{Peak-intensity images of the molecular species detected toward hot core 14A. The maps are ordered based on the frequency of the transition from left to right, top to bottom. The 1.2 mm continuum contours with contour levels of three, six, and nine times the $\sigma_{rms}$ and the 50\% 1.2 mm continuum contour are overlaid in each image for reference. Top is north and right is west.}
\label{fig:PeakIntensityMaps}
\end{figure*}
%
%
\subsubsection{New hot core in the LMC}\label{sec:newhotcore}

The newly identified hot core 14A is located at the position of the YSO 0523333.40$-$693712.1. The color composite image from the 1.2~mm continuum (dust emission, in red), \ce{CS} (dense gas tracer, in green), and K-band (tracer of stellar population and possible \ce{H2} emission, in blue) in the lower insert panel of Fig.~\ref{fig:LMCRGB} shows the peak emission toward the position of the hot core. Figure~\ref{fig:PeakIntensityMaps} shows the integrated intensity maps of different molecular species toward the hot core 14A, with the 1.2~mm continuum emission shown in contours as reference. Apart from the bright emission toward the hot core position, the CS map also reveals prominent filamentary structures. The emission from \ce{CH3OH}, \ce{SO}, \ce{H^13CO^+}, and \ce{H2CS} show faint elongation in the north-south direction. Moreover, \ce{HNCO} and \ce{HDCO} show a very faint emission, offset from the hot core, and \ce{H2CS} emission also peaks offset to the hot core, where the \ce{HNCO} emission peaks.
%
%
\subsubsection{Comparison of hot cores common to other studies}\label{sec:comparisonothersurveys}

In addition to 14A, which is detected for the first time in this work, the other three hot cores (4A, 6A, and 6B) have previously been identified in other studies based on independent observational datasets. In the following, we compare our results with those in the literature for the temperatures and abundances of species detected toward all the cores, including \ce{CH3OH}, \ce{CH3CN}, \ce{HNCO}, \ce{H^13CO^+}, \ce{SO2}, \ce{SO}, and \ce{SiO}. 

\begin{figure*}
\begin{tabular}{c c}
\includegraphics[scale=0.88]{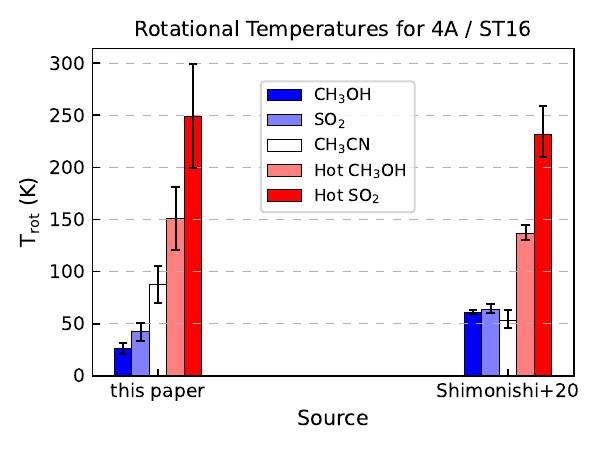} &
\includegraphics[scale=0.88]{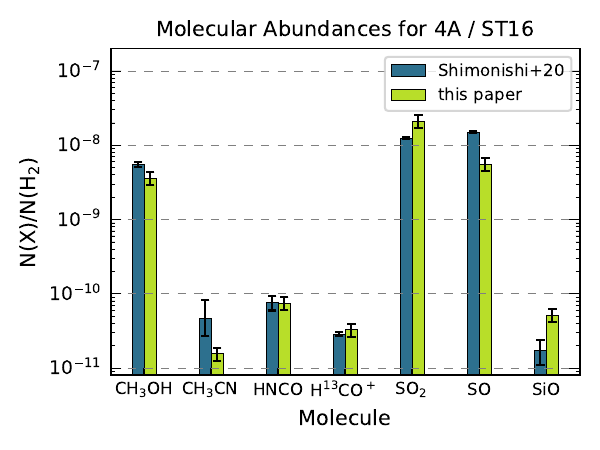}\\
\includegraphics[scale=0.88]{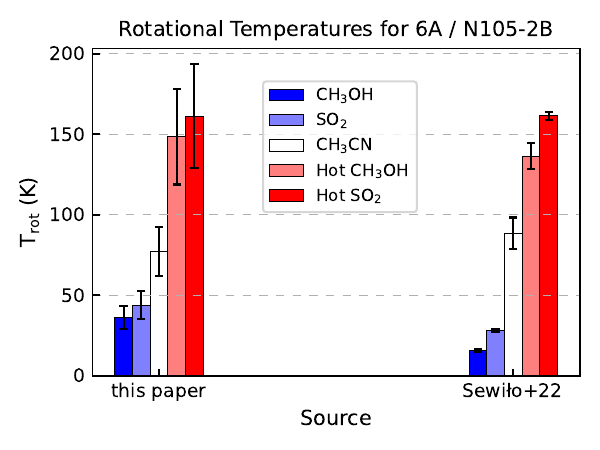} &
\includegraphics[scale=0.88]{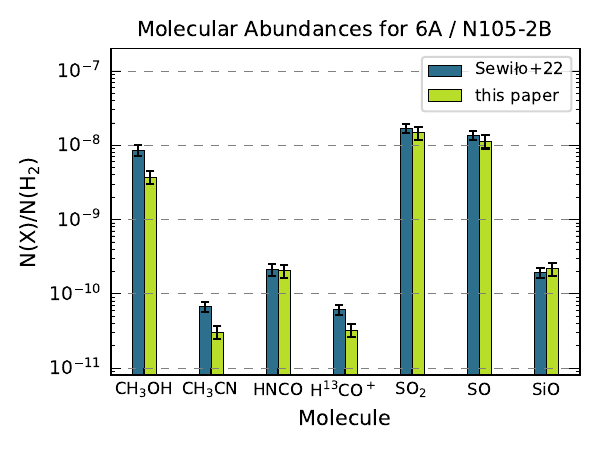}\\
\includegraphics[scale=0.88]{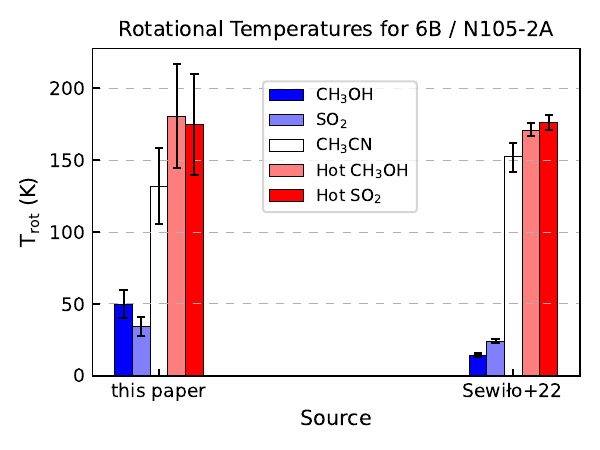} &
\includegraphics[scale=0.88]{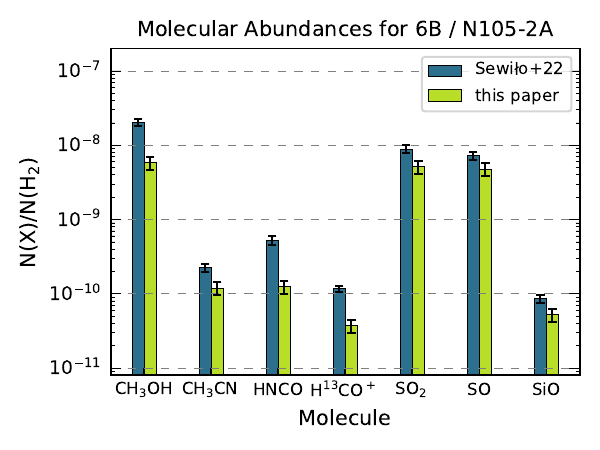}\\
\end{tabular}
\caption{Comparisons of the derived temperatures (left) and abundances (right) from the current study and the literature. The top panels show a comparison of 4A and ST16 \protect\citep{2020ApJ...891..164S}, the middle panels show 6A and N105-2B, and the bottom panels show 6B and N105-2A \protect\citep{2022ApJ...931..102S}. An error range of 40\% is depicted for the temperatures and column densities from the current paper.}
\label{fig:commonHotCores}
\end{figure*}

\cite{2020ApJ...891..164S} reported the detection of a hot molecular core based on ALMA Band 6 and 7 observations of the embedded high-mass infrared source ST16. Our study results in the detection of the same hot core that we name 4A following the adopted nomenclature. The hot components of \ce{CH3OH} and \ce{SO2} have the same temperatures in both studies considering uncertainties (see Fig.~\ref{fig:commonHotCores}, top-left panel). However, the cold component of both species has slightly lower temperatures in our study, and the derived temperature from \ce{CH3CN} is larger by a factor of 1.5. The top-right panel of Fig.~\ref{fig:commonHotCores} compares the abundances from the two studies, after accounting for the difference in the gas-to-dust mass ratio assumed in our study and that of \cite{2020ApJ...891..164S}.\footnote{\cite{2020ApJ...891..164S} assumed a dust-to-gas mass ratio of 0.0027 for ST16, which corresponds to a factor 1.15 lower compared to the value we use. In order to compare the abundances, we applied this factor to the molecular abundances reported by \cite{2020ApJ...891..164S}.} While the abundances for \ce{SO2}, and \ce{SiO} are slightly higher in the current work, \ce{CH3OH}, \ce{CH3CN} and \ce{SO} show lower abundances. The higher temperature and lower abundance for \ce{CH3CN} in the current work could be due to a possible degeneracy between these two parameters when fitting the spectral data, suggesting that these two sets of parameters can reproduce the observed lines. Further observations can help in better constraining both the temperature and density for this hot core.
%
%
\cite{2022ApJ...931..102S} found two hot cores, N105-2\,A and N105-2\,B, from Cycle~7 Band~6 observations of three fields in the N105 star-forming region. We note that the hot cores 6B and 6A in the current study correspond to the objects N105-2\,A and N105-2\,B, respectively. The temperature estimations are in good agreement for each pair of sources (see Fig.~\ref{fig:commonHotCores}, middle- and bottom-left panels), with larger variations happening for the cold components of \ce{CH3OH} and \ce{SO2} in 6B compared to N105-2\,A. The middle- and bottom-right panels of Fig.~\ref{fig:commonHotCores} compare the abundances from the two studies. \cite{2022ApJ...931..102S} assumed a gas-to-dust mass ratio of 316, very close to the value we use, 320. The abundances of nearly all the molecules in 6A compared to N105-2\,B fall in the same range except \ce{CH3OH} and \ce{CH3CN}, which are slightly higher in N105-2\,B. Similar results were obtained for 6B and N105-2\,A, with abundances falling in the same range. However \ce{CH3OH}, \ce{HNCO} and \ce{H^13CO^+} show abundances approximately two to five times lower in 6B (this work). The observed differences are likely due to the slightly different temperatures used to calculate the column densities in both studies. Finally, we searched for detections of other species, such as \ce{HDCO}, \ce{CH2CO}, and \ce{NH2CHO}, that had been previously detected by \cite{2022ApJ...931..102S}. We could not confirm their detection because the expected brightness of the species is less than the sensitivity of the current dataset. Only a tentative feature was seen for \ce{NH2CHO} toward source 6A (see Fig.~\ref{fig:spectraField06}).

Overall, the differences in the molecular abundances between the different works are within a factor of less than five. This factor is acceptable considering all possible errors and uncertainties in the analysis process. For example, different interferometric filtering during the observations as well as weather conditions or calibration effects, procedures followed to extract the spectra, or methodology used for fitting the lines may affect the derived temperatures and abundances by factors commonly in the range of two to ten. Therefore, we consider that our derived abundances are in agreement with previous studies, and we used them in the following discussion.
%
%
\subsection{Properties of hot cores in the LMC}\label{sec:hotcoresLMC}
%
%
\subsubsection{Comparison of the hot cores in the LMC}\label{sec:comparisonhotcores_LMC}

Apart from the four hot cores discussed in the previous section, three more have been reported in the LMC. \citet{2018ApJ...853L..19S} discovered the first two chemically rich hot cores in the N113 star-forming region, N113\,A1 and N113\,B3, with COMs with up to nine atoms being detected in extra-galactic hot cores for the first time. The abundance of the molecular species for these objects, after scaling for metallicity, falls on the lower end of the Galactic values. In contrast, the hot core ST11 reported by \citet{2016ApJ...827...72S} was detected based on the presence of hot \ce{SO2}, although no reliable COMs were detected. The absence of COMs in this hot core was explained by a hot dust temperature resulting in inefficient CO hydrogenation to form \ce{CH3OH} \citep{2016ApJ...827...72S}. 

\begin{figure*}
\centering
\includegraphics[scale=0.70]{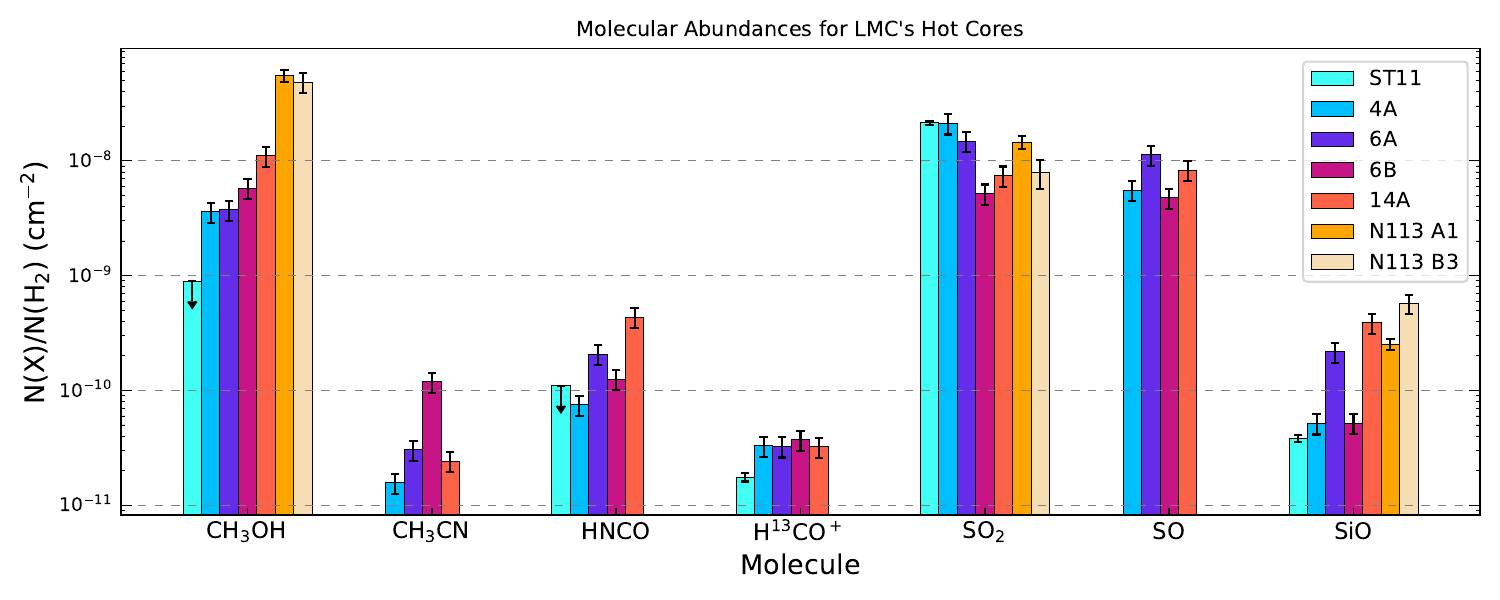}  
\caption{Abundances of different molecular species for all the known hot cores in the LMC after scaling to the gas-to-dust mass ratio. The LMC hot cores are ST11 \protect\citep{2016A&A...585A.107S}, 4A, 6A, 6B and 14A (this work), and N113~A1 and N113~B \protect\citep{2022ApJ...931..102S}. Black arrows show an upper limit estimation. An error range of 40\% is shown on the abundances from the current paper.}
\label{fig:Molecular_Abundances_hot_cores}
\end{figure*}

Figure~\ref{fig:Molecular_Abundances_hot_cores} shows a comparison of the molecular abundances in the seven hot molecular cores detected to date in the LMC after scaling to the gas-to-dust mass ratio used in the current paper (see Section~\ref{sec:comparisonothersurveys}). Each bar stands for one hot core, with error bars and upper limits also depicted in the plot. The non-detections of \ce{CH3CN}, \ce{HNCO}, \ce{H^13CO^+}, and \ce{SO} toward N113\,A and N113\,B are due to a lack of bright transitions of these species in the observational setup, whereas the upper limits for ST11 show the absence or underabundance of this species, which was in the observed frequency ranges but not detected. The methanol abundance varies between \num{3.6E-9} for 4A and 6A and \num{5.3E-8} for N113\,A1, where it is detected. The abundance of \ce{SO2} varies between \num{5.0E-9} for 6B and \num{2.1E-8} for 4A and ST11.

The abundances shown in Fig.~\ref{fig:Molecular_Abundances_hot_cores} reveal some trends for the hot cores in the LMC. We found an ascending pattern of an order of magnitude in the abundance of \ce{CH3OH} from the ST11 to N113 region, while the abundance of \ce{SO2} shows small variations among the hot cores. At a second level, there seems to exist a correlation between \ce{CH3OH} and \ce{SiO}. If \ce{SiO} is produced in shocks, likely associated with molecular outflows \citep[e.g.][]{1997A&A...321..293S, 2013A&A...557A..94S}, this correlation may suggest that the higher abundance of \ce{CH3OH} for some sources may have a shock origin. These correlations will be explored in a forthcoming work evaluating the spatial correlation of multiple molecular species.
%
%
%
\subsubsection{Comparison with chemical models}\label{sec:comparisonmodels}

The presence of hot \ce{SO2} in ST11 has been taken as a signature of a hot core, with the absence of COMs in this source strengthening the idea of a warm dust chemistry in the LMC \citep[see e.g.][]{2016ApJ...827...72S}. Although only one hot core of this kind has been detected in the LMC, we took advantage of the hot core models by \cite{2018ApJ...859...51A} to investigate how the warm dust hypothesis for low-metallicity environments compares to the current observations of hot cores in the LMC.

\cite{2018ApJ...859...51A} simulated the stages of star formation leading to hot cores and their lower-mass counterparts, hot corinos, under the conditions of the LMC (and the SMC). They used a gas-to-dust mass ratio of 175, a cosmic-ray ionization rate of \num{1.3e-17}~s$^{-1}$, and a radiation field strength similar to the Galactic values. The star formation was modeled in three stages. The first stage is an isothermal cloud collapse under free fall with fixed gas and dust temperatures of 10, 15, 20, and 25~K and a visual extinction $A_\mathrm{v}$ of 1.64~mag. The collapse is stopped when the density reaches 10$^7$~cm$^{-3}$, a process that lasts \num{9.3e5}~yr, during which $A_\mathrm{v}$ increases up to $\approx450$~mag. In the second stage, the density is fixed at \num{e7}~cm$^{-3}$, while the temperature increases linearly with time up to 100 or 200~K, resembling the temperature of hot cores. Two different times for the heating process are considered, reproducing expected conditions for high-mass (\num{5e4}~yr) and low-mass (\num{e6}~yr) star formation. Once 100 or 200~K is reached, the warm-up ceases, and the newly formed hot core is allowed to evolve until a final total time of \num{e7}~yr. This model does not include the effects of shocks in the chemical evolution.

\begin{figure*}
\centering
\begin{tabular}{c}
\includegraphics[scale=0.7]{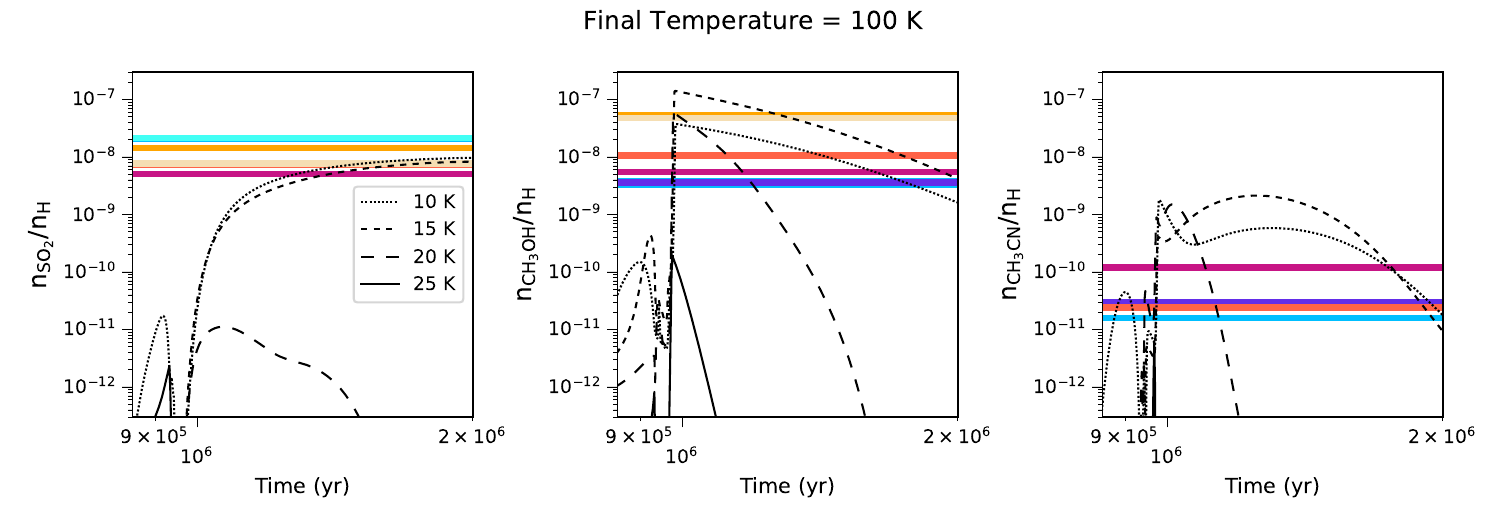} \\
\includegraphics[scale=0.7]{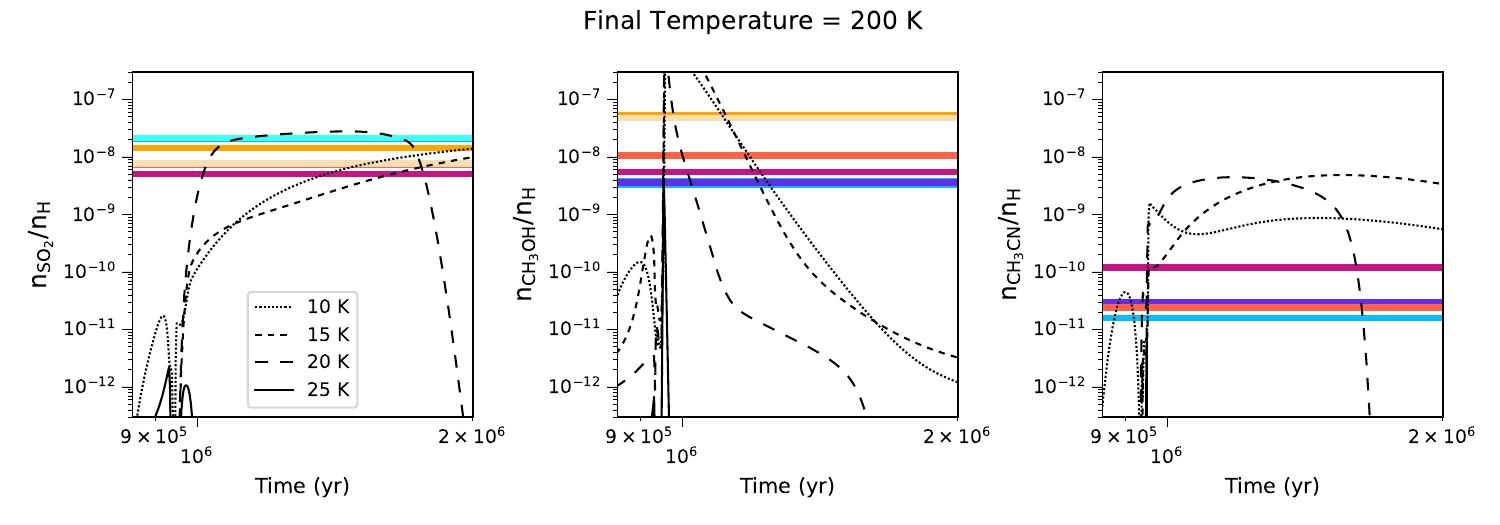}
\end{tabular}
\caption{Comparisons of abundances for \ce{SO2}, \ce{CH3OH}, and \ce{CH3CN} from observations of the seven hot cores detected in this galaxy and the LMC hot core models for the final hot core temperature of 100 K ({\it top}) and 200~K ({\it bottom}). The black curves show abundances over time for an initial dust temperature of 10 (dotted curve), 15 (dashed curve), 20 (larger dashed curve), and 25 K (the solid curve). The horizontal colored line shows the single abundance value for each hot core and follows the same color code as Fig.~\ref{fig:Molecular_Abundances_hot_cores} (see the text for more details).}
\label{fig:Molecular_Abundances_model_3mol}
\end{figure*}

Figure~\ref{fig:Molecular_Abundances_model_3mol} shows the time evolution of \ce{SO2}, \ce{CH3OH}, and \ce{CH3CN} abundances for the hot core model described above for the 100 and 200~K final temperatures (for other molecules see Figures~\ref{appfig:Molecular_Abundances_model1} and \ref{appfig:Molecular_Abundances_model2}). The observed abundances for the seven hot cores in the LMC are overlaid as horizontal lines in each panel following the same color scheme as in Fig.~\ref{fig:Molecular_Abundances_model_3mol}. Interestingly, only the models where the initial dust temperature is assumed to be between 10 and 20~K reproduce the observed molecular abundances. 
The chemical model with an initial highest dust temperature of 25~K results in molecular abundances that are much lower than the observed ones. Based on the comparison with the chemical models of \cite{2018ApJ...859...51A}, a high initial dust temperature does not seem to be a reason for a possible lack of COMs in the hot cores of the LMC since chemical models with initial dust temperatures in the range 10 to 20~K (similar to star-forming regions in the MW) can better reproduce the molecular abundances observed in the LMC.
%
%
%
\subsection{Hot cores in the LMC, SMC, and MW}\label{sec:comparisonMW}

A logical next step when studying the chemistry in low-metallicity environments is to compare the molecular abundances from the hot cores in the LMC, SMC, and outer Galaxy to those found in star-forming regions in the inner Galaxy. To achieve this goal, we collected a sample of Galactic hot cores studied in the literature and added them to the seven hot cores found in the LMC (see Section~\ref{sec:hotcoresLMC}), the two hot cores recently discovered in the SMC \citep{2023ApJ...946L..41S}, and the WB\,89$-$789~SMM1 hot core found in the outer regions of the Galaxy \citep{2021ApJ...922..206S}. In the case of WB\,89$-$789~SMM1, we used the abundances calculated for a source with a size of 0.1~pc (from Table~4 of \cite{2021ApJ...922..206S}). In the interest of matching spatial scales, we focused on single-dish observations of Galactic hot cores. In particular, we considered the Orion Hot Core \citep{1995ApJS...97..455S}, Sgr\,B2(N) \citep{2000ApJS..128..213N} and W3(\ce{H2O}) \citep{1997A&AS..124..205H} to represent the standard conditions and properties of hot cores in the MW.

\begin{figure*}
\centering
\includegraphics[scale=0.75]{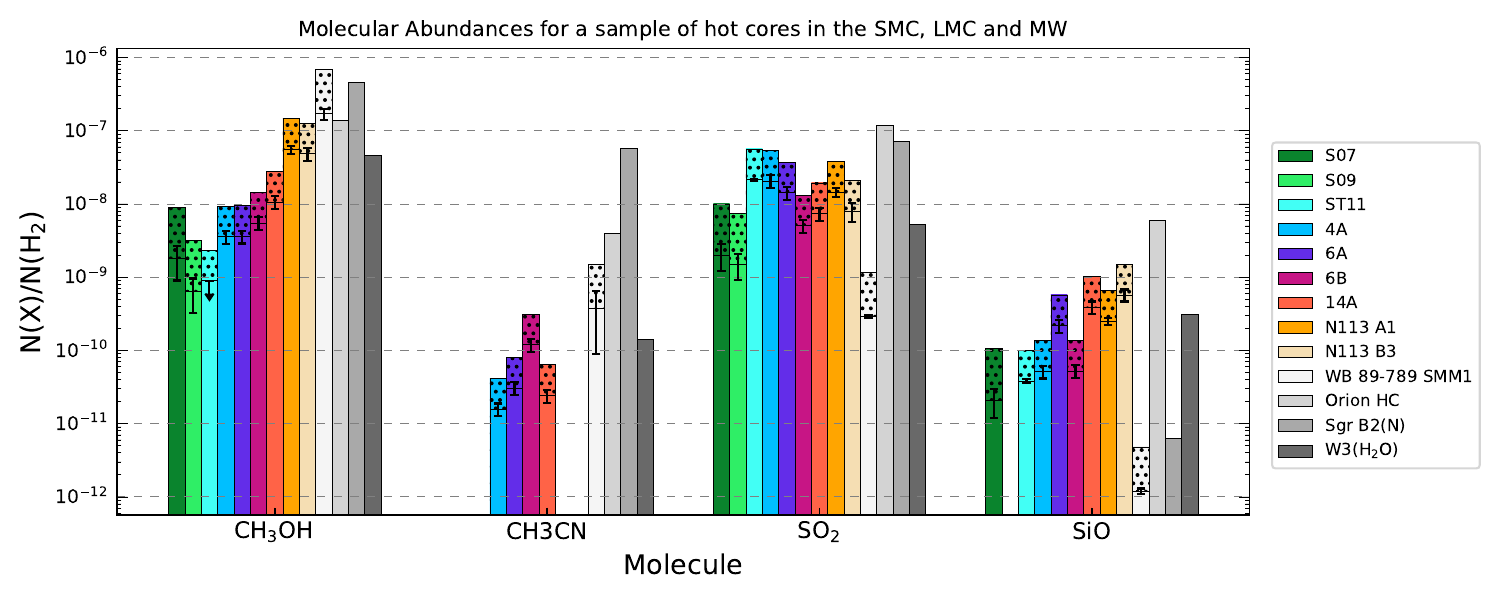}  
\caption{Abundances of \ce{CH3OH}, \ce{CH3CN}, \ce{SO2}, and \ce{SiO} for all known hot cores in the SMC and LMC and for a selection of Galactic hot cores. The SMC hot cores are included, S07 and S09 \protect\citep{2023ApJ...946L..41S}. The LMC hot cores are ST11 \protect\citep{2016A&A...585A.107S}, 4A, 6A, 6B and 14A (this work), and N113~A1 and N113~B \protect\citep{2022ApJ...931..102S}.
We selected four Galactic hot cores: an outer Galaxy hot core \protect\citep{2020ApJ...891..164S}, Orion Hot Core (HC) \protect\citep{1995ApJS...97..455S}, Sgr B2(N) \protect\citep{2000ApJS..128..213N}, and W3(\ce{H2O}) \protect\citep{1997A&AS..124..205H}.  The extent of each bar with a dotted pattern represents the abundances scaled to the Galactic metallicity.}
\label{fig:Molecular_Abundances_LMC_MW_SMC}
\end{figure*}

In Fig. ~\ref{fig:Molecular_Abundances_LMC_MW_SMC}, we present a comparison of the abundances of four molecular species corresponding to the selected list of hot cores. For the LMC hot cores, we first applied a correction factor to account for the different gas-to-dust mass ratio every study assumes (see Section~\ref{sec:comparisonothersurveys}). Then we applied a correction factor of 5, 2.6, and 4 to the abundances of the hot cores in the SMC, LMC, and extreme outer Galaxy\footnote{We adopted Z$_\mathrm{Gal}$/Z$_\mathrm{SMC}$ = 5, Z$_\mathrm{Gal}$/Z$_\mathrm{LMC}$ = 2.6, Z$_\mathrm{Gal}$/Z$_\mathrm{ExtOutGal}$ = 4. We note that metallicity decreases with the Galactocentric radius in the MW, resulting in varying metallicities across the outer Galaxy \citep[e.g.,][]{2021MNRAS.502..225A}. The metallicity value we present for the outer Galaxy corresponds to the position of WB 89-789.} to account for differences in metallicity. The original values are shown with single color bars in Fig.~\ref{fig:Molecular_Abundances_LMC_MW_SMC}, while the scaled values are depicted as an excess with dotted patterns.  

The methanol abundances vary by an order of magnitude among the hot cores in the LMC, except for ST11, for which only an upper limit is estimated. While the methanol abundances show an ascending trend from the low metallicity of the SMC to the LMC and to the inner regions of the MW, the only hot core in the outer region of the Galaxy, with a factor of four lower metallicity, shows the highest methanol abundance. We note that \ce{CH3CN} was not detected in all the hot cores, making the comparison less robust. However, one can see that the abundance of \ce{CH3CN} in the hot core in the outskirts of the Galaxy is higher than those in the LMC. In contrast to higher abundances of \ce{CH3OH} and \ce{CH3CN} in the outer Galaxy compared to the LMC and SMC, \ce{SO2} and \ce{SiO} are less abundant. This raises the question of whether the abundance of different molecular species is only a function of metallicity. We note that the outskirts of the Galaxy are more quiescent due to the low star-formation activity, while the star-forming regions of the LMC and the SMC are more active. While comparisons of the chemistry in different conditions still need better observational statistics, we acknowledge the necessity of chemical modeling to further investigate the effect of metallicity, strength of UV radiation, cosmic ionization rate, and shock chemistry on the observed molecular abundances.

\cite{2020ApJ...891..164S} classified the LMC hot cores as "organic-poor" and "organic-rich" based on the comparison of only four hot cores in the LMC. \cite{2022ApJ...931..102S} placed N105-2~A and N105-2~B in the organic-rich class, suggesting a larger sample to verify this classification. We also acknowledge the necessity of larger samples for a better understanding of hot core chemistry in the LMC. At the same time, we revise the classification of organic-poor and organic-rich based on the findings of this study. \cite{2020ApJ...891..164S} classified ST11 and ST16 as organic-poor and N113\,A1 and N113\,B3 as organic-rich hot cores to emphasize the detection of more complex compounds with comparable scaled abundances to the Galactic hot cores. Our survey study, however, demonstrates a different picture. All the hot cores in our sample show comparable chemical complexity. A slightly different chemistry may relate to the evolutionary stage of the hot core and surrounding gas. For example, we detected isotopologues of \ce{SO2}, \ce{^33SO2}, and \ce{^34SO2} in the hot core 4A but had no detection of deuterated molecules. In contrast, hot core 14A shows emission lines from \ce{HDCO} but no emission from \ce{SO2} isotopologues. The common association of deuterated species with early evolutionary stages \citep[e.g.][]{2011A&A...529L...7F, 2014A&A...569A..19T} and the connection of \ce{SO2} with the presence of shocks and ionized gas \citep[e.g.][]{2016ApJ...824...99M, 2021A&A...648A..66G} suggest that different evolutionary stages and physical conditions are likely relevant factors of the chemical content of the hot cores in the LMC. Within this picture from the hot cores in the LMC, the only organic-poor hot core that remains is ST11.

\begin{figure}
\centering
\begin{tabular}{cc}
\vspace{-0.4cm}
\hspace{-0.5cm}
\includegraphics[scale = 0.6]{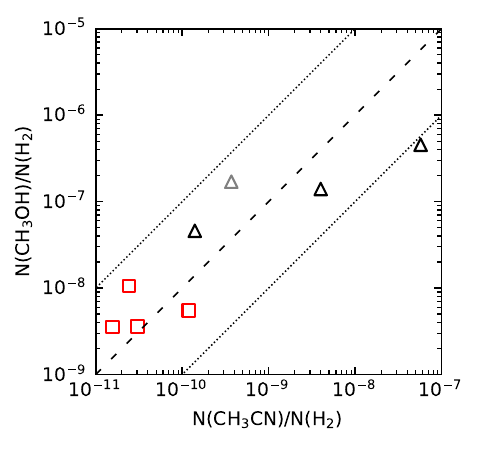} & \hspace{-0.75cm}
\includegraphics[scale = 0.6]{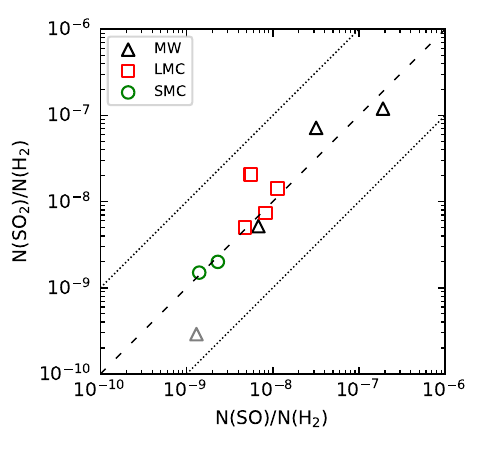} \\
\hspace{-0.5cm}
\vspace{-0.4cm}
\includegraphics[scale = 0.6]{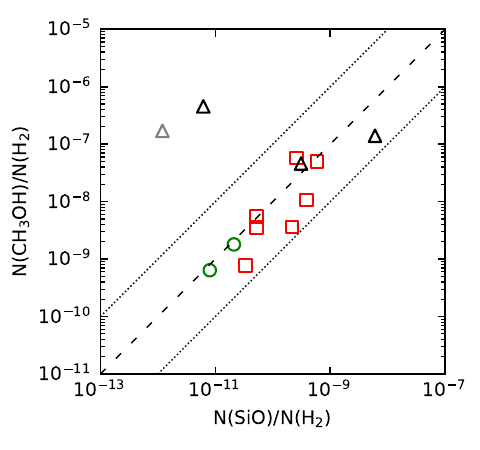} &
\hspace{-0.75cm}
\includegraphics[scale = 0.6]{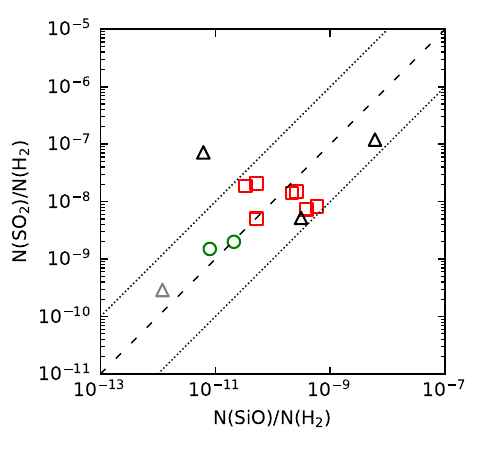} \\
\end{tabular}
\caption{Comparisons of the abundances of \ce{CH3OH} and \ce{CH3CN} and \ce{SO2} and \ce{SO} for the hot cores in the SMC (open green circles), LMC (red rectangles), and MW (black triangles for inner Galactic hot cores and gray triangle for outer Galactic hot core). Dashed lines show abundance relations of 1:1 for the top-right panel and 1:100 for the other panels. The dotted lines show a variation of a factor of ten above and below the relations described by the dashed lines.}
\label{fig:Abundance_HotCores_2by2}
\end{figure}

In Fig.~\ref{fig:Abundance_HotCores_2by2}, we present pair-wise comparisons of the abundances of \ce{CH3CN}, \ce{CH3OH}, \ce{SO2}, \ce{SO}, and \ce{SiO}. Based on these comparisons, we observed that the hot cores in the LMC appear to show lower abundances of \ce{CH3CN} relative to \ce{CH3OH} compared to hot cores in the MW but higher abundances of \ce{SO2} relative to \ce{SO}. Although this study should be extended to a larger sample of targets and evolutionary effects may play a role in the observed abundances, the first results favor the idea that hot cores in the LMC may be brighter in \ce{SO2} lines compared to other molecular species commonly found in hot cores. The comparison and correlation between \ce{CH3OH} and \ce{SO2} with \ce{SiO} suggest that shocks may play an important role in the observed abundances of these species.
%
%
%
\subsection{Location of hot cores in the LMC}\label{sec:hotcoresLocation}

\begin{figure}
\centering 
\includegraphics[scale = 0.33]{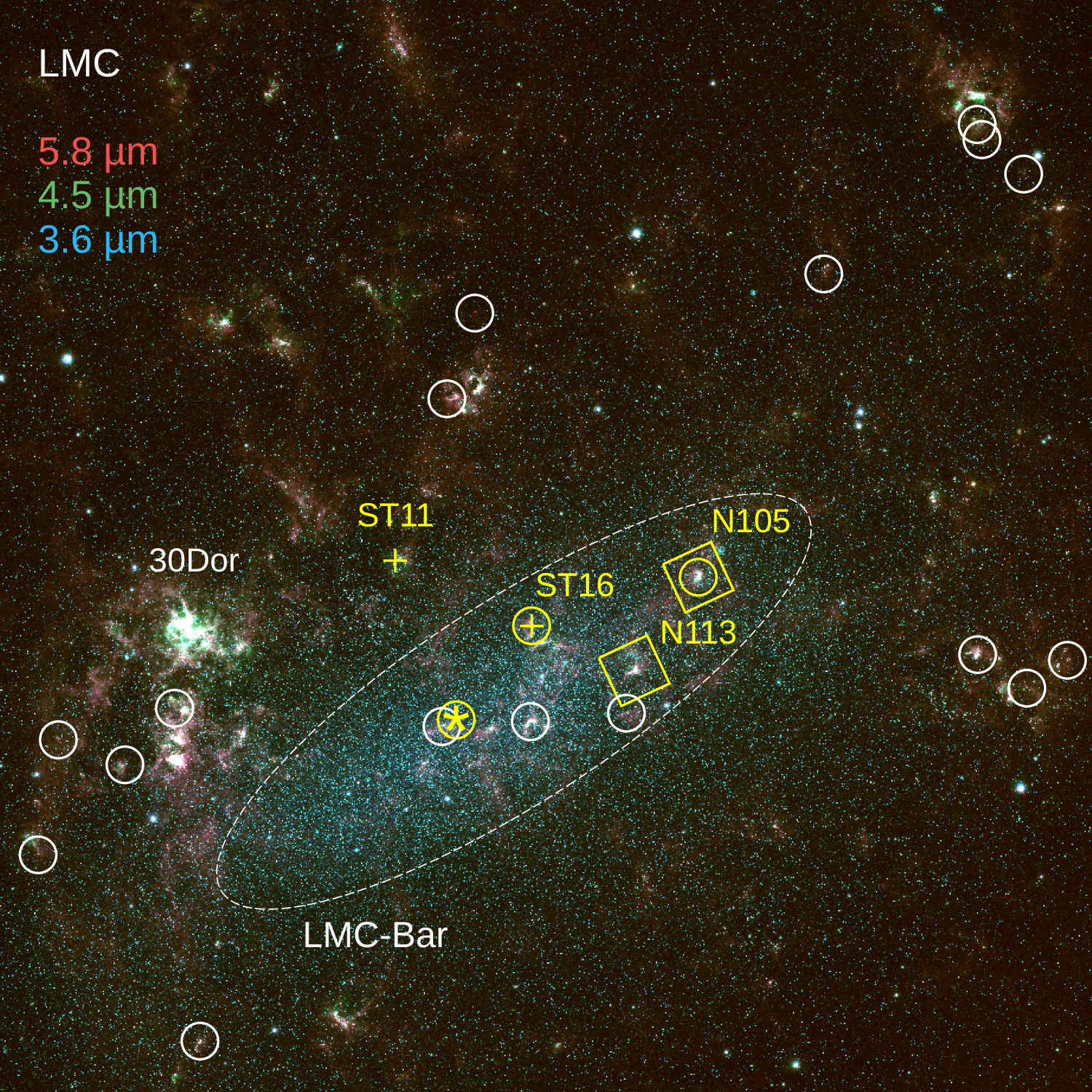} 
\vspace{0.2cm}
\caption{Three color composite image of the LMC combining 3.6 $\mu$m (blue), 4.5 $\mu$m (green), and 5.8 $\mu$m \protect\citep[red][]{2006AJ....132.2268M} images to highlight the position of the stellar bar. The dashed ellipse marks the bar region. The 20 ALMA fields from the current study are overlaid with white circles 15 times larger than the field of view of our ALMA observations. The fields accommodating hot cores are highlighted in yellow, including the YSO ST16 and N105 star-forming region. The field marked with a star shows the position of the new hot core detected in our survey. The location of the hot cores in N105 \protect\citep{2022ApJ...931..102S} and N113 \protect\citep{2018ApJ...853L..19S} are marked with yellow rectangles. The hot cores ST11 \protect\citep{2016ApJ...827...72S} and ST16 \protect\citep{2020ApJ...891..164S} are shown with yellow plus signs. The hot core ST11 with no COMs is the only hot core detected out of the LMC bar.}
\label{fig:HotCoreInTheLMCBar}
\end{figure}
%
In Fig.~\ref{fig:HotCoreInTheLMCBar}, we show the location of all the detected hot cores in the LMC. So far, all the hot cores associated with COMs are located in the central region of the galaxy, which is known as the LMC bar. This region has a mildly higher ($\sim$ 2\%) metallicity compared to the surrounding regions, as suggested by narrow-band near-IR observations \citep{2021MNRAS.507.4752C}. Inspection of the atomic\citep{1998ApJ...503..674K, 2000ASPC..221...83S}, molecular gas \citep[e.g.][]{1999IAUS..190...61F}, and dust maps \citep{2013AJ....146...62M} of the LMC revealed that the dense cold gas and dust are located within a zigzag shape in the bar region and surrounded by bubbles of dispersed dust and gas \citep[known as \hi{} holes;][]{2003MNRAS.339...87S}. This implies that expanding material from different sides results in compressed material into this zigzag region. Based on this observational evidence, we speculate that star formation in the bar must have been triggered at some point when the interstellar medium in the bar was enriched and the metallicity was increased relative to other areas of the LMC. This might explain the preferential detection of hot cores with complex chemistry only toward the LMC bar.

For this scenario to be plausible, one needs to consider if the ignition of the new generation of stars that resulted in the currently detected hot cores took place after the chemical enrichment of the interstellar medium of the LMC bar. Different works in the literature explore the origin of the current distribution of gas and dust in the bar region as well as the existence of possible events in the history of the LMC that may have caused an increase of metallicity in the bar and the triggering of a new generation of stars. In the following, we summarize the main findings and connect them to the presence of hot cores in the LMC bar.

Hydrodynamic simulations \citep{1986A&A...170..107T} infer the creation of \hi{} holes and shells as a result of the interaction between high-velocity clouds and the galactic disk. At the same time, the study of the morphology and kinematics of stellar populations and \hi\ gas in the LMC, SMC, Magellanic Bridge (MB), and the Leading Arm (LA) suggests the interaction between the LMC and SMC \citep[e.g.][and references therein]{2019ApJ...874...78Z}. Furthermore, numerical simulations of the evolution of the Magellanic Clouds favor the sporadic replenishment of low-metallicity gas from the SMC to the LMC's disk over the last 2~Gyr as a result of their tidal interaction \citep[known as Magellanic Squall;][]{2007MNRAS.381L..16B}. This stripped high-velocity gas can be responsible for some of the \hi{} holes observable in the LMC. \cite{2007MNRAS.381L..16B} argues that the triggering of the star formation could be a direct impact of clouds of the SMC interacting with those of the LMC. In addition, numerical models of \cite{2012MNRAS.421.2109B} investigate the history of interaction between the SMC, LMC, and MW. \cite{2012MNRAS.421.2109B} argue that the off-center warped stellar bar of the LMC and its one-armed spiral can be explained by a recent direct collision with the SMC. Given the uncertainties of the model parameters, the exact time of the collision and its impact on star formation cannot be strictly defined. However, a time range from 100 to 300~Myr seems a reasonable approximation \citep{2012MNRAS.421.2109B}. Analysis of the proper motions of stars from Gaia data supports the direct collision and suggests that this event was not a grazing interaction but that the SMC went directly through the LMC, allowing the interstellar medium of both galaxies to interact \citep{2019ApJ...874...78Z}. Furthermore, \cite{2021MNRAS.507.4752C} presumed that the mixing of the metallicity of the two galaxies, caused by tidal interaction, could be a reason for the asymmetry metallicity distribution observed in the LMC. With all this, and given the timescale of formation of hot molecular cores ($\sim$~\num{e5}~yr, \citealt{2001ApJ...550L..81W, 2005ApJ...624..827F, 2009ARA&A..47..427H}), the detected hot cores in the LMC bar region have most likely been formed in the inhomogeneous and slightly metallicity-richer interstellar gas that emerged after the severe interactions of the LMC and the SMC.
%
%
\section{Conclusions}\label{sec:conclusions}

We have observed 20 star-forming sites distributed throughout the LMC with ALMA in four frequency ranges between 241 and 260~GHz and at an angular resolution of  $\approx0\farcs35$ (corresponding to 0.08~pc). We detected a total of 65 continuum cores whose physical properties (e.g., mass, density, size) are presented in Paper~II. In this work, we have studied the chemical content of these compact cores. The main findings are summarized as follows:

\begin{enumerate}

\item The extracted spectra at the location of the cores show different levels of chemical complexity in the cores. There are cores with no molecular line emission and cores with single-line detection of dense or warm gas tracers, such as \ce{CS}, \ce{SO}, or \ce{H^13CO^+}. These include 28\% of all the cores. In addition, other cores contain \ce{CH3OH}. These cores can be put into two distinct groups: cores with emission lines from simple molecules, such as \ce{SO2},\ce{H2CS}, \ce{H^13CN}, \ce{HC^15N}, and \ce{SiO}, and sources with detection of simple species and COMs, such as \ce{CH3CN}, or tentative detection of \ce{NH2CHO}, \ce{CH3OCHO}, and \ce{CH3OCH3}.

\item The detection of multiple transitions from a single molecule allowed for the determination of the temperature. The \ce{CH3OH} molecule, detected in 72\% of the cores, is our main temperature tracer in this study. We estimated the temperature with both LTE and non-LTE assumptions and compared the two temperatures with the \ce{SO2} temperature, where it was reliably detected. Our temperature determination was further confirmed by estimating the \ce{CH3CN} temperature where it was detected.

\item Molecular line width is a tracer of the gas kinematics. We investigated the line width of different molecular species (fit with XCLASS), and \ce{CH3OH} showed the lowest line width value in 51\% of the cores. However, only a small sample of cores showed an average line width above 5 km/s.

\item Comparing the abundance of different molecular species in the 65 cores showed that S-bearing molecules are in general more present and abundant than N- and O-bearing species. Therefore, the detection of N- and O-bearing species, especially the COMs, needs higher sensitivity in future observations. 

\item Four hot molecular cores were detected in this study, confirmed by their size ($\sim$ 0.1pc), chemical richness, temperatures above 100 K, and high abundances of \ce{CH3OH} and \ce{SO2}. 
\end{enumerate}

The four hot cores in our sample reveal the detection of one new hot core, bringing the number of hot cores in the LMC to seven. \cite{2023ApJ...946L..41S} suggest \ce{SO2} as a tracer of hot cores in a low-metallicity environment. Our survey study supports this suggestion by detecting strong lines of \ce{SO2} and its isotopologues only in the hot cores. We note that (six out of the seven) hot cores with detections of methanol or other COMs are preferentially located in the LMC bar. The metallicity-scaled abundances of different molecular species in these hot cores are comparable with those in the MW. \cite{2020ApJ...891..164S} have suggested the initial high dust temperature to be responsible for the lower abundances of COMs in the low metallicity of the LMC. However, comparisons of our observed abundances with the LMC hot core chemical models of \cite{2018ApJ...859...51A} do not support high initial dust temperatures. 
%
%
\begin{acknowledgements}
R.HG.\ acknowledges the German Deutsche Forschungsgemeinschaft (DFG, German Research Foundation) for supporting this research through project SCHI 561/1-1. A.S.-M.\ acknowledges support from the RyC2021-032892-I grant funded by MCIN/AEI/10.13039/501100011033 and by the European Union `Next GenerationEU’/PRTR, as well as the program Unidad de Excelencia María de Maeztu CEX2020-001058-M, and support from the PID2020-117710GB-I00 (MCI-AEI-FEDER, UE). The material is based upon work supported by NASA under award number 80GSFC21M0002 (M.S.). G.A.F. gratefully acknowledges the Collaborative Research Center 1601 (SFB 1601 sub-project A1) funded by the DFG – 500700252. The authors acknowledge the comments provided by the anonymous referee which helped to improve the manuscript.
\end{acknowledgements}
%
%
\bibliographystyle{aa} 
\bibliography{aabib} 
%
%

\clearpage
\onecolumn

\begin{appendix}
%
%
\section{Spectra and XCLASS fits for 1.2 mm continuum cores}\label{app:spectra}
\begin{figure*}
    \centering
    \subfloat[][]{\includegraphics[width=0.95\textwidth]{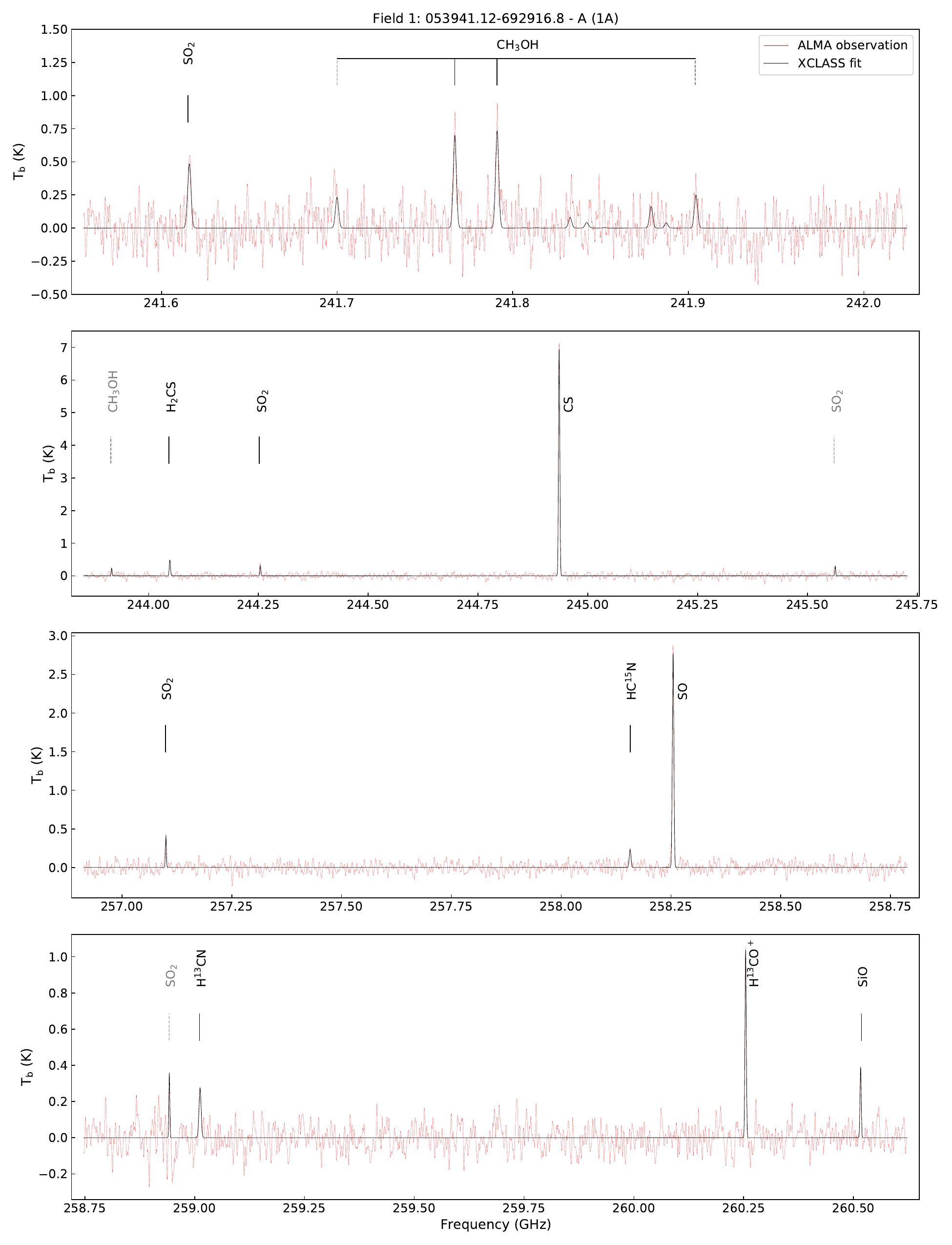}}
    \caption{ALMA Band 6 observed spectra toward source 1A overlaid with the best-fit XCLASS model. Molecular transitions with definite detections ($\mathrm{S/N} > 5$) are represented by solid black lines, while those with tentative detections ($3 < \mathrm{S/N} < 5$) are denoted by dotted gray lines. Dotted red lines indicate transitions of molecules for which abundance upper limits are provided. (See Section~\ref{sec:spectra} for more details.)}
    \label{fig:spectraField01}
\end{figure*}

\begin{figure*}
    \centering
    \subfloat[][]{\includegraphics[width=0.95\textwidth]{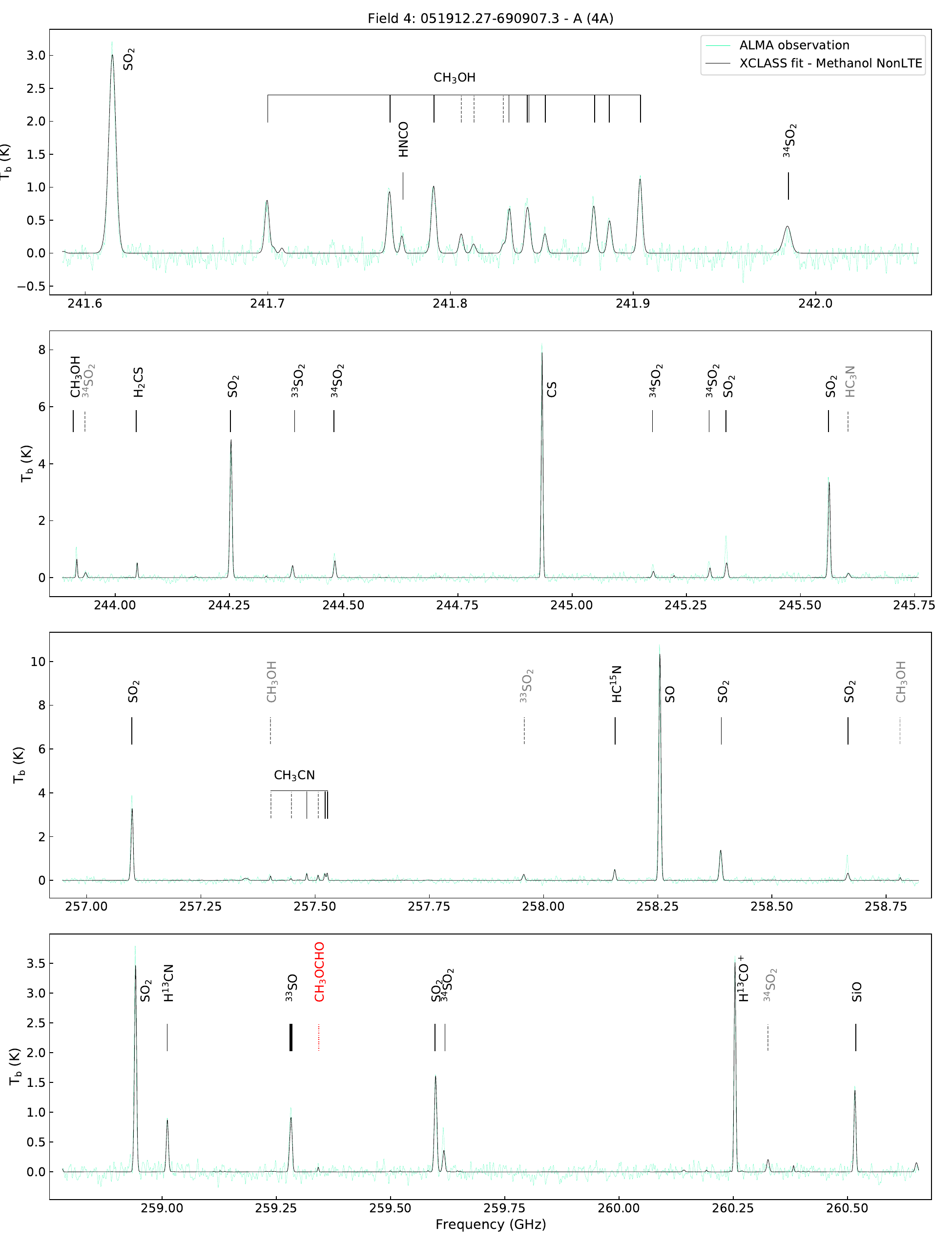}}
    \caption{Same as Fig.~\ref{fig:spectraField01} but for source 4A.}
    \label{fig:spectraField04}
\end{figure*}

\begin{figure*}
    \centering
    \subfloat[][]{\includegraphics[width=0.95\textwidth]{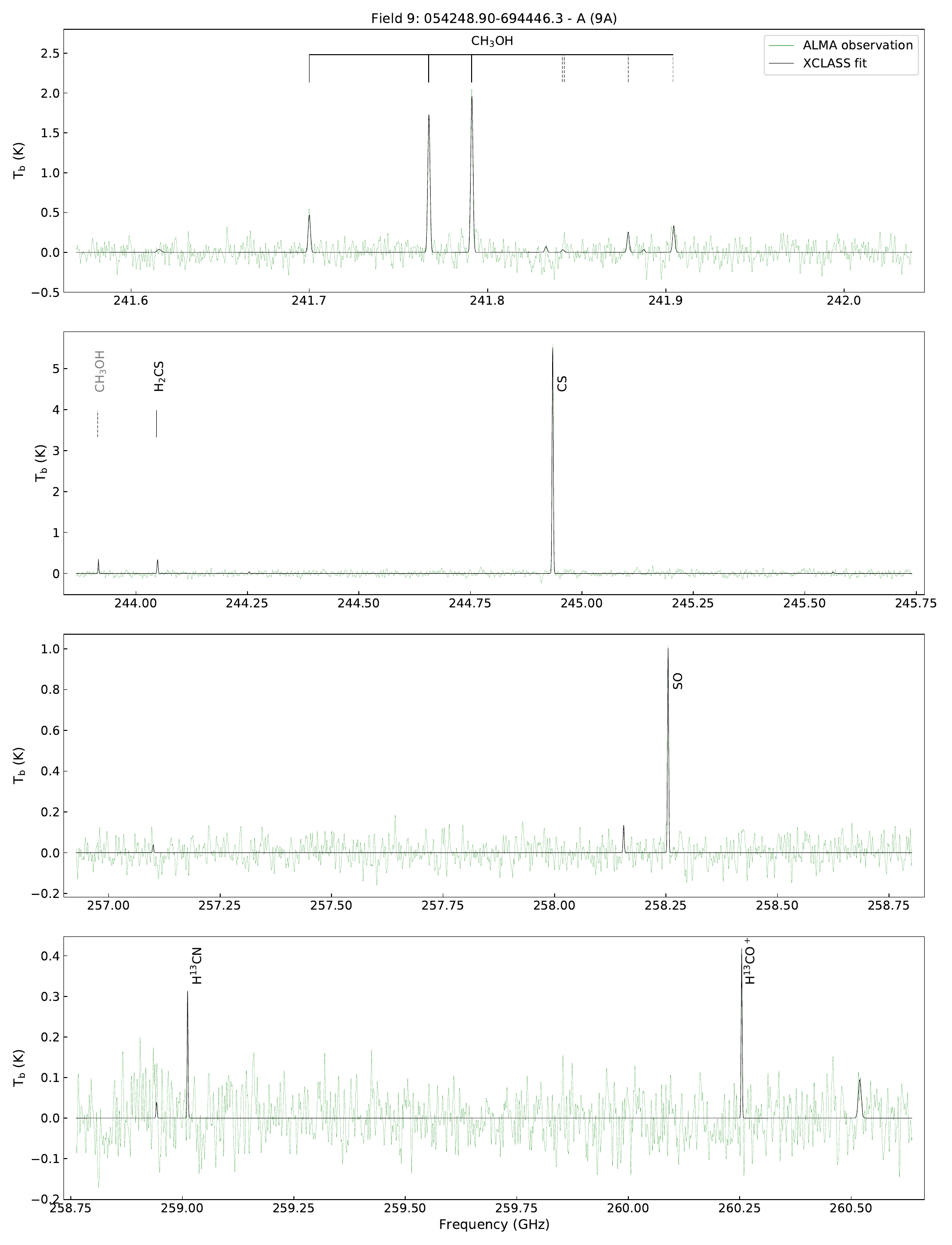}}
    \caption{Same as Fig.~\ref{fig:spectraField01} but for source 9A.}
    \label{fig:spectraField09}
\end{figure*}

\begin{figure*}
    \centering
    \subfloat[][]{\includegraphics[width=0.95\textwidth]{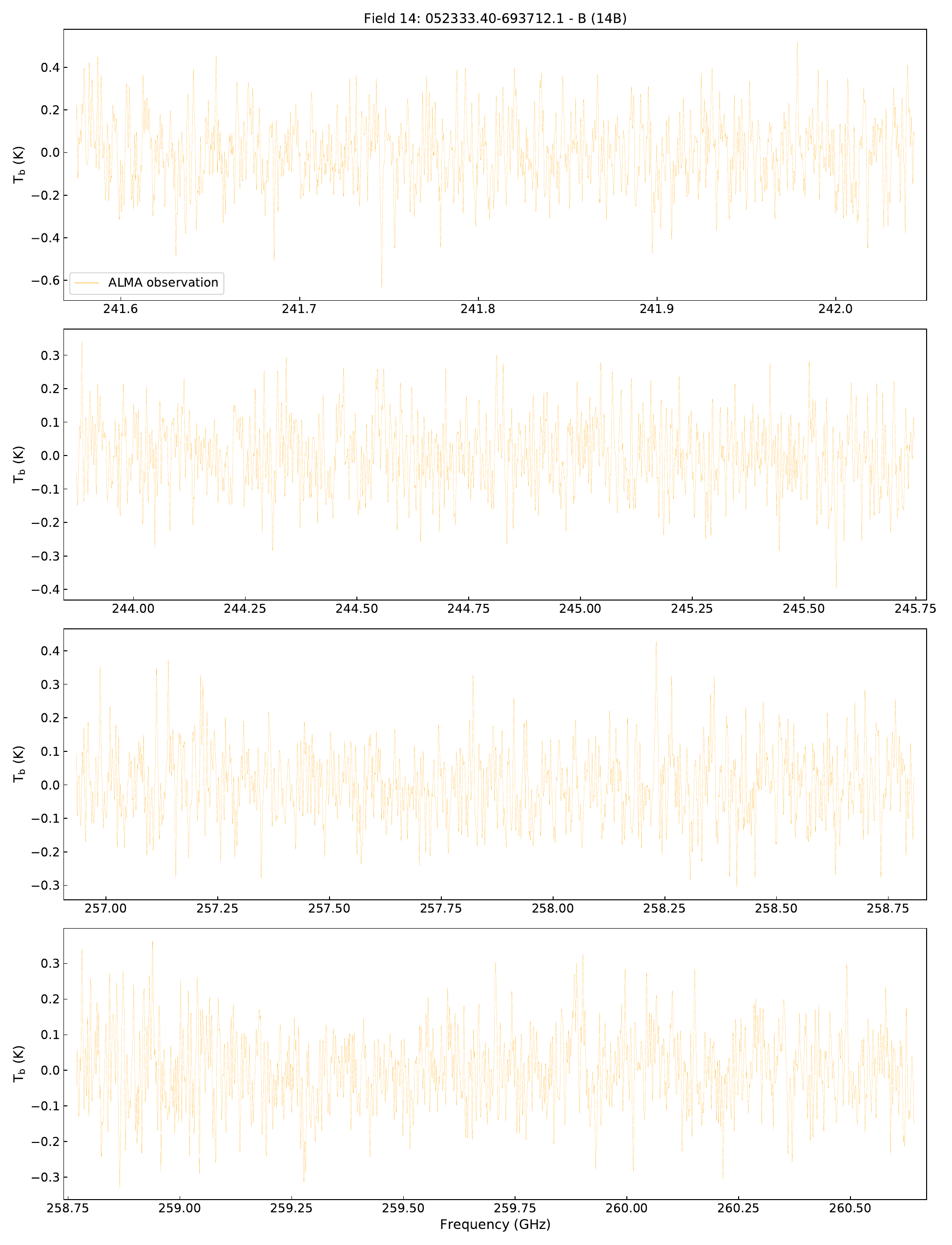}}
    \caption{ALMA Band 6 observed spectra toward source 14B  with no emission line detections. (See Section~\ref{sec:spectra} for more details.)}
    \label{fig:spectraField14}
\end{figure*}

\begin{figure*}
    \centering
    \subfloat[][]{\includegraphics[width=0.95\textwidth]{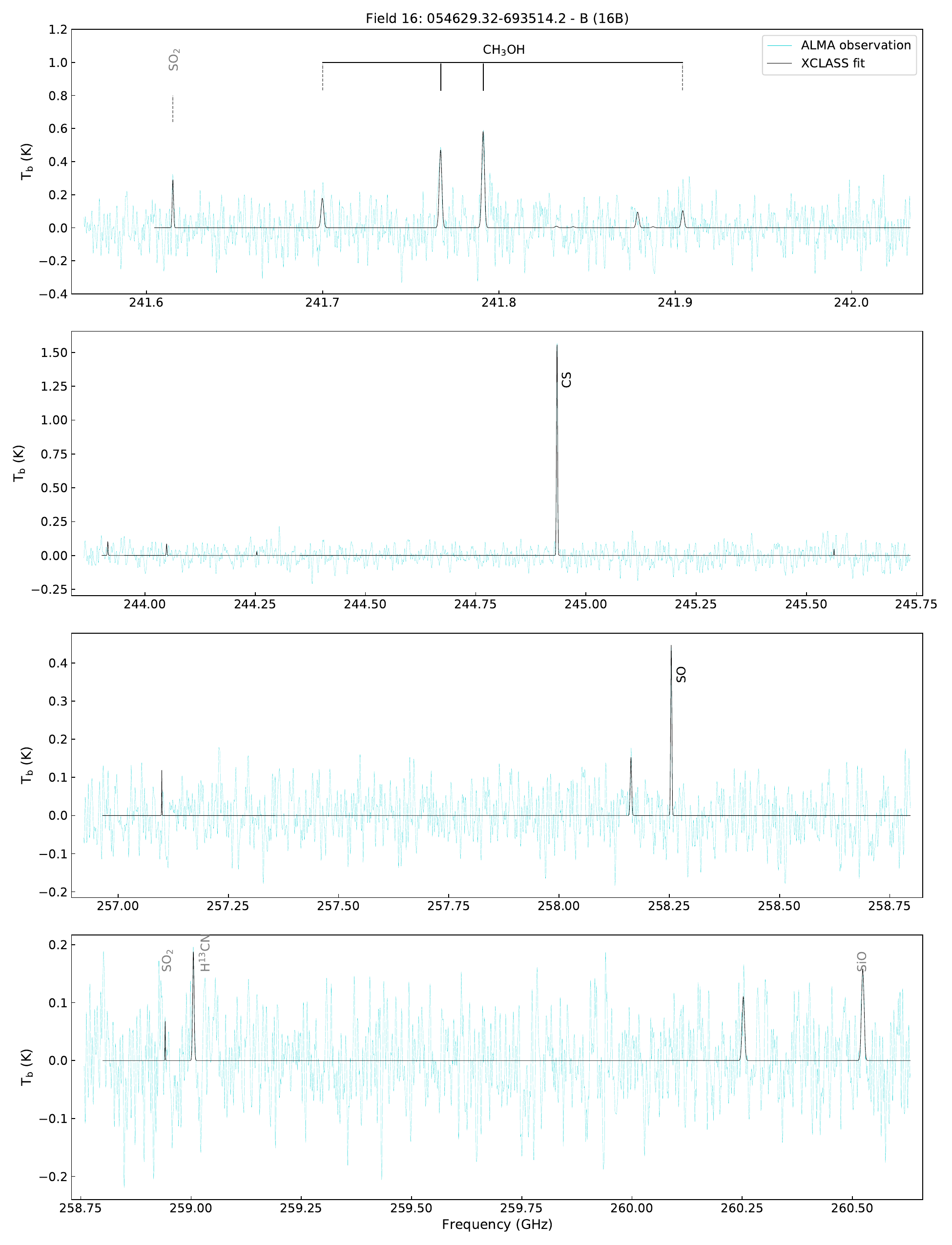}}
    \caption{Same as Fig.~\ref{fig:spectraField01} but for source 16B.}
    \label{fig:spectraField16}
\end{figure*}

\begin{figure*}
    \centering
    \subfloat[][]{\includegraphics[width=0.95\textwidth]{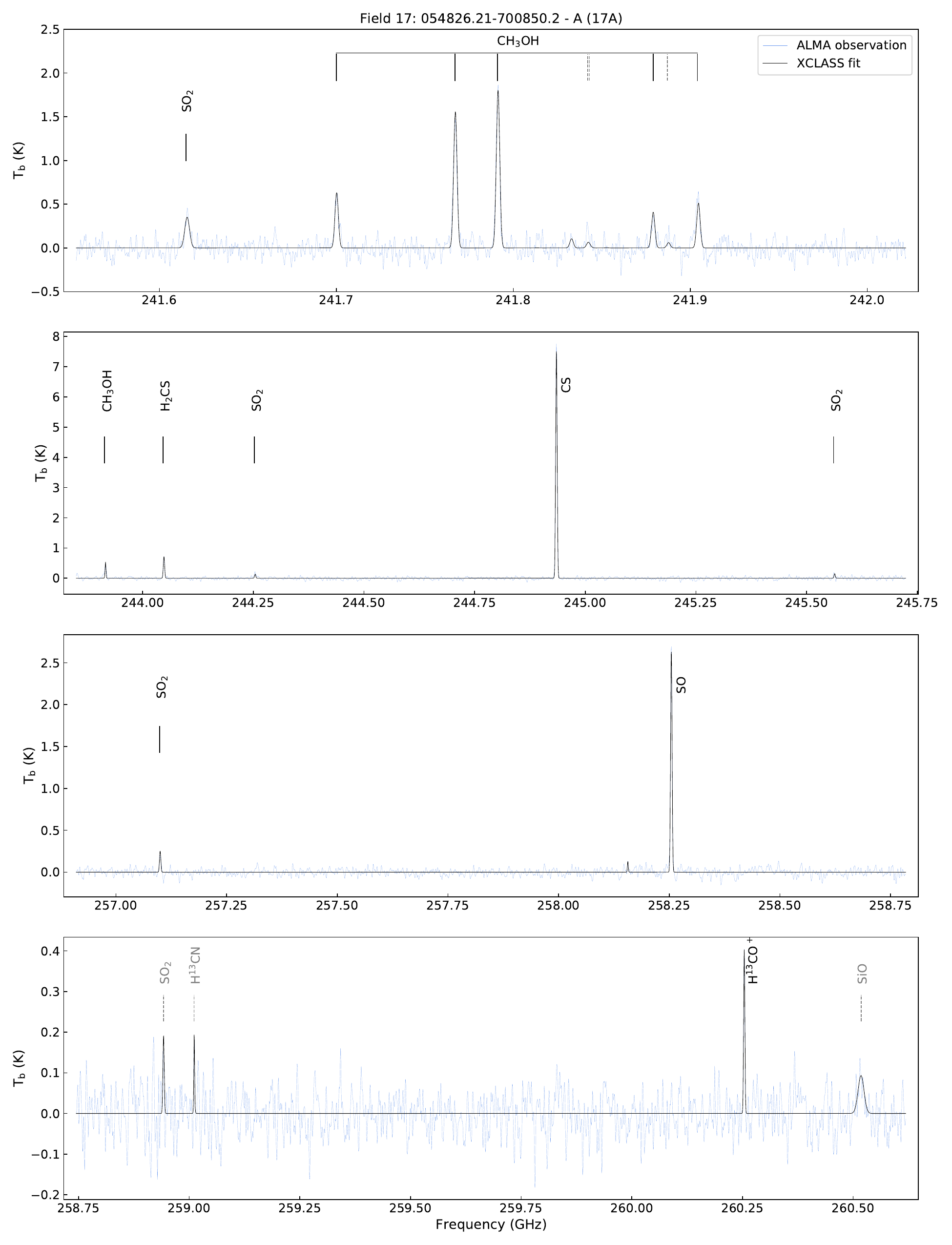}}
    \caption{Same as Fig.~\ref{fig:spectraField01} but for source 17A.}
    \label{fig:spectraField17}
\end{figure*}

\begin{figure*}
    \centering
    \subfloat[][]{\includegraphics[width=0.95\textwidth]{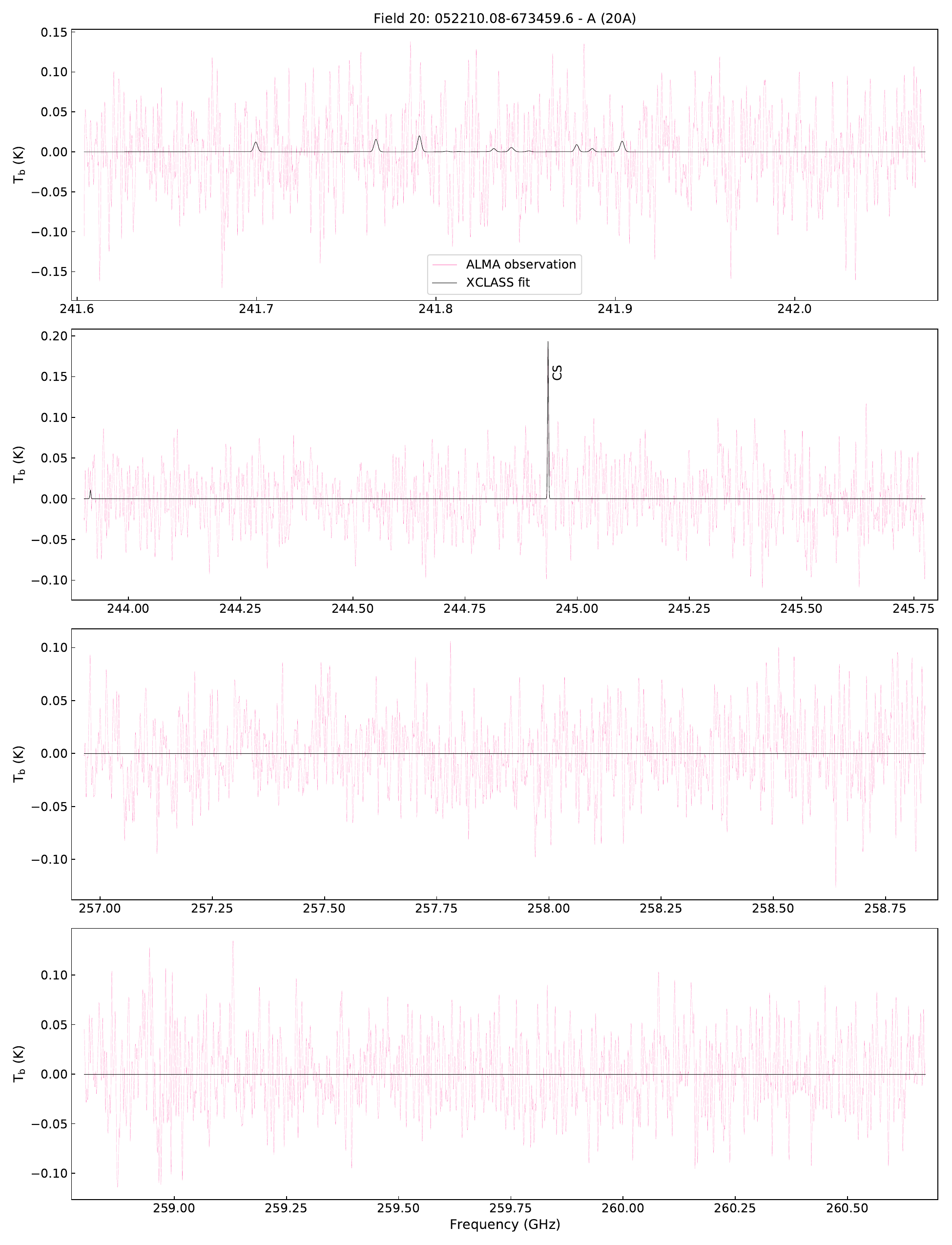}}
    \caption{Same as Fig.~\ref{fig:spectraField01} but for source 20A.}
    \label{fig:spectraField20}
\end{figure*}

Figures~\ref{fig:spectraField01} to \ref{fig:spectraField20} show the ALMA Band 6 spectra observed toward the all cores with identified molecular lines (except the core 14B) in the 20 fields observed throughout the LMC. The observed spectra with line detections were fit using the XCLASS software (see Section~\ref{sec:spectra}), and the results are shown as a black solid line on top of the observed spectra. The molecular species detected in the spectra are marked in black, while the tentative detections are marked in gray.

\clearpage
%
%

%
%
\begin{figure*}
    \centering
    \subfloat[][]{\includegraphics[width=0.95\textwidth]{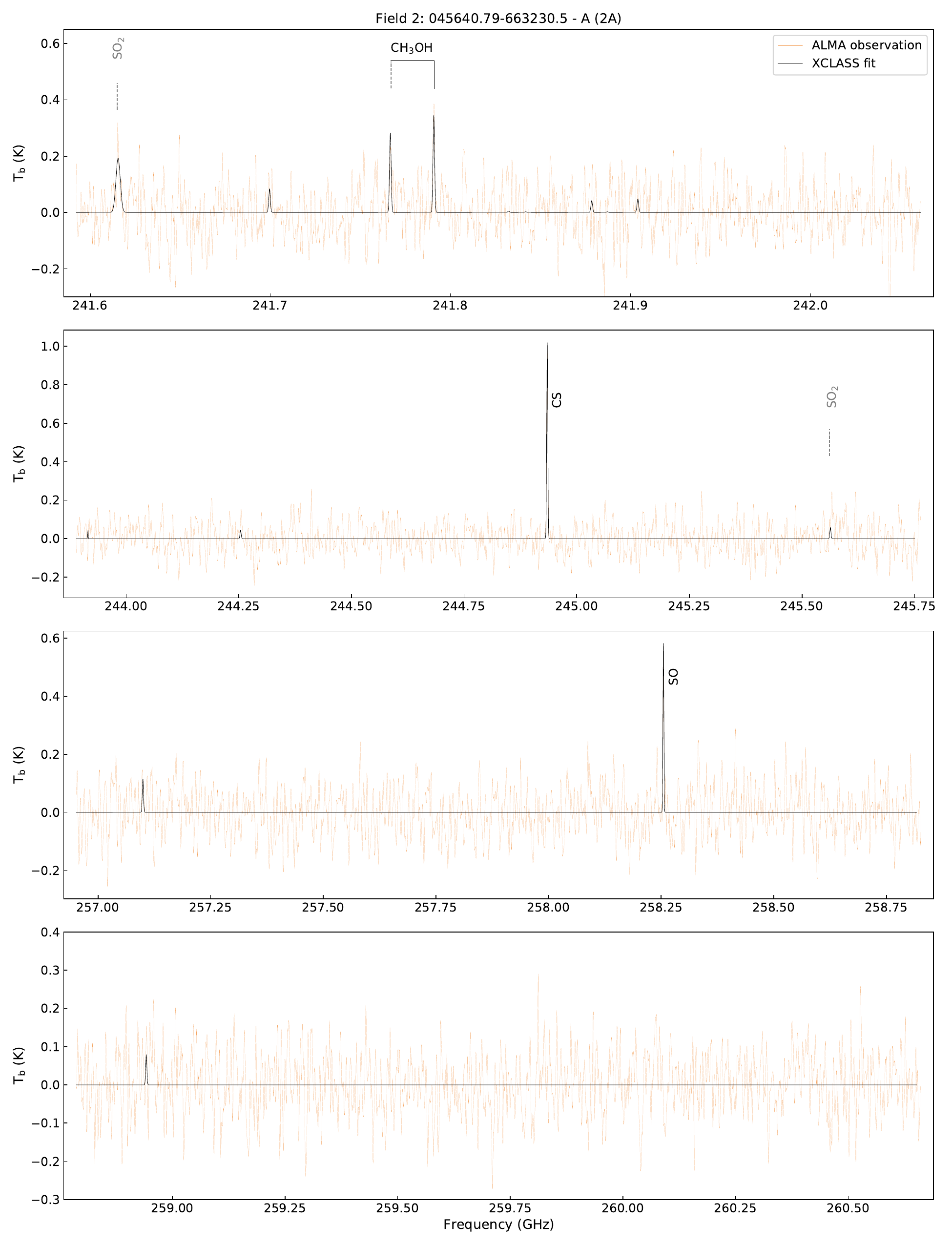}}
    \caption{Same as Fig.~\ref{fig:spectraField01} but for source 2A.}
\end{figure*}

\begin{figure*}
    \ContinuedFloat
    \centering
    \subfloat[][]{\includegraphics[width=0.95\textwidth]{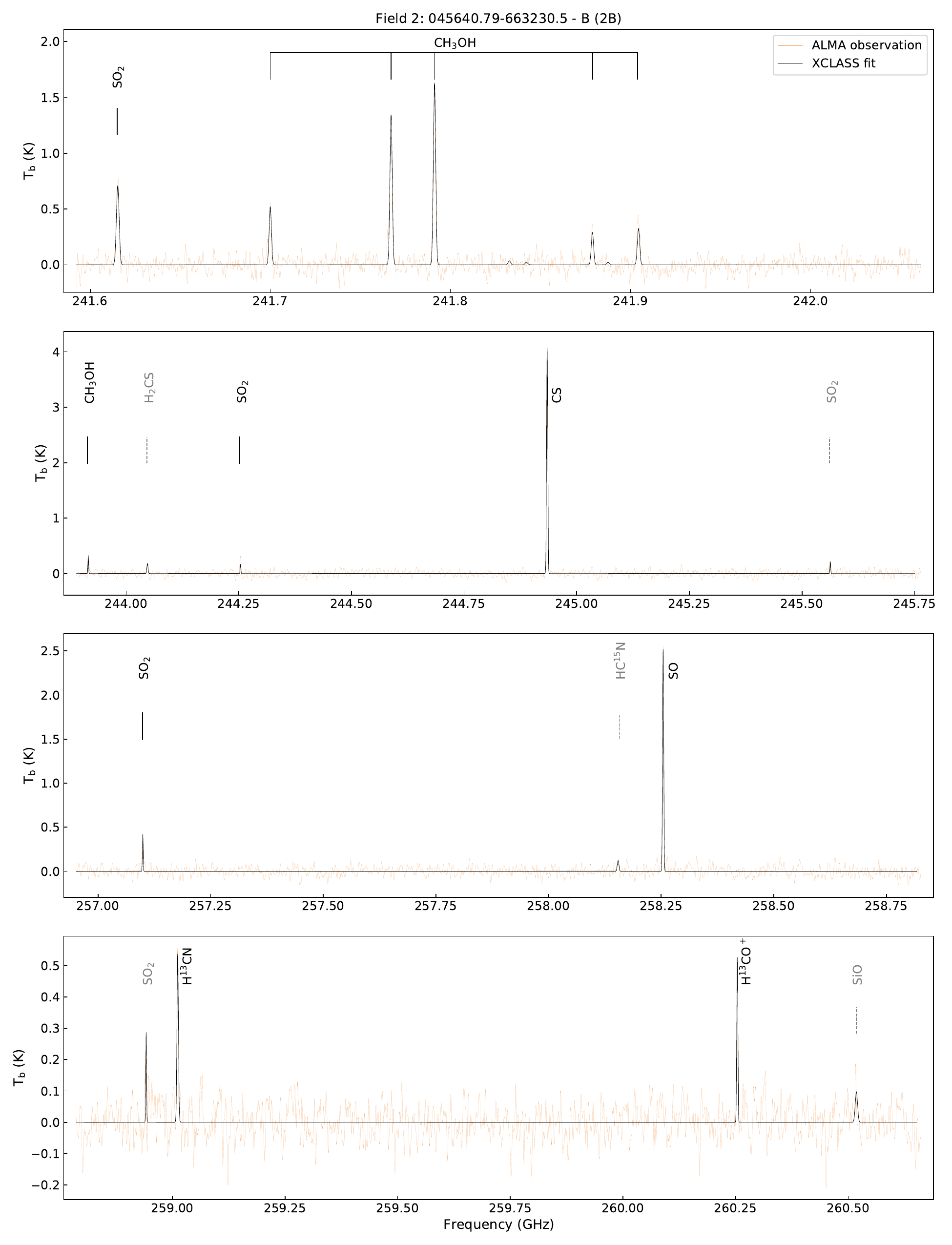}}
    \caption{Same as Fig.~\ref{fig:spectraField01} but for source 2B.}
\end{figure*}

\begin{figure*}
    \ContinuedFloat
    \centering
    \subfloat[][]{\includegraphics[width=0.95\textwidth]{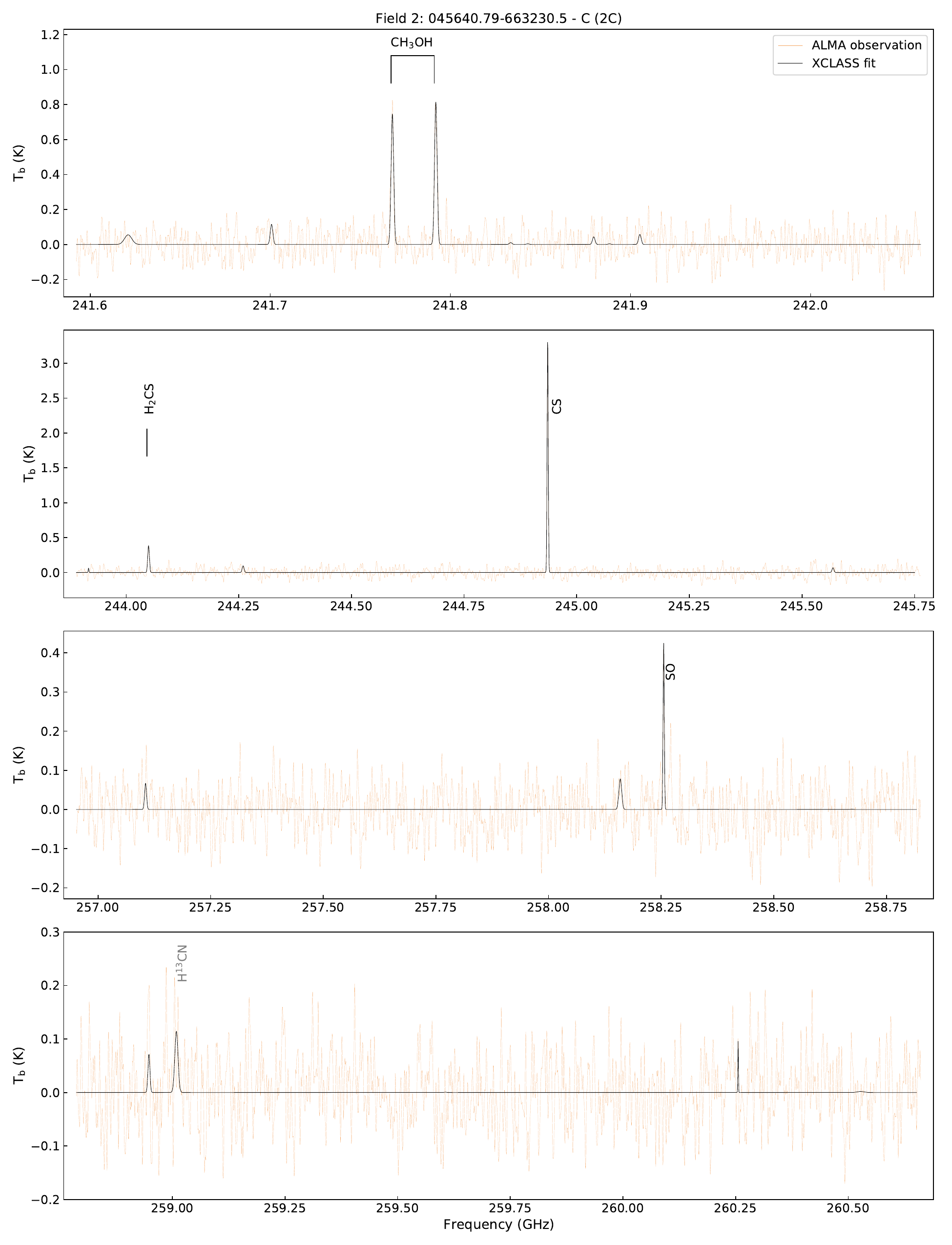}}
    \caption{Same as Fig.~\ref{fig:spectraField01} but for source 2C.}
\end{figure*}

\begin{figure*}
    \ContinuedFloat
    \centering
    \subfloat[][]{\includegraphics[width=0.95\textwidth]{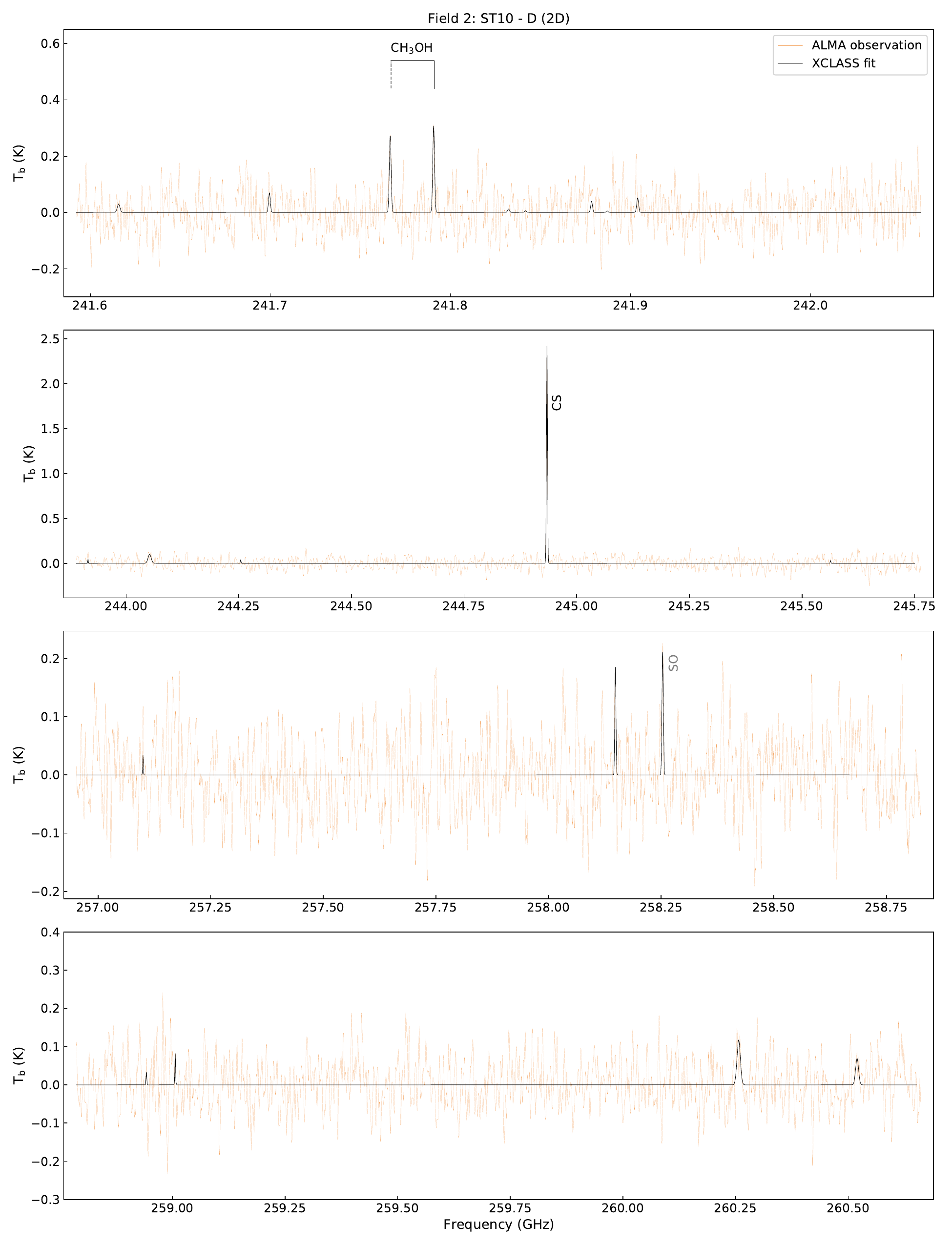}}
    \caption{Same as Fig.~\ref{fig:spectraField01} but for source 2D.}    
\end{figure*}

\begin{figure*}
    \ContinuedFloat
    \centering
    \subfloat[][]{\includegraphics[width=0.95\textwidth]{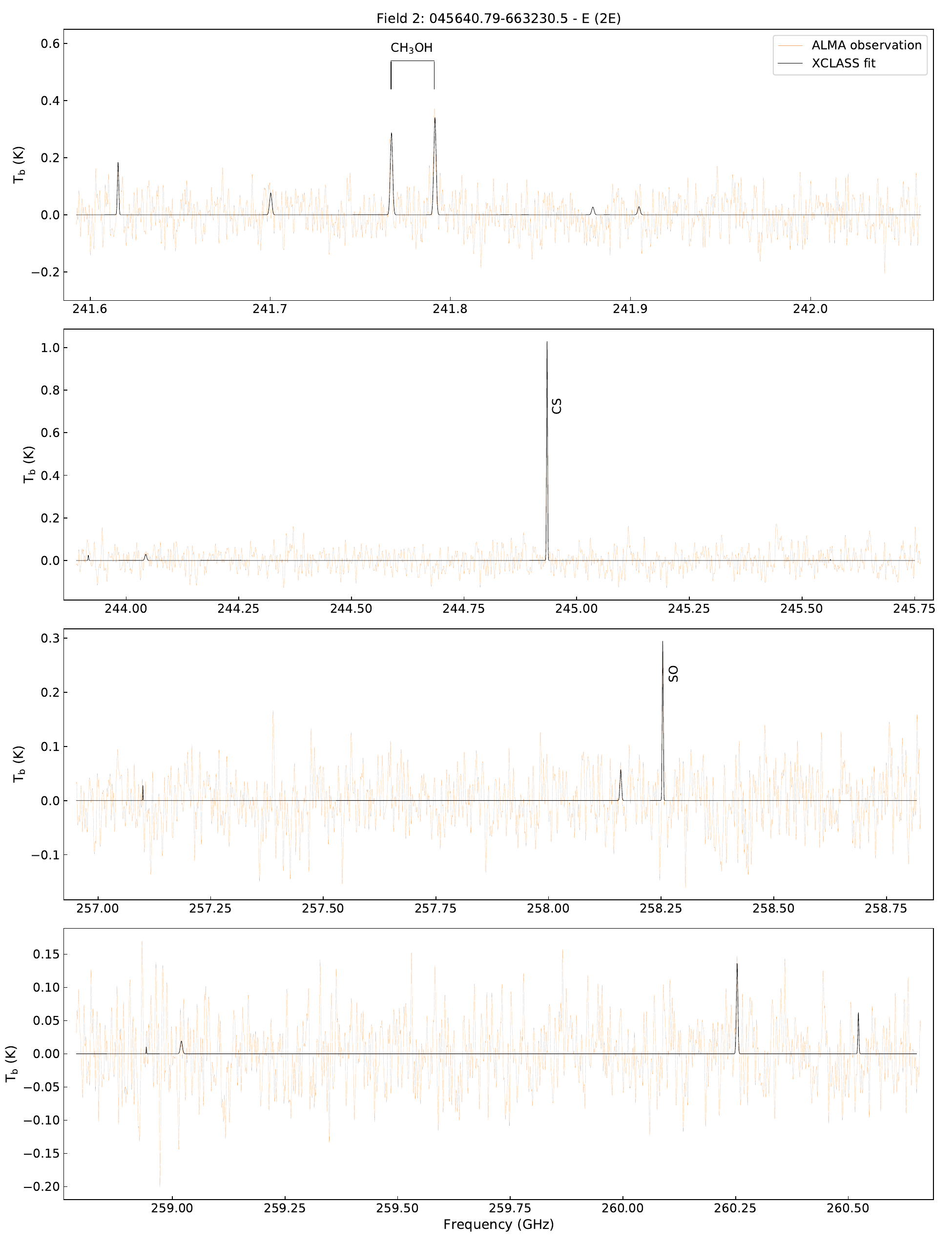}}
    \caption{Same as Fig.~\ref{fig:spectraField01} but for source 2E.}
\end{figure*}

\begin{figure*}
    \ContinuedFloat
    \centering
    \subfloat[][]{\includegraphics[width=0.95\textwidth]{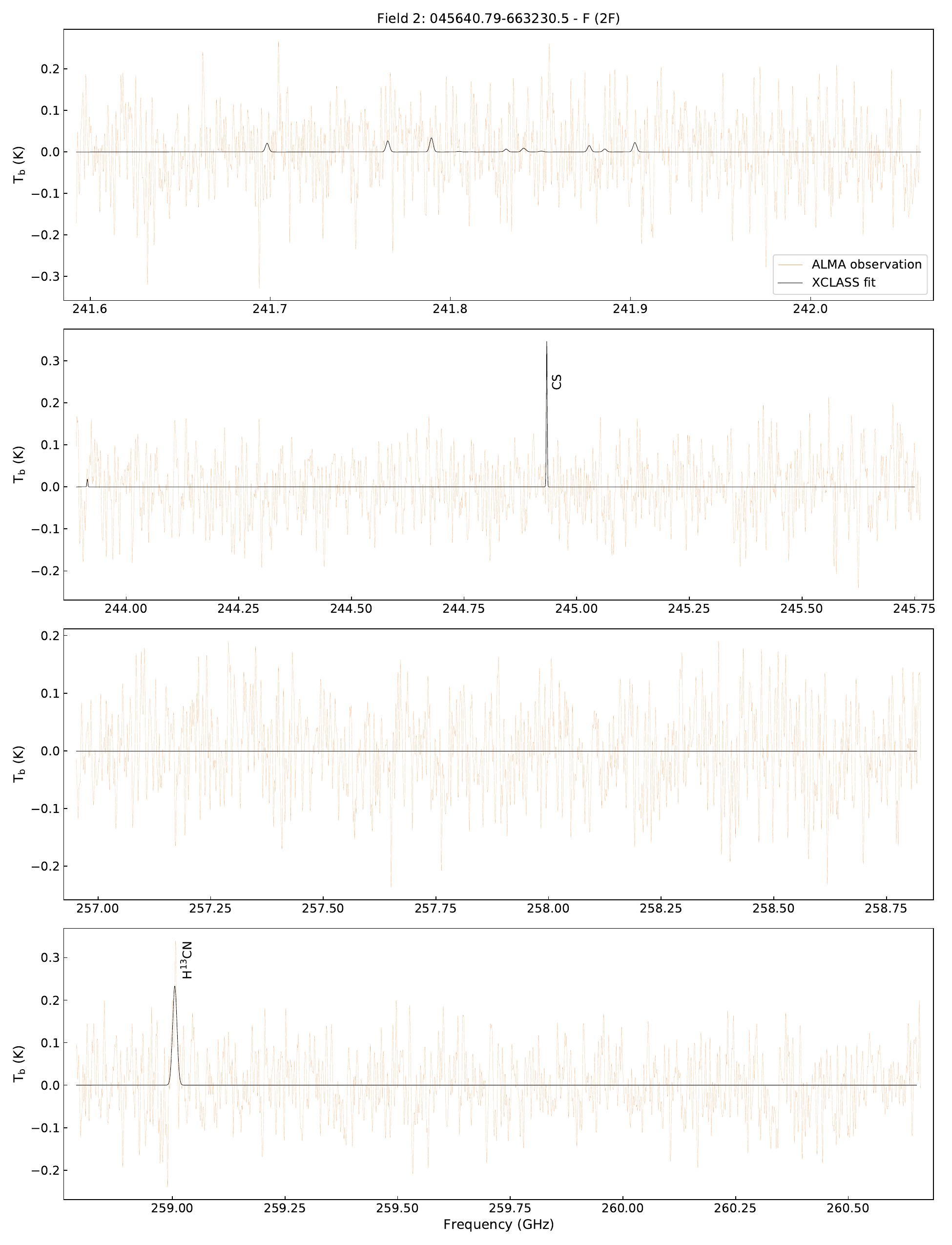}}
    \caption{Same as Fig.~\ref{fig:spectraField01} but for source 2F.}
    \label{fig:spectraField02}
\end{figure*}

%
%
\begin{figure*}
    \centering
    \subfloat[][]{\includegraphics[width=0.95\textwidth]{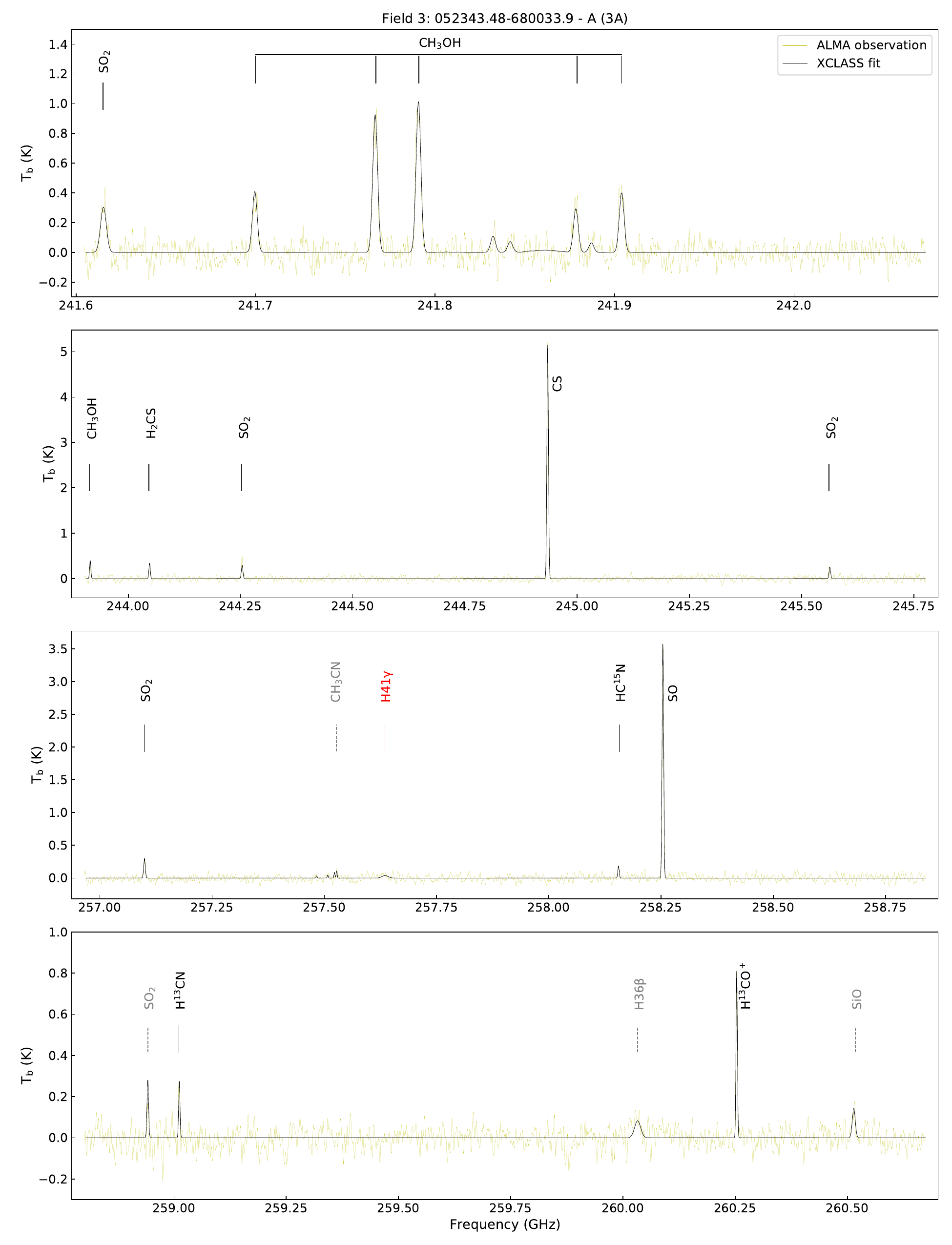}}
    \caption{Same as Fig.~\ref{fig:spectraField01} but for source 3A.}
\end{figure*}

\begin{figure*}
    \ContinuedFloat
    \centering
    \subfloat[][]{\includegraphics[width=0.95\textwidth]{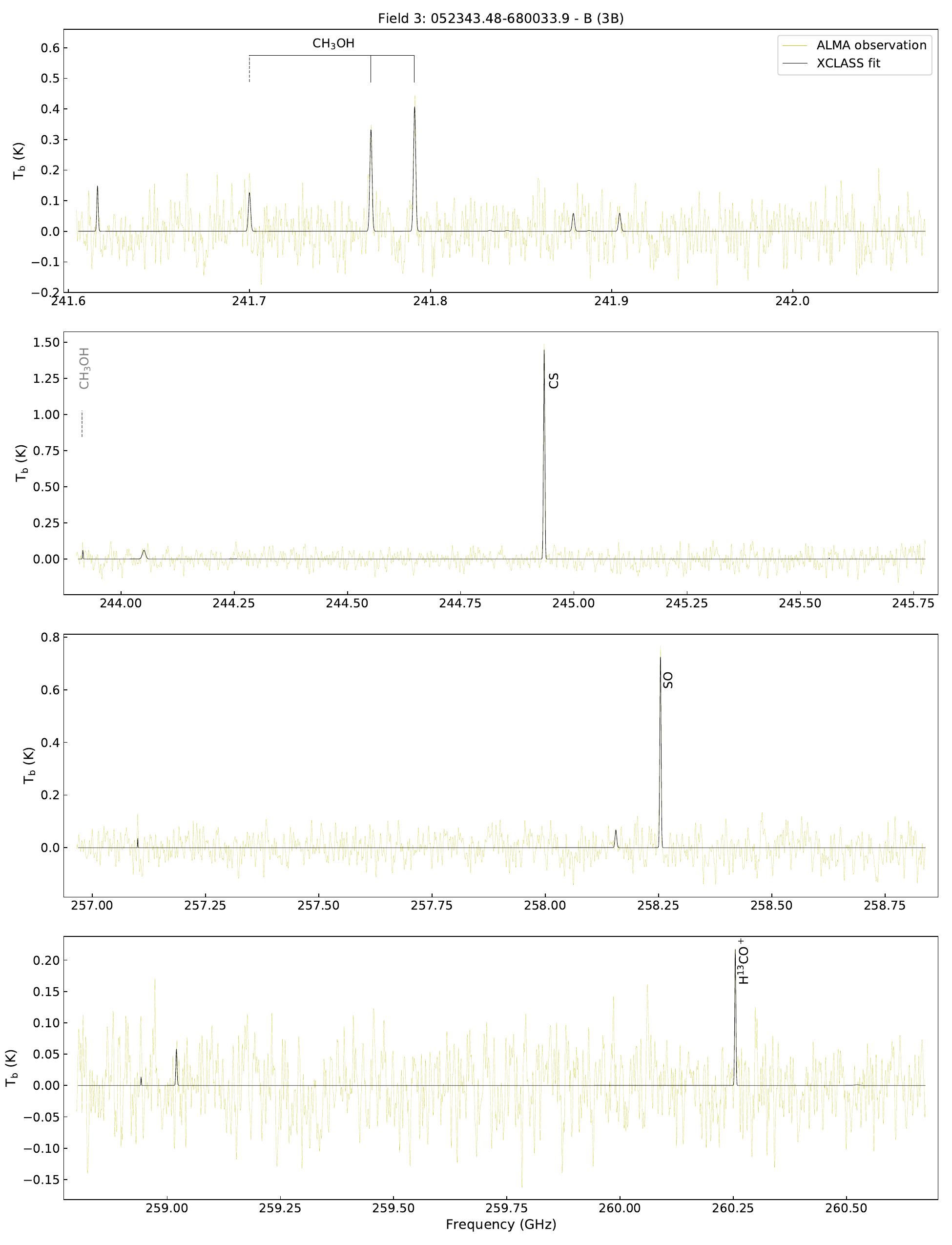}}
    \caption{Same as Fig.~\ref{fig:spectraField01} but for source 3B.}
\end{figure*}

\begin{figure*}
    \ContinuedFloat
    \centering
    \subfloat[][]{\includegraphics[width=0.95\textwidth]{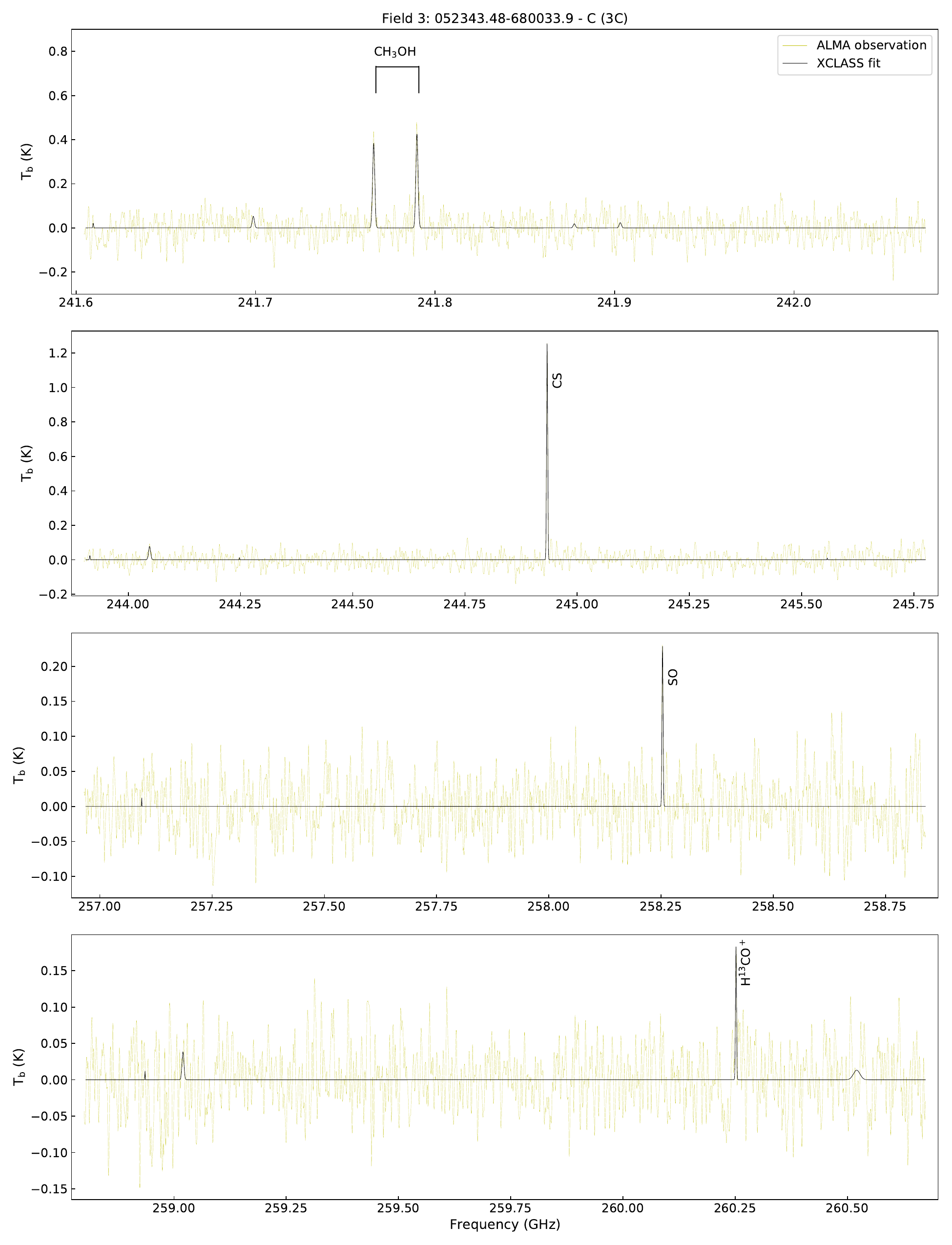}}
    \caption{Same as Fig.~\ref{fig:spectraField01} but for source 3C.}
\end{figure*}

\begin{figure*}
    \ContinuedFloat
    \centering
    \subfloat[][]{\includegraphics[width=0.95\textwidth]{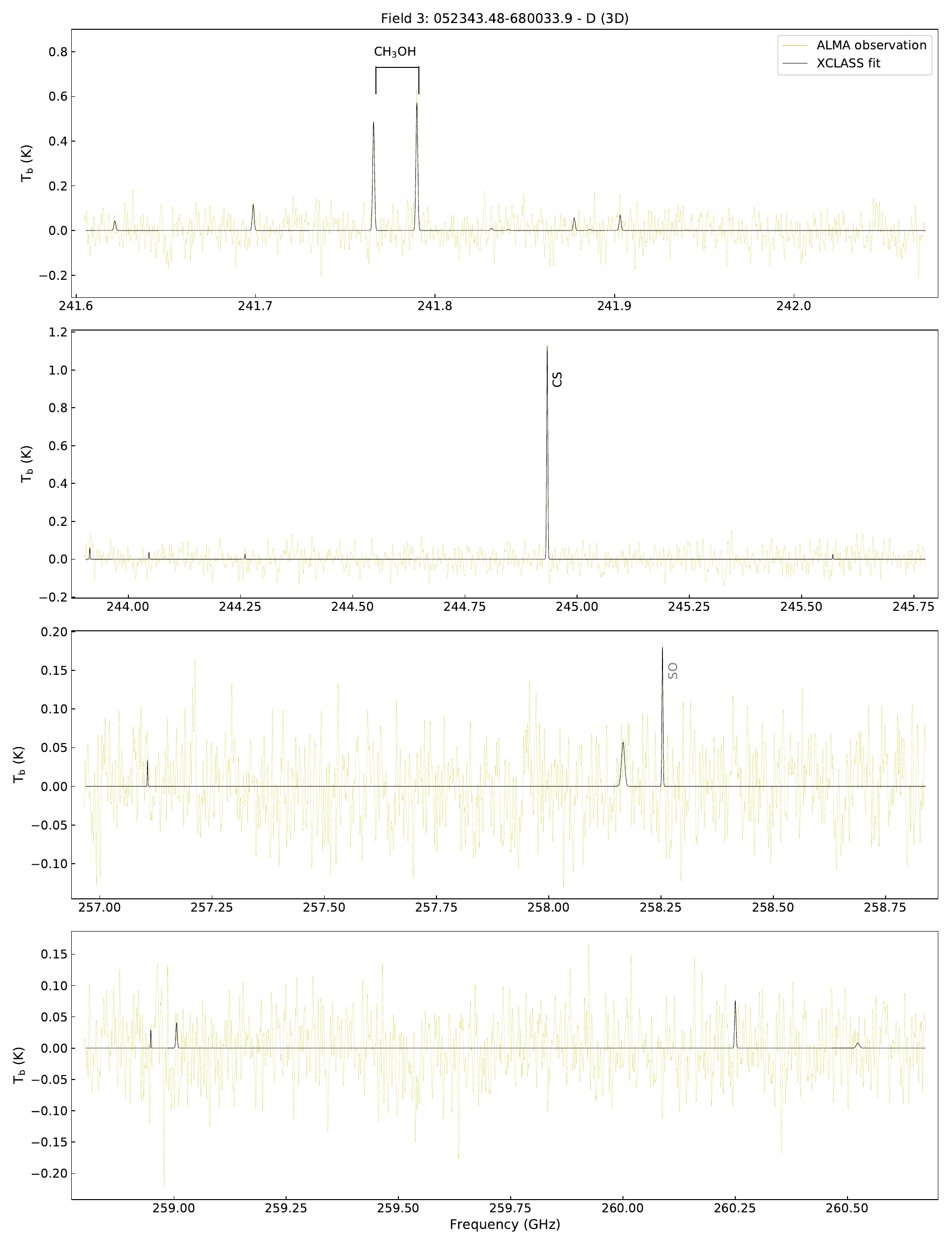}}
    \caption{Same as Fig.~\ref{fig:spectraField01} but for source 3D.}
    \label{fig:spectraField03}
\end{figure*}

%
%

%
%
\begin{figure*}
    \centering
    \subfloat[][]{\includegraphics[width=0.95\textwidth]{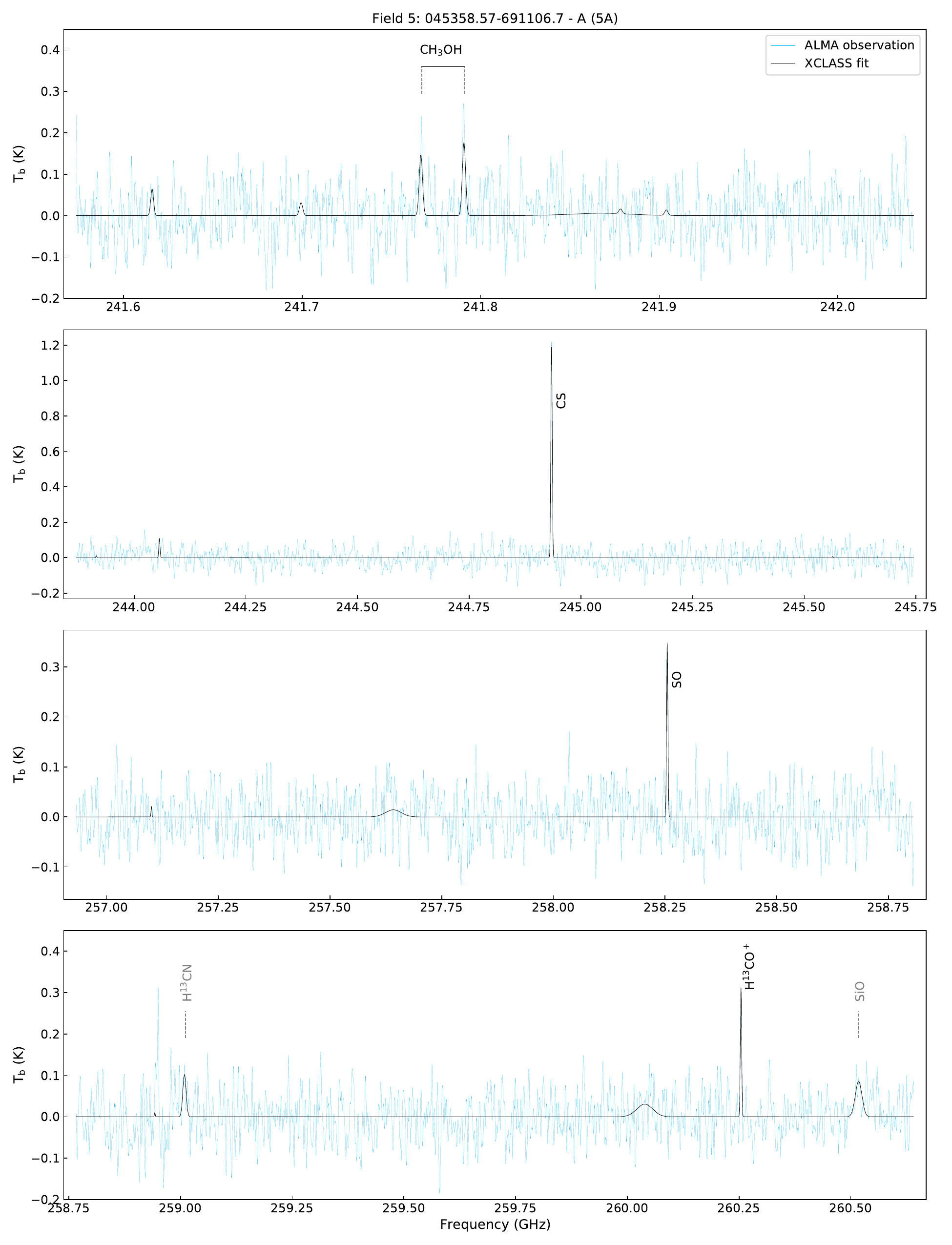}}
    \caption{Same as Fig.~\ref{fig:spectraField01} but for source 5A.}
\end{figure*}

\begin{figure*}
    \ContinuedFloat
    \centering
    \subfloat[][]{\includegraphics[width=0.95\textwidth]{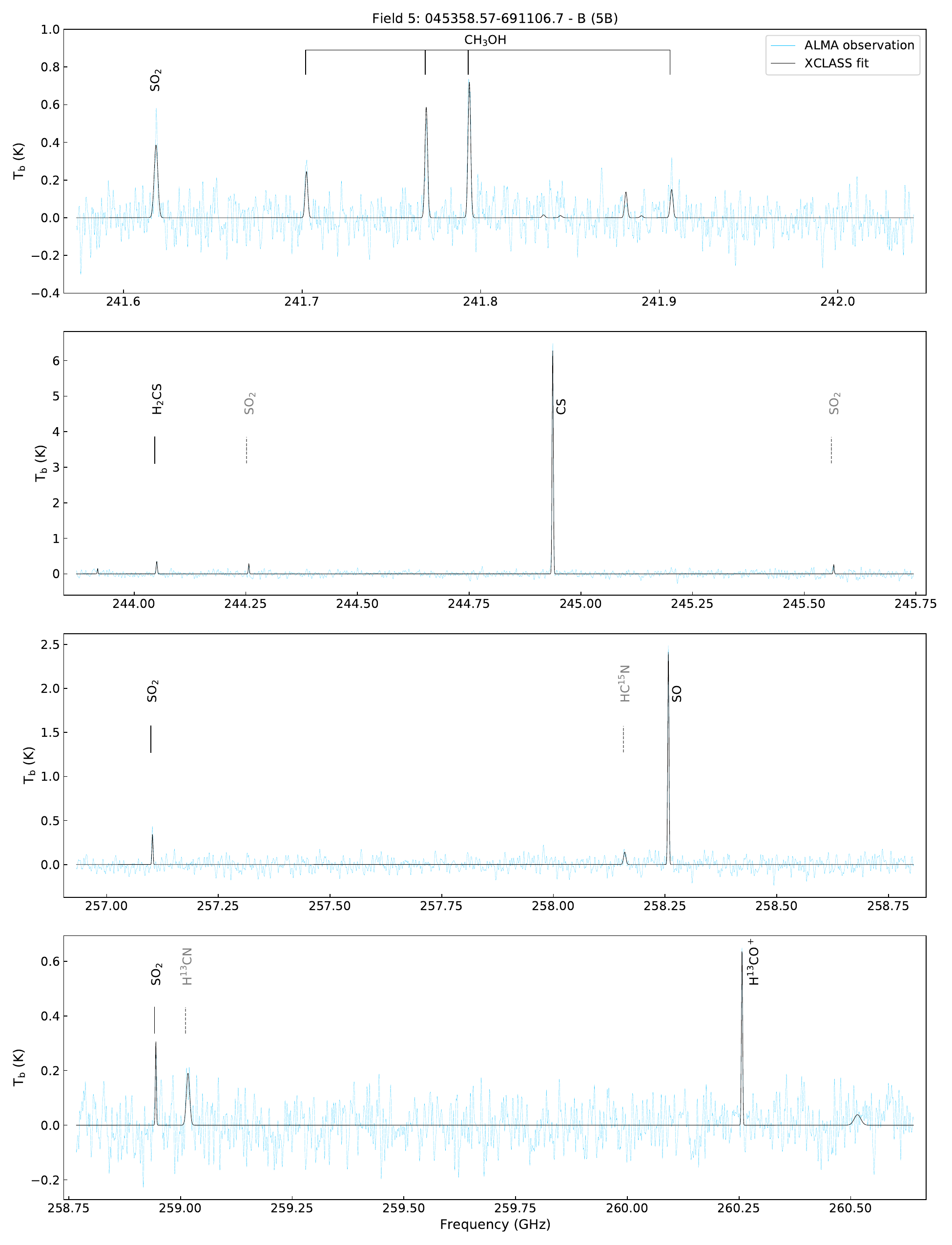}}
    \caption{Same as Fig.~\ref{fig:spectraField01} but for source 5B.}
    \label{fig:spectraField05}
\end{figure*}

%
%
\begin{figure*}
    \centering
    \subfloat[][]{\includegraphics[width=0.95\textwidth]{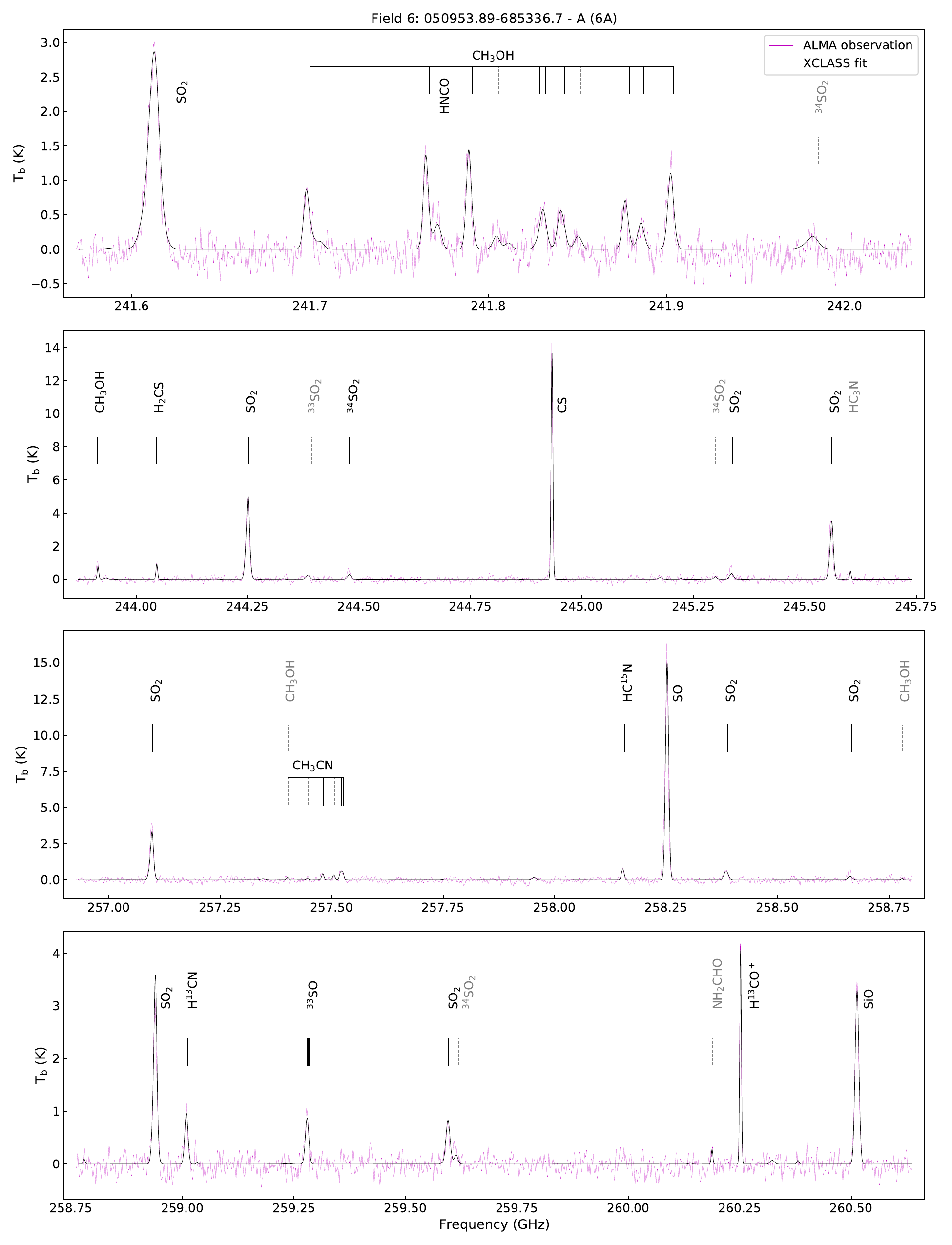}}
    \caption{Same as Fig.~\ref{fig:spectraField01} but for source 6A.}
\end{figure*}

\begin{figure*}
    \ContinuedFloat
    \centering
    \subfloat[][]{\includegraphics[width=0.95\textwidth]{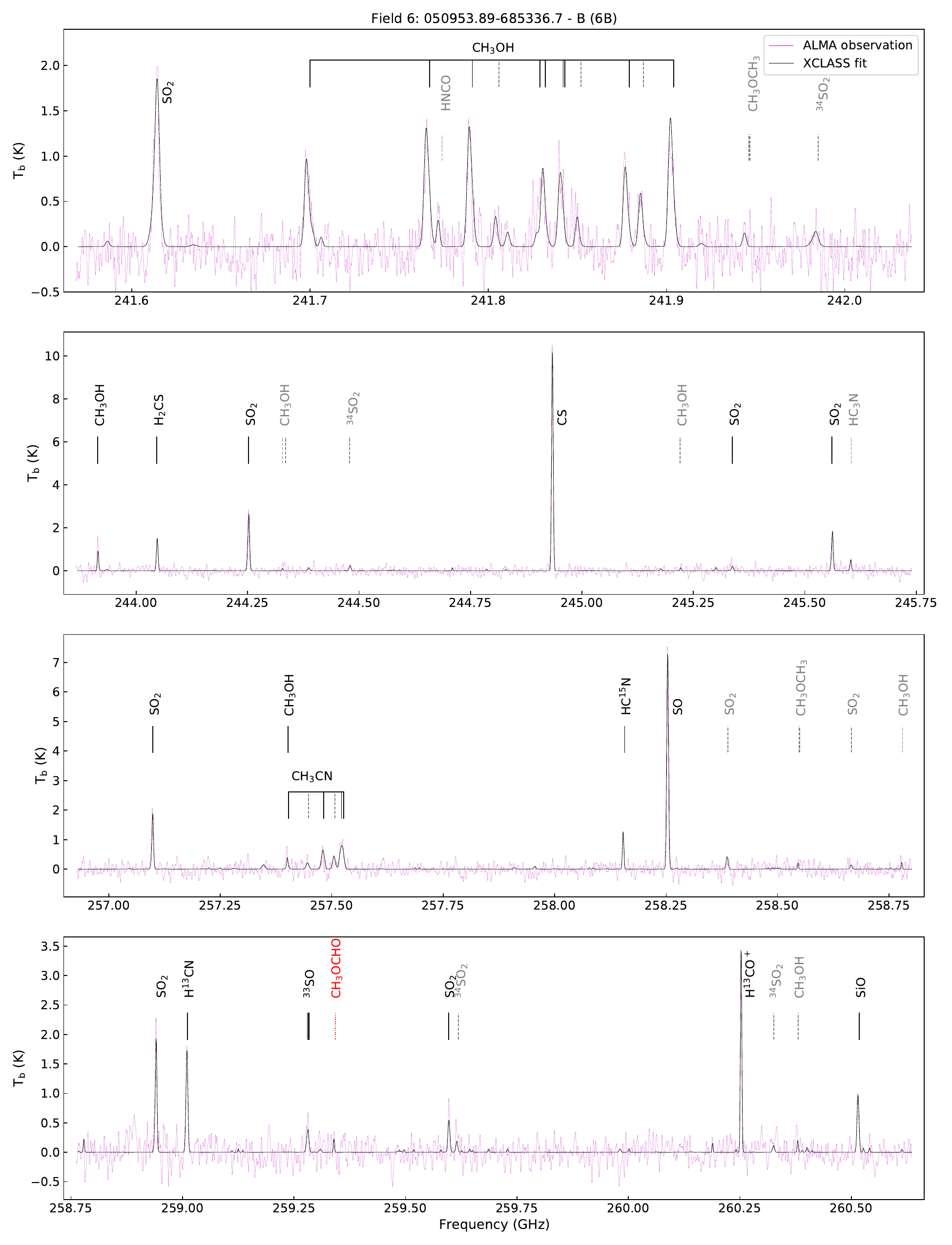}}
    \caption{Same as Fig.~\ref{fig:spectraField01} but for source 6B.}
\end{figure*}

\begin{figure*}
    \ContinuedFloat
    \centering
    \subfloat[][]{\includegraphics[width=0.95\textwidth]{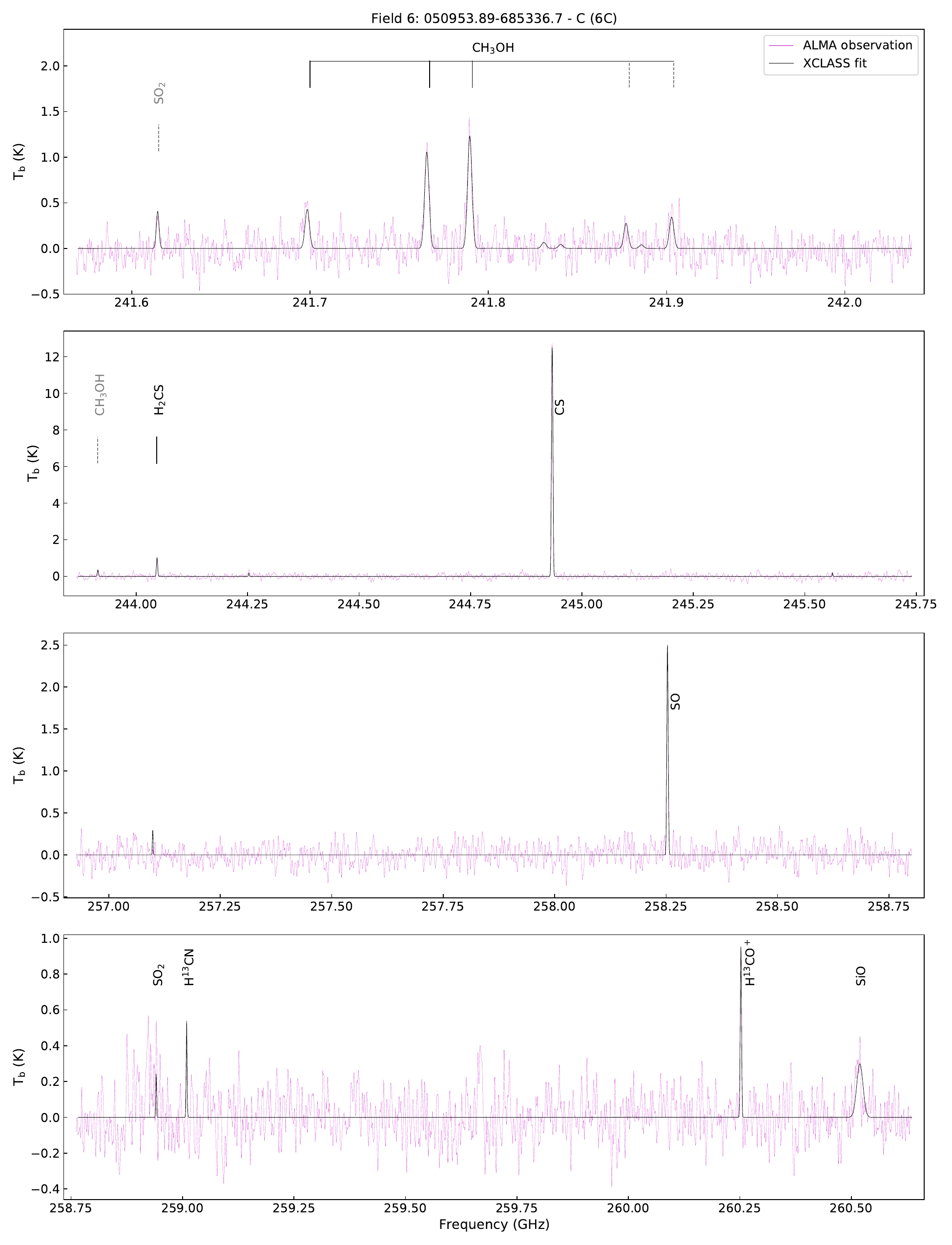}}
    \caption{Same as Fig.~\ref{fig:spectraField01} but for source 6C.}
\end{figure*}

\begin{figure*}
    \ContinuedFloat
    \centering
    \subfloat[][]{\includegraphics[width=0.95\textwidth]{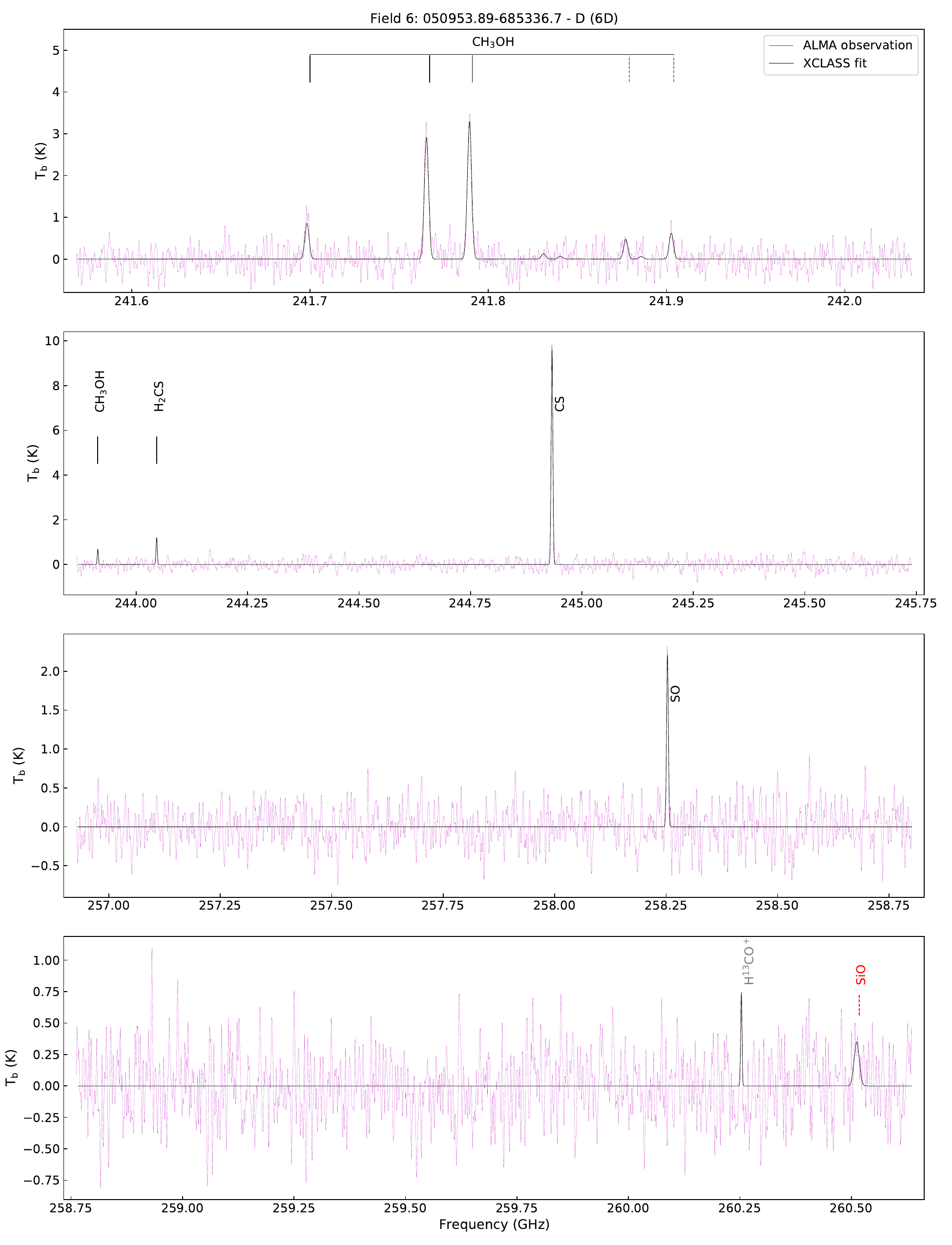}}
    \caption{Same as Fig.~\ref{fig:spectraField01} but for source 6D.}
\end{figure*}

\begin{figure*}
    \ContinuedFloat
    \centering
    \subfloat[][]{\includegraphics[width=0.95\textwidth]{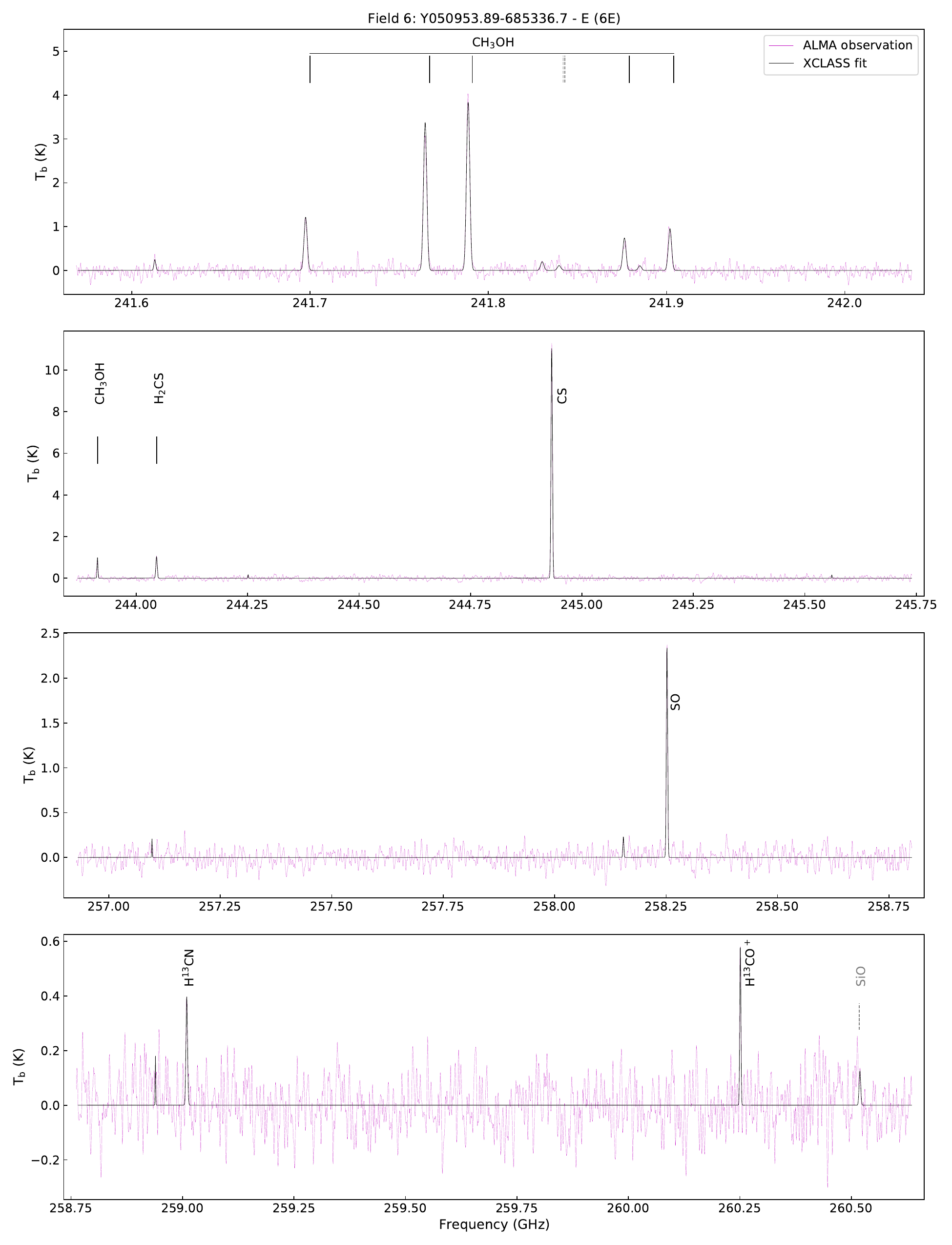}}
    \caption{Same as Fig.~\ref{fig:spectraField01} but for source 6E.}
\end{figure*}

\begin{figure*}
    \ContinuedFloat
    \centering
    \subfloat[][]{\includegraphics[width=0.95\textwidth]{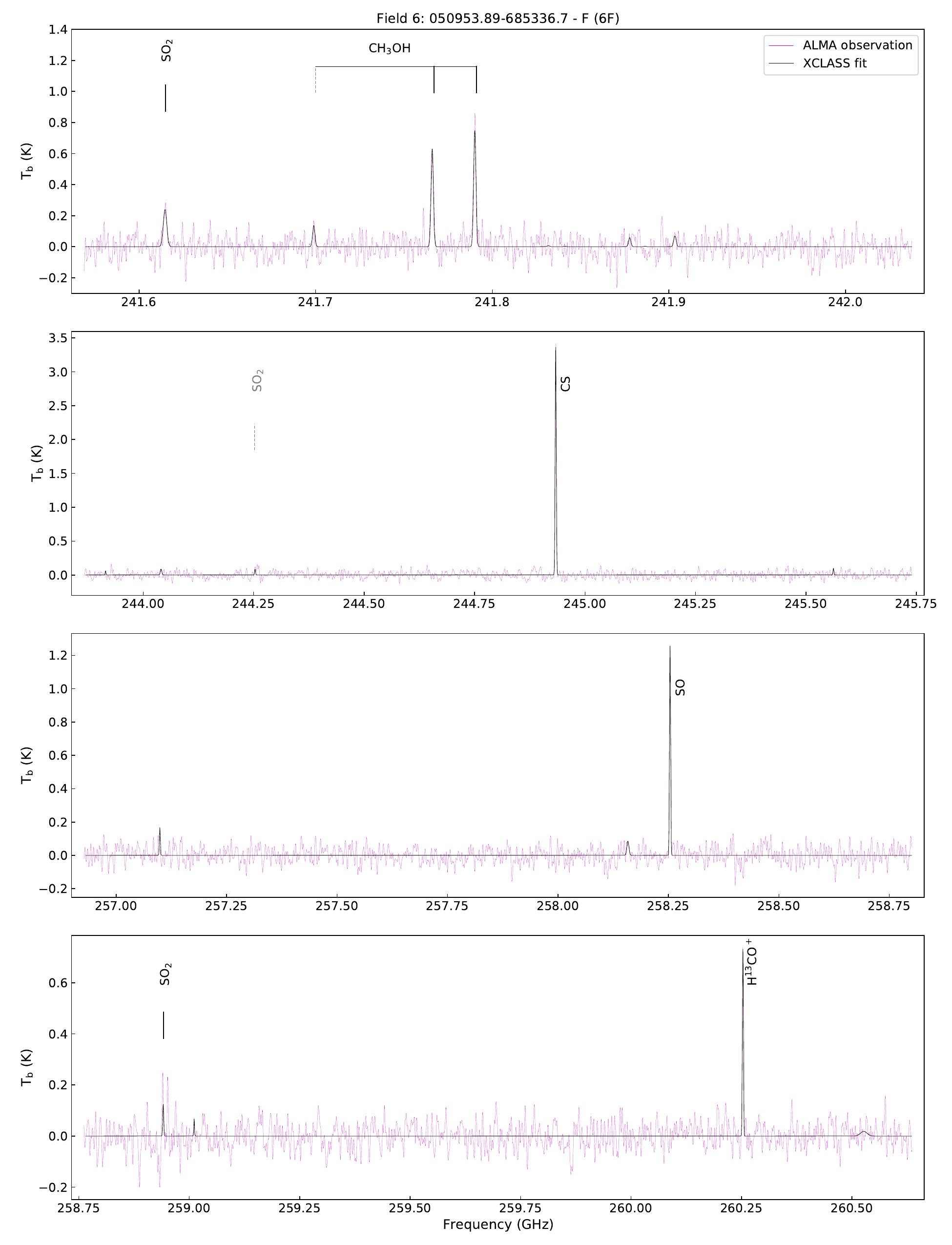}}
    \caption{Same as Fig.~\ref{fig:spectraField01} but for source 6F.}
    \label{fig:spectraField06}
\end{figure*}

%
%
\begin{figure*}
    \centering
    \subfloat[][]{\includegraphics[width=0.95\textwidth]{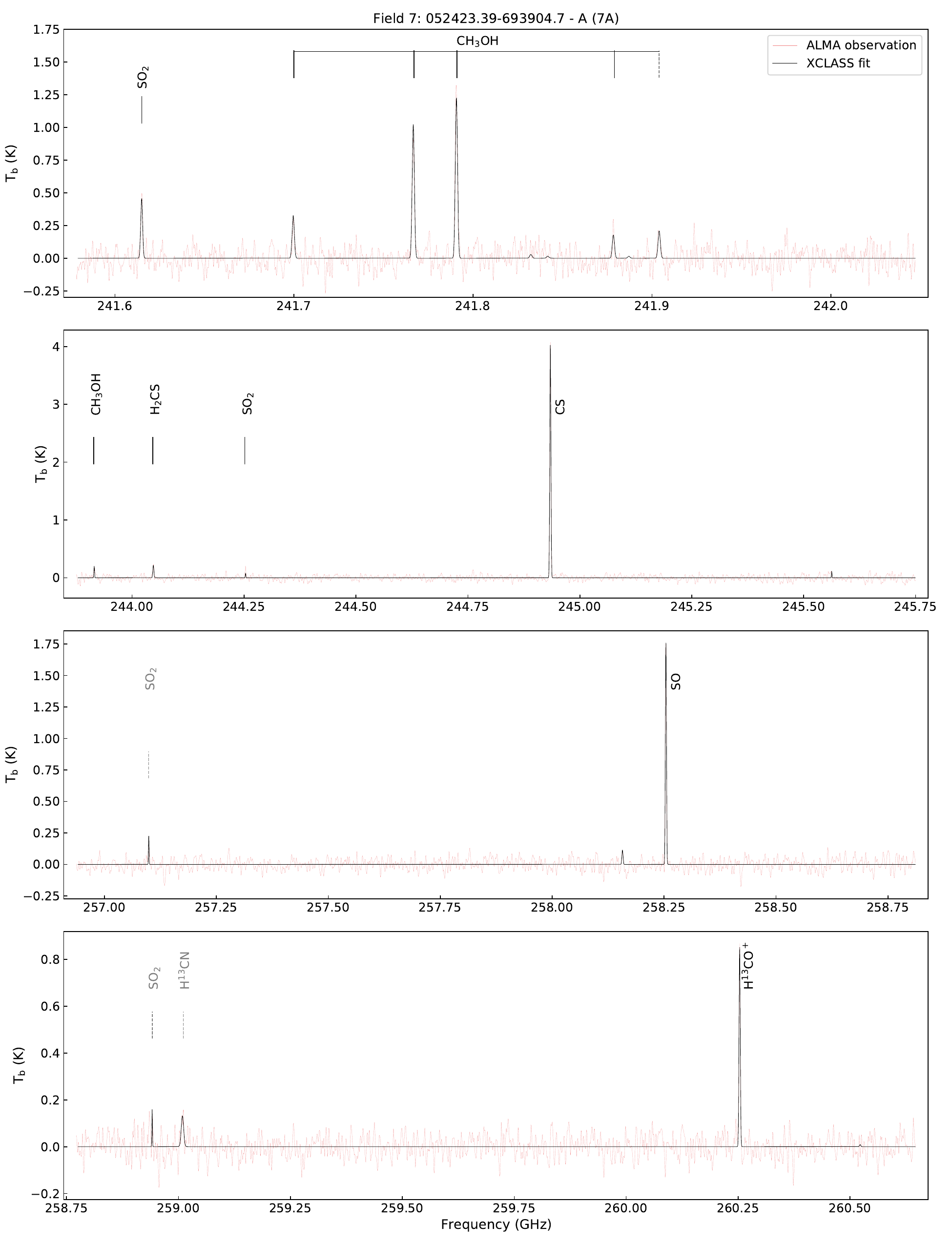}}
    \caption{Same as Fig.~\ref{fig:spectraField01} but for source 7A.}
\end{figure*}

\begin{figure*}
    \ContinuedFloat
    \centering
    \subfloat[][]{\includegraphics[width=0.95\textwidth]{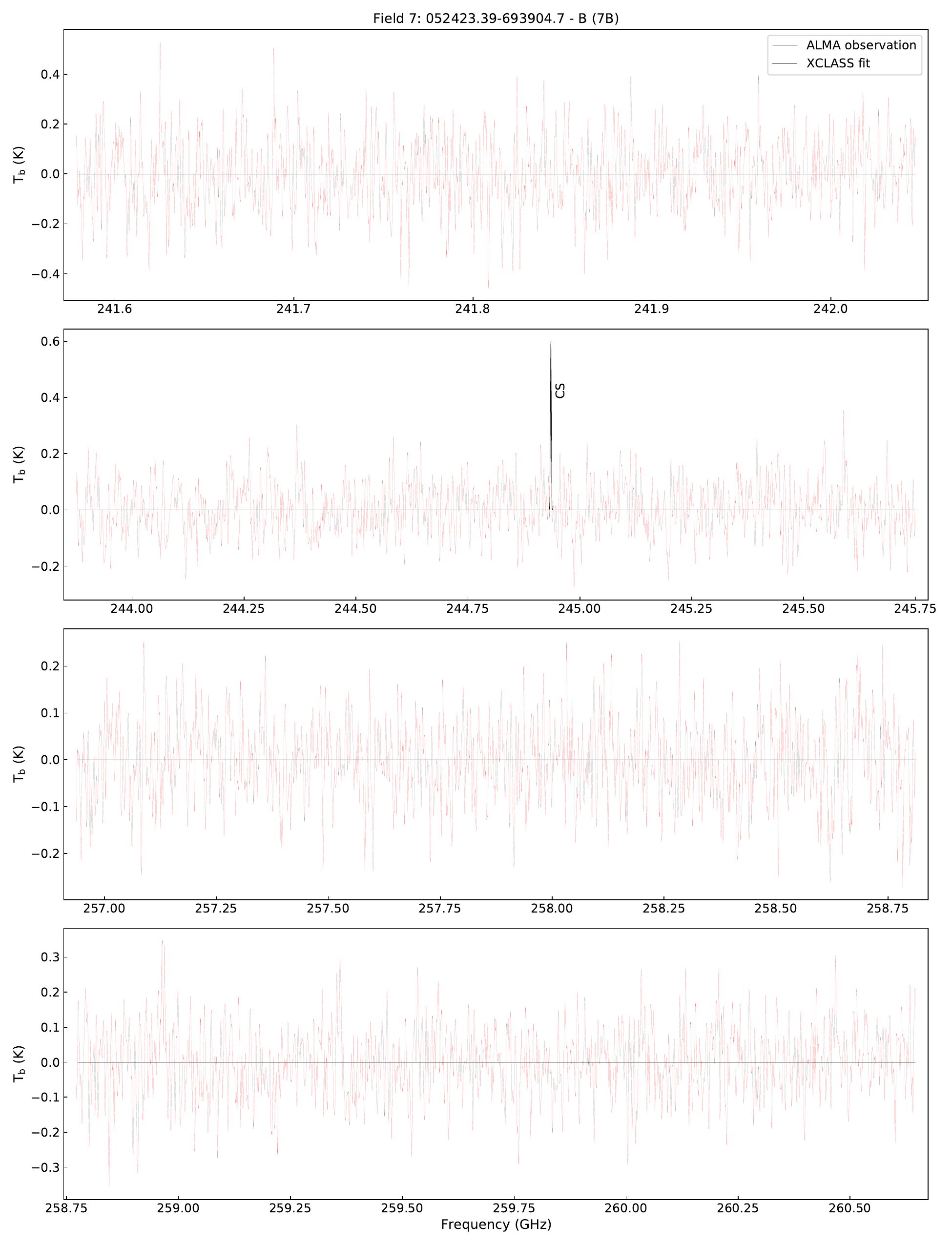}}
    \caption{Same as Fig.~\ref{fig:spectraField01} but for source 7B.}
\end{figure*}

\begin{figure*}
    \ContinuedFloat
    \centering
    \subfloat[][]{\includegraphics[width=0.95\textwidth]{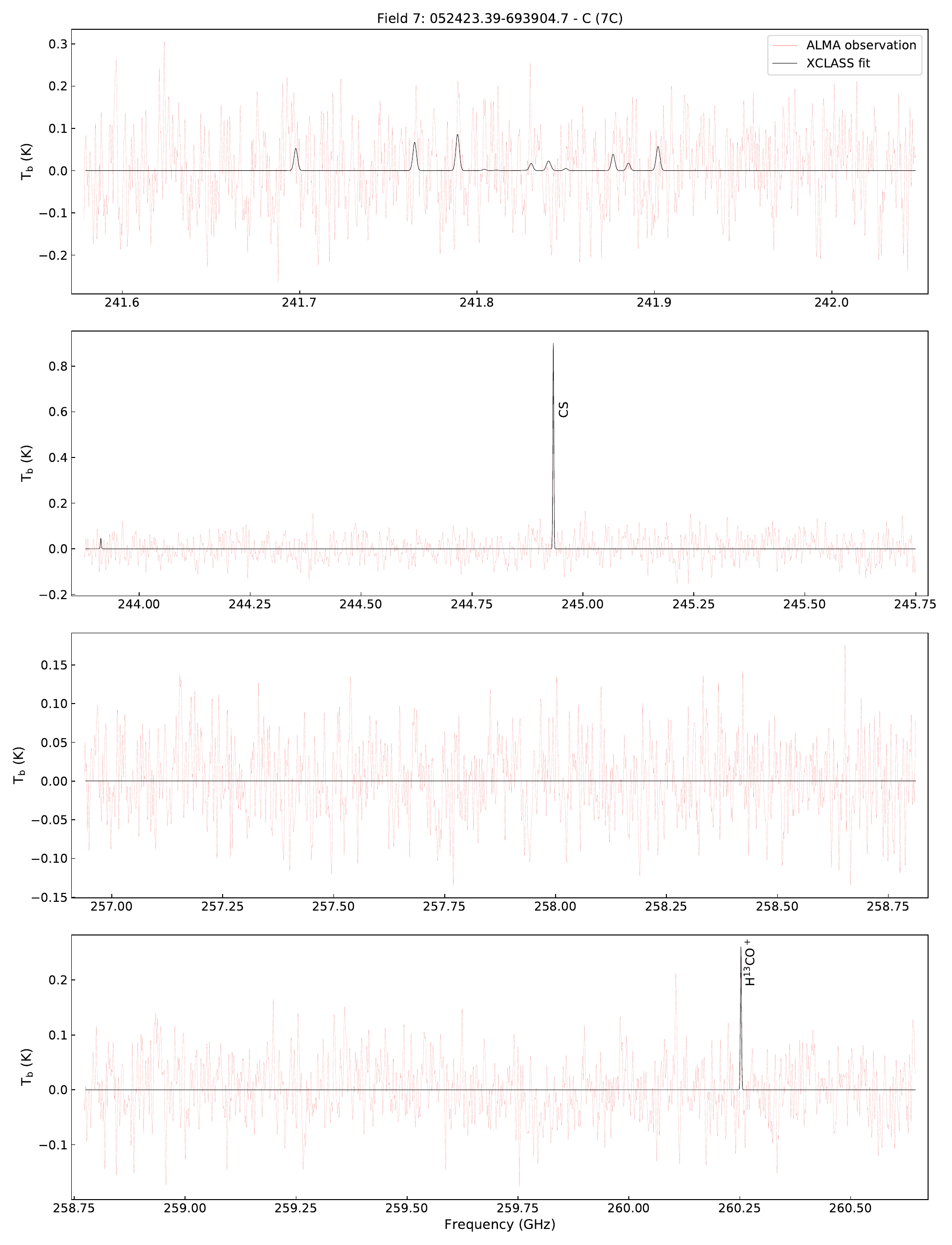}}
    \caption{Same as Fig.~\ref{fig:spectraField01} but for source 7C.}
    \label{fig:spectraField07}
\end{figure*}

%
%
\begin{figure*}
    \centering
    \subfloat[][]{\includegraphics[width=0.95\textwidth]{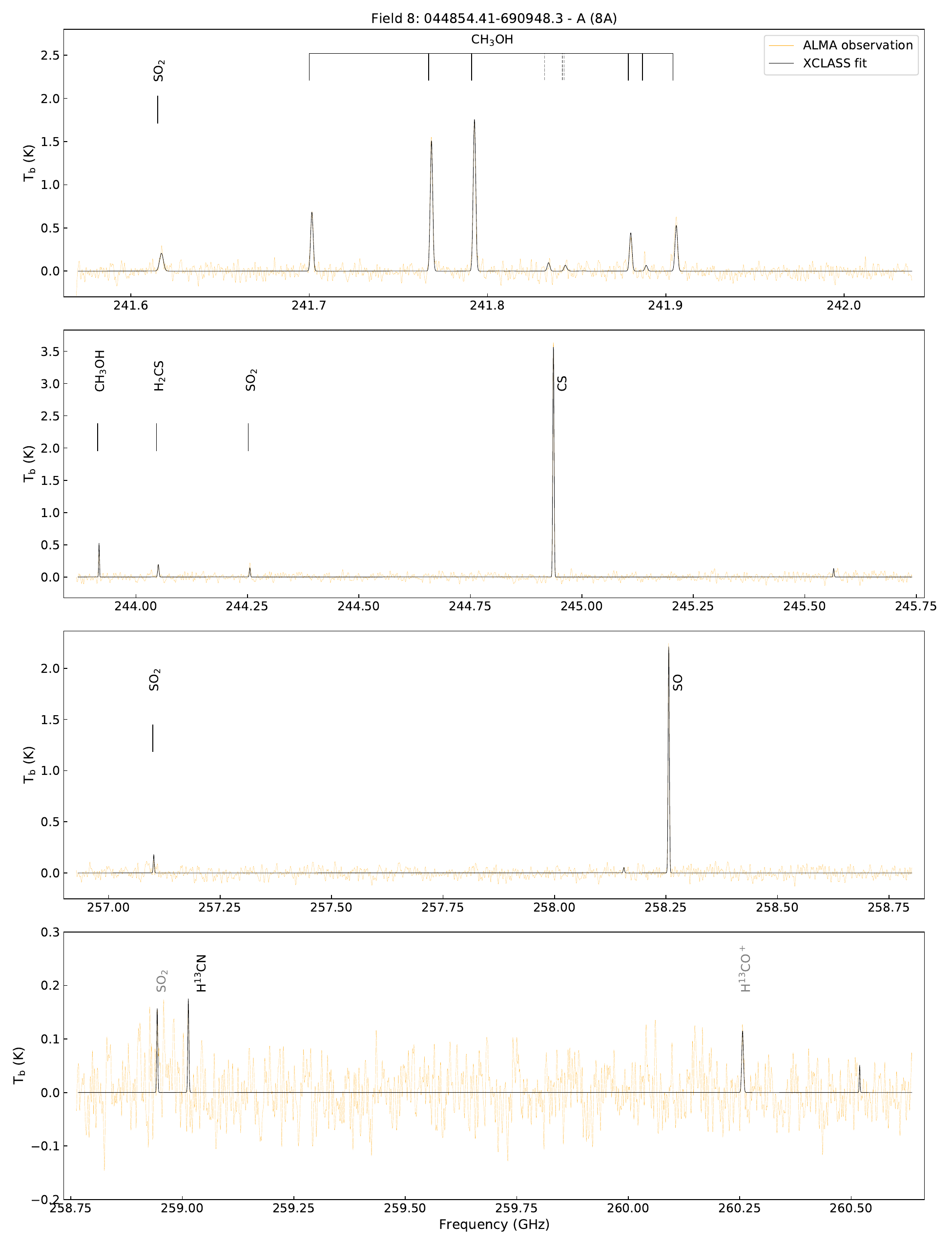}}
    \caption{Same as Fig.~\ref{fig:spectraField01} but for source 8A.}
\end{figure*}


\begin{figure*}
    \ContinuedFloat
    \centering
    \subfloat[][]{\includegraphics[width=0.95\textwidth]{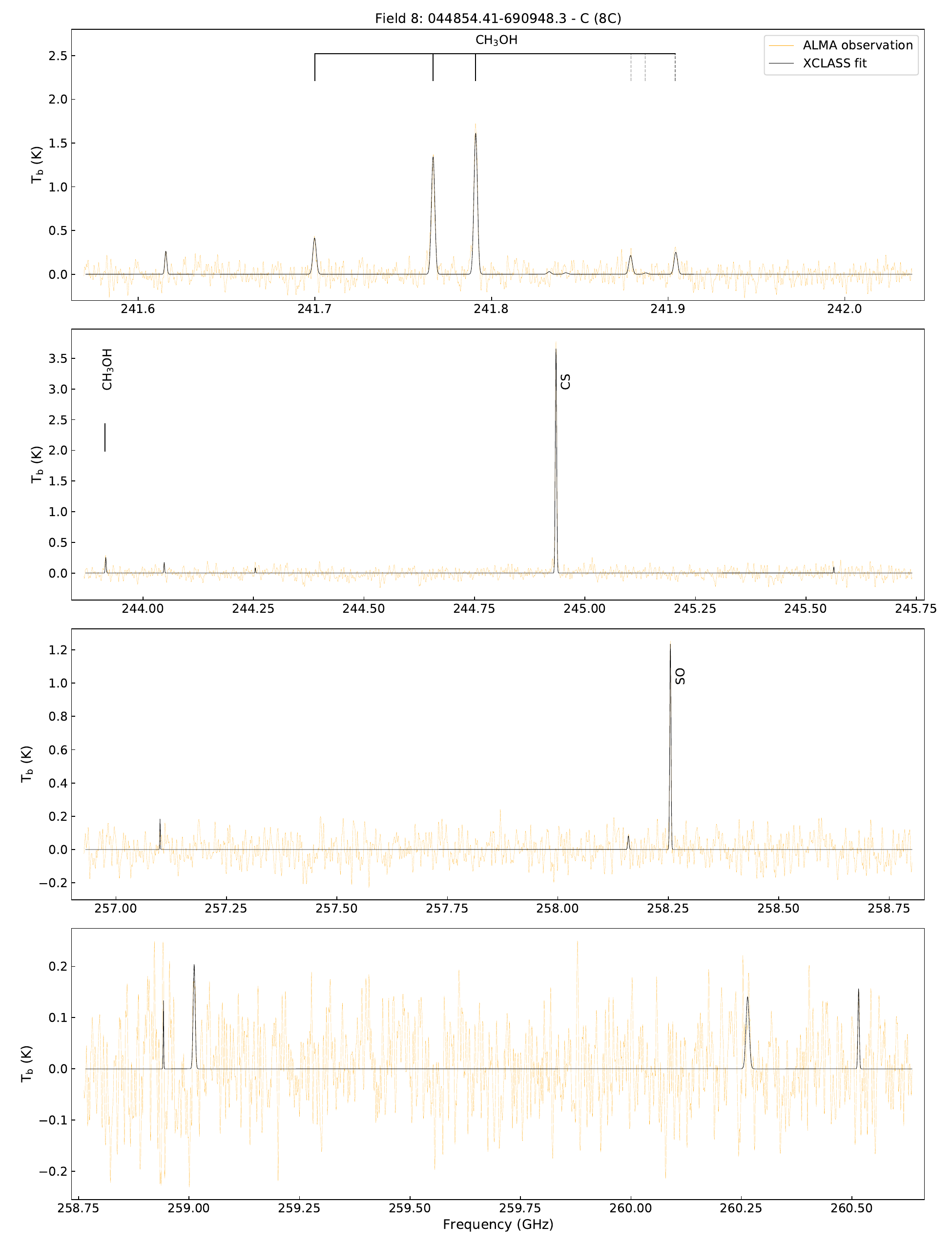}}
    \caption{Same as Fig.~\ref{fig:spectraField01} but for source 8C.}
    \label{fig:spectraField08}
\end{figure*}

%
%

%
%
\begin{figure*}
    \centering
    \subfloat[][]{\includegraphics[width=0.95\textwidth]{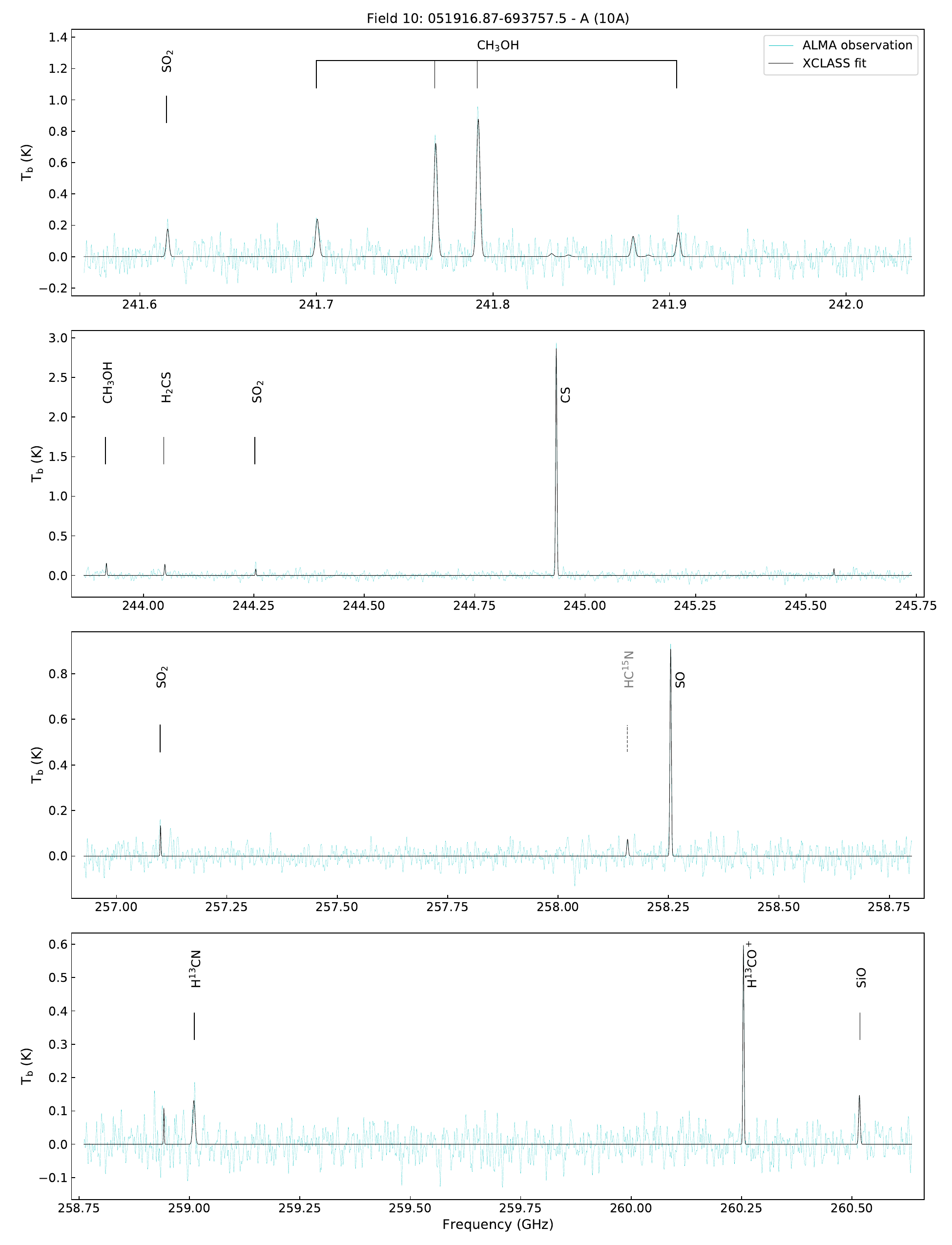}}
    \caption{Same as Fig.~\ref{fig:spectraField01} but for source 10A.}
\end{figure*}

\begin{figure*}
    \ContinuedFloat
    \centering
    \subfloat[][]{\includegraphics[width=0.95\textwidth]{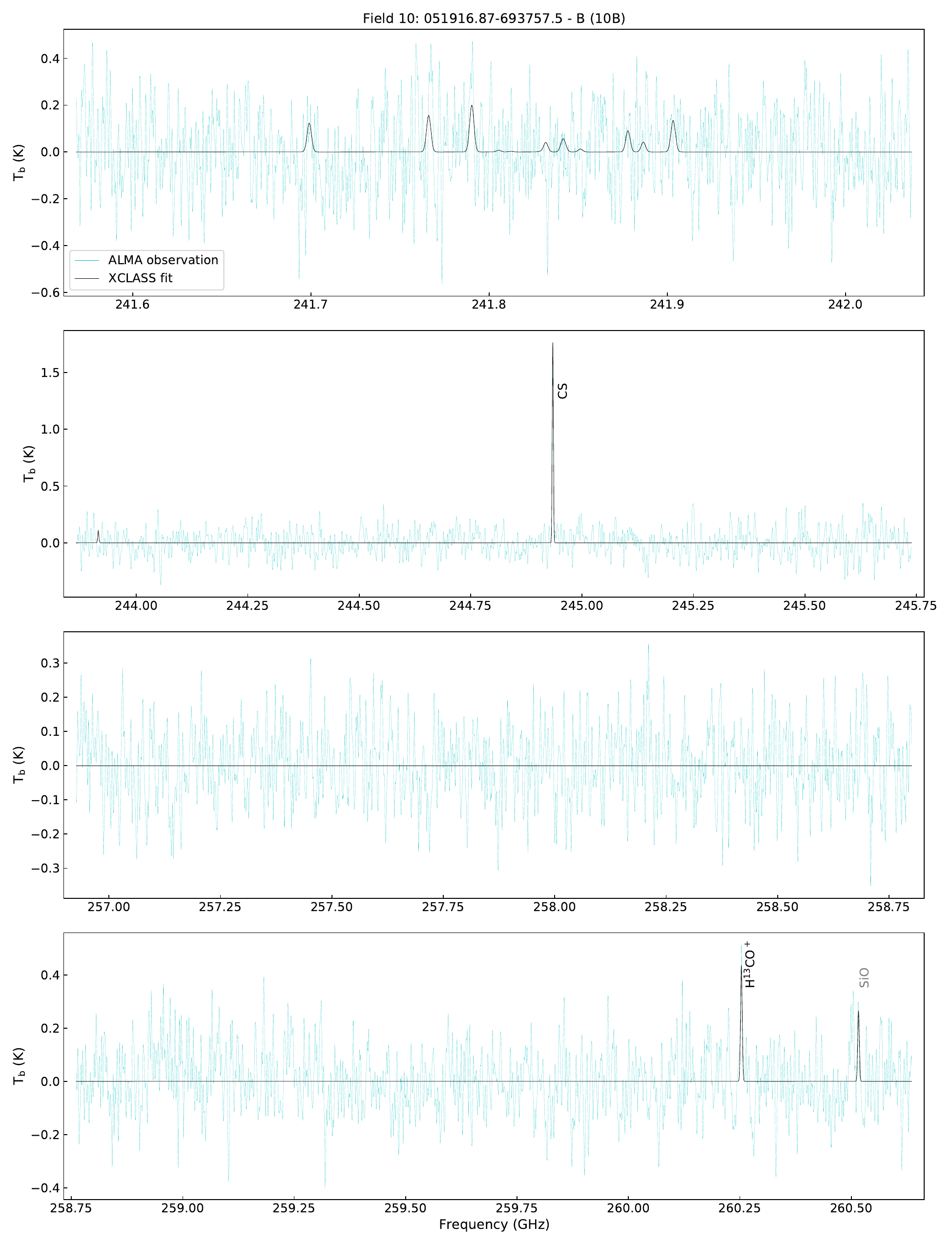}}
    \caption{Same as Fig.~\ref{fig:spectraField01} but for source 10B.}
\end{figure*}

\begin{figure*}
    \ContinuedFloat
    \centering
    \subfloat[][]{\includegraphics[width=0.95\textwidth]{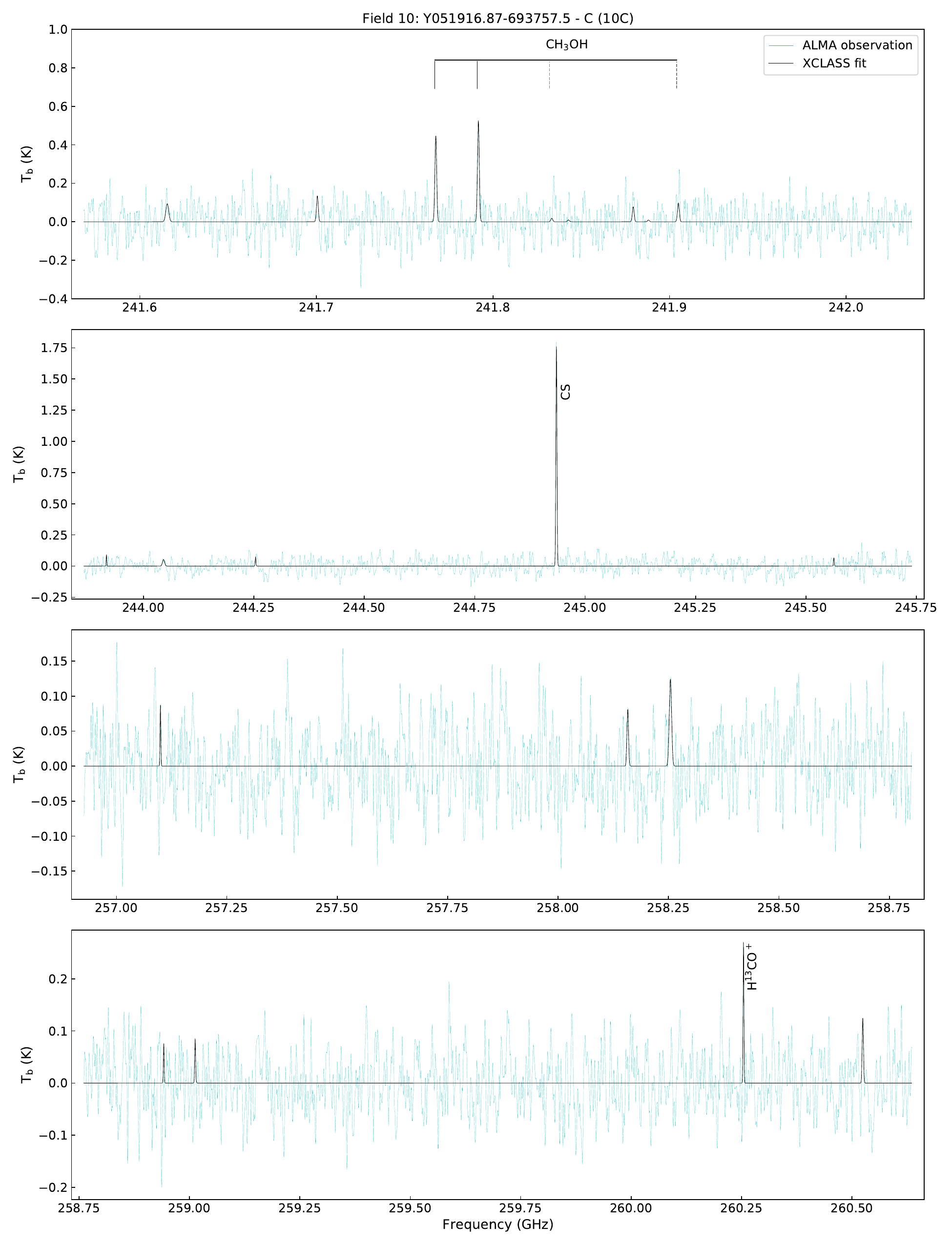}}
    \caption{Same as Fig.~\ref{fig:spectraField01} but for source 10C.}
\end{figure*}

\begin{figure*}
    \ContinuedFloat
    \centering
    \subfloat[][]{\includegraphics[width=0.95\textwidth]{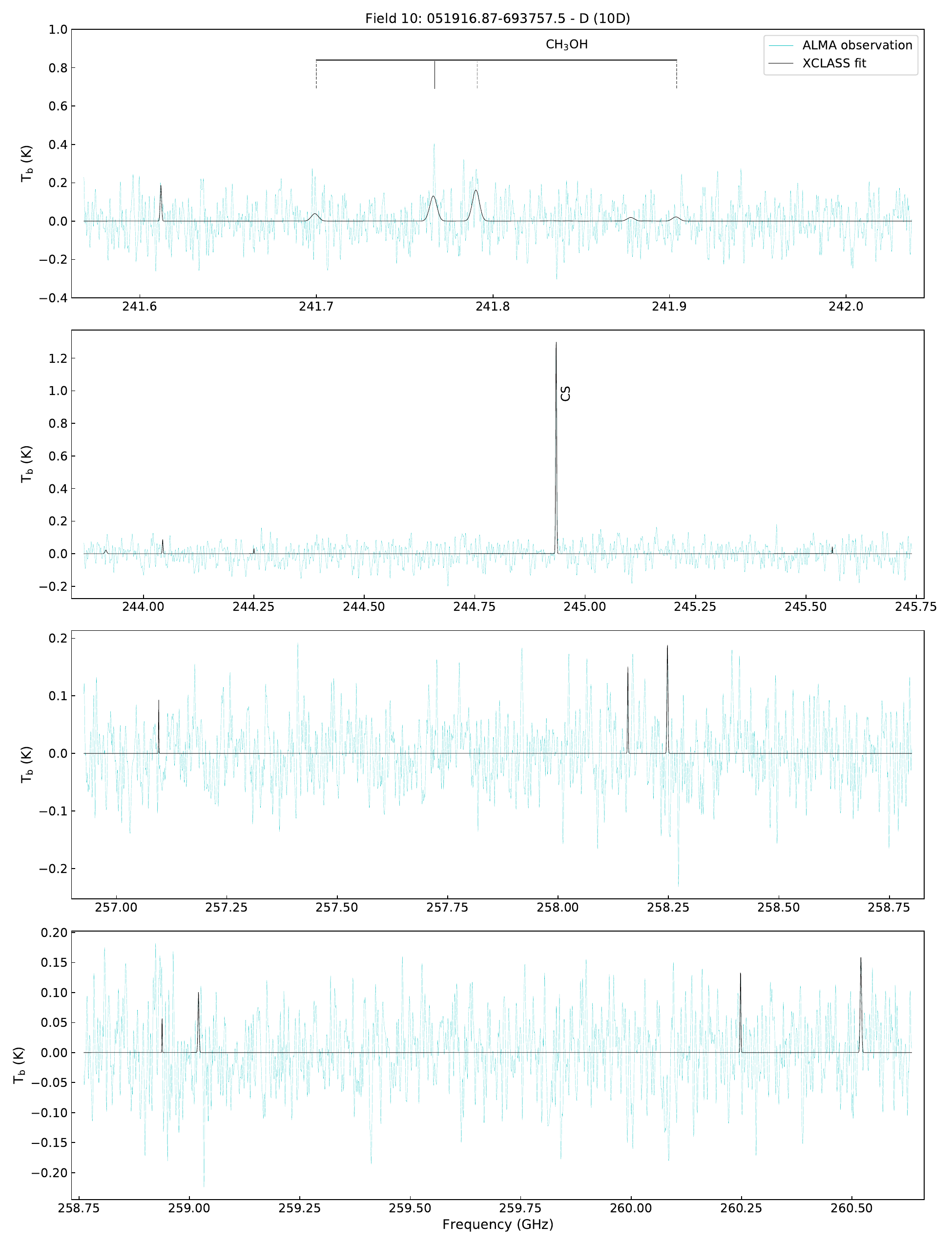}}
    \caption{Same as Fig.~\ref{fig:spectraField01} but for source 10D.}
    \label{fig:spectraField10}
\end{figure*}

%
%

\begin{figure*}
    \centering
    \subfloat[][]{\includegraphics[width=0.95\textwidth]{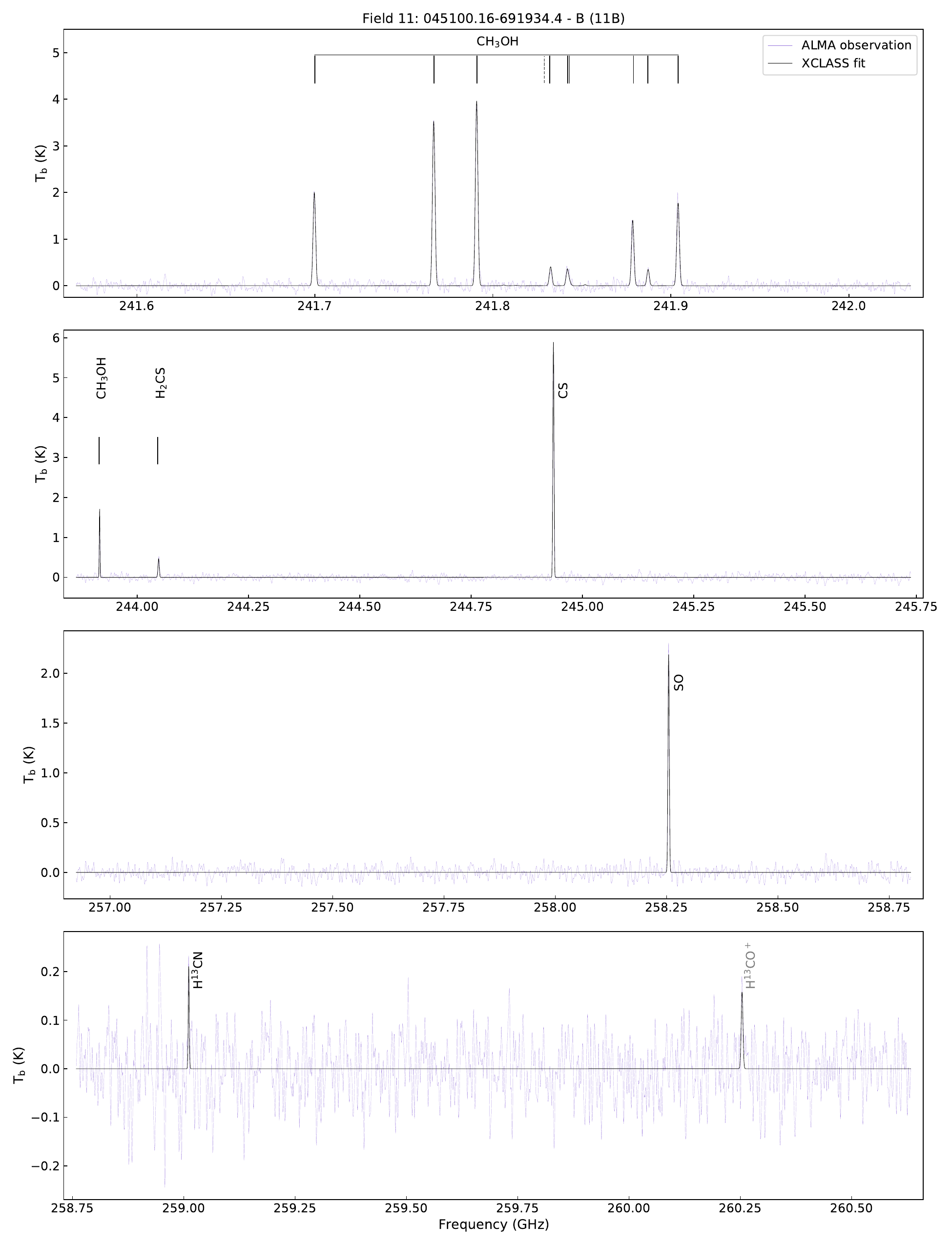}}
    \caption{Same as Fig.~\ref{fig:spectraField01} but for source 11B.}
\end{figure*}

\begin{figure*}
    \ContinuedFloat
    \centering
    \subfloat[][]{\includegraphics[width=0.95\textwidth]{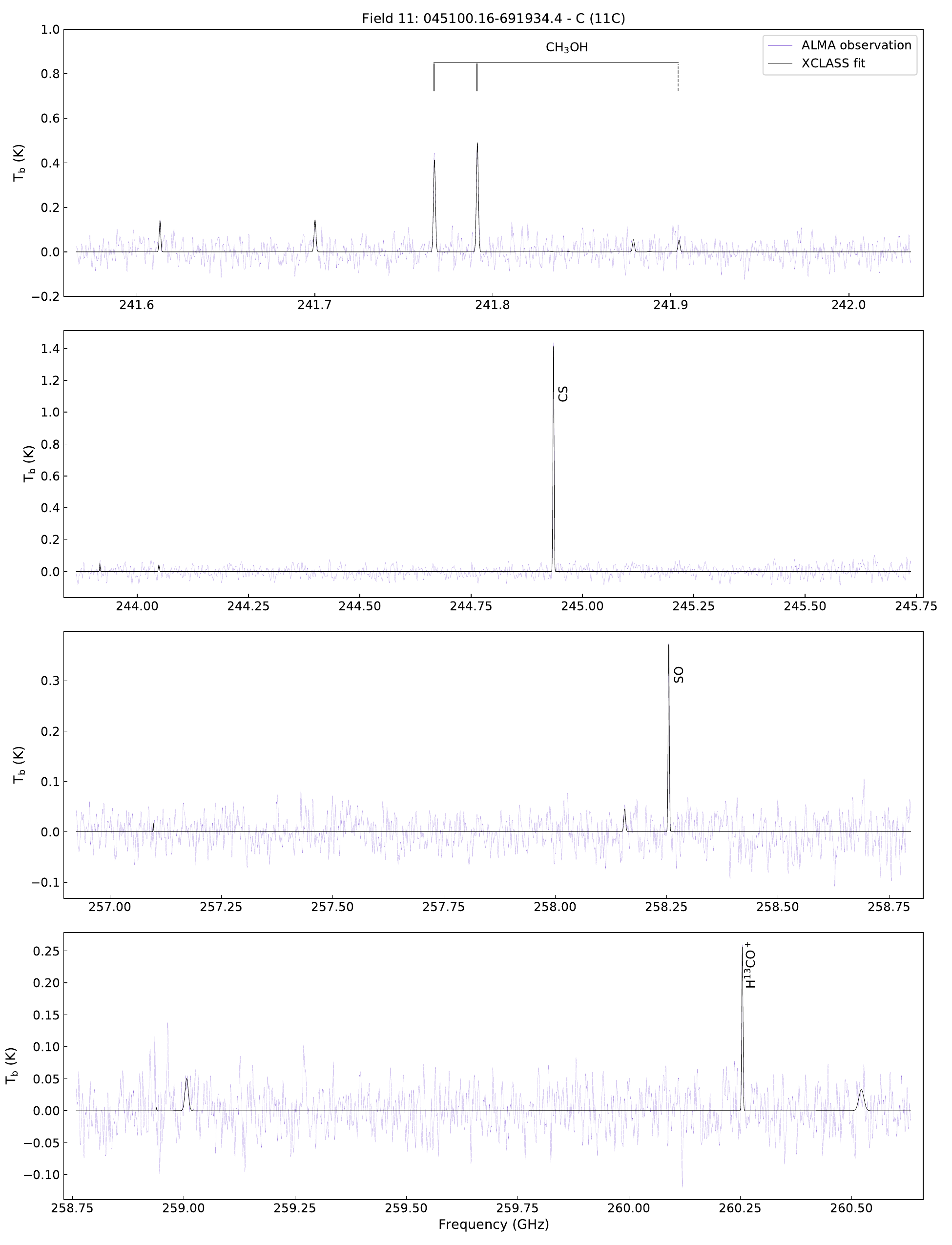}}
    \caption{Same as Fig.~\ref{fig:spectraField01} but for source 11C.}
\end{figure*}

\begin{figure*}
    \ContinuedFloat
    \centering
    \subfloat[][]{\includegraphics[width=0.95\textwidth]{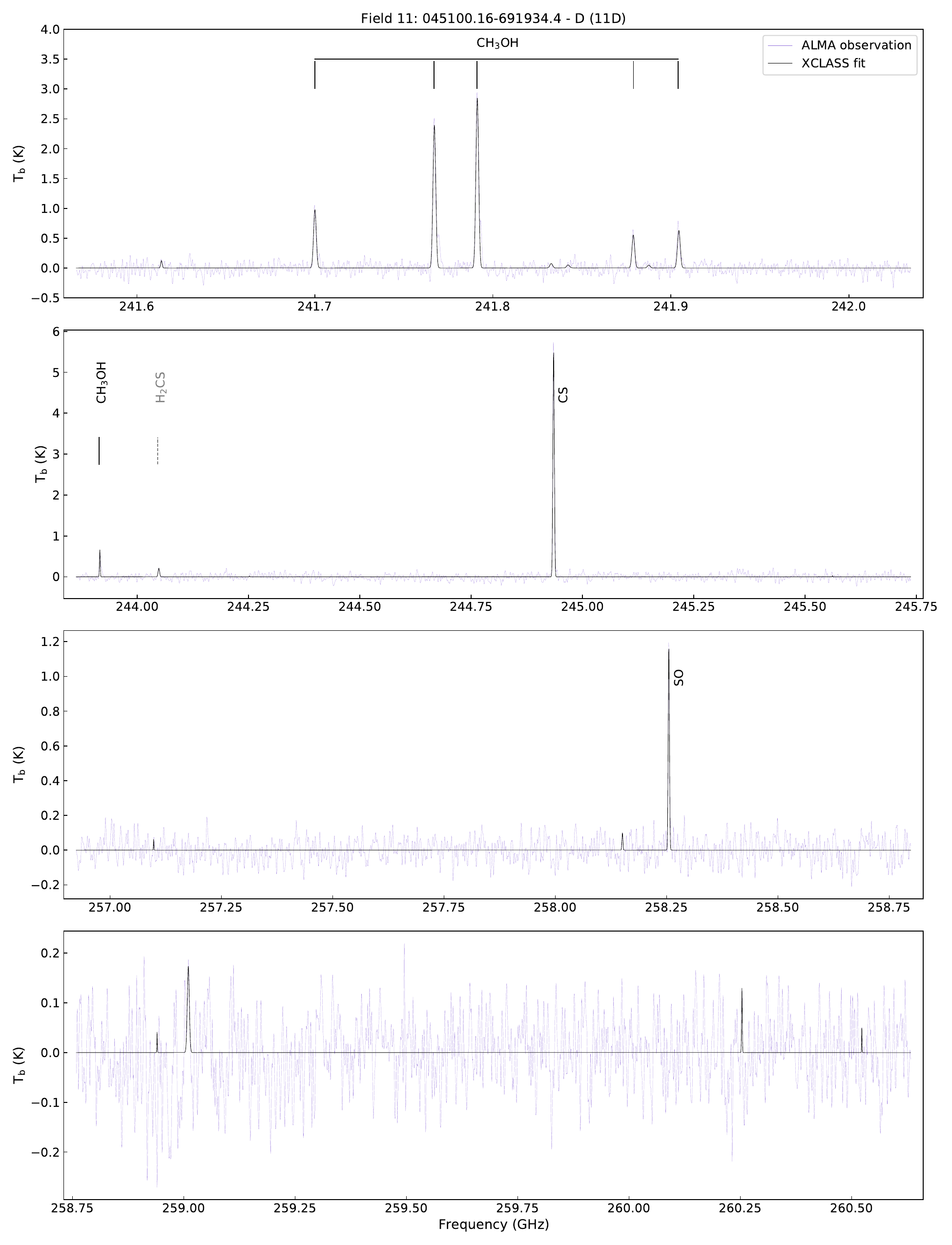}}
    \caption{Same as Fig.~\ref{fig:spectraField01} but for source 11D.}
    \label{fig:spectraField11}
\end{figure*}

%
%
\begin{figure*}
    \centering
    \subfloat[][]{\includegraphics[width=0.95\textwidth]{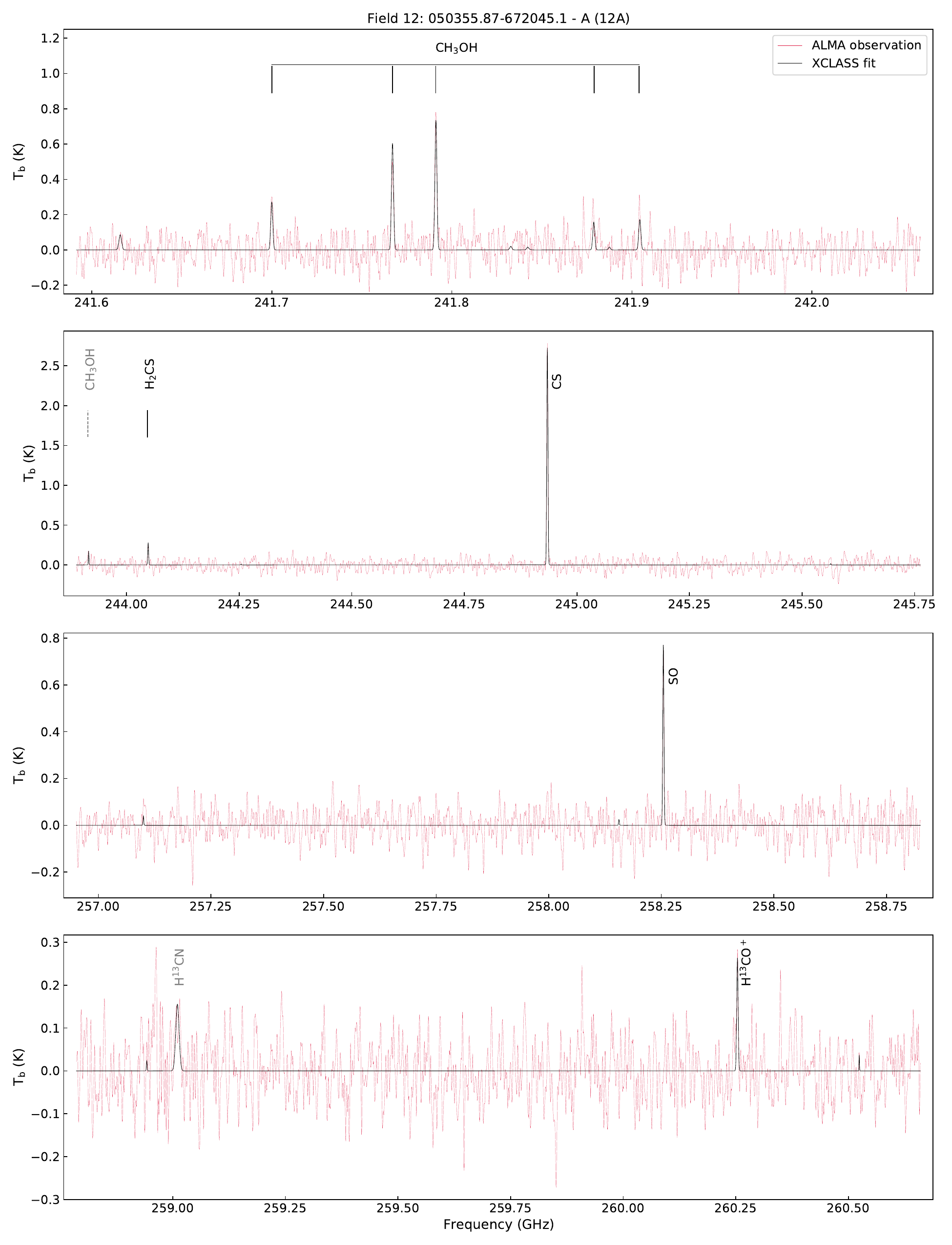}}
    \caption{Same as Fig.~\ref{fig:spectraField01} but for source 12A.}
\end{figure*}

\begin{figure*}
    \ContinuedFloat
    \centering
    \subfloat[][]{\includegraphics[width=0.95\textwidth]{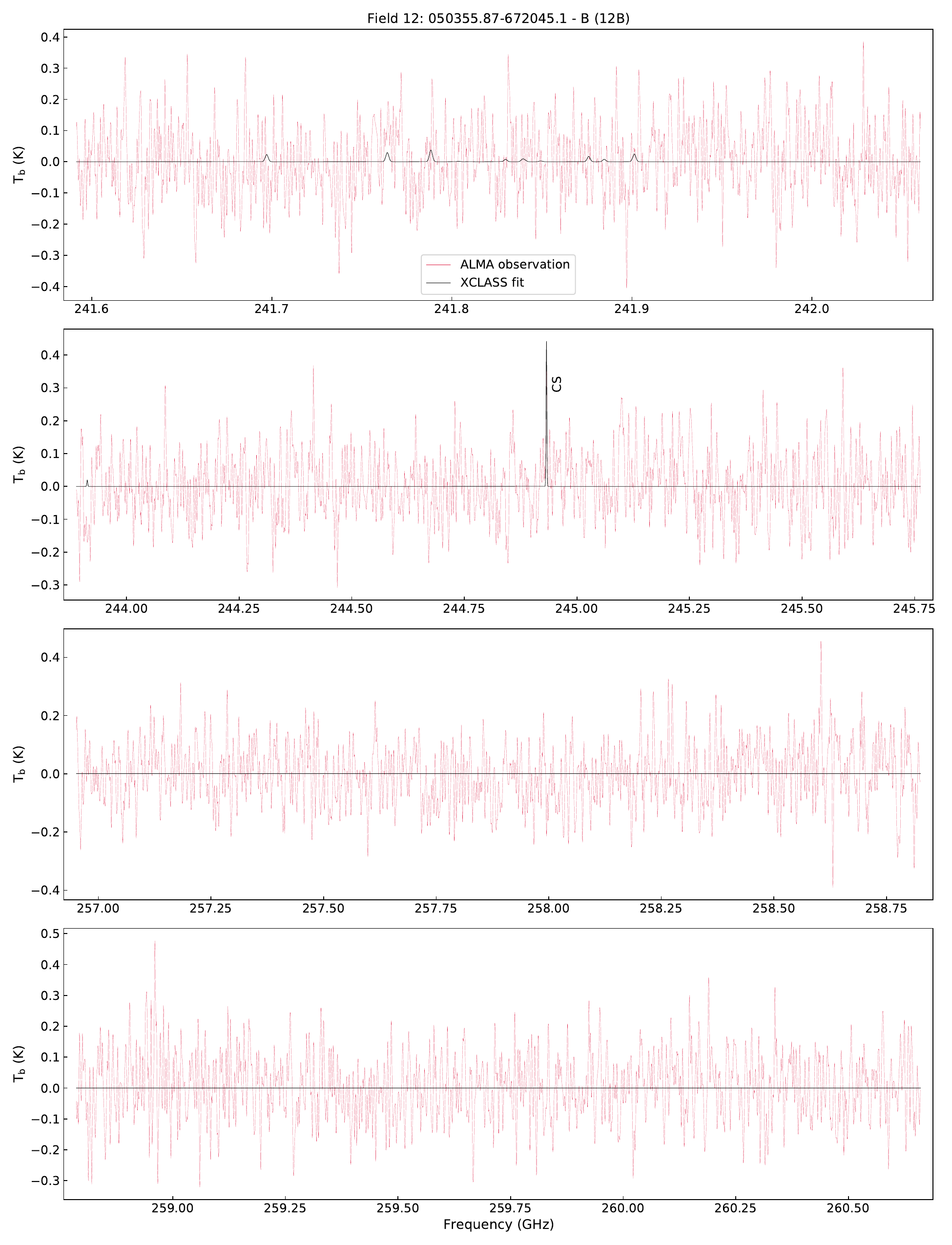}}
    \caption{Same as Fig.~\ref{fig:spectraField01} but for source 12B.}
\end{figure*}

\begin{figure*}
    \ContinuedFloat
    \centering
    \subfloat[][]{\includegraphics[width=0.95\textwidth]{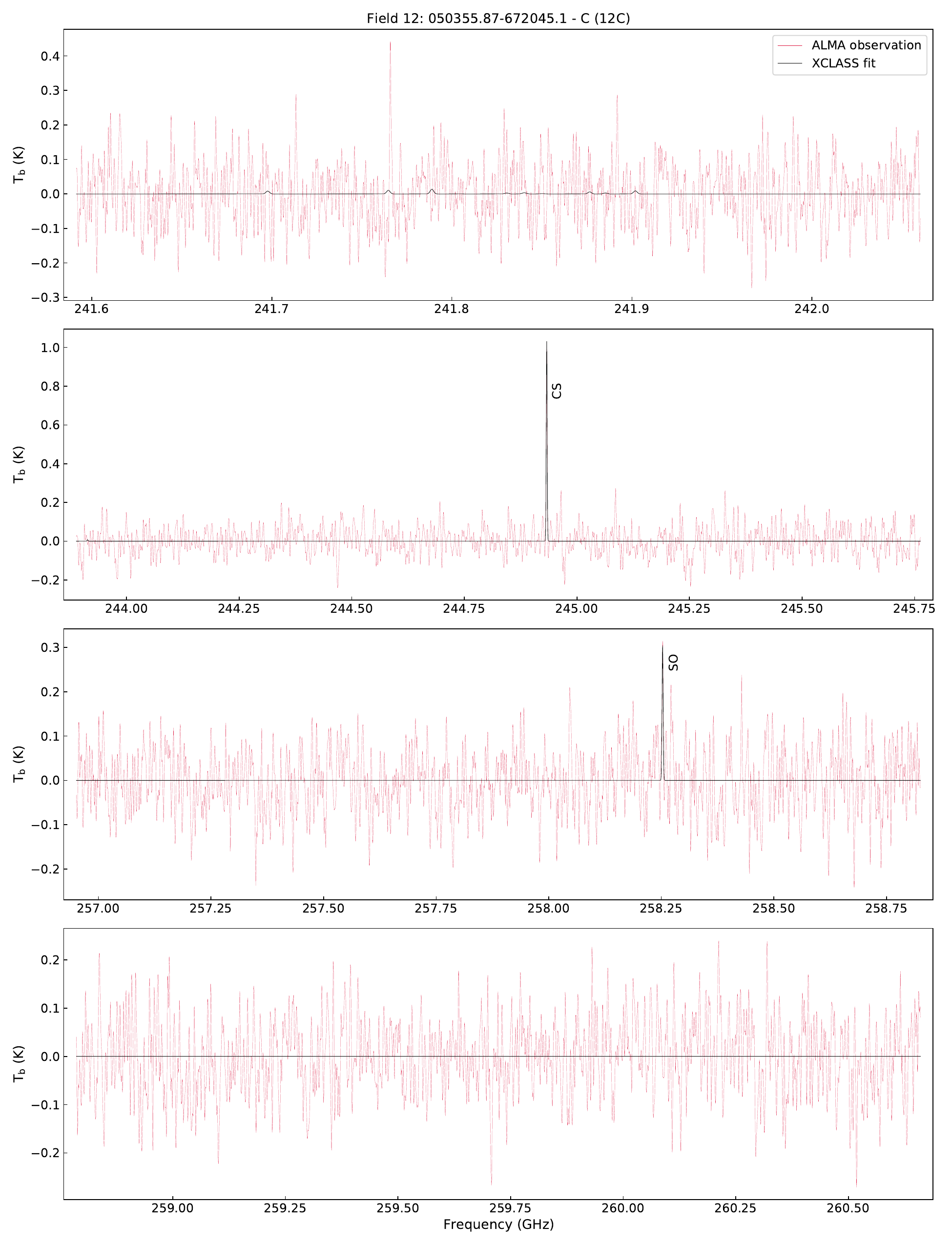}}
    \caption{Same as Fig.~\ref{fig:spectraField01} but for source 12C.}
\end{figure*}

\begin{figure*}
    \ContinuedFloat
    \centering
    \subfloat[][]{\includegraphics[width=0.95\textwidth]{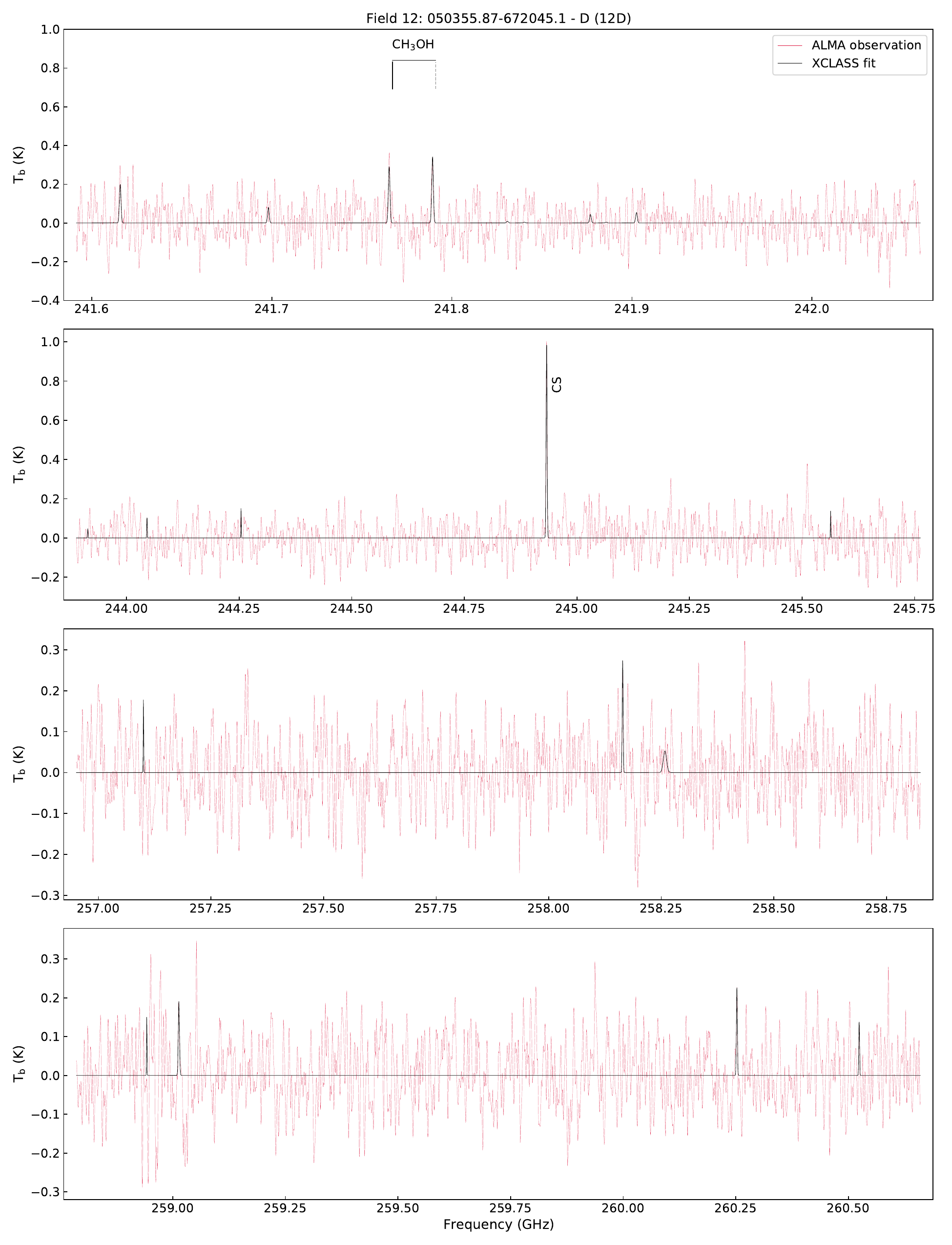}}
    \caption{Same as Fig.~\ref{fig:spectraField01} but for source 12D.}
    \label{fig:spectraField12}
\end{figure*}

%
%
\begin{figure*}
    \centering
    \subfloat[][]{\includegraphics[width=0.95\textwidth]{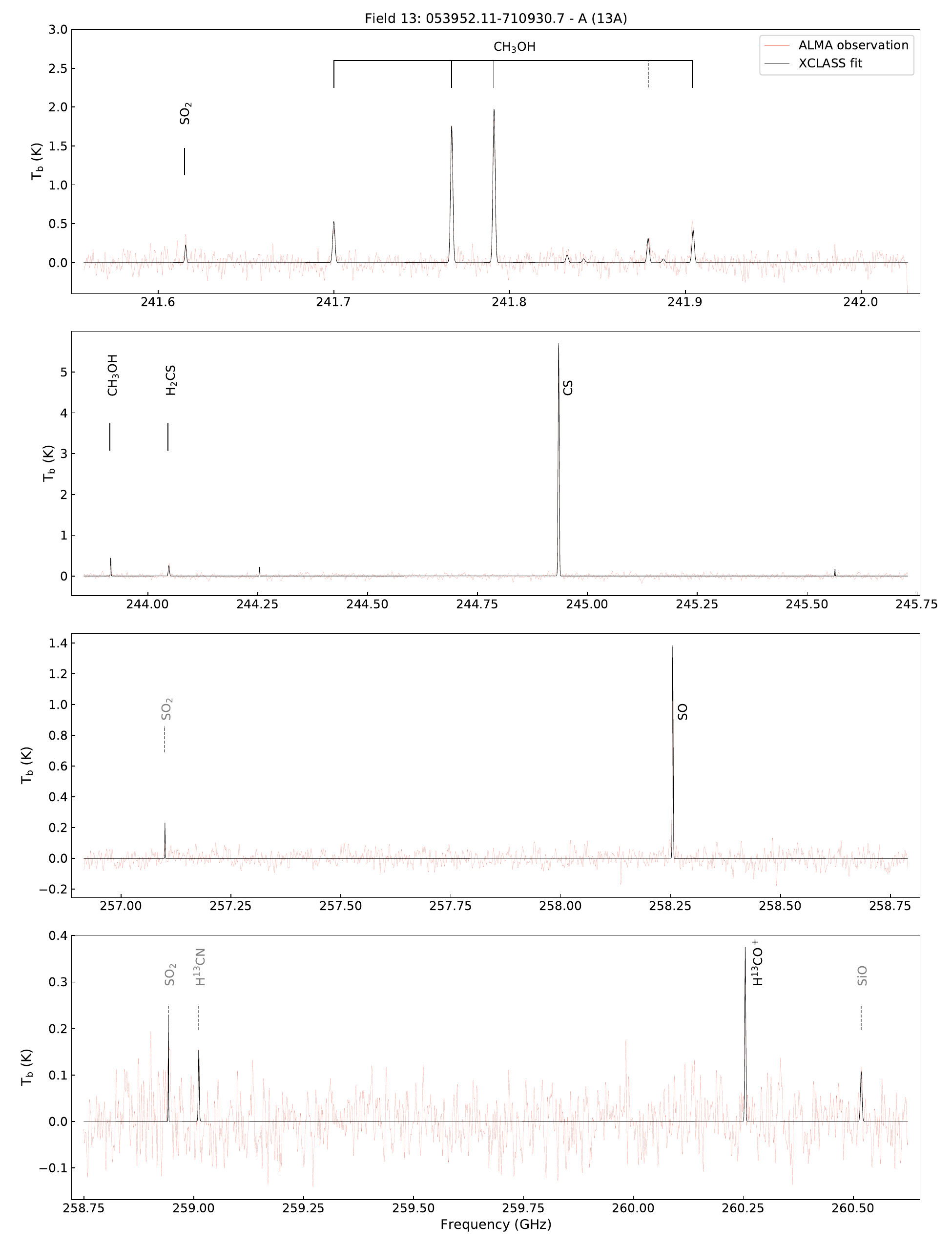}}
    \caption{Same as Fig.~\ref{fig:spectraField01} but for source 13A.}
\end{figure*}


\begin{figure*}
    \ContinuedFloat
    \centering
    \subfloat[][]{\includegraphics[width=0.95\textwidth]{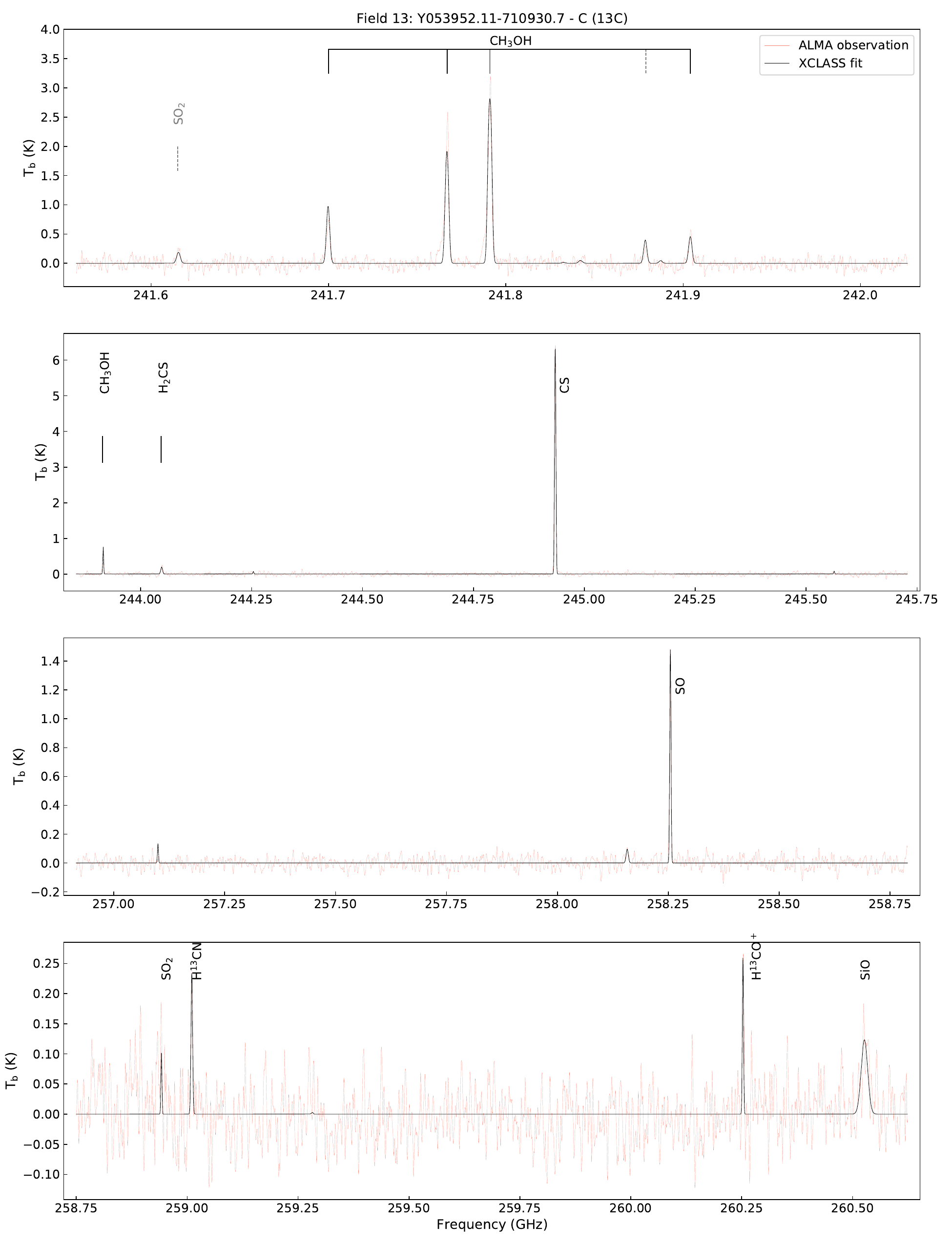}}
    \caption{Same as Fig.~\ref{fig:spectraField01} but for source 13C.}
\end{figure*}

\begin{figure*}
    \ContinuedFloat
    \centering
    \subfloat[][]{\includegraphics[width=0.95\textwidth]{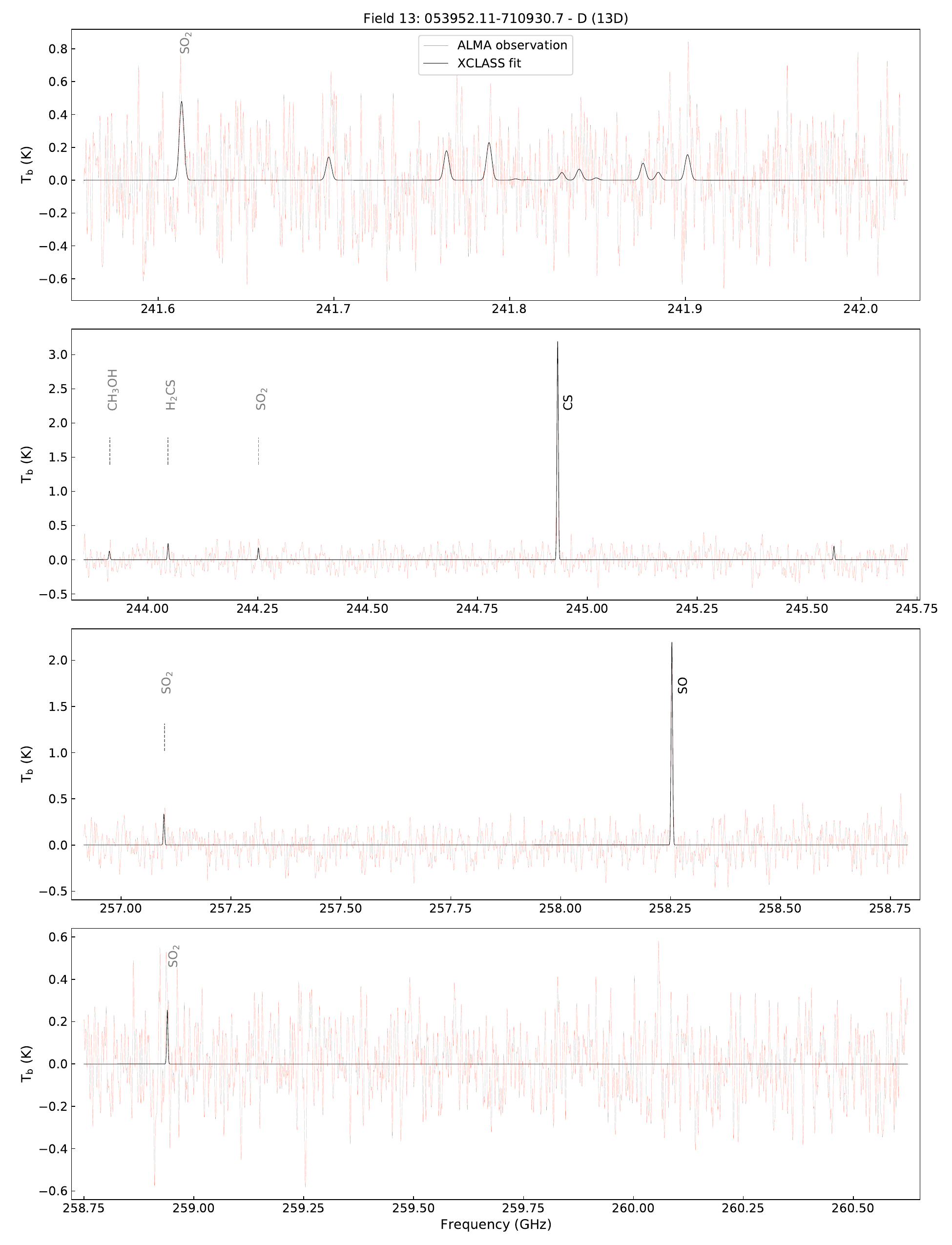}}
    \caption{Same as Fig.~\ref{fig:spectraField01} but for source 13D.}
\end{figure*}

\begin{figure*}
    \ContinuedFloat
    \centering
    \subfloat[][]{\includegraphics[width=0.95\textwidth]{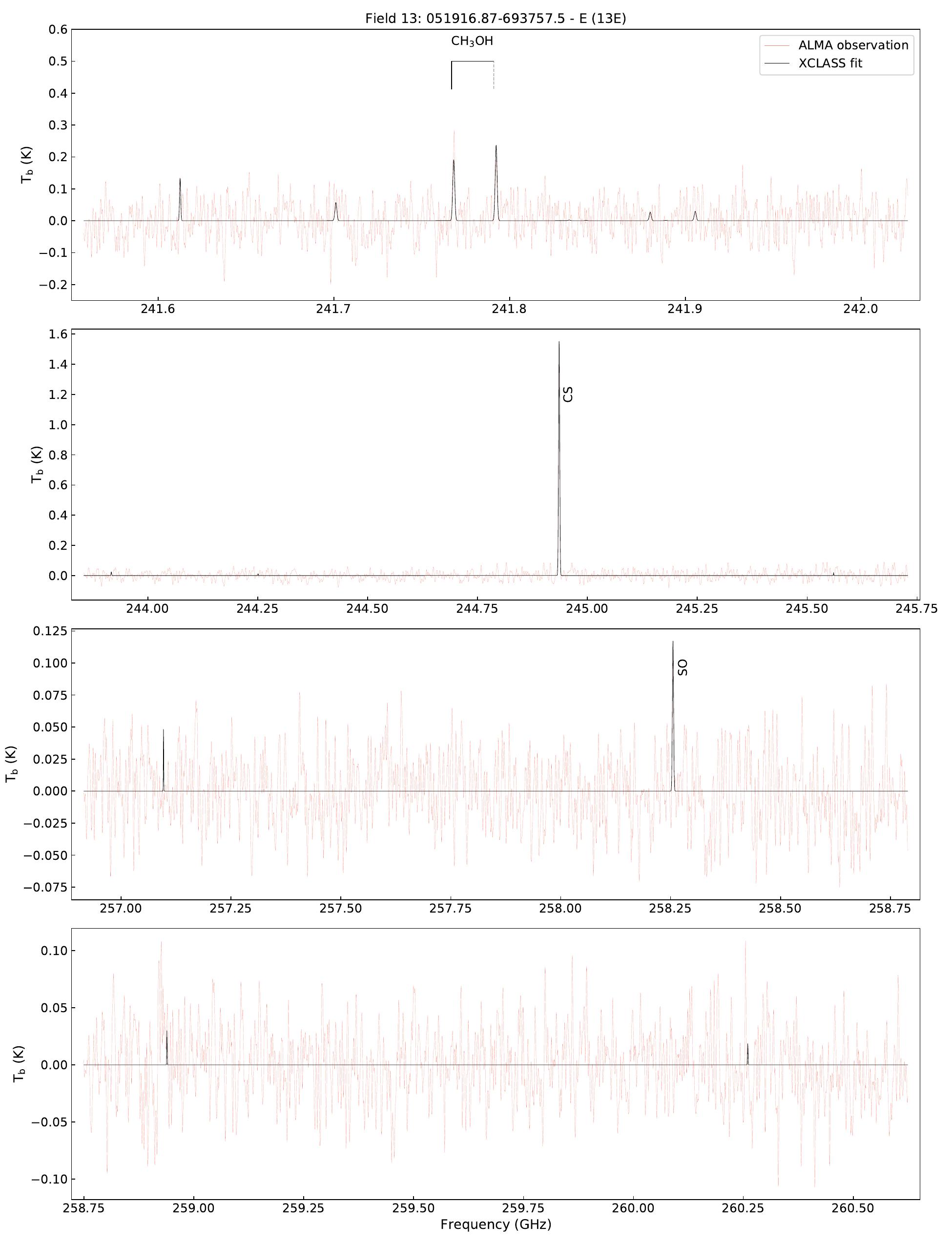}}
    \caption{Same as Fig.~\ref{fig:spectraField01} but for source 13E.}
\end{figure*}

\begin{figure*}
    \ContinuedFloat
    \centering
    \subfloat[][]{\includegraphics[width=0.95\textwidth]{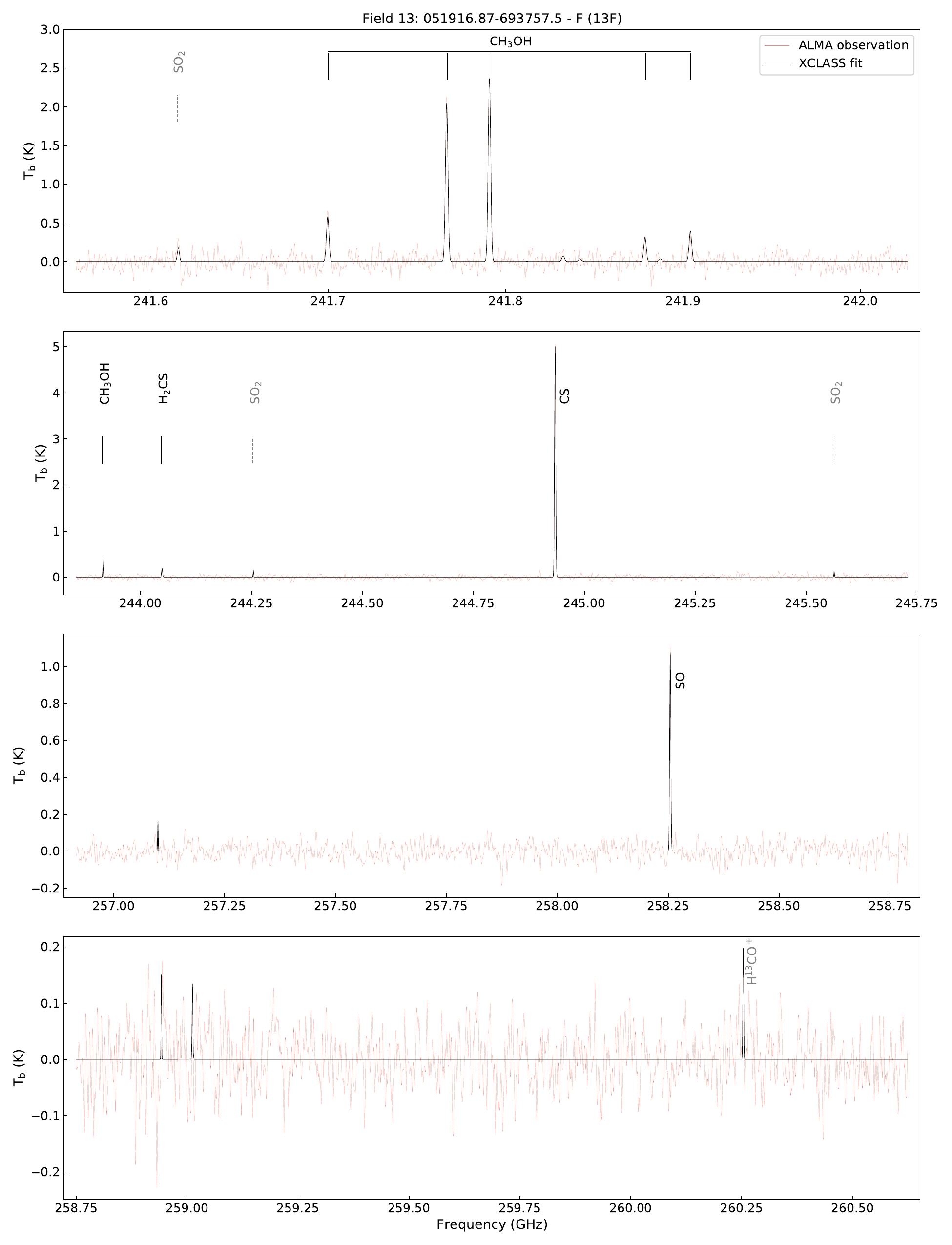}}
    \caption{Same as Fig.~\ref{fig:spectraField01} but for source 13F.}
\end{figure*}

\begin{figure*}
    \ContinuedFloat
    \centering
    \subfloat[][]{\includegraphics[width=0.95\textwidth]{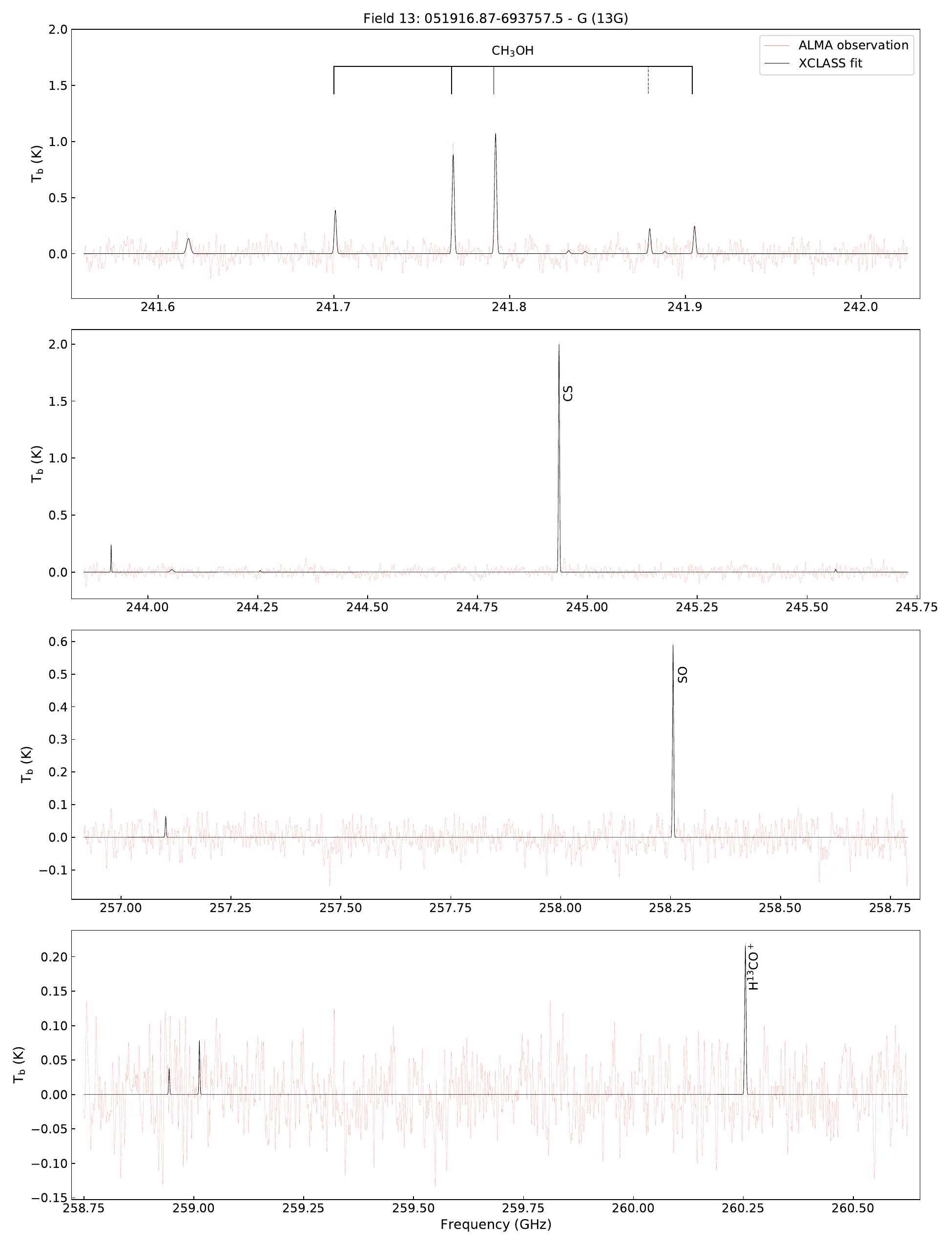}}
    \caption{Same as Fig.~\ref{fig:spectraField01} but for source 13G.}
\end{figure*}

\begin{figure*}
    \ContinuedFloat
    \centering
    \subfloat[][]{\includegraphics[width=0.95\textwidth]{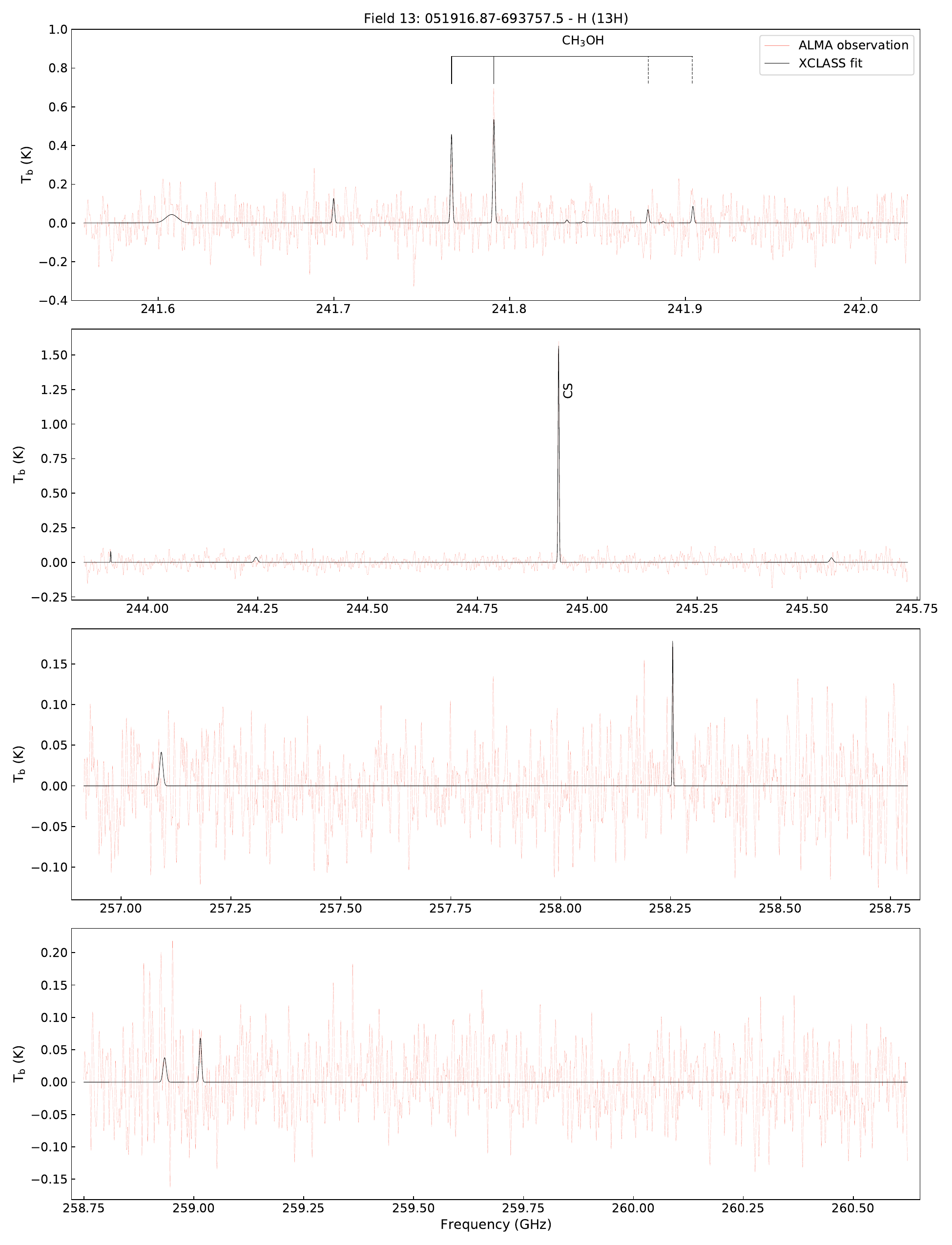}}
    \caption{Same as Fig.~\ref{fig:spectraField01} but for source 13H.}
\end{figure*}

\begin{figure*}
    \ContinuedFloat
    \centering
    \subfloat[][]{\includegraphics[width=0.95\textwidth]{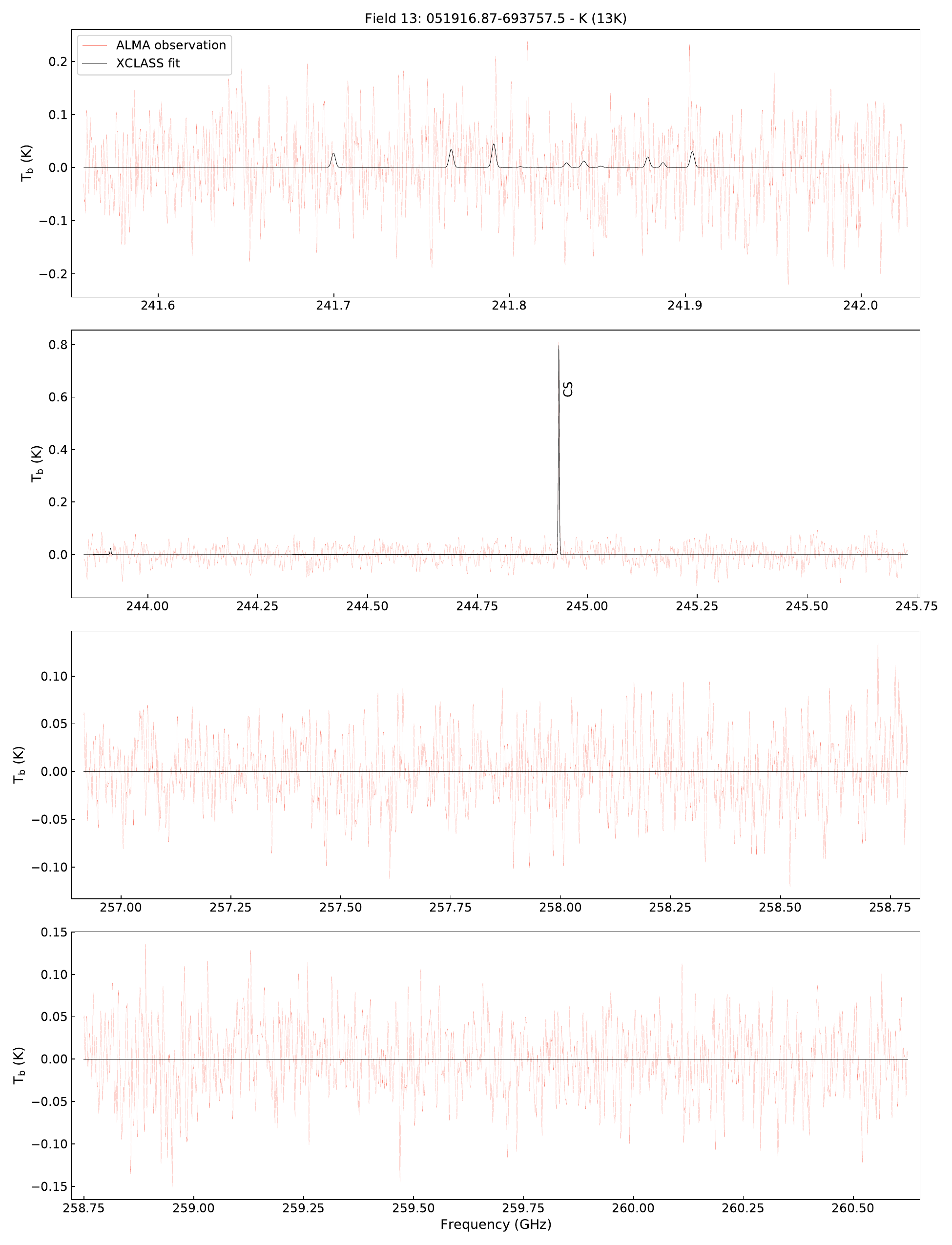}}
    \caption{Same as Fig.~\ref{fig:spectraField01} but for source 13K.}
    \label{fig:spectraField13}
\end{figure*}

%
%

\clearpage
%
%
\begin{figure*}
    \centering
    \subfloat[][]{\includegraphics[width=0.95\textwidth]{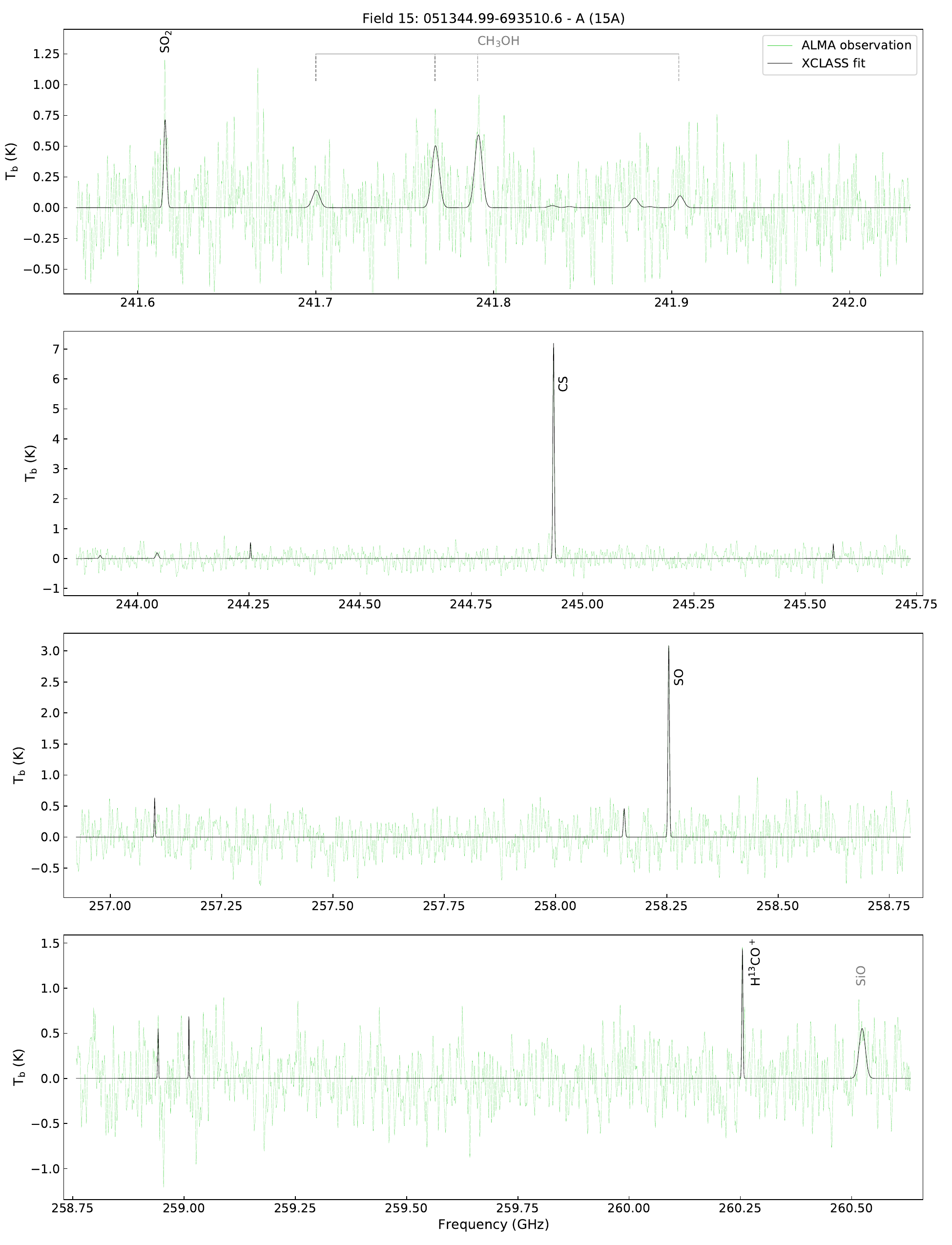}}
    \caption{Same as Fig.~\ref{fig:spectraField01} but for source 15A.}
\end{figure*}

\begin{figure*}
    \ContinuedFloat
    \centering
    \subfloat[][]{\includegraphics[width=0.95\textwidth]{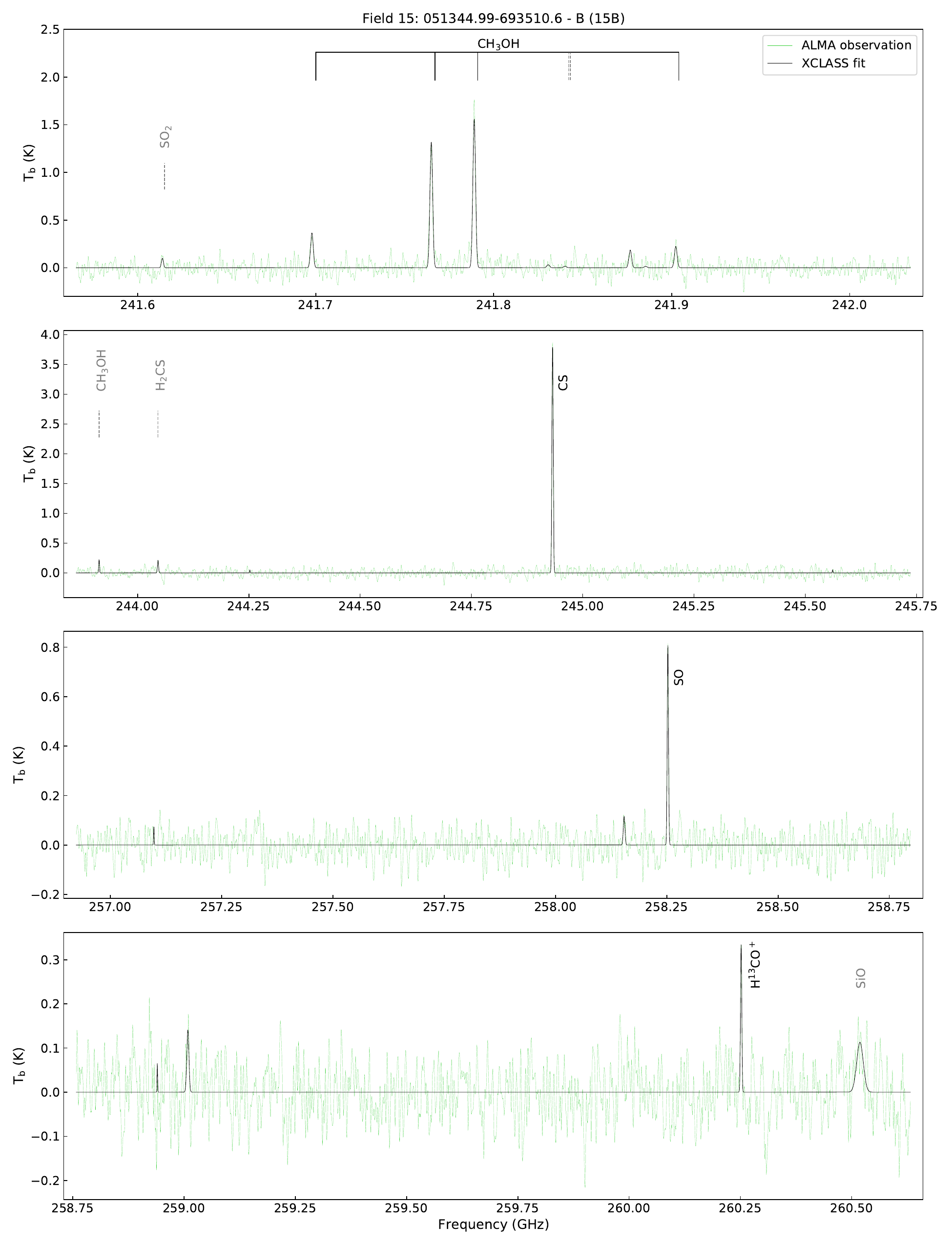}}
    \caption{Same as Fig.~\ref{fig:spectraField01} but for source 15B.}
\end{figure*}

\begin{figure*}
    \ContinuedFloat
    \centering
    \subfloat[][]{\includegraphics[width=0.95\textwidth]{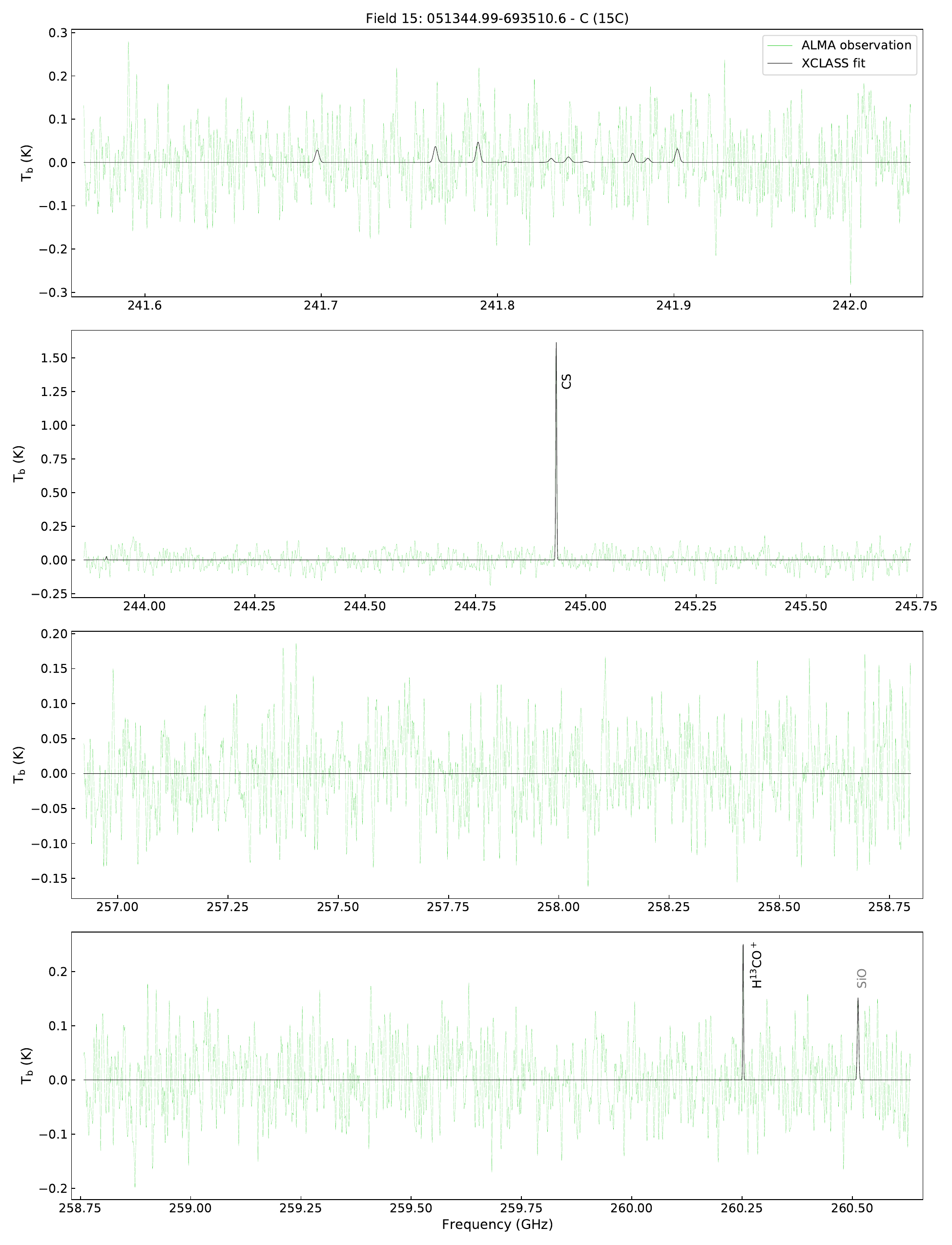}}
    \caption{Same as Fig.~\ref{fig:spectraField01} but for source 15C.}
    \label{fig:spectraField15}
\end{figure*}

%
%

%
%


%
%
\begin{figure*}
    \centering
    \subfloat[][]{\includegraphics[width=0.95\textwidth]{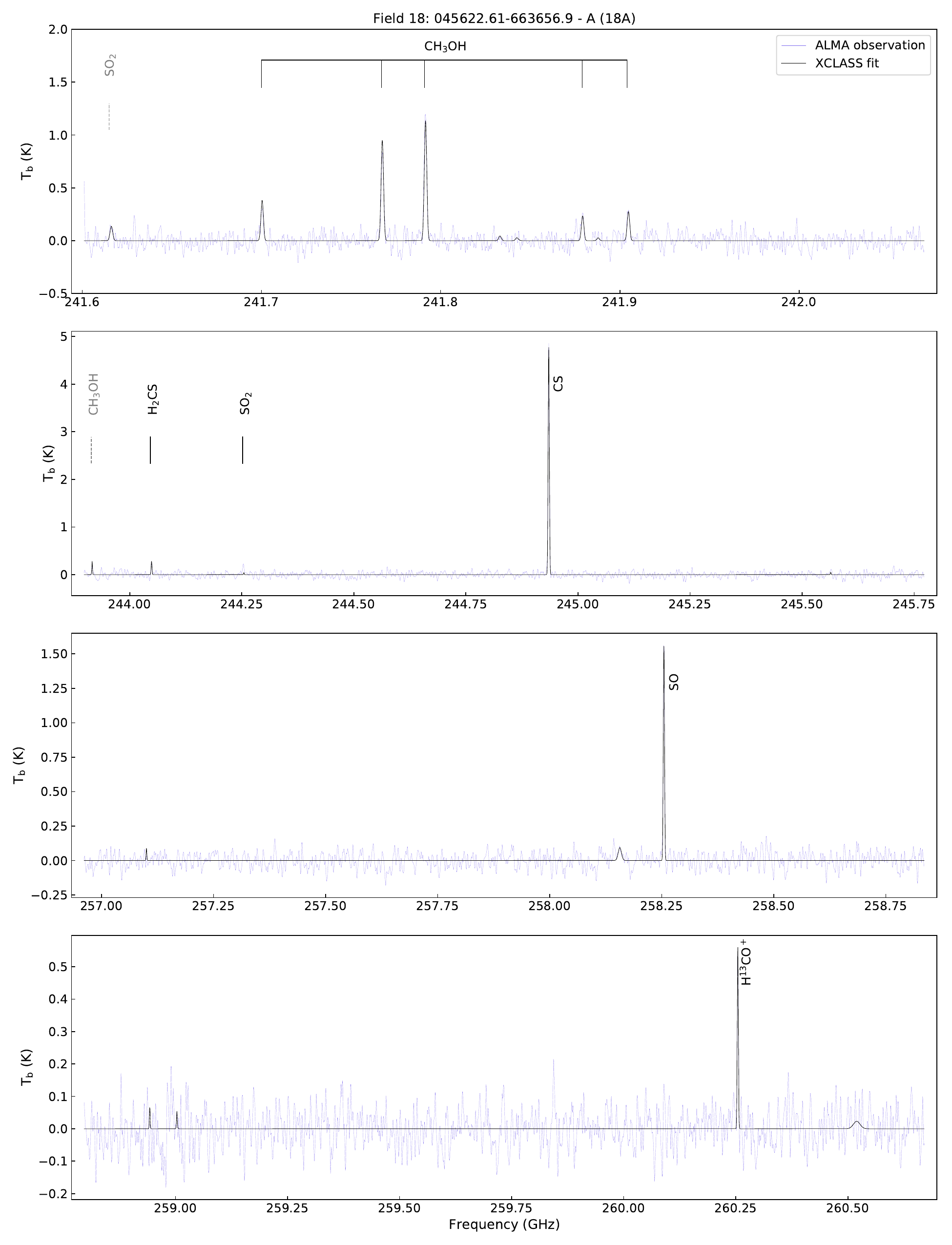}}
    \caption{Same as Fig.~\ref{fig:spectraField01} but for source 18A.}
\end{figure*}

\begin{figure*}
    \ContinuedFloat
    \centering
    \subfloat[][]{\includegraphics[width=0.95\textwidth]{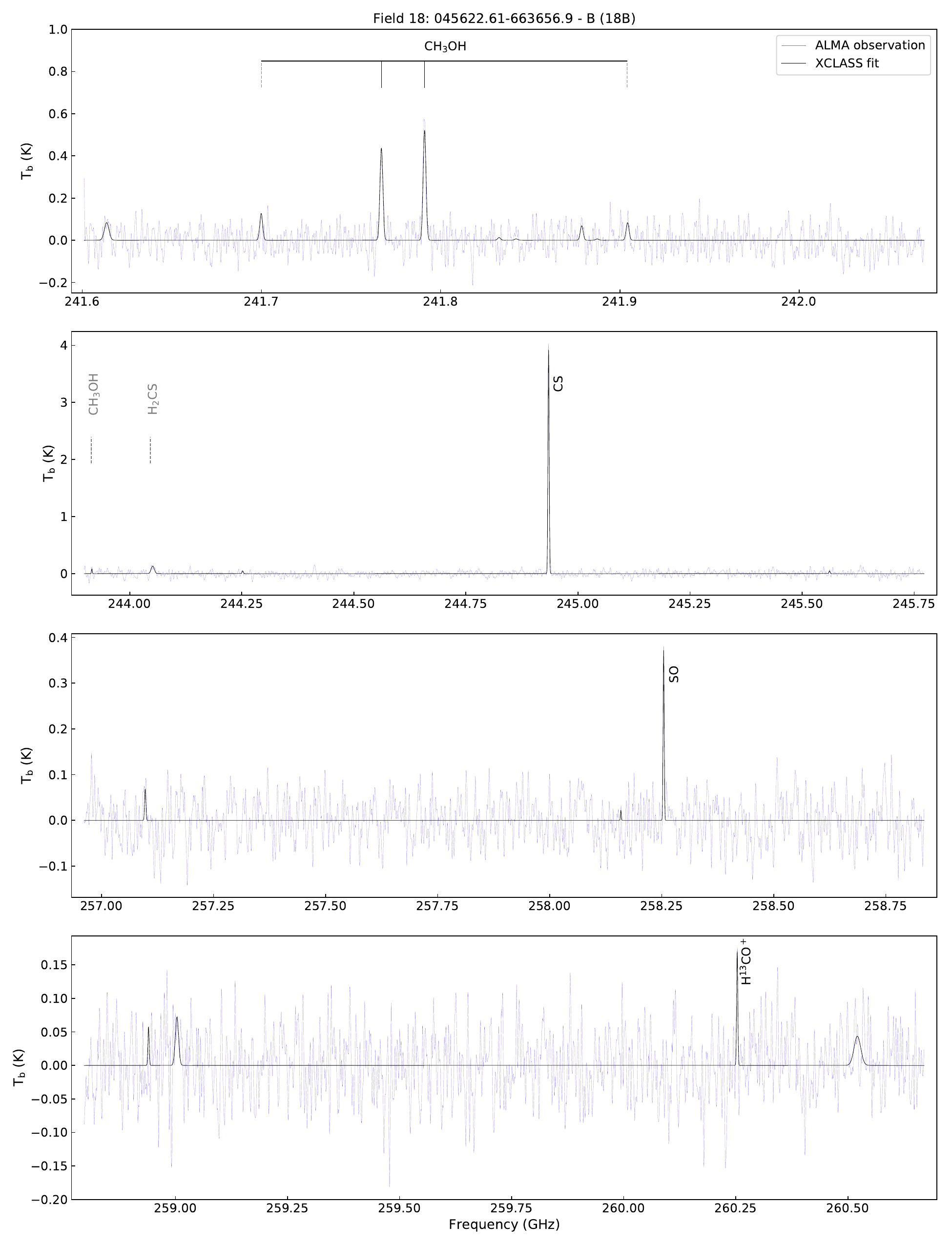}}
    \caption{Same as Fig.~\ref{fig:spectraField01} but for source 18B.}
\end{figure*}

\begin{figure*}
    \ContinuedFloat
    \centering
    \subfloat[][]{\includegraphics[width=0.95\textwidth]{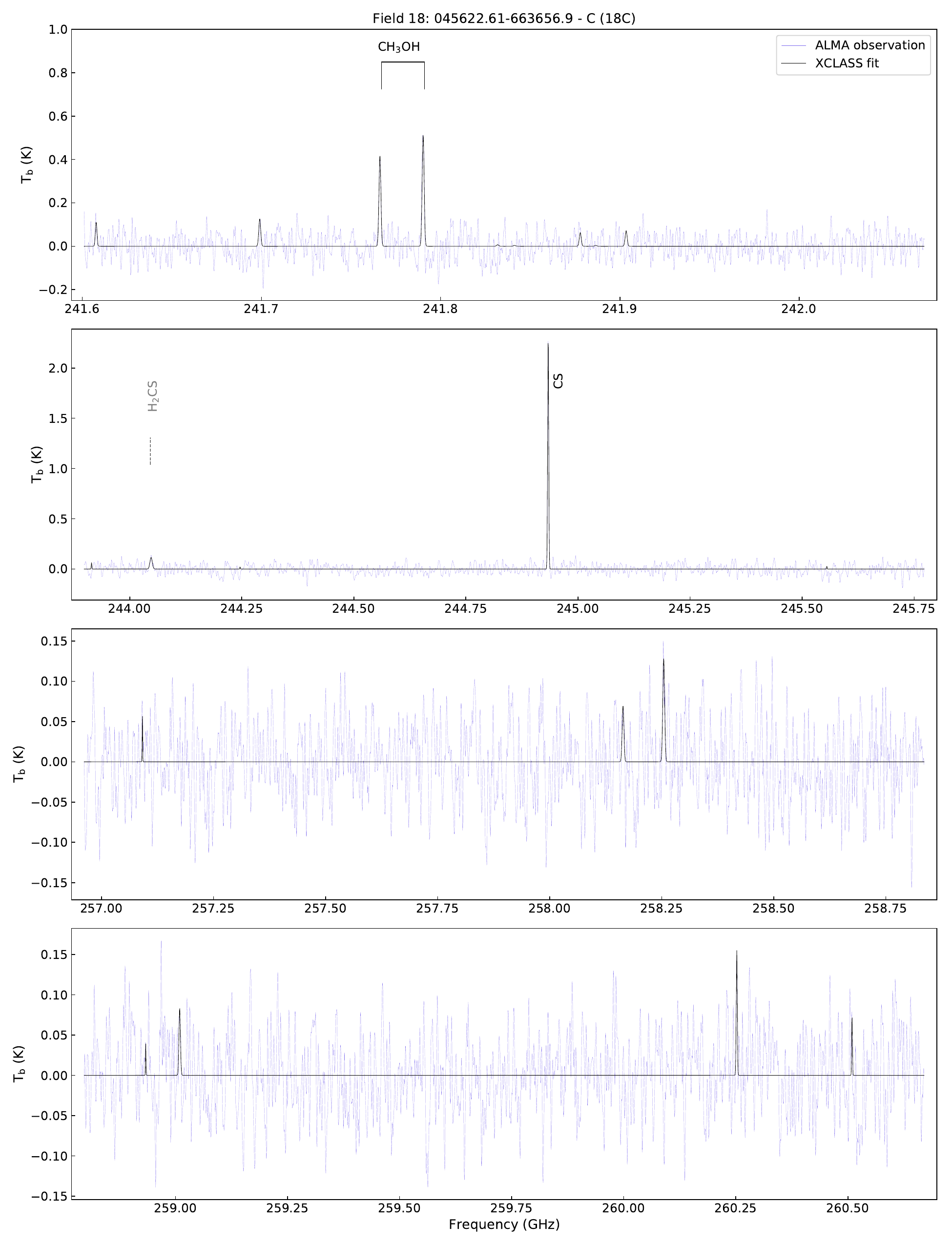}}
    \caption{Same as Fig.~\ref{fig:spectraField01} but for source 18C.}
    \label{fig:spectraField18}
\end{figure*}

%
%
\begin{figure*}
    \centering
    \subfloat[][]{\includegraphics[width=0.95\textwidth]{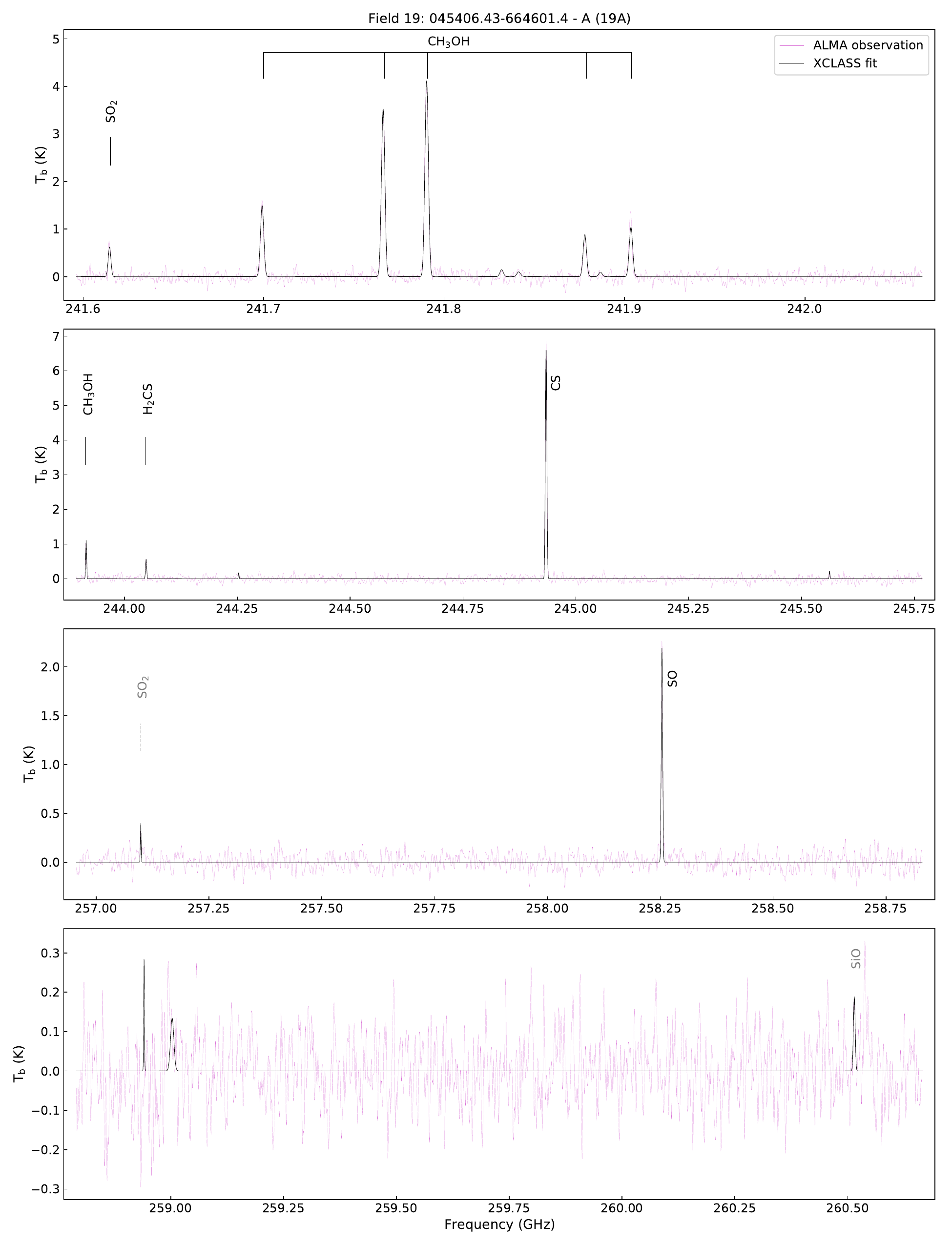}}
    \caption{Same as Fig.~\ref{fig:spectraField01} but for source 19A.}
\end{figure*}


\begin{figure*}
    \ContinuedFloat
    \centering
    \subfloat[][]{\includegraphics[width=0.95\textwidth]{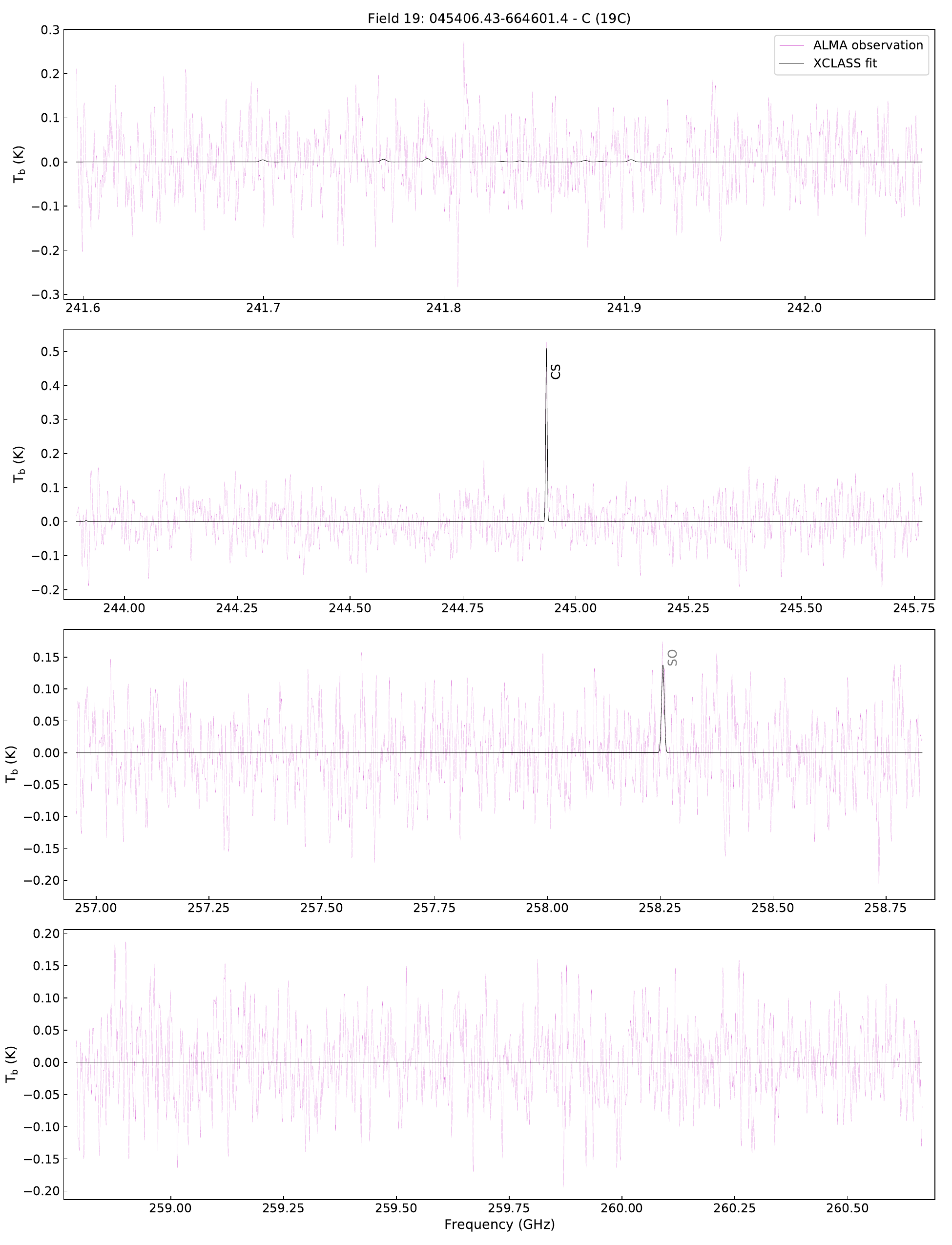}}
    \caption{Same as Fig.~\ref{fig:spectraField01} but for source 19C.}
\end{figure*}

\begin{figure*}
    \ContinuedFloat
    \centering
    \subfloat[][]{\includegraphics[width=0.95\textwidth]{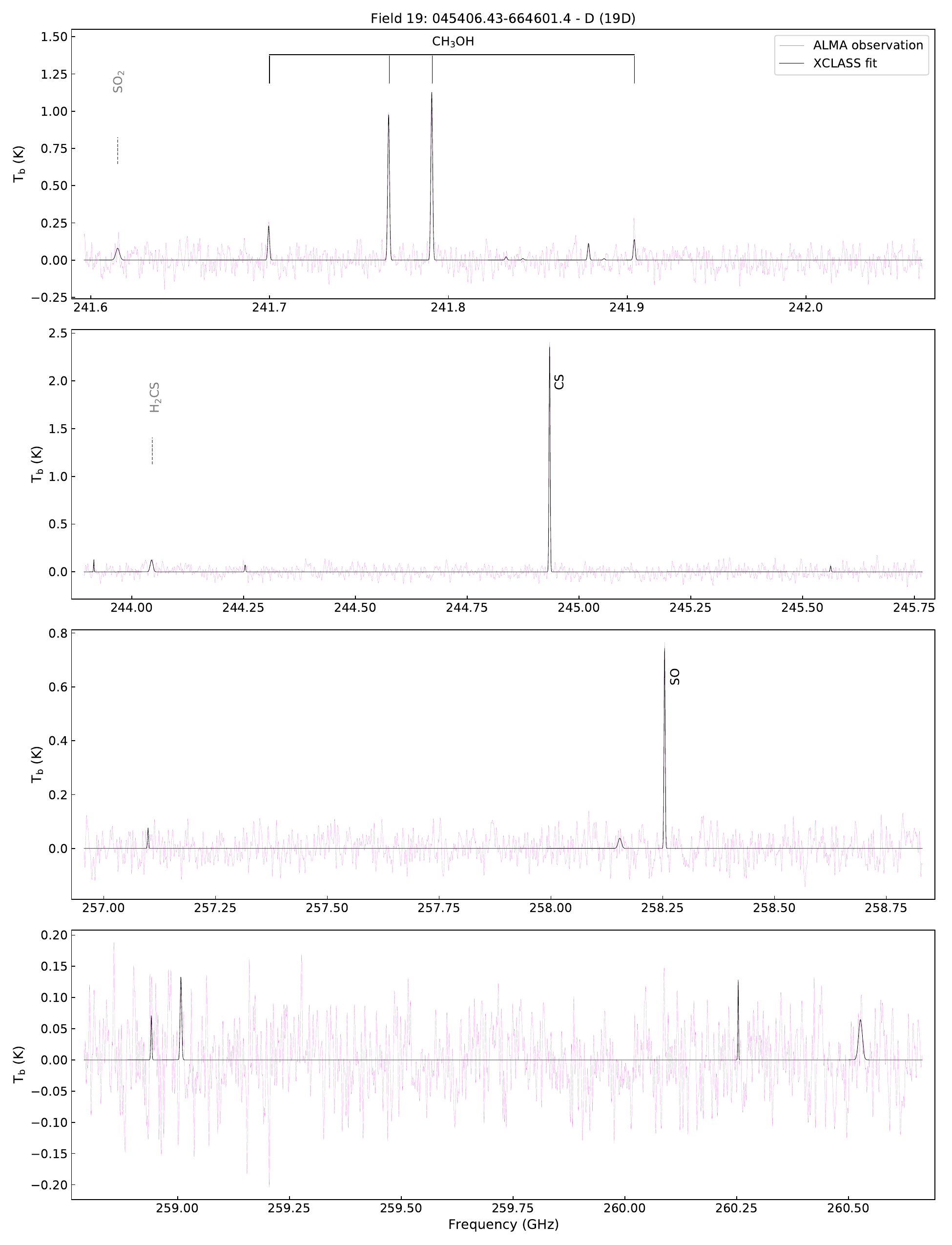}}
    \caption{Same as Fig.~\ref{fig:spectraField01} but for source 19D.}
    \label{fig:spectraField19}
\end{figure*}

%
%
%
%
\clearpage
\section{XCLASS fits parameters}\label{app:XCLASS_params}

\begin{figure}[h!]
\centering
\includegraphics[scale=0.55]{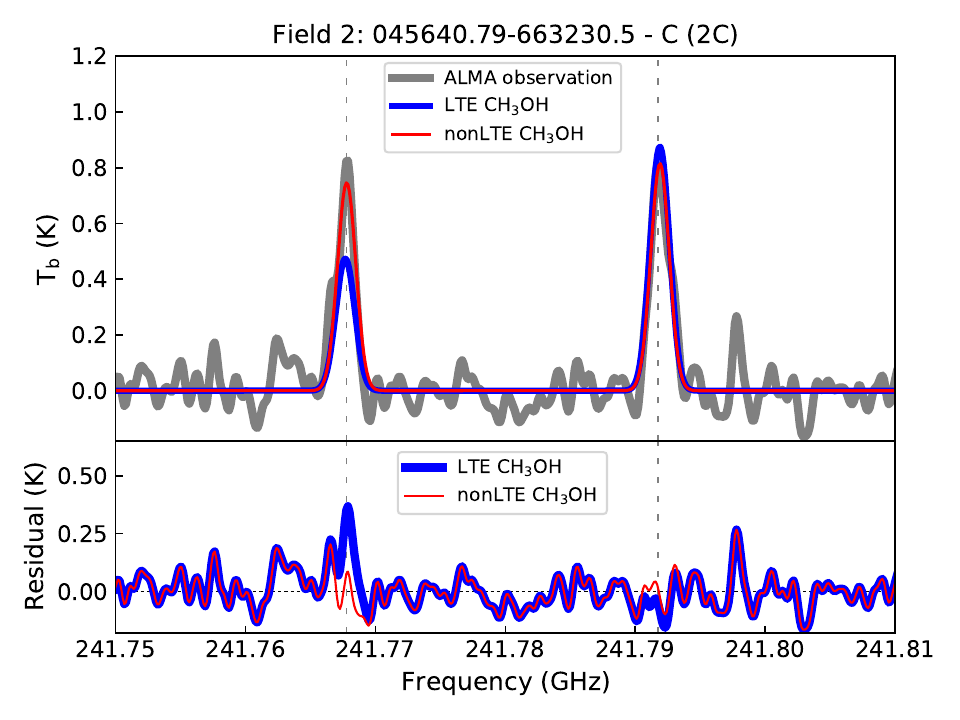}
\includegraphics[scale=0.55]{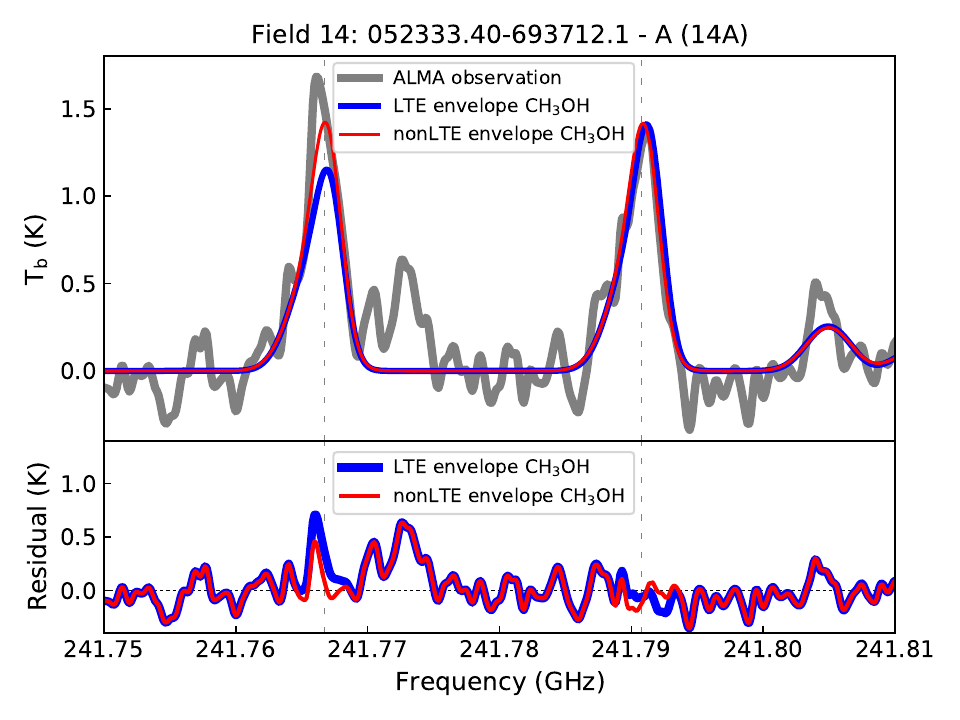}
\caption{Comparisons of the LTE and non-LTE fit for two \ce{CH3OH} lines at 241.767 and 241.791 GHz.}
\label{fig:residual_spectra}
\end{figure}

In Table~\ref{tab:XCLASS_LTE}, we list the parameters assuming LTE conditions for the \ce{CH3OH}, and Table~\ref{tab:XCLASS_params} collects the fit parameters for \ce{CH3OH} in non-LTE. We used the non-LTE temperature for methanol to fit other molecules. Therefore, the fit parameters for other molecules are only presented in Table~\ref{tab:XCLASS_params}. (For more details, see Section~\ref{sec:spectra}.) Figure~\ref{fig:residual_spectra} compares the LTE and non-LTE fit for two methanol lines at 241.767 and 241.792 for two sources, 2C and 14A. While methanol is fit with one component in 2C, 14A gets the best-fitting results with two components for methanol, where the cold foreground component (the envelope) is assumed to be in non-LTE.

\begin{table}[h!]
\centering
\caption{Parameters of the LTE fits for \ce{CH3OH}.} 
\label{tab:XCLASS_LTE}
\begin{tabular}{c c c c c c}
\hline\hline \noalign{\smallskip}
& ${\rm T}$
& ${\rm N(X)}$
& $\Delta$v$_\mathrm{FWHM}$
& v$_\mathrm{LSR}$
\\
Source
& (K)
& (cm$^{-2}$)
& (km~s$^{-1}$)
& (km~s$^{-1}$)
\\
\hline \noalign{\smallskip}
	1A  & 16 & \num{2.09E+14} & 4.3 & 224.8 \\
	2A  & 12 & \num{5.29E+13} & 1.3 & 269.2 \\
	2B  & 12 & \num{3.48E+14} & 1.8 & 268.7 \\
	2C  & 9  & \num{3.50E+14} & 2.1 & 267.8 \\
	2D  & 16 & \num{4.51E+13} & 3.5 & 268.4 \\
	2E  & 10 & \num{1.09E+14} & 2.0 & 269.0 \\
	3A  & 16 & \num{3.72E+14} & 3.9 & 285.2 \\
	3B  & 11 & \num{9.17E+13} & 1.8 & 284.7 \\
	3C  & 10 & \num{1.15E+14} & 2.1 & 286.3 \\
	3D  & 10 & \num{1.39E+14} & 1.6 & 286.3 \\
    4A  & 144 & \num{4.88E+15} & 3.17 & 264.3 \\
        & 20 & \num{1.35E+14} & 4.49 & 264.2 \\
	5A  & 5  & \num{1.11E+15} & 2.0 & 246.1 \\
	5B  & 15 & \num{1.41E+14} & 2.3 & 242.3 \\
    6A  & 170 & \num{6.56E+15} & 5.74 & 242.3 \\
        & 19 & \num{4.17E+14} & 3.80& 242.9 \\
    6B  & 228 & \num{9.15E+15} & 3.21 & 243.3 \\
        & 33 & \num{3.27E+14} & 3.47 & 241.2 \\
	6C  & 12 & \num{5.35E+14} & 3.7 & 242.0 \\
	6D  & 8  & \num{5.36E+15} & 2.7 & 242.1 \\
	6E  & 12 & \num{1.54E+15} & 2.5 & 243.0 \\
	6F  & 9  & \num{2.39E+14} & 1.8 & 241.3 \\
	7A  & 10 & \num{3.84E+14} & 1.7 & 252.8 \\
	8A  & 15 & \num{4.04E+14} & 1.9 & 238.0 \\
	8C  & 10 & \num{7.57E+14} & 2.6 & 239.9 \\
	9A  & 11 & \num{5.31E+14} & 1.8 & 241.6 \\
	10A & 10 & \num{3.77E+14} & 2.9 & 239.3 \\
	10C & 11 & \num{1.08E+14} & 1.6 & 239.3 \\
	10D & 10 & \num{8.89E+13} & 2.8 & 239.3 \\
	11B & 17 & \num{8.15E+14} & 1.9 & 236.0 \\
	11C & 10 & \num{1.15E+14} & 1.6 & 235.5 \\
	11D & 11 & \num{9.39E+14} & 2.0 & 235.7 \\
	12A & 14 & \num{1.17E+14} & 1.6 & 267.9 \\
	12D & 12 & \num{5.28E+13} & 1.6 & 270.3 \\
	13A & 12 & \num{4.64E+14} & 1.8 & 227.1 \\
	13C & 8  & \num{3.19E+15} & 2.6 & 227.3 \\
	13E & 14 & \num{3.25E+13} & 1.5 & 225.7 \\
	13F & 10 & \num{8.35E+14} & 2.5 & 227.0 \\
	13G & 10 & \num{3.03E+14} & 3.0 & 226.8 \\
	13H & 10 & \num{1.27E+14} & 1.6 & 226.8 \\
    14A & 160 & \num{6.48E+15} & 4.70 & 250.0 \\
        & 26 & \num{2.96E+14} & 3.03 & 248.0 \\
	15A & 16 & \num{2.70E+14} & 5.5 & 235.4 \\
	15B & 10 & \num{5.80E+14} & 2.2 & 238.5 \\
	16B & 12 & \num{1.48E+14} & 2.0 & 236.1 \\
	17A & 14 & \num{5.11E+14} & 2.9 & 221.4 \\
	18A & 12 & \num{2.61E+14} & 2.0 & 279.0 \\
	18B & 11 & \num{1.62E+14} & 2.6 & 279.8 \\
	18C & 10 & \num{1.26E+14} & 1.7 & 280.7 \\
	19A & 10 & \num{3.01E+15} & 2.3 & 274.5 \\
	19D & 10 & \num{2.61E+14} & 1.4 & 274.1 \\
\hline
\end{tabular}
\tablefoot{The final best-fit models for \ce{CH3OH} are listed in Table~\ref{tab:XCLASS_params} (see Section~\ref{sec:spectra} for more details).}
\end{table}

\begin{center}
\begin{longtable}{c c c c c c c c}
\caption{Parameters of the best-fit model for all detected molecules and their abundances.}
\label{tab:XCLASS_params} \\
\hline\hline \noalign{\smallskip}

Source
& Species, X
& T
& N(X)
& $\Delta$v$_\mathrm{FWHM}$
& v$_\mathrm{LSR}$
& N(X)/N(H$_2$)
& n$_\mathrm{col}$\tablefootmark{$\ast$}
\\

& 
& (K)
& (cm$^{-2}$)
& (km~s$^{-1}$)
& (km~s$^{-1}$)
& 
& (cm$^{-3}$)
\\
\hline \noalign{\smallskip}
\endhead 
\hline \noalign{\smallskip}
\multicolumn{8}{r}{Continued \ldots}
\endfoot
\hline \noalign{\smallskip}
\endlastfoot
1A & \ce{CH3OH}\tablefootmark{$\star$} & 58 & \num{1.16E+14} & 2.6 & 224.8 & \num{6.34E-10} & \num{4.9E+06} \\ 
	 & \ce{SO2} & 35 & \num{8.40E+13} & 2.8 & 224.5 & \num{4.59E-10} \\ 
	 & \ce{H2CS} & 58\tablefootmark{$\dagger$} & \num{3.55E+13} & 3.8 & 224.5 & \num{1.94E-10} \\ 
	 & \ce{CS} & " & \num{1.06E+14} & 4.0 & 225.4 & \num{5.79E-10} \\ 
	 & \ce{HC^15N} & " & \num{1.42E+12} & 5.4 & 224.5 & \num{7.76E-12} \\ 
	 & \ce{SO} & " & \num{2.61E+14} & 4.6 & 225.7 & \num{1.43E-09} \\ 
	 & \ce{H^13CN} & " & \num{2.02E+12} & 6.3 & 224.0 & \num{1.10E-11} \\ 
	 & \ce{H^13CO^+} & " & \num{1.37E+12} & 4.0 & 225.5 & \num{7.48E-12} \\ 
	 & \ce{SiO} & " & \num{2.07E+12} & 3.7 & 226.5 & \num{1.13E-11} \\ 
	 \\
2A & \ce{CH3OH}\tablefootmark{$\star$} & 23 & \num{2.09E+13} & 1.3 & 269.1 & \num{3.10E-11} & \num{2.0E+06} \\ 
	 & \ce{SO2} & 23\tablefootmark{$\dagger$} & \num{3.12E+13} & 3.5 & 268.8 & \num{4.63E-11} \\ 
	 & \ce{CS} & " & \num{1.30E+13} & 3.4 & 269.6 & \num{1.93E-11} \\ 
	 & \ce{SO} & " & \num{4.65E+13} & 2.5 & 269.4 & \num{6.90E-11} \\ 
	 \\
2B & \ce{CH3OH}\tablefootmark{$\star$} & 22 & \num{1.70E+14} & 1.8 & 268.7 & \num{4.33E-10} & \num{3.7E+06} \\ 
	 & \ce{SO2} & 23 & \num{7.33E+13} & 2.2 & 269.0 & \num{1.87E-10} \\ 
	 & \ce{H2CS} & 22\tablefootmark{$\dagger$} & \num{1.63E+13} & 3.7 & 270.0 & \num{4.15E-11} \\ 
	 & \ce{CS} & " & \num{5.96E+13} & 3.5 & 269.8 & \num{1.52E-10} \\ 
	 & \ce{HC^15N} & " & \num{5.13E+11} & 4.7 & 271.5 & \num{1.31E-12} \\ 
	 & \ce{SO} & " & \num{2.88E+14} & 3.3 & 270.0 & \num{7.34E-10} \\ 
	 & \ce{H^13CN} & " & \num{2.08E+12} & 4.1 & 268.6 & \num{5.30E-12} \\ 
	 & \ce{H^13CO^+} & " & \num{4.41E+11} & 3.1 & 270.3 & \num{1.12E-12} \\ 
	 & \ce{SiO} & " & \num{1.25E+12} & 6.8 & 268.3 & \num{3.18E-12} \\ 
	 \\
2C & \ce{CH3OH}\tablefootmark{$\star$} & 67 & \num{5.56E+13} & 2.1 & 267.8 & \num{6.87E-10} & \num{6.2E+05} \\ 
	 & \ce{H2CS} & 67\tablefootmark{$\dagger$} & \num{4.11E+13} & 5.2 & 266.9 & \num{5.08E-10} \\ 
	 & \ce{CS} & " & \num{4.52E+13} & 3.5 & 268.4 & \num{5.59E-10} \\ 
	 & \ce{SO} & " & \num{3.10E+13} & 3.5 & 268.6 & \num{3.83E-10} \\ 
	 & \ce{H^13CN} & " & \num{1.35E+12} & 9.6 & 271.9 & \num{1.67E-11} \\ 
	 \\
2D & \ce{CH3OH}\tablefootmark{$\star$} & 55 & \num{1.91E+13} & 1.4 & 269.3 & \num{2.81E-10} & \num{2.0E+06} \\ 
	 & \ce{CS} & 55\tablefootmark{$\dagger$} & \num{2.82E+13} & 3.3 & 270.4 & \num{4.15E-10} \\ 
	 & \ce{SO} & " & \num{1.69E+13} & 4.0 & 271.4 & \num{2.49E-10} \\ 
	 \\
2E & \ce{CH3OH}\tablefootmark{$\star$} & 11 & \num{7.80E+13} & 1.8 & 268.4 & \num{2.07E-10} & \num{2.0E+06} \\ 
	 & \ce{CS} & 11\tablefootmark{$\dagger$} & \num{3.14E+13} & 2.8 & 270.1 & \num{8.33E-11} \\ 
	 & \ce{SO} & " & \num{1.54E+14} & 2.9 & 271.2 & \num{4.08E-10} \\ 
	 \\
2F & \ce{CH3OH}\tablefootmark{$\star$} & 27\tablefootmark{$\ddagger$} & < \num{8.28E+12} & 2.8\tablefootmark{$\equiv$} & 270.9\tablefootmark{$\equiv$} & < \num{1.11E-10} \\ 
	 & \ce{CS} & " & \num{3.31E+12} & 2.8 & 270.9 & \num{4.46E-11} \\ 
	 & \ce{H^13CN} & " & \num{2.68E+12} & 13.1 & 276.0 & \num{3.61E-11} \\ 
	 \\
3A & \ce{CH3OH}\tablefootmark{$\star$} & 35 & \num{2.46E+14} & 3.9 & 285.2 & \num{1.83E-10} & \num{9.0E+06} \\ 
	 & \ce{SO2} & 43 & \num{1.06E+14} & 4.7 & 285.0 & \num{7.89E-11} \\ 
	 & \ce{H2CS} & 35\tablefootmark{$\dagger$} & \num{2.59E+13} & 4.3 & 285.5 & \num{1.93E-11} \\ 
	 & \ce{CS} & " & \num{8.60E+13} & 4.8 & 285.8 & \num{6.40E-11} \\ 
	 & \ce{CH3CN} & " & \num{2.97E+12} & 2.9 & 284.1 & \num{2.21E-12} \\ 
	 & \ce{HC^15N} & " & \num{6.82E+11} & 4.0 & 286.4 & \num{5.08E-13} \\ 
	 & \ce{SO} & " & \num{3.66E+14} & 4.5 & 286.3 & \num{2.73E-10} \\ 
	 & \ce{H^13CN} & " & \num{1.00E+12} & 3.8 & 284.3 & \num{7.45E-13} \\ 
	 & \ce{H^13CO^+} & " & \num{8.83E+11} & 4.0 & 286.5 & \num{6.57E-13} \\ 
	 & \ce{SiO} & " & \num{1.79E+12} & 8.9 & 288.7 & \num{1.33E-12} \\ 
  \\
3B & \ce{CH3OH}\tablefootmark{$\star$} & 13 & \num{6.50E+13} & 1.7 & 284.6 & \num{8.08E-11} & \num{4.9E+06} \\ 
	 & \ce{CS} & 13\tablefootmark{$\dagger$} & \num{4.29E+13} & 4.0 & 285.9 & \num{5.33E-11} \\ 
	 & \ce{SO} & " & \num{2.79E+14} & 3.7 & 286.2 & \num{3.47E-10} \\ 
	 & \ce{H^13CO^+} & " & \num{2.60E+11} & 3.4 & 285.5 & \num{3.23E-13} \\ 
	 \\
3C & \ce{CH3OH}\tablefootmark{$\star$} & 51 & \num{2.38E+13} & 1.8 & 286.2 & \num{3.77E-10} & \num{2.0E+06} \\ 
	 & \ce{CS} & 51\tablefootmark{$\dagger$} & \num{1.45E+13} & 3.4 & 287.4 & \num{2.30E-10} \\ 
	 & \ce{SO} & " & \num{1.48E+13} & 3.2 & 287.2 & \num{2.34E-10} \\ 
	 & \ce{H^13CO^+} & " & \num{1.74E+11} & 3.0 & 288.3 & \num{2.76E-12} \\ 
	 \\
3D & \ce{CH3OH}\tablefootmark{$\star$} & 37 & \num{3.25E+13} & 1.5 & 286.3 & \num{4.48E-10} & \num{1.3E+06} \\ 
	 & \ce{CS} & 37\tablefootmark{$\dagger$} & \num{1.28E+13} & 3.5 & 287.1 & \num{1.77E-10} \\ 
	 & \ce{SO} & " & \num{1.22E+13} & 3.2 & 287.4 & \num{1.68E-10} \\ 
	 \\
4A & \ce{CH3OH} & 151 & \num{1.29E+15} & 3.2 & 264.3 & \num{3.55E-09} \\ 
	 & \ce{CH3OH}\tablefootmark{$\star$} & 26 & \num{1.25E+14} & 4.5 & 264.3 & \num{4.86E-11} & \num{4.6E+07} \\ 
	 & \ce{SO2} & 249 & \num{4.47E+15} & 7.1 & 265.5 & \num{2.06E-08} \\ 
	 &  & 42 & \num{8.07E+14} & 5.0 & 264.3 & \num{5.57E-10} \\ 
	 & \ce{^33SO2} & 249 & \num{3.73E+14} & 7.1 & 265.5 & \num{1.72E-09} \\ 
	 &  & 42 & \num{6.72E+13} & 5.0 & 264.3 & \num{4.63E-11} \\ 
	 & \ce{^34SO2} & 249 & \num{5.05E+14} & 7.1 & 265.5 & \num{2.33E-09} \\ 
	 &  & 42 & \num{9.10E+13} & 5.0 & 264.3 & \num{6.28E-11} \\ 
	 & \ce{SO} & 151\tablefootmark{$\dagger$} & \num{2.01E+15} & 6.0 & 265.7 & \num{5.55E-09} \\ 
	 & \ce{^33SO} & " & \num{1.68E+14} & 6.0 & 265.7 & \num{4.62E-10} \\ 
	 & \ce{CH3CN} & 88 & \num{1.02E+13} & 3.9 & 265.3 & \num{1.57E-11} \\ 
	 & \ce{CS} & 151\tablefootmark{$\dagger$} & \num{2.60E+14} & 5.0 & 265.1 & \num{7.15E-10} \\ 
	 & \ce{H2CS} & " & \num{8.29E+13} & 3.6 & 264.0 & \num{2.28E-10} \\ 
	 & \ce{H^13CO^+} & " & \num{1.20E+13} & 5.1 & 265.6 & \num{3.30E-11} \\ 
	 & \ce{SiO} & " & \num{1.88E+13} & 5.8 & 265.7 & \num{5.16E-11} \\ 
	 & \ce{H^13CN} & " & \num{1.20E+13} & 6.0 & 264.0 & \num{3.31E-11} \\ 
	 & \ce{HC^15N} & " & \num{6.48E+12} & 5.7 & 264.8 & \num{1.79E-11} \\ 
	 & \ce{HNCO} & " & \num{2.71E+13} & 2.6 & 264.1 & \num{7.47E-11} \\ 
	 & \ce{HC3N} & " & \num{6.62E+12} & 8.0 & 264.9 & \num{1.82E-11} \\ 
	 & \ce{HDCO} & " & < \num{4.17E+09} & 3.2\tablefootmark{$\equiv$} & 264.3\tablefootmark{$\equiv$} & < \num{1.15E-14} \\ 
	 & \ce{NH2CHO} & " & < \num{5.60E+10} & " & " & < \num{1.54E-13} \\ 
	 & \ce{H2CCO} & " & < \num{5.23E+12} & " & " & < \num{1.44E-11} \\ 
	 & \ce{CH3CHO} & " & < \num{1.20E+12} & " & " & < \num{3.31E-12} \\ 
	 & \ce{CH3OCHO} & " & < \num{5.35E+13} & " & " & < \num{1.47E-10} \\ 
	 & \ce{CH3OCH3} & " & < \num{5.07E+08} & " & " & < \num{1.40E-15} \\ 
	 \\ 
5A & \ce{CH3OH}\tablefootmark{$\star$} & 15 & \num{3.16E+13} & 2.7 & 246.0 & \num{2.12E-11} & \num{1.1E+06} \\ 
	 & \ce{CS} & 15\tablefootmark{$\dagger$} & \num{2.91E+13} & 4.2 & 246.6 & \num{1.95E-11} \\ 
	 & \ce{SO} & " & \num{9.08E+13} & 3.9 & 246.3 & \num{6.08E-11} \\ 
	 & \ce{H^13CN} & " & \num{1.07E+12} & 10.0 & 248.4 & \num{7.17E-13} \\ 
	 & \ce{H^13CO^+} & " & \num{3.84E+11} & 3.9 & 246.0 & \num{2.57E-13} \\ 
	 & \ce{SiO} & " & \num{5.64E+12} & 20.0 & 245.3 & \num{3.78E-12} \\
  \\
5B & \ce{CH3OH}\tablefootmark{$\star$} & 19 & \num{9.76E+13} & 2.1 & 242.2 & \num{1.53E-10} & \num{5.4E+06} \\ 
	 & \ce{SO2} & 38 & \num{7.48E+13} & 3.0 & 242.0 & \num{1.17E-10} \\ 
	 & \ce{H2CS} & 19\tablefootmark{$\dagger$} & \num{3.94E+13} & 3.8 & 242.3 & \num{6.18E-11} \\ 
	 & \ce{CS} & " & \num{1.27E+14} & 3.7 & 243.3 & \num{1.99E-10} \\ 
	 & \ce{HC^15N} & " & \num{9.31E+11} & 7.1 & 241.9 & \num{1.46E-12} \\ 
	 & \ce{SO} & " & \num{3.92E+14} & 3.7 & 243.3 & \num{6.15E-10} \\ 
	 & \ce{H^13CN} & " & \num{1.76E+12} & 10.0 & 239.2 & \num{2.76E-12} \\ 
	 & \ce{H^13CO^+} & " & \num{6.35E+11} & 3.5 & 243.4 & \num{9.96E-13} \\ 
	 \\
6A & \ce{CH3OH} & 149 & \num{1.46E+15} & 5.7 & 242.4 & \num{3.60E-09} \\ 
	 & \ce{CH3OH}\tablefootmark{$\star$} & 36 & \num{2.62E+14} & 3.7 & 242.9 & \num{1.38E-10} & \num{1.3E+07} \\ 
	 & \ce{SO2} & 161 & \num{5.28E+15} & 12.5 & 245.5 & \num{1.42E-08} \\ 
	 &  & 44 & \num{9.75E+14} & 7.1 & 243.7 & \num{6.50E-10} \\ 
	 & \ce{^33SO2} & 161 & \num{2.55E+14} & 12.5 & 245.5 & \num{6.88E-10} \\ 
	 &  & 44 & \num{4.74E+13} & 7.1 & 243.7 & \num{3.16E-11} \\ 
	 & \ce{^34SO2} & 161 & \num{2.80E+14} & 12.5 & 245.5 & \num{7.55E-10} \\ 
	 &  & 44 & \num{5.19E+13} & 7.1 & 243.7 & \num{3.46E-11} \\ 
	 & \ce{SO} & 149\tablefootmark{$\dagger$} & \num{4.58E+15} & 9.5 & 244.9 & \num{1.13E-08} \\ 
	 & \ce{^33SO} & " & \num{2.23E+14} & 9.5 & 244.9 & \num{5.50E-10} \\ 
	 & \ce{CH3CN} & 77 & \num{2.46E+13} & 6.8 & 243.7 & \num{3.04E-11} \\ 
	 & \ce{CS} & 149\tablefootmark{$\dagger$} & \num{4.99E+14} & 5.5 & 243.6 & \num{1.23E-09} \\ 
	 & \ce{H2CS} & " & \num{2.05E+14} & 5.3 & 242.8 & \num{5.07E-10} \\ 
	 & \ce{H^13CO^+} & " & \num{1.32E+13} & 4.9 & 244.3 & \num{3.25E-11} \\ 
	 & \ce{SiO} & " & \num{8.80E+13} & 11.5 & 246.0 & \num{2.17E-10} \\ 
	 & \ce{H^13CN} & " & \num{2.13E+13} & 9.8 & 243.6 & \num{5.27E-11} \\ 
	 & \ce{HC^15N} & " & \num{1.31E+13} & 7.5 & 244.7 & \num{3.25E-11} \\ 
	 & \ce{HNCO} & " & \num{8.35E+13} & 5.8 & 242.8 & \num{2.06E-10} \\ 
	 & \ce{HC3N} & " & \num{1.24E+13} & 4.7 & 245.1 & \num{3.07E-11} \\ 
	 & \ce{NH2CHO} & " & \num{4.09E+13} & 4.8 & 241.5 & \num{1.01E-10} \\ 
	 & \ce{HDCO} & " & < \num{6.74E+12} & 5.7\tablefootmark{$\equiv$} & 242.4\tablefootmark{$\equiv$} & < \num{1.66E-11} \\ 
	 & \ce{H2CCO} & " & < \num{5.00E+08} & " & " & < \num{1.23E-15} \\ 
	 & \ce{CH3CHO} & " & < \num{5.02E+08} & " & " & < \num{1.24E-15} \\ 
	 & \ce{CH3OCHO} & " & < \num{5.03E+08} & " & " & < \num{1.24E-15} \\ 
	 & \ce{CH3OCH3} & " & < \num{8.51E+08} & " & " & < \num{2.10E-15} \\ 
	 \\ 
6B & \ce{CH3OH} & 181 & \num{1.60E+15} & 2.9 & 243.1 & \num{5.50E-09} \\ 
	 & \ce{CH3OH}\tablefootmark{$\star$} & 50 & \num{3.25E+14} & 4.4 & 241.6 & \num{2.83E-10} & \num{2.8E+07} \\ 
	 & \ce{SO2} & 175 & \num{1.53E+15} & 6.0 & 243.0 & \num{5.03E-09} \\ 
	 &  & 34 & \num{2.68E+14} & 3.3 & 241.7 & \num{1.49E-10} \\ 
	 & \ce{^34SO2} & 175 & \num{1.30E+14} & 6.0 & 243.0 & \num{4.28E-10} \\ 
	 &  & 34 & \num{2.28E+13} & 3.3 & 241.7 & \num{1.27E-11} \\ 
	 & \ce{SO} & 181\tablefootmark{$\dagger$} & \num{1.39E+15} & 5.2 & 243.2 & \num{4.75E-09} \\ 
	 & \ce{^33SO} & " & \num{7.48E+13} & 5.2 & 243.2 & \num{2.56E-10} \\ 
	 & \ce{CH3CN} & 132 & \num{4.86E+13} & 8.7 & 243.4 & \num{1.19E-10} \\ 
	 & \ce{CS} & 181\tablefootmark{$\dagger$} & \num{3.85E+14} & 5.0 & 242.4 & \num{1.32E-09} \\ 
	 & \ce{H2CS} & " & \num{4.10E+14} & 5.2 & 241.5 & \num{1.41E-09} \\ 
	 & \ce{H^13CO^+} & " & \num{1.08E+13} & 4.0 & 242.9 & \num{3.72E-11} \\ 
	 & \ce{SiO} & " & \num{1.51E+13} & 5.8 & 243.5 & \num{5.19E-11} \\ 
	 & \ce{H^13CN} & " & \num{2.65E+13} & 5.6 & 242.1 & \num{9.08E-11} \\ 
	 & \ce{HC^15N} & " & \num{1.46E+13} & 4.3 & 243.7 & \num{4.99E-11} \\ 
	 & \ce{HNCO} & " & \num{3.64E+13} & 2.5 & 242.3 & \num{1.25E-10} \\ 
	 & \ce{HC3N} & " & \num{1.41E+13} & 4.5 & 243.9 & \num{4.83E-11} \\
	 &\ce{HDCO} & " & < \num{1.64E+10} & 2.9\tablefootmark{$\equiv$} & 243.1\tablefootmark{$\equiv$} & < \num{5.61E-14} \\ 
	 & \ce{NH2CHO} & " & < \num{2.77E+09} & " & " & < \num{9.49E-15} \\ 
	 & \ce{H2CCO} & " & < \num{3.45E+13} & " & " & < \num{1.18E-10} \\ 
	 & \ce{CH3CHO} & " & < \num{4.60E+13} & " & " & < \num{1.58E-10} \\ 
	 & \ce{CH3OCHO} & " & < \num{1.67E+14} & " & " & < \num{5.71E-10} \\ 
	 & \ce{CH3OCH3} & " & < \num{2.39E+14} & " & " & < \num{8.19E-10} \\ 
	 \\ 
6C & \ce{CH3OH}\tablefootmark{$\star$} & 30 & \num{2.38E+14} & 3.6 & 241.9 & \num{3.23E-10} & \num{5.3E+06} \\ 
	 & \ce{SO2} & 30\tablefootmark{$\dagger$} & \num{4.90E+13} & 2.2 & 241.3 & \num{6.66E-11} \\ 
	 & \ce{H2CS} & " & \num{7.05E+13} & 3.6 & 241.7 & \num{9.58E-11} \\ 
	 & \ce{CS} & " & \num{2.34E+14} & 4.1 & 242.5 & \num{3.18E-10} \\ 
	 & \ce{SO} & " & \num{2.56E+14} & 4.2 & 242.7 & \num{3.48E-10} \\ 
	 & \ce{H^13CN} & " & \num{1.25E+12} & 2.3 & 242.6 & \num{1.70E-12} \\ 
	 & \ce{H^13CO^+} & " & \num{9.47E+11} & 3.8 & 243.0 & \num{1.29E-12} \\ 
	 & \ce{SiO} & " & \num{8.50E+12} & 18.9 & 238.2 & \num{1.16E-11} \\ 
	 \\
6D & \ce{CH3OH}\tablefootmark{$\star$} & 46 & \num{5.08E+14} & 3.4 & 242.1 & \num{1.54E-09} & \num{2.1E+06} \\ 
	 & \ce{H2CS} & 46\tablefootmark{$\dagger$} & \num{8.41E+13} & 3.9 & 242.7 & \num{2.55E-10} \\ 
	 & \ce{CS} & " & \num{1.88E+14} & 5.2 & 243.0 & \num{5.69E-10} \\ 
	 & \ce{SO} & " & \num{2.47E+14} & 5.5 & 242.8 & \num{7.48E-10} \\ 
	 & \ce{H^13CO^+} & " & \num{8.91E+11} & 4.0 & 241.8 & \num{2.70E-12} \\ 
	 & \ce{SiO} & " & < \num{7.51E+12} & 15.6 & 246.3 & \num{2.27E-11} \\ 
	 \\
6E & \ce{CH3OH}\tablefootmark{$\star$} & 38 & \num{5.19E+14} & 2.6 & 243.0 & \num{1.28E-09} & \num{3.3E+06} \\ 
	 & \ce{H2CS} & 38\tablefootmark{$\dagger$} & \num{7.75E+13} & 4.2 & 243.0 & \num{1.91E-10} \\ 
	 & \ce{CS} & " & \num{1.82E+14} & 4.2 & 243.7 & \num{4.48E-10} \\ 
	 & \ce{SO} & " & \num{1.92E+14} & 3.8 & 244.0 & \num{4.73E-10} \\ 
	 & \ce{H^13CN} & " & \num{1.78E+12} & 4.6 & 242.4 & \num{4.38E-12} \\ 
	 & \ce{H^13CO^+} & " & \num{5.54E+11} & 3.4 & 244.5 & \num{1.36E-12} \\ 
	 & \ce{SiO} & " & \num{8.59E+11} & 4.9 & 238.1 & \num{2.12E-12} \\ 
	 \\
6F & \ce{CH3OH}\tablefootmark{$\star$} & 27 & \num{5.82E+13} & 1.9 & 241.2 & \num{1.33E-10} & \num{1.0E+06} \\ 
	 & \ce{SO2} & 27\tablefootmark{$\dagger$} & \num{3.54E+13} & 2.9 & 240.9 & \num{8.08E-11} \\ 
	 & \ce{CS} & " & \num{4.22E+13} & 3.4 & 241.9 & \num{9.63E-11} \\ 
	 & \ce{SO} & " & \num{1.08E+14} & 3.3 & 242.1 & \num{2.47E-10} \\ 
	 & \ce{H^13CO^+} & " & \num{6.01E+11} & 3.1 & 242.1 & \num{1.37E-12} \\ 
	 \\
7A & \ce{CH3OH}\tablefootmark{$\star$} & 29 & \num{9.92E+13} & 1.8 & 252.8 & \num{3.05E-10} & \num{2.4E+06} \\ 
	 & \ce{SO2} & 22 & \num{3.25E+13} & 1.6 & 253.1 & \num{9.98E-11} \\ 
	 & \ce{H2CS} & 29\tablefootmark{$\dagger$} & \num{1.40E+13} & 3.3 & 253.3 & \num{4.30E-11} \\ 
	 & \ce{CS} & " & \num{4.91E+13} & 3.4 & 253.8 & \num{1.51E-10} \\ 
	 & \ce{SO} & " & \num{1.36E+14} & 3.1 & 254.1 & \num{4.18E-10} \\ 
	 & \ce{H^13CN} & " & \num{8.42E+11} & 7.2 & 255.4 & \num{2.59E-12} \\ 
	 & \ce{H^13CO^+} & " & \num{8.17E+11} & 3.6 & 254.2 & \num{2.51E-12} \\ 
	 \\
7B & \ce{CH3OH} & 27\tablefootmark{$\ddagger$} & < \num{1.00E+10} & 2.8\tablefootmark{$\equiv$} & 252.4\tablefootmark{$\equiv$} & < \num{8.11E-14} \\ 
	 & \ce{CS} & " & \num{5.83E+12} & 2.8 & 252.4 & \num{4.73E-11} \\ 
	 \\
7C & \ce{CH3OH} & 27\tablefootmark{$\ddagger$} & < \num{2.20E+13} & 2.9\tablefootmark{$\equiv$} & 255.0\tablefootmark{$\equiv$} & < \num{1.74E-10} \\ 
	 & \ce{CS} & 27\tablefootmark{$\dagger$} & \num{9.13E+12} & 2.9 & 255.0 & \num{7.24E-11} \\ 
	 & \ce{H^13CO^+} & " & \num{2.19E+11} & 3.2 & 255.3 & \num{1.74E-12} \\ 
	 \\
8A & \ce{CH3OH}\tablefootmark{$\star$} & 25 & \num{2.14E+14} & 2.0 & 238.0 & \num{6.30E-10} & \num{7.3E+06} \\ 
	 & \ce{SO2} & 36 & \num{4.00E+13} & 3.1 & 237.9 & \num{1.18E-10} \\ 
	 & \ce{H2CS} & 25\tablefootmark{$\dagger$} & \num{1.68E+13} & 4.0 & 237.4 & \num{4.94E-11} \\ 
	 & \ce{CS} & " & \num{5.23E+13} & 3.8 & 238.6 & \num{1.54E-10} \\ 
	 & \ce{SO} & " & \num{2.33E+14} & 3.6 & 239.0 & \num{6.86E-10} \\ 
	 & \ce{H^13CN} & " & \num{5.13E+11} & 3.1 & 237.5 & \num{1.51E-12} \\ 
	 & \ce{H^13CO^+} & " & \num{1.55E+11} & 5.3 & 238.2 & \num{4.56E-13} \\ 
	 \\ 
8B\tablefootmark{$\S$} \\
\\
8C & \ce{CH3OH}\tablefootmark{$\star$} & 27 & \num{2.08E+14} & 2.8 & 239.9 & \num{1.13E-09} & \num{2.0E+06} \\ 
	 & \ce{CS} & 27\tablefootmark{$\dagger$} & \num{5.67E+13} & 4.2 & 240.8 & \num{3.09E-10} \\ 
	 & \ce{SO} & " & \num{1.21E+14} & 3.7 & 240.8 & \num{6.59E-10} \\ 
	 \\
9A & \ce{CH3OH}\tablefootmark{$\star$} & 50 & \num{1.53E+14} & 1.8 & 241.6 & \num{1.16E-09} & \num{1.9E+06} \\ 
	 & \ce{H2CS} & 50\tablefootmark{$\dagger$} & \num{2.08E+13} & 3.4 & 241.3 & \num{1.57E-10} \\ 
	 & \ce{CS} & " & \num{7.34E+13} & 3.7 & 242.4 & \num{5.55E-10} \\ 
	 & \ce{SO} & " & \num{6.96E+13} & 3.4 & 242.6 & \num{5.26E-10} \\ 
	 & \ce{H^13CN} & " & \num{8.95E+11} & 2.5 & 241.2 & \num{6.77E-12} \\ 
	 & \ce{H^13CO^+} & " & \num{4.01E+11} & 3.1 & 242.5 & \num{3.03E-12} \\ 
	 \\ 
10A & \ce{CH3OH}\tablefootmark{$\star$} & 26 & \num{1.20E+14} & 2.9 & 239.0 & \num{3.48E-10} & \num{2.5E+06} \\ 
	 & \ce{SO2} & 30 & \num{2.19E+13} & 2.3 & 239.4 & \num{6.36E-11} \\ 
	 & \ce{H2CS} & 26\tablefootmark{$\dagger$} & \num{9.92E+12} & 3.4 & 239.3 & \num{2.88E-11} \\ 
	 & \ce{CS} & " & \num{4.20E+13} & 4.0 & 240.4 & \num{1.22E-10} \\ 
	 & \ce{HC^15N} & " & \num{2.97E+11} & 4.5 & 239.0 & \num{8.62E-13} \\ 
	 & \ce{SO} & " & \num{1.01E+14} & 4.1 & 240.5 & \num{2.93E-10} \\ 
	 & \ce{H^13CN} & " & \num{8.44E+11} & 7.2 & 241.2 & \num{2.45E-12} \\ 
	 & \ce{H^13CO^+} & " & \num{5.35E+11} & 3.4 & 240.6 & \num{1.55E-12} \\ 
	 & \ce{SiO} & " & \num{1.13E+12} & 4.7 & 240.7 & \num{3.28E-12} \\ 
	 \\ 
10B & \ce{CH3OH} & 27\tablefootmark{$\ddagger$} & < \num{6.25E+13} & 3.6\tablefootmark{$\equiv$} & 240.7\tablefootmark{$\equiv$} & < \num{3.43E-10} \\ 
	 & \ce{CS} & " & \num{2.21E+13} & 3.6 & 240.7 & \num{1.21E-10} \\ 
	 & \ce{H^13CO^+} & " & \num{5.72E+11} & 5.1 & 241.7 & \num{3.14E-12} \\ 
	 & \ce{SiO} & " & \num{2.01E+12} & 4.7 & 241.5 & \num{1.10E-11} \\ 
	 \\
10C & \ce{CH3OH}\tablefootmark{$\star$} & 39 & \num{3.35E+13} & 1.5 & 238.9 & \num{5.54E-10} & \num{2.5E+06} \\ 
	 & \ce{CS} & 39\tablefootmark{$\dagger$} & \num{1.96E+13} & 3.4 & 240.0 & \num{3.24E-10} \\ 
	 & \ce{H^13CO^+} & " & \num{1.88E+11} & 2.4 & 240.4 & \num{3.11E-12} \\ 
	 \\
10D & \ce{CH3OH}\tablefootmark{$\star$} & 22 & \num{4.59E+13} & 5.9 & 240.7 & \num{3.99E-10} & \num{2.0E+06} \\ 
	 & \ce{CS} & 22\tablefootmark{$\dagger$} & \num{1.55E+13} & 3.1 & 240.7 & \num{1.35E-10} \\ 
	 \\
11A \\ 
  \\ 
11B & \ce{CH3OH}\tablefootmark{$\star$} & 24 & \num{6.28E+14} & 2.0 & 236.0 & \num{1.87E-09} & \num{2.0E+07} \\ 
	 & \ce{H2CS} & 24\tablefootmark{$\dagger$} & \num{4.05E+13} & 3.9 & 236.3 & \num{1.21E-10} \\ 
	 & \ce{CS} & " & \num{8.37E+13} & 3.3 & 236.7 & \num{2.49E-10} \\ 
	 & \ce{SO} & " & \num{2.56E+14} & 3.8 & 237.1 & \num{7.63E-10} \\ 
	 & \ce{H^13CN} & " & \num{5.41E+11} & 2.6 & 236.1 & \num{1.61E-12} \\ 
	 & \ce{H^13CO^+} & " & \num{2.19E+11} & 5.4 & 237.1 & \num{6.52E-13} \\ 
	 \\
11C & \ce{CH3OH}\tablefootmark{$\star$} & 10 & \num{1.21E+14} & 1.5 & 235.6 & \num{2.11E-10} & \num{4.9E+06} \\ 
	 & \ce{CS} & 10\tablefootmark{$\dagger$} & \num{6.58E+13} & 3.3 & 236.5 & \num{1.15E-10} \\ 
	 & \ce{SO} & " & \num{3.48E+14} & 3.5 & 236.8 & \num{6.06E-10} \\ 
	 & \ce{H^13CO^+} & " & \num{4.70E+11} & 3.6 & 236.4 & \num{8.18E-13} \\ 
	 \\
11D & \ce{CH3OH}\tablefootmark{$\star$} & 22 & \num{3.74E+14} & 2.2 & 235.6 & \num{2.97E-09} & \num{3.9E+06} \\ 
	 & \ce{H2CS} & 22\tablefootmark{$\dagger$} & \num{2.42E+13} & 4.8 & 235.5 & \num{1.92E-10} \\ 
	 & \ce{CS} & " & \num{1.06E+14} & 4.4 & 236.1 & \num{8.41E-10} \\ 
	 & \ce{SO} & " & \num{1.40E+14} & 3.6 & 236.6 & \num{1.11E-09} \\ 
	 \\
12A & \ce{CH3OH}\tablefootmark{$\star$} & 19 & \num{7.88E+13} & 1.6 & 267.9 & \num{1.91E-10} & \num{7.5E+06} \\ 
	 & \ce{H2CS} & 19\tablefootmark{$\dagger$} & \num{2.35E+13} & 2.7 & 268.2 & \num{5.70E-11} \\ 
	 & \ce{CS} & " & \num{4.35E+13} & 3.5 & 268.9 & \num{1.06E-10} \\ 
	 & \ce{SO} & " & \num{1.08E+14} & 3.4 & 269.1 & \num{2.62E-10} \\ 
	 & \ce{H^13CN} & " & \num{1.44E+12} & 10.0 & 269.4 & \num{3.49E-12} \\ 
	 & \ce{H^13CO^+} & " & \num{3.04E+11} & 4.2 & 269.4 & \num{7.38E-13} \\ 
	 \\
12B & \ce{CH3OH} & 27\tablefootmark{$\ddagger$} & < \num{7.99E+12} & 2.4\tablefootmark{$\equiv$} & 271.4\tablefootmark{$\equiv$} & < \num{5.84E-11} \\ 
	 & \ce{CS} & " & \num{3.73E+12} & 2.4 & 271.4 & \num{2.73E-11} \\ 
	 \\
12C & \ce{CH3OH} & 27\tablefootmark{$\ddagger$} & < \num{3.33E+12} & 2.8 & 270.7 & < \num{2.85E-11} \\ 
	 & \ce{CS} & " & \num{9.99E+12} & 2.8 & 270.7 & \num{8.55E-11} \\ 
	 & \ce{SO} & " & \num{2.61E+13} & 3.4 & 271.0 & \num{2.23E-10} \\ 
	 \\
12D & \ce{CH3OH}\tablefootmark{$\star$} & 38 & \num{1.85E+13} & 1.3 & 270.2 & \num{3.06E-10} & \num{2.0E+06} \\ 
	 & \ce{CS} & 38\tablefootmark{$\dagger$} & \num{1.02E+13} & 3.1 & 271.0 & \num{1.69E-10} \\ 
	 \\
13A & \ce{CH3OH}\tablefootmark{$\star$} & 49 & \num{1.71E+14} & 1.8 & 227.0 & \num{9.63E-10} & \num{2.6E+06} \\ 
	 & \ce{SO2} & 49\tablefootmark{$\dagger$} & \num{2.32E+13} & 1.2 & 227.0 & \num{1.31E-10} \\ 
	 & \ce{H2CS} & " & \num{1.75E+13} & 3.8 & 227.5 & \num{9.86E-11} \\ 
	 & \ce{CS} & " & \num{7.01E+13} & 3.4 & 227.9 & \num{3.95E-10} \\ 
	 & \ce{SO} & " & \num{7.19E+13} & 2.5 & 227.9 & \num{4.05E-10} \\ 
	 & \ce{H^13CN} & " & \num{4.80E+11} & 2.7 & 227.5 & \num{2.70E-12} \\ 
	 & \ce{H^13CO^+} & " & \num{3.49E+11} & 3.0 & 228.0 & \num{1.97E-12} \\ 
	 & \ce{SiO} & " & \num{7.28E+11} & 4.9 & 226.7 & \num{4.10E-12} \\ 
	 \\
13B\tablefootmark{$\S$} \\ 
	 \\ 
13C & \ce{CH3OH}\tablefootmark{$\star$} & 27 & \num{3.91E+14} & 3.0 & 227.3 & \num{1.26E-09} & \num{1.4E+06} \\ 
	 & \ce{SO2} & 27\tablefootmark{$\dagger$} & \num{2.86E+13} & 2.3 & 227.1 & \num{9.18E-11} \\ 
	 & \ce{H2CS} & " & \num{2.61E+13} & 5.3 & 227.2 & \num{8.38E-11} \\ 
	 & \ce{CS} & " & \num{1.03E+14} & 4.1 & 228.2 & \num{3.31E-10} \\ 
	 & \ce{SO} & " & \num{1.50E+14} & 4.0 & 228.1 & \num{4.82E-10} \\ 
	 & \ce{H^13CN} & " & \num{1.01E+12} & 4.8 & 228.5 & \num{3.24E-12} \\ 
	 & \ce{H^13CO^+} & " & \num{3.18E+11} & 4.5 & 229.1 & \num{1.02E-12} \\ 
	 & \ce{SiO} & " & \num{3.92E+12} & 20.0 & 216.8 & \num{1.26E-11} \\ 
	 \\
13D & \ce{CH3OH} & 27\tablefootmark{$\ddagger$} & < \num{8.62E+13} & 4.3\tablefootmark{$\equiv$} & 230.7\tablefootmark{$\equiv$} & < \num{3.04E-10} \\ 
	 & \ce{SO2} & " & \num{9.27E+13} & 3.8 & 229.8 & \num{3.27E-10} \\ 
	 & \ce{H2CS} & " & \num{1.94E+13} & 4.0 & 229.6 & \num{6.85E-11} \\ 
	 & \ce{CS} & " & \num{4.98E+13} & 4.3 & 230.7 & \num{1.76E-10} \\ 
	 & \ce{SO} & " & \num{2.68E+14} & 4.5 & 230.2 & \num{9.46E-10} \\ 
	 \\
13E & \ce{CH3OH}\tablefootmark{$\star$} & 19 & \num{2.06E+13} & 1.6 & 225.6 & \num{8.11E-11} & \num{2.0E+06} \\ 
	 & \ce{CS} & 19\tablefootmark{$\dagger$} & \num{2.41E+13} & 3.6 & 226.5 & \num{9.49E-11} \\ 
	 & \ce{SO} & " & \num{1.64E+13} & 3.5 & 227.3 & \num{6.46E-11} \\ 
	 \\
13F & \ce{CH3OH}\tablefootmark{$\star$} & 42 & \num{2.09E+14} & 2.1 & 227.6 & \num{2.51E-09} & \num{1.9E+06} \\ 
	 & \ce{SO2} & 42\tablefootmark{$\dagger$} & \num{2.23E+13} & 1.7 & 227.3 & \num{2.68E-10} \\ 
	 & \ce{H2CS} & " & \num{1.06E+13} & 3.2 & 227.2 & \num{1.27E-10} \\ 
	 & \ce{CS} & " & \num{6.38E+13} & 3.7 & 228.5 & \num{7.65E-10} \\ 
	 & \ce{SO} & " & \num{7.82E+13} & 3.5 & 228.6 & \num{9.38E-10} \\ 
	 & \ce{H^13CO^+} & " & \num{1.45E+11} & 2.5 & 228.9 & \num{1.74E-12} \\ 
	 \\
13G & \ce{CH3OH}\tablefootmark{$\star$} & 19 & \num{1.22E+14} & 1.7 & 226.0 & \num{6.38E-10} & \num{6.4E+06} \\ 
	 & \ce{CS} & 19\tablefootmark{$\dagger$} & \num{3.00E+13} & 3.4 & 226.9 & \num{1.57E-10} \\ 
	 & \ce{SO} & " & \num{8.09E+13} & 3.4 & 227.1 & \num{4.23E-10} \\ 
	 & \ce{H^13CO^+} & " & \num{2.51E+11} & 4.3 & 227.7 & \num{1.31E-12} \\ 
	 \\
13H & \ce{CH3OH}\tablefootmark{$\star$} & 40 & \num{3.41E+13} & 1.5 & 227.2 & \num{4.75E-10} & \num{2.0E+06} \\ 
	 & \ce{CS} & 40\tablefootmark{$\dagger$} & \num{1.82E+13} & 3.5 & 228.1 & \num{2.54E-10} \\ 
	 \\
13K & \ce{CH3OH} & 27\tablefootmark{$\ddagger$} & < \num{1.27E+13} & 3.3\tablefootmark{$\equiv$} & 227.4\tablefootmark{$\equiv$} & < \num{1.68E-10} \\ 
	 & \ce{CS} & " & \num{8.86E+12} & 3.3 & 227.4 & \num{1.17E-10} \\ 
	 \\
14A & \ce{CH3OH} & 163 & \num{1.79E+15} & 4.9 & 249.8 & \num{1.06E-08} \\ 
	 & \ce{CH3OH}\tablefootmark{$\star$} & 50 & \num{2.63E+14} & 3.1 & 248.1 & \num{4.38E-10} & \num{1.6E+07} \\ 
	 & \ce{SO2} & 138 & \num{1.48E+15} & 8.0 & 250.6 & \num{7.36E-09} \\ 
	 &  & 19 & \num{1.32E+14} & 2.8 & 248.8 & \num{6.77E-11} \\ 
	 & \ce{SO} & 163 & \num{1.40E+15} & 7.3 & 250.1 & \num{8.26E-09} \\ 
	 & \ce{CH3CN} & 42 & \num{1.78E+13} & 4.2 & 249.9 & \num{2.42E-11} \\ 
	 & \ce{CS} & 163\tablefootmark{$\dagger$} & \num{3.70E+14} & 5.4 & 249.2 & \num{2.18E-09} \\ 
	 & \ce{H2CS} & " & \num{1.19E+14} & 4.0 & 248.2 & \num{7.04E-10} \\ 
	 & \ce{H^13CO^+} & " & \num{5.46E+12} & 4.2 & 249.6 & \num{3.23E-11} \\ 
	 & \ce{SiO} & " & \num{6.57E+13} & 34.9 & 256.4 & \num{3.88E-10} \\ 
	 & \ce{HDCO} & " & \num{2.54E+13} & 4.0 & 247.5 & \num{1.50E-10} \\ 
	 & \ce{H^13CN} & " & \num{1.77E+13} & 6.8 & 249.4 & \num{1.04E-10} \\ 
	 & \ce{HC^15N} & " & \num{1.01E+13} & 5.8 & 250.8 & \num{5.97E-11} \\ 
	 & \ce{HNCO} & " & \num{7.35E+13} & 4.6 & 249.5 & \num{4.34E-10} \\ 
	 & \ce{NH2CHO} & " & < \num{5.09E+08} & 4.9\tablefootmark{$\equiv$} & 249.8\tablefootmark{$\equiv$} & < \num{3.00E-15} \\ 
	 & \ce{H2CCO} & " & < \num{5.10E+08} & " & " & < \num{3.01E-15} \\ 
	 & \ce{CH3CHO} & " & < \num{5.09E+08} & " & " & < \num{3.00E-15} \\ 
	 & \ce{CH3OCHO} & " & < \num{9.43E+13} & " & " & < \num{5.57E-10} \\ 
	 & \ce{CH3OCH3} & " & < \num{1.18E+14} & " & " & < \num{6.97E-10} \\ 
	 \\ 
14B\tablefootmark{$\S$} \\ 
	 \\ 
15A & \ce{CH3OH}\tablefootmark{$\star$} & 38 & \num{1.47E+14} & 6.0 & 235.5 & \num{2.77E-10} & \num{2.0E+06} \\ 
	 & \ce{SO2} & 38\tablefootmark{$\dagger$} & \num{1.11E+14} & 2.4 & 236.0 & \num{2.09E-10} \\ 
	 & \ce{CS} & " & \num{1.10E+14} & 4.2 & 236.7 & \num{2.08E-10} \\ 
	 & \ce{SO} & " & \num{2.99E+14} & 4.5 & 237.1 & \num{5.64E-10} \\ 
	 & \ce{H^13CO^+} & " & \num{1.42E+12} & 3.5 & 236.4 & \num{2.68E-12} \\ 
	 & \ce{SiO} & " & \num{1.49E+13} & 19.3 & 229.3 & \num{2.81E-11} \\ 
	 \\
15B & \ce{CH3OH}\tablefootmark{$\star$} & 33 & \num{1.42E+14} & 2.2 & 238.5 & \num{7.48E-10} & \num{1.7E+06} \\ 
	 & \ce{SO2} & 33\tablefootmark{$\dagger$} & \num{9.49E+12} & 1.6 & 238.0 & \num{5.00E-11} \\ 
	 & \ce{H2CS} & " & \num{1.33E+13} & 3.4 & 238.7 & \num{7.00E-11} \\ 
	 & \ce{CS} & " & \num{5.30E+13} & 4.0 & 239.4 & \num{2.79E-10} \\ 
	 & \ce{SO} & " & \num{6.85E+13} & 3.8 & 239.4 & \num{3.61E-10} \\ 
	 & \ce{H^13CO^+} & " & \num{3.39E+11} & 3.8 & 239.8 & \num{1.79E-12} \\ 
	 & \ce{SiO} & " & \num{3.25E+12} & 20.0 & 234.9 & \num{1.71E-11} \\ 
	 \\
15C & \ce{CH3OH} & 27\tablefootmark{$\ddagger$} & < \num{1.29E+13} & 3.1\tablefootmark{$\equiv$} & 238.6\tablefootmark{$\equiv$} & < \num{1.02E-10} \\ 
	 & \ce{CS} & " & \num{1.78E+13} & 3.1 & 238.6 & \num{1.41E-10} \\ 
	 & \ce{H^13CO^+} & " & \num{1.58E+11} & 2.3 & 239.0 & \num{1.25E-12} \\ 
	 & \ce{SiO} & " & \num{1.04E+12} & 4.3 & 241.4 & \num{8.25E-12} \\ 
	 \\
16A\tablefootmark{$\S$} \\ 
  \\ 
16B & \ce{CH3OH}\tablefootmark{$\star$} & 19 & \num{7.37E+13} & 2.1 & 236.1 & \num{4.49E-10} & \num{3.7E+06} \\ 
	 & \ce{SO2} & 19\tablefootmark{$\dagger$} & \num{1.36E+13} & 1.1 & 236.5 & \num{8.28E-11} \\ 
	 & \ce{CS} & " & \num{2.34E+13} & 3.4 & 236.9 & \num{1.43E-10} \\ 
	 & \ce{SO} & " & \num{6.18E+13} & 3.4 & 237.1 & \num{3.76E-10} \\ 
	 & \ce{H^13CN} & " & \num{8.00E+11} & 4.4 & 243.1 & \num{4.87E-12} \\ 
	 & \ce{SiO} & " & \num{2.78E+12} & 7.8 & 229.4 & \num{1.69E-11} \\ 
	 \\
17A & \ce{CH3OH}\tablefootmark{$\star$} & 31 & \num{2.76E+14} & 2.8 & 221.3 & \num{4.93E-10} & \num{5.1E+06} \\ 
	 & \ce{SO2} & 28 & \num{7.08E+13} & 3.9 & 221.4 & \num{1.26E-10} \\ 
	 & \ce{H2CS} & 31\tablefootmark{$\dagger$} & \num{5.60E+13} & 4.2 & 221.8 & \num{9.99E-11} \\ 
	 & \ce{CS} & " & \num{1.30E+14} & 4.6 & 222.4 & \num{2.32E-10} \\ 
	 & \ce{SO} & " & \num{2.82E+14} & 4.4 & 222.5 & \num{5.03E-10} \\ 
	 & \ce{H^13CN} & " & \num{4.49E+11} & 2.3 & 221.8 & \num{8.01E-13} \\ 
	 & \ce{H^13CO^+} & " & \num{3.92E+11} & 3.7 & 222.9 & \num{7.00E-13} \\ 
	 & \ce{SiO} & " & \num{2.48E+12} & 18.1 & 221.6 & \num{4.43E-12} \\ 
	 \\
17B\tablefootmark{$\S$} \\ 
  \\ 
18A & \ce{CH3OH}\tablefootmark{$\star$} & 26 & \num{1.21E+14} & 2.0 & 278.9 & \num{3.05E-10} & \num{4.7E+06} \\ 
	 & \ce{SO2} & 26\tablefootmark{$\dagger$} & \num{1.46E+13} & 2.2 & 278.6 & \num{3.67E-11} \\ 
	 & \ce{H2CS} & " & \num{1.58E+13} & 2.7 & 278.8 & \num{3.98E-11} \\ 
	 & \ce{CS} & " & \num{6.62E+13} & 3.6 & 279.9 & \num{1.67E-10} \\ 
	 & \ce{SO} & " & \num{1.53E+14} & 3.6 & 280.3 & \num{3.85E-10} \\ 
	 & \ce{H^13CO^+} & " & \num{4.66E+11} & 3.2 & 280.1 & \num{1.17E-12} \\ 
	 \\
18B & \ce{CH3OH}\tablefootmark{$\star$} & 33 & \num{5.11E+13} & 2.3 & 279.6 & \num{4.66E-10} & \num{2.1E+06} \\ 
	 & \ce{H2CS} & 33\tablefootmark{$\dagger$} & \num{2.36E+13} & 10.0 & 275.6 & \num{2.15E-10} \\ 
	 & \ce{CS} & " & \num{4.99E+13} & 3.7 & 280.4 & \num{4.55E-10} \\ 
	 & \ce{SO} & " & \num{2.65E+13} & 3.2 & 281.0 & \num{2.42E-10} \\ 
	 & \ce{H^13CO^+} & " & \num{1.58E+11} & 3.5 & 281.6 & \num{1.44E-12} \\ 
	 \\
18C & \ce{CH3OH}\tablefootmark{$\star$} & 23 & \num{3.80E+13} & 1.6 & 280.6 & \num{3.96E-10} & \num{2.0E+06} \\ 
	 & \ce{H2CS} & 23\tablefootmark{$\dagger$} & \num{1.96E+13} & 7.5 & 280.1 & \num{2.04E-10} \\ 
	 & \ce{CS} & " & \num{2.93E+13} & 3.3 & 281.5 & \num{3.06E-10} \\ 
	 \\
19A & \ce{CH3OH}\tablefootmark{$\star$} & 25 & \num{6.73E+14} & 2.7 & 274.4 & \num{1.89E-09} & \num{4.1E+06} \\ 
	 & \ce{SO2} & 25\tablefootmark{$\dagger$} & \num{6.59E+13} & 2.2 & 274.6 & \num{1.85E-10} \\ 
	 & \ce{H2CS} & " & \num{4.32E+13} & 3.5 & 273.9 & \num{1.22E-10} \\ 
	 & \ce{CS} & " & \num{1.23E+14} & 4.4 & 274.8 & \num{3.46E-10} \\ 
	 & \ce{SO} & " & \num{2.53E+14} & 4.0 & 275.6 & \num{7.12E-10} \\ 
	 & \ce{SiO} & " & \num{1.67E+12} & 5.2 & 277.1 & \num{4.70E-12} \\ 
	 \\
19B\tablefootmark{$\S$} \\ 
  \\ 
19C & \ce{CH3OH} & 27\tablefootmark{$\ddagger$} & < \num{3.07E+12} & 4.4\tablefootmark{$\equiv$} & 274.1\tablefootmark{$\equiv$} & < \num{2.39E-11} \\ 
	 & \ce{CS} & " & \num{7.51E+12} & 4.4 & 274.1 & \num{5.86E-11} \\ 
	 & \ce{SO} & " & \num{2.81E+13} & 8.1 & 273.2 & \num{2.19E-10} \\ 
	 \\
19D & \ce{CH3OH}\tablefootmark{$\star$} & 42 & \num{6.23E+13} & 1.4 & 274.0 & \num{8.66E-10} & \num{1.3E+06} \\ 
	 & \ce{SO2} & 42\tablefootmark{$\dagger$} & \num{1.75E+13} & 3.1 & 274.1 & \num{2.43E-10} \\ 
	 & \ce{H2CS} & " & \num{1.87E+13} & 8.6 & 278.2 & \num{2.60E-10} \\ 
	 & \ce{CS} & " & \num{2.79E+13} & 3.5 & 274.8 & \num{3.88E-10} \\ 
	 & \ce{SO} & " & \num{5.19E+13} & 3.4 & 275.2 & \num{7.21E-10} \\ 
	 \\
20A & \ce{CH3OH} & 27\tablefootmark{$\ddagger$} & < \num{5.32E+12} & 3.0\tablefootmark{$\equiv$} & 284.5\tablefootmark{$\equiv$} & < \num{5.36E-11} \\ 
	 & \ce{CS} & " & \num{1.99E+12} & 3.0 & 284.5 & \num{2.00E-11} & 
\end{longtable}
\tablefoot{
N(H$_2$) was derived in Paper~II with the temperatures derived in the current paper and assuming all of the 1.3 mm continuum emission originated in optically thin dust with an opacity of 1.04~cm$^2$~gr$^{-1}$ and a gas-to-dust mass ratio of 320. We consider that the error values range from 40\% to 60\% for the temperatures, 20\% to 40\% for the column densities, and 10\% to 30\% for the line widths and velocity shifts.\\
\tablefoottext{$\ast$}{The collision partner number density.}\\
\tablefoottext{$\star$}{The fit parameters were derived with the non-LTE assumption.}\\
\tablefoottext{$\dagger$}{We adopted this value from the temperature for the \ce{CH3OH}. Where there were two components, the temperature we adopted was from the hot component.}\\
\tablefoottext{$\ddagger$}{The median value of the fit temperature for \ce{CH3OH} from all the sources, where we could fit.}\\
\tablefoottext{<}{The upper limit estimations.}\\
\tablefoottext{$\equiv$}{The upper limit on column density and abundance was estimated by fixing the line width and line shift. In the case of estimating the upper limits for \ce{CH3OH}, we used the best-fit parameters for \ce{CS}, and for the other molecules, we used the best-fit parameters for \ce{CH3OH}.}\\
\tablefoottext{$\S$}{No emission line is detected.}}
\end{center}
%
%
\section{Comparisons of molecular abundances in hot cores with chemical models}

\begin{figure*}
\centering
\includegraphics[scale=0.69]{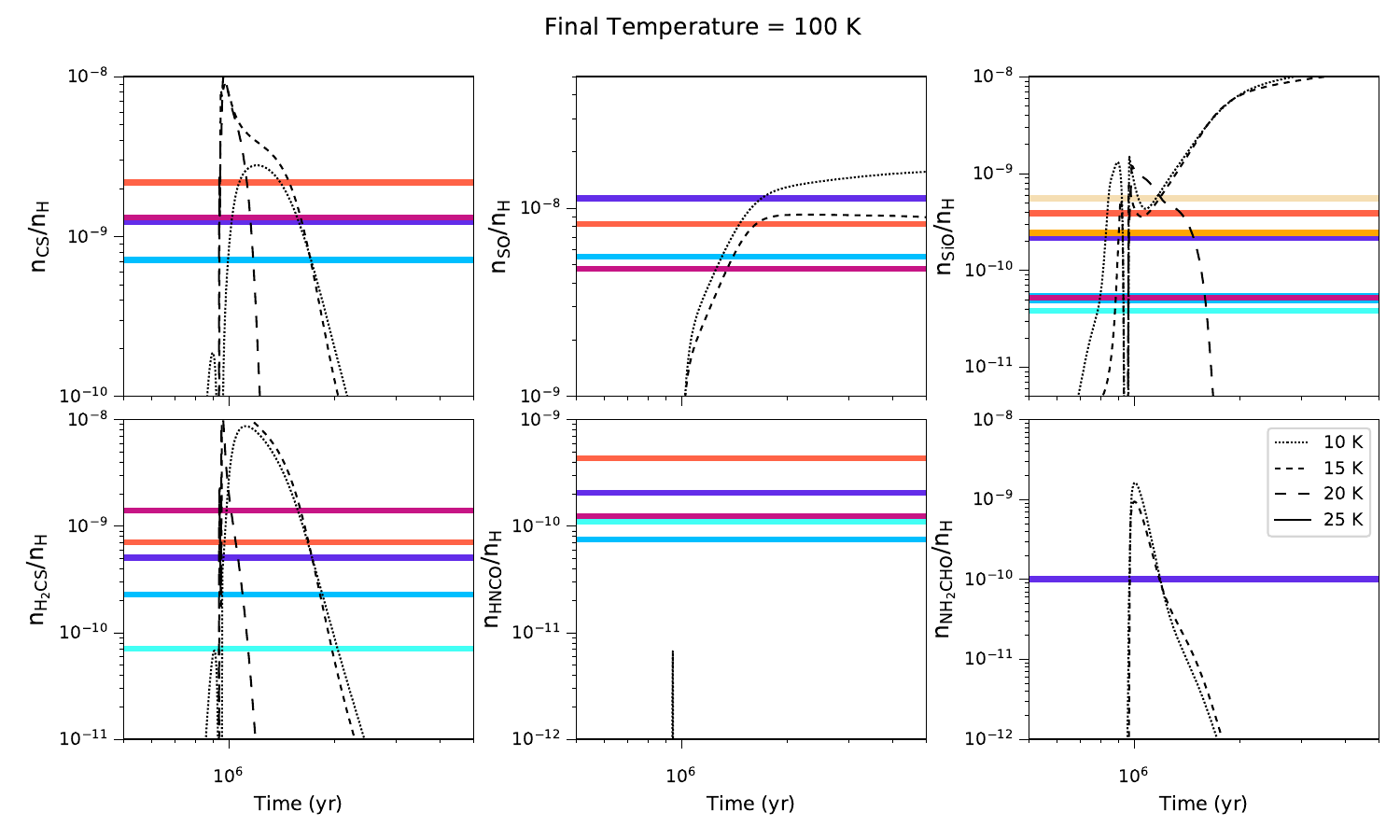}
\caption{Comparisons of abundances for \ce{CS}, \ce{SO}, \ce{SiO}, \ce{H2CS}, \ce{HNCO}, and \ce{NH2CHO} from observations of the seven hot cores detected in the LMC and the LMC hot core models for the final hot core temperature of 100 K. The black curves show abundances over time for initial dust temperatures of 10 (dotted line), 15 (dashed curve), 20 (larger dashed curve), and 25 K (solid curve; this curve does not necessarily appear in the range of abundances in the plot). The horizontal colored line shows the single abundance value for each hot core and follows the same color code as Fig.~\ref{fig:Molecular_Abundances_hot_cores} (see the text for more details).}
\label{appfig:Molecular_Abundances_model1}
\end{figure*}

\begin{figure*}
\centering
\includegraphics[scale=0.69]{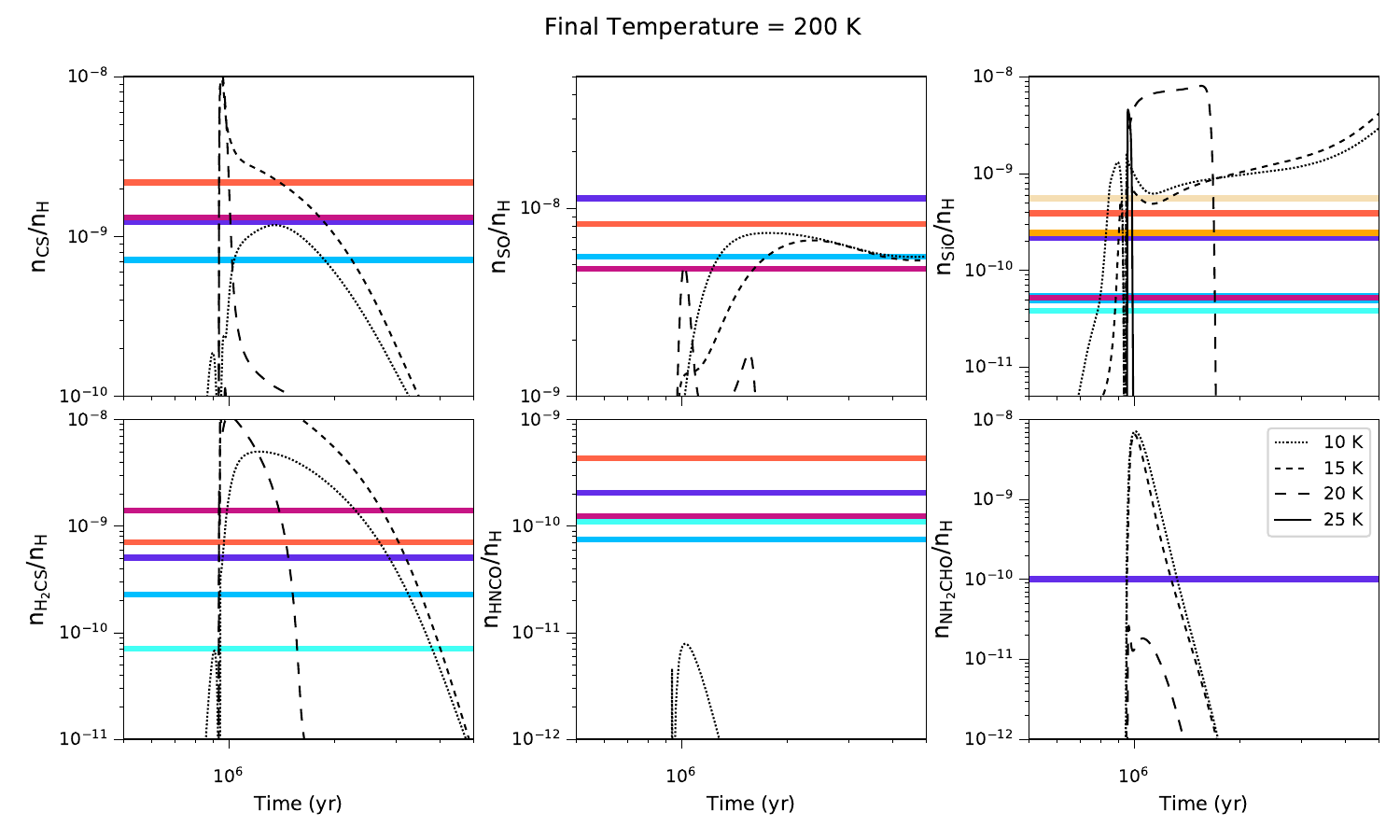}
\caption{Same as Fig.~\ref{appfig:Molecular_Abundances_model1} but for the LMC hot core models for the final hot core temperature of 200 K.}
\label{appfig:Molecular_Abundances_model2}
\end{figure*}

Section~\ref{sec:comparisonmodels} discusses how the observed abundances of the hot cores in the LMC compare with the only hot core modeling for LMC's condition \citep{2018ApJ...859...51A}. We present the comparisons for three molecules (\ce{CH3OH}, \ce{SO2}, and \ce{CH3CN}) in Section~\ref{sec:comparisonmodels}, and we expand this comparison in this appendix. Figures~\ref{appfig:Molecular_Abundances_model1} and \ref{appfig:Molecular_Abundances_model2} show the comparisons of observed abundance for the seven hot cores in the LMC with the chemical models for the final hot core temperature of 100 and 200 K, respectively. It is evident from the plots that the solid curve representing the warm dust initial condition (T$_d$ = 25 K) does not appear in any of the subplots. The only exception is \ce{SiO}, for which a sharp increase in the model at a time of $\approx$\num{e6}~yr could also reproduce the observed abundances. As the modelings do not include shocks, this increase in the \ce{SiO} is not reliable. Moreover, the observed abundance of \ce{HNCO} could not be reproduced by models; the peak abundance of the models is smaller by more than an order of magnitude. The underestimation of abundances is also apparent in the case of \ce{SO2} and \ce{SO}. These three species (HNCO, \ce{SO2}, and \ce{SO}) are also commonly found to be associated with shocks in star-forming regions. The higher observed abundances could imply the presence of shocks in the hot cores of the LMC, which is not included in the modelings.

\end{appendix}
\end{document}